\pdfoutput=1

\documentclass[11pt,twoside,a4paper,cmspaper,final,collab]{cms-tdr}
\def\svnVersion{491901P}\def\cmsCernNoTag{CERN-EP-2018-252}\def\cmsCernDate{\today}\def\cmsMessage{Published in the Journal of High Energy Physics as \href{http://dx.doi.org/10.1007/JHEP02(2019)149}{\doi{10.1007/JHEP02(2019)149.}}}

\begin{document}\cmsNoteHeader{TOP-17-014}

\hyphenation{had-ron-i-za-tion}
\hyphenation{cal-or-i-me-ter}
\hyphenation{de-vices}
\RCS$HeadURL: svn+ssh://svn.cern.ch/reps/tdr2/papers/TOP-17-014/trunk/TOP-17-014.tex $
\RCS$Id: TOP-17-014.tex 489072 2019-02-14 15:19:43Z keaveney $

\newlength\cmsFigWidth
\ifthenelse{\boolean{cms@external}}{\setlength\cmsFigWidth{0.85\columnwidth}}{\setlength\cmsFigWidth{0.4\textwidth}}
\ifthenelse{\boolean{cms@external}}{\providecommand{\cmsLeft}{top\xspace}}{\providecommand{\cmsLeft}{left\xspace}}
\ifthenelse{\boolean{cms@external}}{\providecommand{\cmsRight}{bottom\xspace}}{\providecommand{\cmsRight}{right\xspace}}

\providecommand{\cmsTable}[1]{\resizebox{\textwidth}{!}{#1}}
\providecommand{\NA}{\ensuremath{\text{---}}}

\newcommand{\mt}{\ensuremath{m_{\cPqt}}\xspace}
\newcommand{\mz}{\ensuremath{m_{\cPZ}}}
\newcommand{\Wjets}{{\PW}+jets\xspace}
\newcommand{\Zjets}{{\cPZ}+jets\xspace}
\newcommand{\WW}{{\PW}{\PW}\xspace}
\newcommand{\WZ}{{\PW}{\cPZ}\xspace}
\newcommand{\ZZ}{{\cPZ}{\cPZ}\xspace}
\newcommand{\tW}{{\cPqt}{\PW}\xspace}
\newcommand{\emu}{\ensuremath{\Pe^{\pm}\mu^{\mp}}\xspace}
\newcommand{\MadSpin}{\textsc{MadSpin}}
\newcommand{\MGaMCatNLO}{\textsc{mg}5\_\text{a}\textsc{mc@nlo}\xspace}
\newcommand{\RIVET}{\textsc{rivet}}
\newcommand{\lep}{\ensuremath{\ell}}
\newcommand{\lbar}{\ensuremath{\bar{\ell}}}
\newcommand{\llbar}{\ensuremath{\ell\bar{\ell}}}
\newcommand{\Nj}{\ensuremath{N_{\text{jets}}}}
\newcommand{\Nbj}{\ensuremath{N_{\text{\cPqb\ jets}}}}
\newcommand{\act}{\ensuremath{A_\text{c}^{\ttbar}}}
\newcommand{\acl}{\ensuremath{A_\text{c}^{\llbar}}}
\newcommand{\acx}{\ensuremath{A_\text{c}^{\text{x}}}}
\newcommand{\otg}{\ensuremath{O_\text{tG}}}
\newcommand{\ctgl}{\ensuremath{C_\mathrm{tG}/\Lambda^{2}}}
\newcommand{\alphaew}{\alpha_\mathrm{EW}}
\newcommand{\ttbarother}{\ttbar~other}
\newcommand{\lumivalue}{35.9\fbinv}
\newcommand{\xsectheo}{\ensuremath{832\;\substack{+20\\-29}{\,\text{(scale)}} \pm 35\,({\mathrm{PDF}}+\alpS)\unit{pb}}\xspace}

\newcommand{\pttop}{\ensuremath{\pt^{\cPqt}}\xspace}
\newcommand{\ptantitop}{\ensuremath{\pt^{\text{\cPaqt}}}\xspace}
\newcommand{\ytop}{\ensuremath{y_{\cPqt}}\xspace}
\newcommand{\yantitop}{\ensuremath{y_{\cPaqt}}\xspace}
\newcommand{\delytt}{\ensuremath{\Delta\abs{y}(\cPqt,\cPaqt)}\xspace}
\newcommand{\delphitt}{\ensuremath{\Delta\phi(\cPqt,\cPaqt)}\xspace}
\newcommand{\pttt}{\ensuremath{\pt^{\ttbar}}\xspace}
\newcommand{\ytt}{\ensuremath{y_{\ttbar}}\xspace}
\newcommand{\mtt}{\ensuremath{m_{\ttbar}}\xspace}
\newcommand{\ptlep}{\ensuremath{\pt^{\lep}}\xspace}
\newcommand{\ptalep}{\ensuremath{\pt^{\lbar}}\xspace}
\newcommand{\etalep}{\ensuremath{\eta_{\lep}}\xspace}
\newcommand{\etaalep}{\ensuremath{\eta_{\lbar}}\xspace}
\newcommand{\ptll}{\ensuremath{\pt^{\llbar}}\xspace}
\newcommand{\mll}{\ensuremath{m_{\llbar}}\xspace}
\newcommand{\delphill}{\ensuremath{\Delta\phi(\lep,\lbar)}\xspace}
\newcommand{\deletall}{\ensuremath{\Delta\eta(\lep,\lbar)}\xspace}
\newcommand{\ptb}{\ensuremath{\pt^{\cPqb}}\xspace}
\newcommand{\etab}{\ensuremath{\eta_{\cPqb}}\xspace}
\newcommand{\ptbb}{\ensuremath{\pt^{\bbbar}}\xspace}
\newcommand{\mbb}{\ensuremath{m_{\bbbar}}\xspace}
\newcommand{\chisq}{\ensuremath{\chi^{2}}\xspace}
\newcommand{\chidof}{\ensuremath{\chi^{2}}/dof\xspace}
\newcommand{\delchi}{\ensuremath{\Delta\chi^{2}}\xspace}
\newcommand{\ttzw}{\ensuremath{\ttbar\text{+\cPZ/\PW}}\xspace}
\newcommand{\irel}{\ensuremath{I_{\text{rel}}}\xspace}
\newcommand{\nnloew}{NNLO+\ensuremath{\alphaew^{3}}\xspace}
\newcommand{\nnlonnllprime}{NNLO+NNLL'\xspace}
\newcommand{\annnlo}{a\ensuremath{\mathrm{N^3LO}}\xspace}
\newcommand{\pwhgpy}{\POWHEG{+}\PYTHIA}
\newcommand{\pwhghpp}{\POWHEG{+}\HERWIGpp}
\newcommand{\mgamcpy}{\MGaMCatNLO{+}\PYTHIA{\textsc{[FxFx]}}\xspace}
\newcommand{\annlo}{aNNLO\xspace}
\newcommand{\etaphispace}{$\eta$--$\phi$~space\xspace}

\cmsNoteHeader{TOP-17-014}

\title{Measurements of \ttbar differential cross sections in proton-proton collisions at $\sqrt{s}=13\TeV$ using events containing two leptons}

\date{\today}

\abstract{Measurements of differential top quark pair \ttbar cross sections using events produced in proton-proton collisions at a centre-of-mass energy of 13\TeV containing two oppositely charged leptons are presented. The data were recorded by the CMS experiment at the CERN LHC in 2016 and correspond to an integrated luminosity of 35.9\fbinv. The differential cross sections are presented as functions of kinematic observables of the top quarks and their decay products, the \ttbar system, and the total number of jets in the event. The differential cross sections are defined both with particle-level objects in a fiducial phase space close to that of the detector acceptance and with parton-level top quarks in the full phase space. All results are compared with standard model predictions from Monte Carlo simulations with next-to-leading-order (NLO) accuracy in quantum chromodynamics (QCD) at matrix-element level interfaced to parton-shower simulations. Where possible, parton-level results are compared to calculations with beyond-NLO precision in QCD. Significant disagreement is observed between data and all predictions for several observables. The measurements are used to constrain the top quark chromomagnetic dipole moment in an effective field theory framework at NLO in QCD and to extract \ttbar and leptonic charge asymmetries.
}

\hypersetup{
pdfauthor={CMS Collaboration},
pdftitle={Measurements of ttbar differential cross sections in proton-proton collisions at sqrt(s) = 13 TeV using events containing two leptons},
pdfsubject={CMS},
pdfkeywords={CMS, physics, ttbar}}

\maketitle

\clearpage

\section{Introduction}
\label{sec:introduction}
The top quark could play a central role in the electroweak symmetry breaking mechanism of the standard model (SM) and in numerous physics models beyond the SM (BSM). Measurements of the differential production cross sections for top quark pair (\ttbar) production test SM predictions, while also probing scenarios of BSM physics. Such measurements also allow determination of the top quark mass (\mt), the strong coupling constant (\alpS), and the parton distribution functions (PDFs) of the proton. In this paper the dilepton decay channel of the \ttbar process is utilised. Although the dilepton channel has a relatively small branching fraction, it has significantly lower backgrounds than the other \ttbar decay channels. As a consequence of the excellent lepton energy resolution, the precise measurement of kinematic observables based on lepton pairs is unique to the dilepton channel. However, because of the presence of two neutrinos in the final state, the measurement of top quark kinematic observables in the dilepton channel requires specialised kinematic reconstruction techniques.

Differential \ttbar cross sections have been measured by the ATLAS and CMS Collaborations at the CERN LHC in proton-proton ($\Pp\Pp$) collisions at centre-of-mass energies of 7, 8, and 13\TeV~\cite{bib:ATLASnew,Aad:2014iaa,bib:TOP-11-013_paper,Khachatryan:2016oou,bib:atlas_8TeV,Aaboud:2017ujq,Aad:2015hna,bib:TOP-12-028_paper,bib:TOP-12-041_paper,Khachatryan:2015fwh,Khachatryan:2016gxp,Sirunyan:2017azo,Aaboud:2017fha,Aaboud:2016xii,bib:Aaboud:2016syx, bib:TOP-16-008,Sirunyan:2018wem, Sirunyan:2017mzl, Sirunyan:2018ptc}. In this paper, measurements of absolute and normalised differential \ttbar cross sections are presented. The measurements are based on either particle-level objects and extrapolated to a fiducial phase space close to that of the detector acceptance or on parton-level top quarks extrapolated to the full phase space. In the absolute measurements, the integrated and differential cross sections are measured simultaneously. In the normalised measurements, the integrated cross sections are not measured and the uncertainties affecting only the normalisations of the differential cross sections in both data and predictions are reduced. The measurements represent the most comprehensive study of \ttbar production in the dilepton channel to date. Furthermore, this work includes a more detailed treatment of the systematic uncertainties compared to previous CMS measurements of the \ttbar differential cross sections in the dilepton channel at centre-of-mass energies of 7 and 8\TeV, and approximately 17 times more data than the previous CMS measurement at 13\TeV\ using data recorded in 2015.

The analysis utilises a 13\TeV data sample recorded by the CMS experiment in 2016, corresponding to an integrated luminosity of \lumivalue~\cite{bib:CMS-PAS-LUM-17-001}. The differential cross sections are presented as a function of kinematic observables of the top quarks, their decay products, the \ttbar system, and the number of jets in the event.

Results presented at the parton level may be compared to state-of-the-art SM calculations with beyond next-to-leading-order (NLO) precision in quantum chromodynamics (QCD)~\cite{bib:difftop,bib:kidonakis_13TeV,bib:mitov,bib:Pecjak:2016nee}. These comparisons facilitate the extraction of \mt, \alpS, and the PDFs. Both parton- and particle-level results may be compared to theoretical predictions from Monte Carlo (MC) generators. The particle-level results do not exhibit the theoretical model dependence of the parton-level results introduced when extrapolating to an experimentally inaccessible phase-space region. This reduced model dependence allows calibration of parton-shower, hadronisation, and underlying-event models within MC generators.

Numerous BSM scenarios that lead to modifications of the \ttbar differential cross sections involve the production of new states, \eg top squarks or \cPZpr\ bosons~\cite{Li:2013uma,Frederix:2007gi,Harris:2011ez}. The absence of signals of such new states in the LHC data recorded so far suggests that BSM physics might only be directly manifested at an energy scale that is large with respect to the typical scales probed at the LHC. In this case, the new states are only produced virtually at the LHC. These virtual effects can modify the rates and kinematic properties of \ttbar events. Theoretically, these effects can be accommodated by adding higher-dimensional operators to the SM Lagrangian in an effective field theory (EFT). The reduced model dependence of the particle-level results makes them particularly suitable to constrain BSM theories. The generation of predictions for the particle-level observables does not require any detector simulation. Hence, such measurements facilitate a future, global analysis of EFT operators using multiple observables measured by more than one experiment to simultaneously constrain all relevant EFT operators. An anomalous top quark chromomagnetic dipole moment (CMDM) is a feature of BSM models such as two-Higgs-doublet models, supersymmetry, technicolor, and top quark compositeness models~\cite{bib:zhang,Martinez:2007qf}. In this paper, the measured particle-level differential \ttbar cross section as a function of the azimuthal angle between the two charged leptons is used to constrain the CMDM in an EFT framework. Signals of BSM physics could also appear in \ttbar production as anomalous top quark or leptonic charge asymmetries. Hence, we extract these quantities from differential \ttbar cross section measurements as a function of the difference in absolute rapidity between the top quark and antiquark, and the difference in absolute pseudorapidity between the charged leptons.

The paper is organised as follows. In Section~\ref{sec:cms}, a brief description of the CMS detector is provided. In Section~\ref{sec:simulation}, the simulation of signal and background processes is detailed, followed by the description of the selection of events at the trigger level and in the offline analysis in Section~\ref{sec:selection}. The sources of systematic uncertainties that affect the measurements are discussed in Section~\ref{sec:errors}, along with the methods employed to estimate the size of their effects. In Section~\ref{sec:diffxsec}, details of the objects and phase-space regions used to define the measured observables are provided, together with a description of the unfolding procedure used to determine the particle- and parton-level data. The unfolded data are presented and compared to theoretical predictions in Section~\ref{sec:results}. In Sections~\ref{sec:eft} and~\ref{sec:ca}, constraints on the top quark CMDM in an EFT framework and the \ttbar and leptonic charge asymmetries are derived from the unfolded data. Finally, the paper is summarised in Section~\ref{sec:summary}.

\section{The CMS detector}
\label{sec:cms}
The central feature of the CMS apparatus is a superconducting solenoid of 6\unit{m} internal diameter, providing a magnetic field of 3.8\unit{T}. Within the solenoid volume are a silicon pixel and strip tracker, a lead tungstate crystal electromagnetic calorimeter (ECAL), and a brass and scintillator hadron calorimeter, each composed of a barrel and two endcap sections. Forward calorimeters extend the pseudorapidity coverage provided by the barrel and endcap detectors. Muons are measured in gas-ionisation detectors embedded in the steel flux-return yoke outside the solenoid. Events of interest are selected using a two-tiered trigger system~\cite{Khachatryan:2016bia}. The first level, composed of custom hardware processors, uses information from the calorimeters and muon detectors to select events at a rate of around 100\unit{kHz} within a time interval of less than 4\mus. The second level, known as the high-level trigger (HLT), consists of a farm of processors running a version of the full event reconstruction software optimised for fast processing, and reduces the event rate to around 1\unit{kHz} before data storage. A more detailed description of the CMS detector, together with a definition of the coordinate system used and the relevant kinematic variables, can be found in Ref.~\cite{bib:Chatrchyan:2008zzk}.

\section{Event simulation}
\label{sec:simulation}
The simulation of physics processes is important in order to estimate event reconstruction and selection efficiencies, resolutions of the event reconstruction, and to provide predictions for the \ttbar signal and backgrounds. This motivates the use of MC generators interfaced to a detector simulation. The default simulation setup for the \ttbar process is provided at NLO in QCD at the matrix-element (ME) level by the \POWHEG\ (v.2)~\cite{Frixione:2007nw,bib:powheg0,bib:powheg,bib:powheg2} generator (\POWHEG). For this setup, the $h_\mathrm{damp}$ parameter of \POWHEG, which regulates the damping of real emissions in the NLO calculation when matching to the parton shower, is set to 1.58 \mt~=~272.72\GeV as a result of the tuning of this parameter from Ref.~\cite{bib:CMS:2016kle}. The generated events are subsequently processed with the \PYTHIA (v.~8.219)~\cite{Sjostrand:2007gs} program (\PYTHIA), with the CUETP8M2T4 tune~\cite{bib:CMS:2016kle,bib:CUETP8tune,Skands:2014pea}, for parton showering and hadronisation. In order to compare the predictive powers of alternative ME, parton shower, and hadronisation models, two additional samples are generated using different generator setups. Firstly, a sample is generated using the \MGvATNLO~\cite{Alwall:2014hca} (v.~2.2.2) (\MGaMCatNLO) generator including up to two extra partons at the ME level with NLO precision. In this setup, referred to as "\mgamcpy", \MadSpin~\cite{bib:madspin} is used to model the decays of the top quarks, while preserving their spin correlation, and events are matched to \PYTHIA\ for parton showering and hadronisation using the FxFx prescription~\cite{Frederix:2012ps}. Secondly, a sample is generated with \POWHEG\ at NLO in QCD at the ME level and interfaced with \HERWIGpp~(v.~2.7.1)~\cite{bib:herwigpp} with the EE5C tune~\cite{bib:EE5Ctune} for parton showering and hadronisation. This setup is referred to as \pwhghpp.

Only \ttbar\ events with two electrons or muons that do not originate from the decays of $\tau$ leptons or semileptonic \cPqb\ hadron decays are considered as signal, with all other \ttbar\ events regarded as a background, which we refer to as "\ttbarother". The largest background contributions originate from \ttbarother, single top quarks produced in association with a {\PW} boson (\tW), {\cPZ}/{$\cPgg^{*}$} bosons produced with additional jets (\Zjets), {\PW} boson production with additional jets (\Wjets), diboson (\WW, \WZ, and \ZZ) events, and the production of a \ttbar pair in association with a {\cPZ} or {\PW} boson (\ttzw). Other backgrounds are negligible in comparison to the uncertainties in the main backgrounds. The \Wjets process is simulated at leading-order (LO) precision using \MGaMCatNLO with up to four additional partons at ME level and matched to \PYTHIA\ using the MLM prescription~\cite{Alwall:2007fs}. The \Zjets process is simulated at NLO precision using \MGaMCatNLO with up to two additional partons at ME level and matched to \PYTHIA\ using the FxFx prescription. The \ttzw processes are simulated with \MGaMCatNLO\ with NLO precision at ME level and matched to \PYTHIA. In the case of $\ttbar\text{+\PW}$, one extra parton is simulated at ME level and the calculation is matched to \PYTHIA\ using the FxFx prescription. Single top quark production is simulated with \POWHEG\ (v.~1)~\cite{bib:powheg1,bib:powheg3} using the CUETP8M2T4 tune in \PYTHIA. Diboson events are simulated with \PYTHIA.

For all \ttbar\ samples, the NNPDF3.0\_nlo\_as\_0118~\cite{bib:NNPDF} PDF set is used. Predictions are normalised based on their theoretical cross sections and the integrated luminosity of the data. The cross sections are calculated at the highest orders of perturbative QCD currently available. This corresponds to next-to-NLO (NNLO) for \Wjets and \Zjets~\cite{Li:2012wna}, approximate NNLO for single top quark in the \tW channel~\cite{bib:twchan}, and NLO calculations for diboson~\cite{bib:mcfm:diboson} and \ttzw~\cite{bib:Maltoni:2015ena}. The \ttbar predictions are normalised to a cross section of \xsectheo\ calculated with the \textsc{Top++}2.0 program~\cite{Czakon:2011xx} at NNLO including resummation of next-to-next-to-leading-logarithmic (NNLL) soft-gluon terms, assuming a top quark mass \mt\ = 172.5\GeV. Additional proton-proton interactions within the same or nearby bunch crossings (pileup) is simulated for all samples. The interactions of particles with the CMS detector is simulated using \GEANTfour (v.~9.4)~\cite{bib:geant}.

\section{Event selection}
\label{sec:selection}
The event selection procedure is designed to select events corresponding to the decay topology where both top quarks decay into a {\PW} boson and a bottom quark ({\cPqb} quark), and each of the {\PW} bosons decays into a muon or an electron, and a neutrino. Three distinct channels based on the flavours of the final-state leptons are defined: the same-flavour channels corresponding to two electrons (\EE) or two muons (\MM), and the different-flavour channel corresponding to one electron and one muon (\emu). The final results are derived by combining the three channels.
At HLT level, events are selected either by single-lepton triggers that require the presence of at least one electron or muon or by dilepton triggers that require the presence of either two electrons, two muons, or an electron and a muon. For the single-electron and single-muon triggers, transverse momentum \pt thresholds of 27 and 24\GeV are applied, respectively. The same-flavour dilepton triggers require either an electron pair with $\pt > 23 (12)\GeV$ for the leading (trailing) electron or a muon pair with $\pt > 17 (8)\GeV$ for the leading (trailing) muon, where leading (trailing) refers to the electron or muon with the highest (second-highest) \pt in the event. The different-flavour dilepton triggers require either a muon with $\pt > 23\GeV$ and an electron with $\pt > 12\GeV$, or an electron with $\pt > 23\GeV$ and a muon with $\pt > 8\GeV$.

The events selected by the trigger are reconstructed offline using a particle-flow algorithm~\cite{bib:Sirunyan:2017ulk}. The particle-flow algorithm aims to reconstruct and identify each individual particle in an event, with an optimised combination of information from the various elements of the CMS detector. Electron candidates are reconstructed from a combination of the track momentum at the main interaction vertex and the corresponding clusters in the ECAL with a Gaussian sum filter algorithm~\cite{1748-0221-10-06-P06005}. The electron candidates are required to have $\pt > 25 (20)\GeV$ for the leading (trailing) candidate and $\abs{\eta} < 2.4$. Electron candidates with ECAL clusters in the region between the barrel and endcap ($1.44 < \abs{\eta_{\mathrm{cluster}}}< 1.57$) are excluded because of less efficient electron reconstruction. A relative isolation criterion $\irel < 0.0588 (0.0571)$ is applied for an electron candidate in the barrel (endcap), where \irel is defined as the sum of the \pt of all neutral hadron, charged hadron, and photon candidates within a distance of 0.3 from the electron in \etaphispace, divided by the \pt of the electron candidate. In addition, electron identification requirements are applied to reject misidentified electron candidates and candidates originating from photon conversions. Muon candidates are reconstructed using the track information from the tracker and the muon system~\cite{Sirunyan:2018fpa}. They are required to have $\pt > 25 (20)\GeV$ for the leading (trailing) candidates and $\abs{\eta}<2.4$. An isolation requirement of $\irel < 0.15$ is applied to muon candidates with particles within 0.4 of the muon in \etaphispace included in the calculation of \irel. In addition, muon identification requirements are applied to reject misidentified muon candidates and candidates originating from decay-in-flight processes. For both electron and muon candidates, a correction is applied to \irel to suppress the residual effect of pileup.

Jets are reconstructed by clustering the particle-flow candidates using the anti-\kt clustering algorithm with a distance parameter of 0.4~\cite{bib:antikt,bib:Cacciari:2011ma}. Jet momentum is determined as the vectorial sum of all particle momenta in the jet, and is found from simulation to be within 5 to 10\% of the true momentum over the whole \pt spectrum and detector acceptance. Pileup can contribute additional tracks and calorimetric energy deposits to the jet momentum. To mitigate this effect, tracks identified to be originating from pileup vertices are discarded, and an offset correction is applied to correct for remaining contributions. Jet energy corrections are derived from simulation to bring the measured response of jets to that of particle-level jets on average. In situ measurements of the momentum imbalance in dijet, photon+jets, \Zjets, and multijet events are used to account for any residual differences in jet energy in data and simulation. Additional selection criteria are applied to remove badly reconstructed jets. Jets are selected if they have $\pt >30\GeV$ and $\abs{\eta}<2.4$. Jets are rejected if the distance in \etaphispace between the jet and the closest lepton, $\Delta R(\mathrm{jet,lepton})$, is less than 0.4. Jets originating from the hadronisation of {\cPqb} quarks ({\cPqb} jets) are identified (\cPqb\ tagged) by combining information related to secondary decay vertices reconstructed within the jets and track-based lifetime information in an algorithm CSV (v.2)~\cite{Sirunyan:2017ezt} that provides a \cPqb\ jet identification efficiency of $\approx$79--87\% and a probability to misidentify light-flavour jets as {\cPqb} jets of $\approx$10\%.

The missing transverse momentum vector \ptvecmiss is defined as the projection on the plane perpendicular to the beams of the negative vector sum of the momenta of all reconstructed particles in an event. Its magnitude is referred to as \ptmiss.

Events are selected offline if they contain exactly two isolated, oppositely charged electrons or muons (\EE, \MM, \emu) and at least two jets. At least one of the jets is required to be \cPqb\ tagged. Events with an invariant mass of the lepton pair (\mll) smaller than 20\GeV are removed in order to suppress contributions from heavy-flavour resonance decays and low-mass Drell--Yan processes. Backgrounds from \Zjets processes in the \EE and \MM channels are further suppressed by requiring $\mll <76\GeV$ or $\mll >106\GeV$, and $\ptmiss > 40\GeV$. The normalisation of the remaining background contribution from \Zjets events, which is large in the \EE and \MM channels, is determined by applying a factor derived from simulation to the number of \Zjets events observed in data in a control region where \mll\ is close to \mz~\cite{bib:TOP-12-028_paper,bib:TOP-15-003_paper}. A correction to account for non-\Zjets backgrounds in the control region is derived from the \emu channel. The shape of the \Zjets background is taken from simulation. Other sources of background such as \tW, diboson, \ttzw, \ttbarother, misidentified leptons, and leptons within jets are estimated from simulation.

The kinematic observables of the top quarks are estimated via a kinematic reconstruction algorithm~\cite{bib:TOP-12-028_paper}. The algorithm examines all combinations of jets and leptons and solves a system of equations based on the following constraints: \ptmiss is assumed to originate solely from the two neutrinos; the invariant mass of the reconstructed {\PW} boson must equal 80.4\GeV~\cite{Patrignani:2016xqp}; and the invariant mass of each reconstructed top quark must equal 172.5\GeV. Effects of detector resolution are accounted for by randomly varying the measured energies and directions of the reconstructed lepton and {\cPqb} jet candidates by their resolutions as measured in simulation. This procedure is referred to in the following to as smearing. In addition, the assumed invariant mass of the {\PW} boson is smeared according to the Breit--Wigner distribution of {\PW} boson masses in simulation. For a given smearing, the solution of the equations for the neutrino momenta yielding the smallest invariant mass of the \ttbar system is chosen. For each solution, a weight is calculated based on the spectrum of the true invariant mass of the lepton and \cPqb\ jet system from simulated top quark decays at particle level. The weights are summed over 100 smearings for each combination, and the kinematic observables of the top quark and antiquark are calculated as a weighted average. The smearing procedure increases the fraction of combinations in which a valid solution to the system of equations is found. Increasing the number of smearings beyond 100 did not significantly increase this fraction further. The top quark and antiquark candidates are distinguished according to the charge of the lepton in the chosen solution. The solution with the most {\cPqb}-tagged jets is chosen to represent the top quark momenta. If multiple combinations with the same number of {\cPqb}-tagged jets are found, the combination that yields the maximum sum of weights is chosen. The efficiency of the kinematic reconstruction, defined as the number of events where a solution is found divided by the total number of selected \ttbar events, is about 90\% in both data and simulation. Events with no valid solution for the neutrino momenta are excluded from further analysis.

\begin{figure*}[htbp]
\centering
\includegraphics[width=0.48\textwidth]{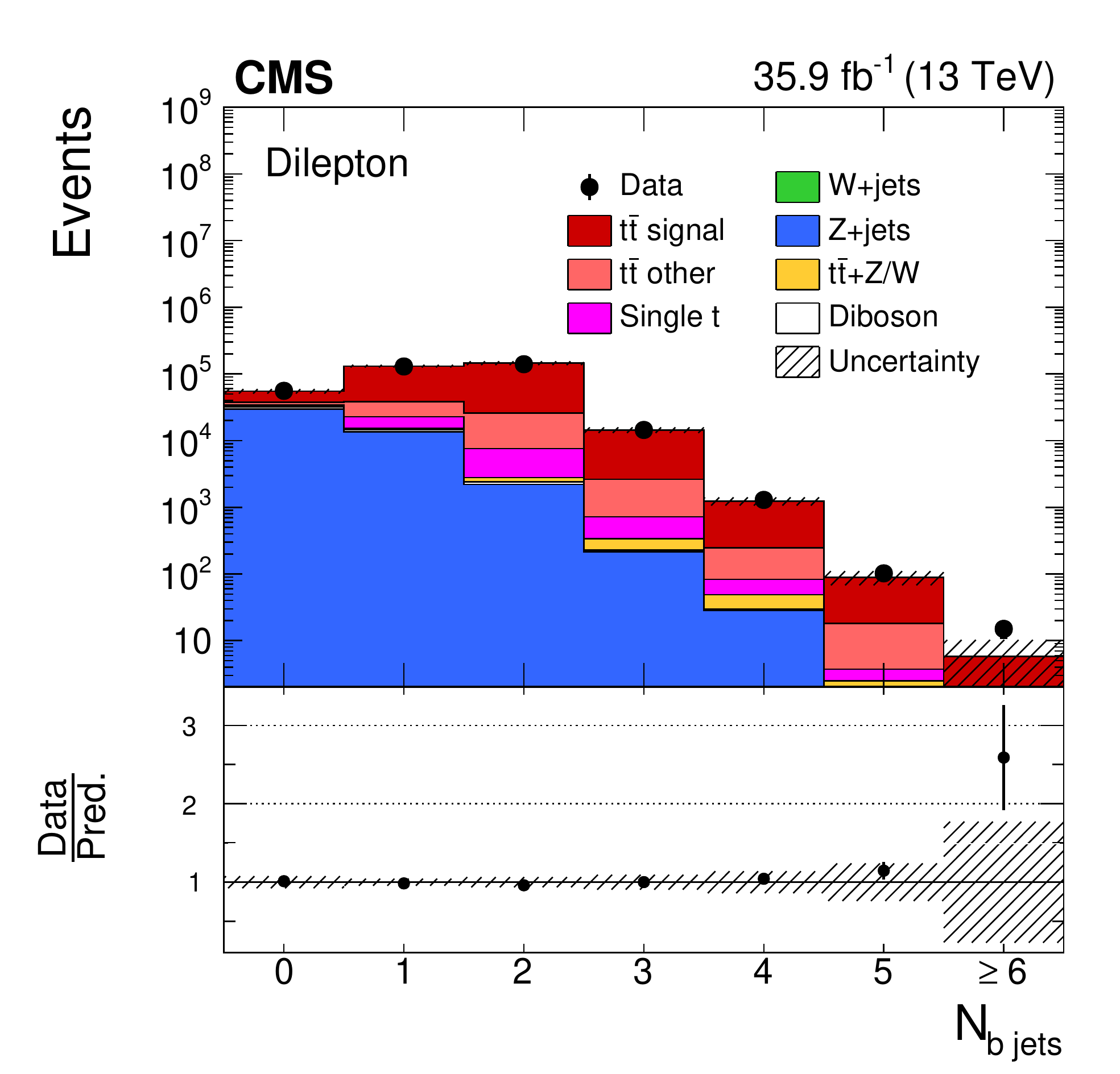}
\includegraphics[width=0.48\textwidth]{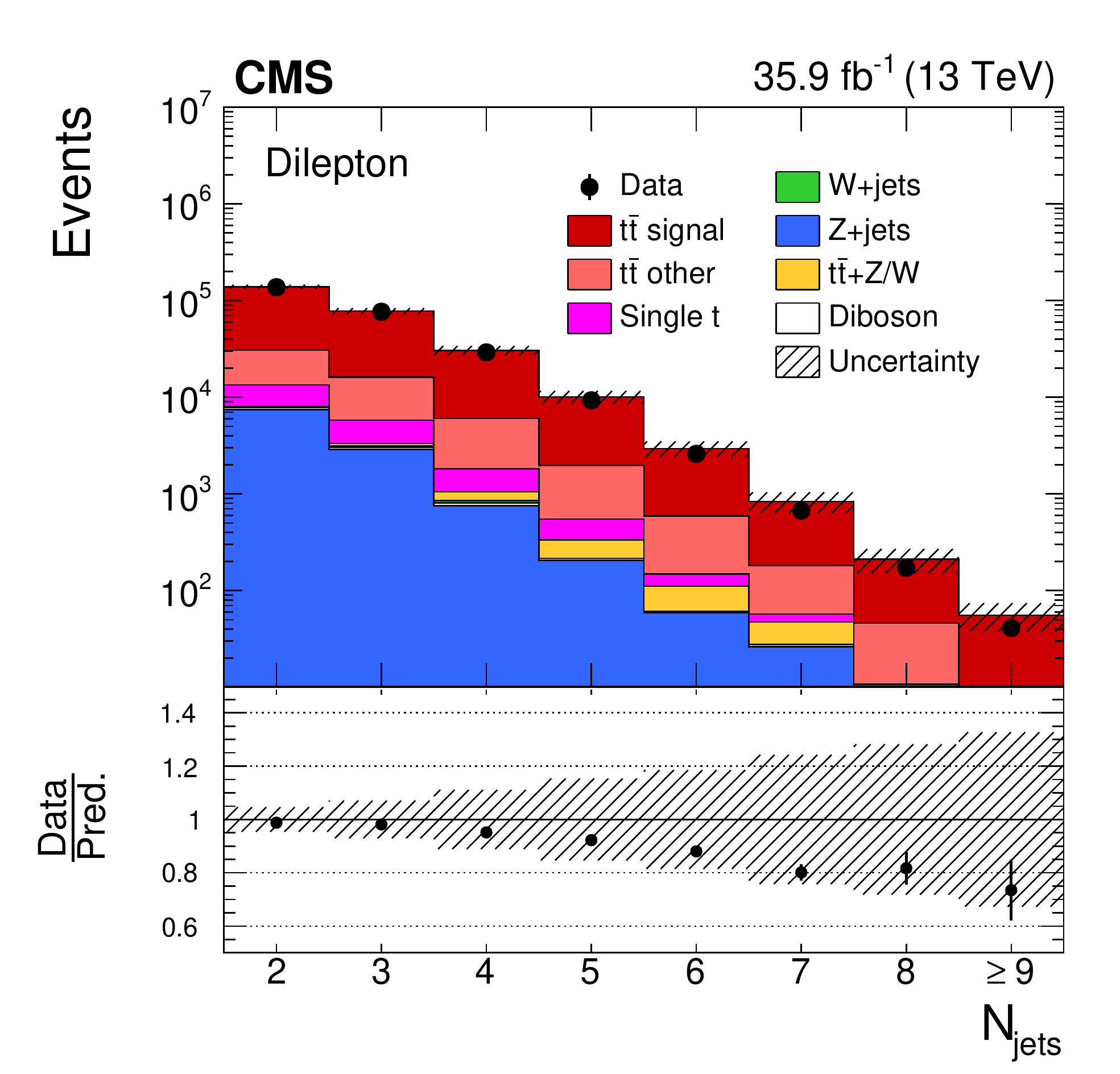}
\includegraphics[width=0.48\textwidth]{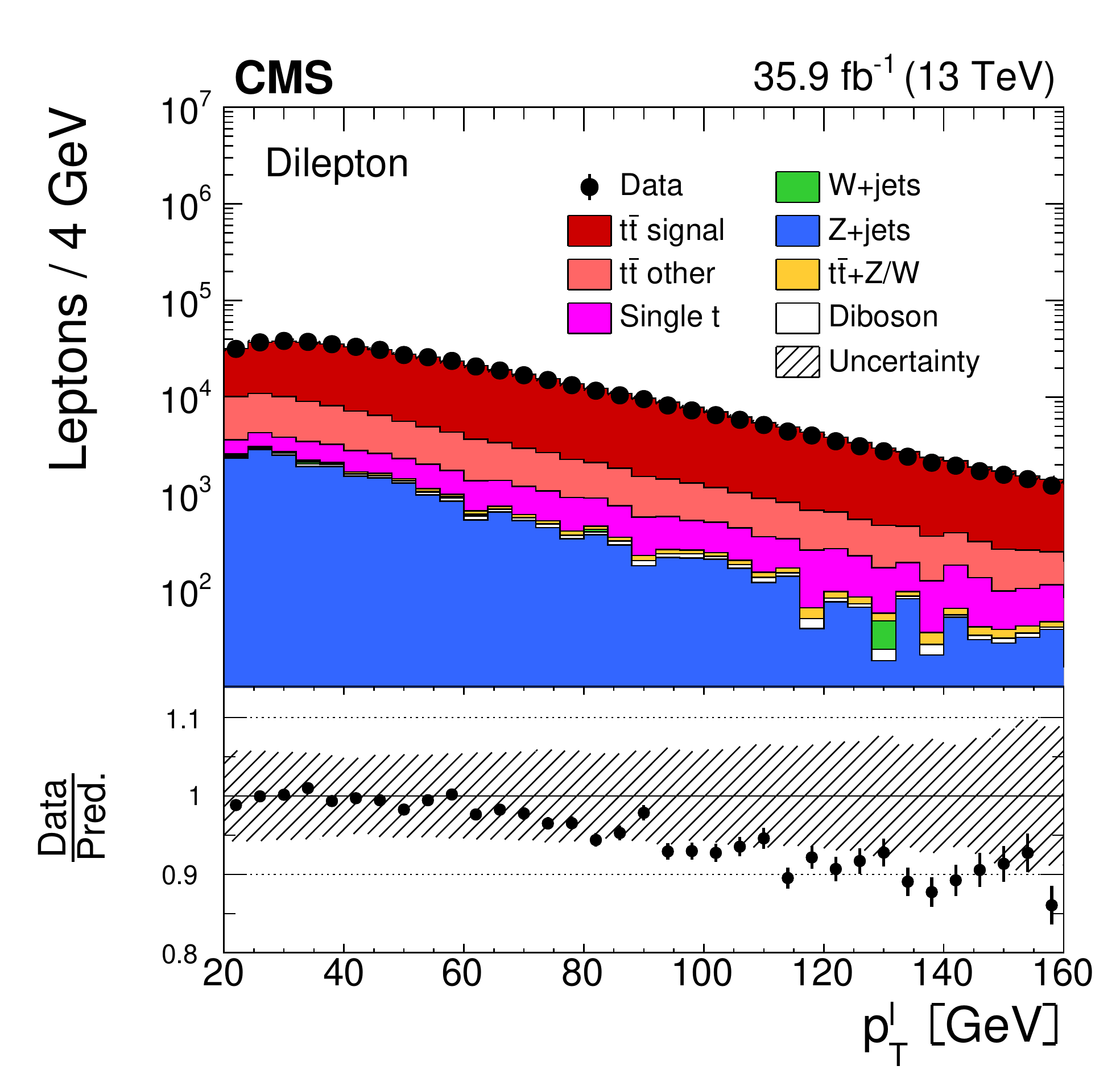}
\includegraphics[width=0.48\textwidth]{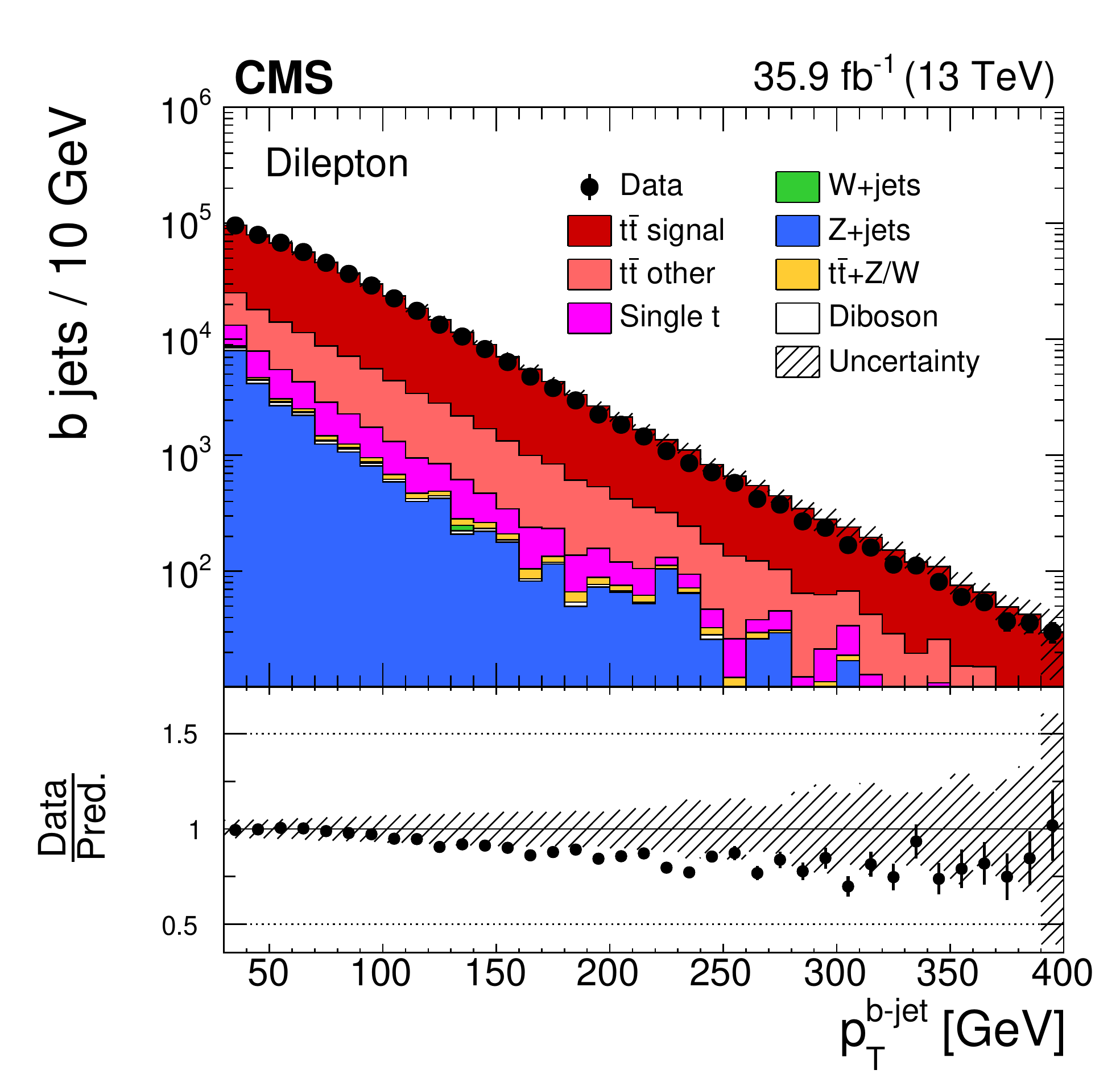}
\setlength{\unitlength}{\textwidth}
\caption{Distributions of the {\cPqb} jet (upper left), and total jet (upper right) multiplicities, and the \pt of the leptons (lower left), and {\cPqb} jets (lower right) are shown for data (points) and simulation (histograms). The \ttbar\ process is simulated with \POWHEG\ + \PYTHIA. The vertical bars on the points represent the statistical uncertainties in the data. The hatched regions correspond to the systematic uncertainties in the signal and backgrounds, as described in Section~\ref{sec:errors}. The lower panel of each plot shows the ratio of the data to the predictions from simulation.}
\label{fig:ctrl:dileptons}
\end{figure*}

\begin{figure*}[htbp]
\centering
\includegraphics[width=0.44\textwidth]{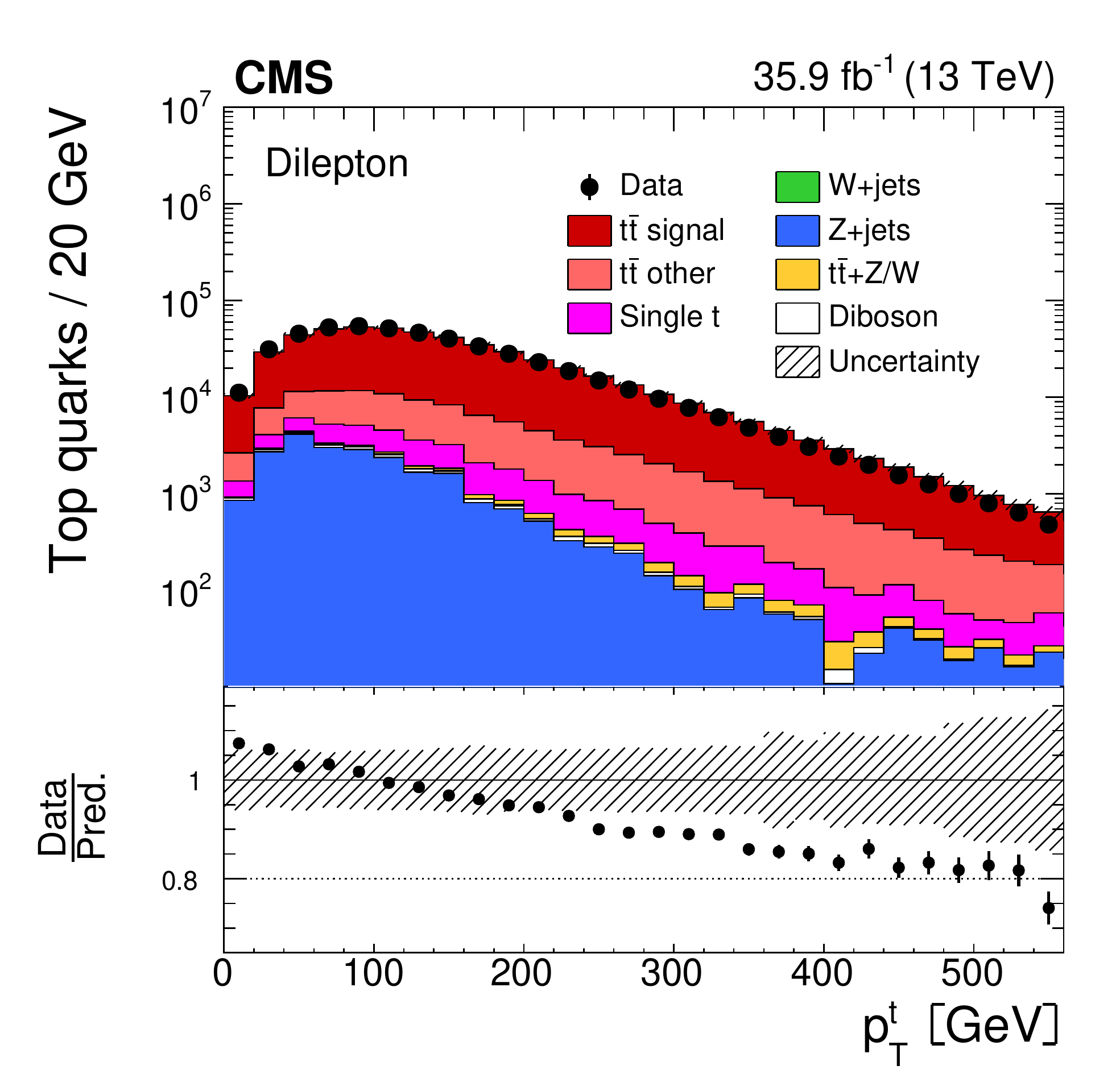}
\includegraphics[width=0.44\textwidth]{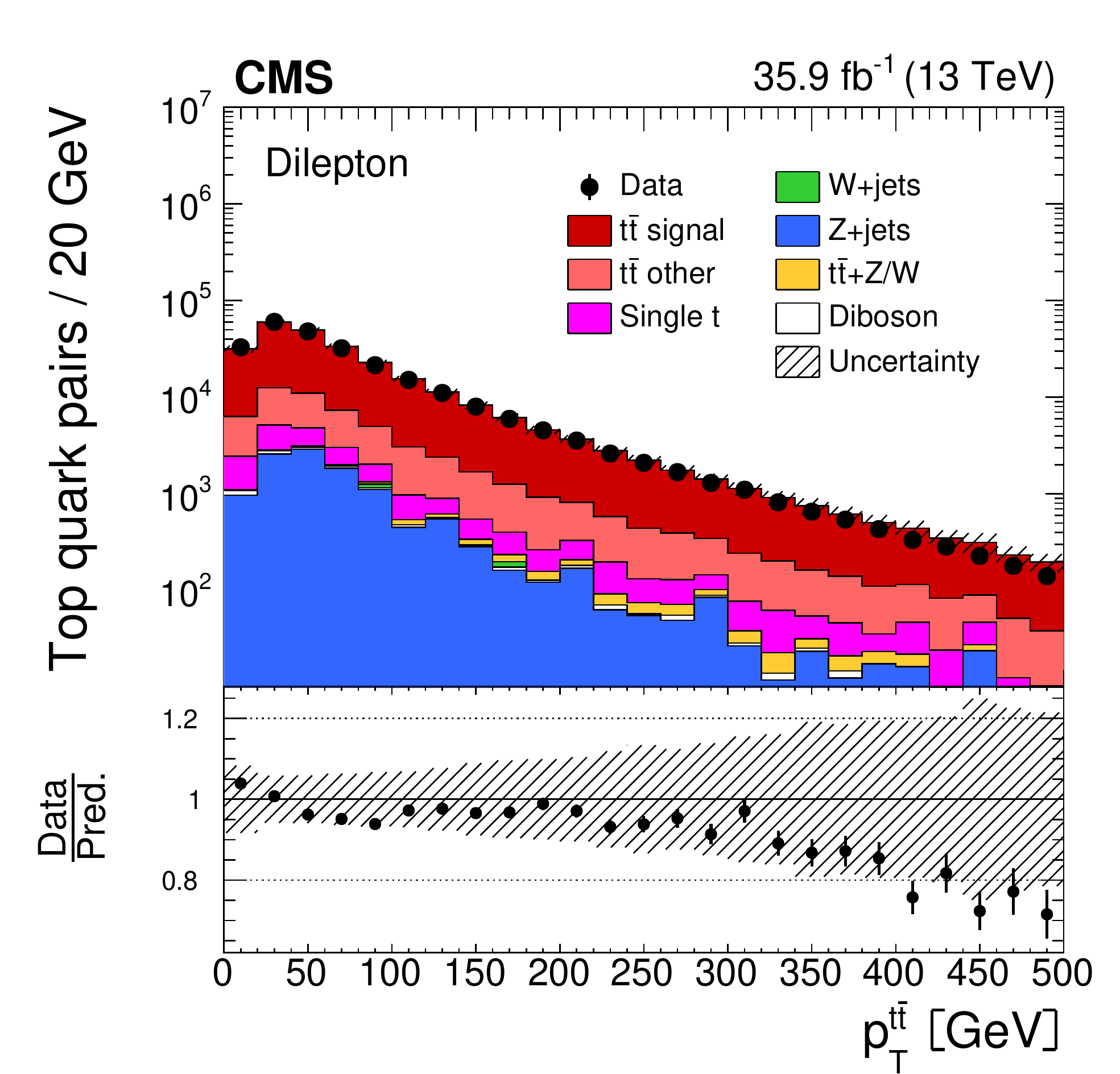}\\
\includegraphics[width=0.44\textwidth]{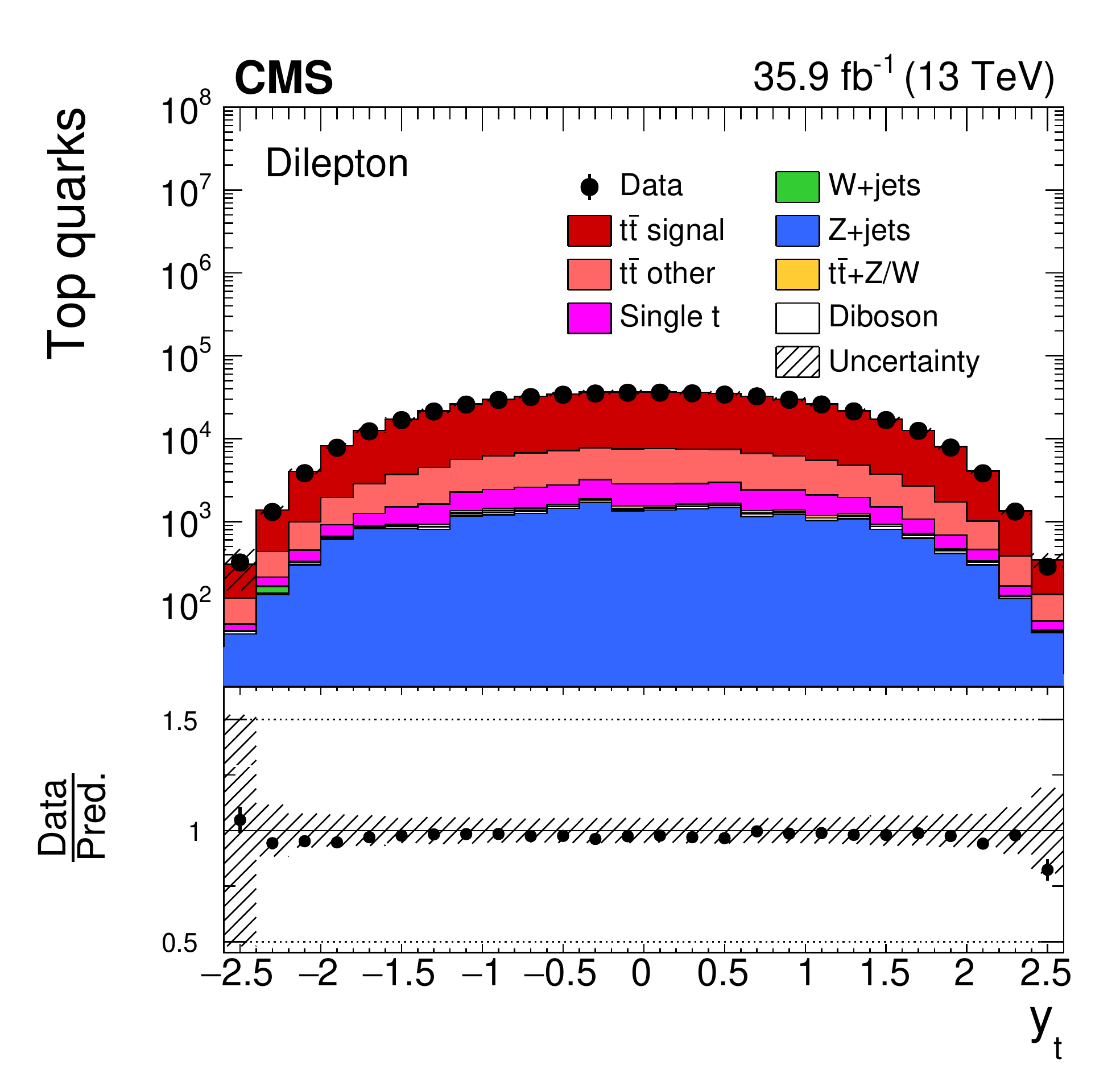}
\includegraphics[width=0.44\textwidth]{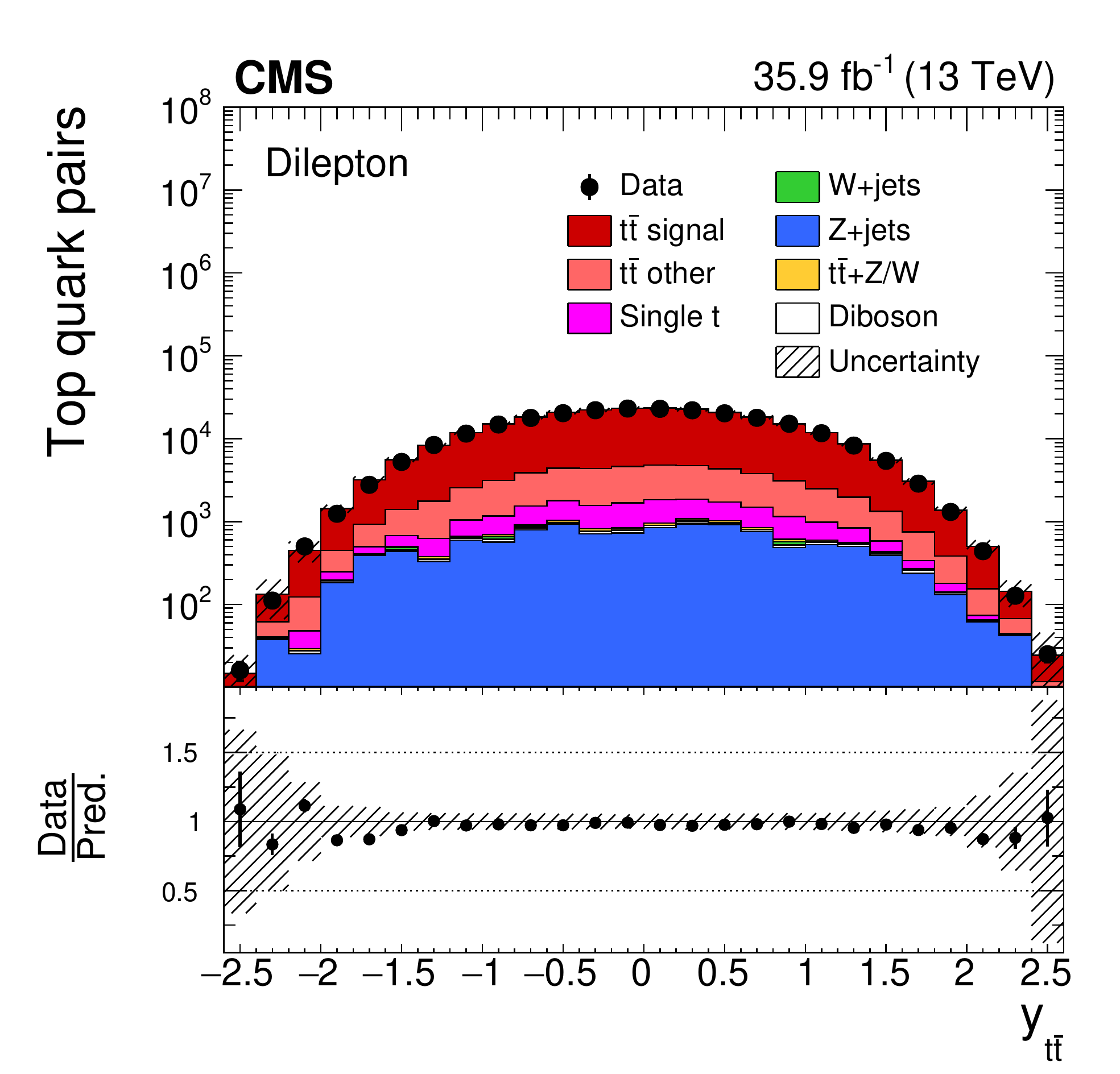}\\
\includegraphics[width=0.44\textwidth]{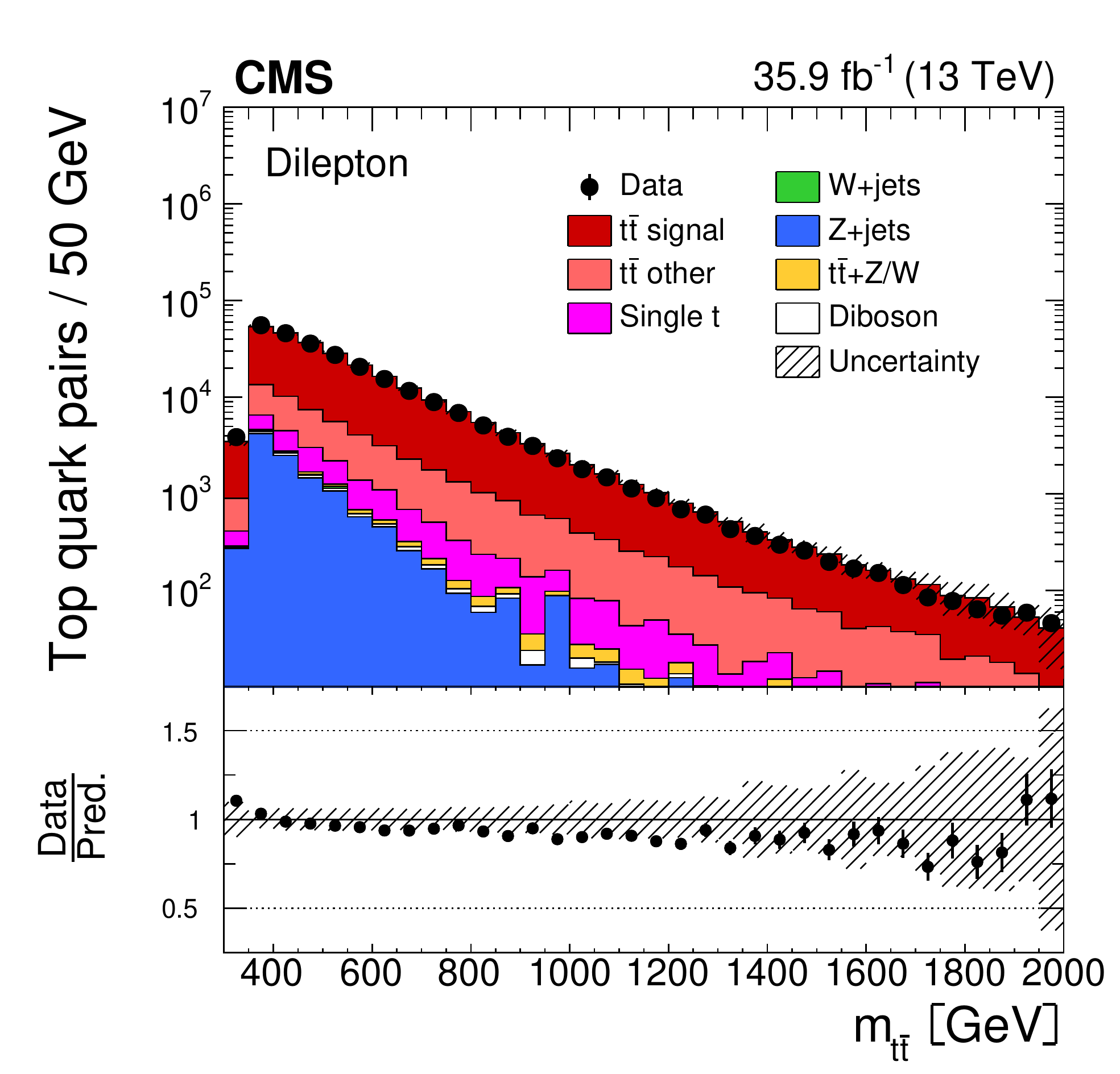}
\caption{Distributions of the \pt (upper row) and rapidities (middle row), at detector level for the top quarks (left column), and \ttbar system (right column), and \mtt (lower plot) are shown for data (points) and simulation (histograms). The \ttbar\ process is simulated with \POWHEG\ + \PYTHIA. The vertical bars on the points represent the statistical uncertainties in the data. The hatched regions correspond to the systematic uncertainties in the signal and backgrounds, as described in Section~\ref{sec:errors}. The lower panel of each plot shows the ratio of the data to the predictions from simulation.}
\label{fig:kinreco:dileptons}
\end{figure*}

After applying the full event selection, 34\,890 events in the \EE channel, 70\,346 events in the \MM channel, and 150\,410 events in the \emu channel are observed. In all decay channels combined, the estimated signal contribution to the data is 80.6\%. In Fig.~\ref{fig:ctrl:dileptons}, selected distributions of the kinematic observables and multiplicities of the selected jets (\Nj) and {\cPqb} jets (\Nbj) are shown. For each distribution, all event selection criteria are applied, with the exception of the \Nbj\ distribution where no {\cPqb}-tagged jets are required. Figure~\ref{fig:kinreco:dileptons} shows the distributions of the top quark or antiquark and \ttbar kinematic observables (the transverse momenta \pttop, \pttt, the rapidities \ytop, \ytt, and the invariant mass of the \ttbar system \mtt). The mismodelling of the data by the simulation, apparent in the tails of the distributions, is accounted for by the corresponding systematic uncertainties, as described in Section~\ref{sec:errors}. Simulation is used to verify that mismodelling of the \pttop distribution does not bias the results for the differential cross section as a function of \pttop.

\section{Differential cross section extraction}
\label{sec:diffxsec}

For a given variable $X$, the absolute differential \ttbar cross section ${\rd\sigma_{i}}/{\rd X}$ is determined via the relation~\cite{bib:TOP-12-028_paper}:
\begin{equation}
\label{eq:diffXsec}
\frac{\rd\sigma_{i}}{\rd X}=\frac{1}{\lumi} \frac{x_{i}}{\Delta^{X}_{i}},
\end{equation}

where \lumi\ is the integrated luminosity of the data, $x_{i}$ is the number of signal events observed in data for bin $i$ after the background subtraction and correction for the detector efficiencies, acceptances, and bin migration, and $\Delta_{i}^{X}$ is the bin width. The normalised differential cross section is obtained by dividing the absolute differential cross section by the measured total cross section $\sigma$ in the same phase space, which is evaluated by summing the binned cross section measurements over all bins of the observable $X$. The background from other \ttbar decays is taken into account, after subtracting all other background components, by correcting the number of signal events in data using the expected signal fraction. The expected signal fraction is defined as the ratio of the number of selected \ttbar signal events to the total number of selected \ttbar events in simulation. This procedure avoids the dependence on the total inclusive \ttbar cross section used in the normalisation of the simulated signal sample.

The finite resolution introduced by the detector response, parton shower, and hadronisation lead to migration of events across bins when correcting the data to both the fiducial phase space based on particle-level objects or the full phase space based on the parton-level top quarks. These effects are accounted for with a regularised unfolding method~\cite{bib:svd,bib:blobel,bib:TOP-12-028_paper}. For each measured distribution, a response matrix that accounts for migrations and efficiencies is calculated using the default \ttbar simulation. For the parton-level measurements in the full phase space, the response matrix also accounts for the branching fraction of \ttbar events into two leptons excluding $\tau$ leptons. The generalised inverse of the response matrix is used to obtain the unfolded distribution from the measured distribution by applying a \chisq minimisation technique. Regularisation is applied to suppress nonphysical fluctuations. The regularisation level is determined individually for each distribution using the average squared global correlation method~\cite{bib:james}.
To keep the bin-to-bin migrations small, the width of the measurement bins are chosen according to their purity and stability. Purity is defined as the fraction of events in a given bin at the detector level that originate from the same bin at the generator level, and stability is defined as the fraction of events in a given bin at the generator level that are reconstructed in the same bin at the detector level. The purities and stabilities are typically ${\approx}50\%$, except in the regions where the distributions are steeply rising or falling, where values of 30\% are typical. The statistical uncertainty is small in comparison to the systematic uncertainties in all bins. The data in the three channels are combined before unfolding in order to model correlations between channels and reduce statistical uncertainties in poorly populated regions of the unfolding matrix.

For some observables, both the absolute and normalised differential cross sections are measured at both the particle level in a fiducial phase space and at the parton level in the full phase space. This leads to four measurements for each of these observables. The observables related to the kinematics and multiplicities of jets and leptons are determined at the particle level only.

\textbf{Object and phase-space definitions}

The definition of the particle-level objects and the kinematic reconstruction procedure employed to estimate the kinematic properties of the particle-level top quarks are described in Ref.~\cite{bib:CMS-NOTE-2017-004}. We detail here the additional event-level requirements that define the fiducial phase-space region in which the particle-level differential cross sections are measured. We require that the {\PW} bosons produced from decays of the top quark and antiquark in a \ttbar event themselves decay to an electron or muon. Events where these {\PW} bosons decay to tau leptons are rejected. The requirements of exactly two selected lepton candidates with opposite charges, a dilepton invariant mass greater than 20\GeV, and at least two {\cPqb} jets are also added.

For the parton-level results, the momenta of the parton-level top quarks are defined after QCD radiation but before the top quark decays. The parton-level results are extrapolated to the full phase space using the default simulation.

\section{Systematic uncertainties}
\label{sec:errors}
The systematic uncertainties in the measured differential cross sections are categorised into experimental uncertainties arising from imperfect modelling of the detector response and conditions and theoretical uncertainties arising from the modelling of the signal and background processes. Each systematic uncertainty is determined separately in each bin of the measured differential cross section via a variation of the corresponding aspect of the simulation setup.

A regularised unfolding method, described in Section \ref{sec:diffxsec}, is used to correct for the migration of events between bins due to the finite detector resolutions and to extrapolate the detector-level data to the fiducial and full phases spaces. The variations are applied both at detector level and in the response matrices that define the unfolding. For each variation, the difference between the varied and nominal results is taken as the systematic uncertainty. The total systematic uncertainty is calculated by adding these differences in quadrature. In this section, each of these applied variations is detailed.

\subsection{Experimental sources of uncertainty}
In order to account for the differences in trigger efficiencies between data and simulation, scale factors, defined as the ratio of the efficiencies measured in data and simulation, are calculated in bins of lepton $\eta$ and \pt and applied to the simulation. The efficiencies of the dilepton triggers in data are measured as the fraction of events passing triggers based on a \ptmiss requirement that also satisfy the dilepton trigger criteria. As the efficiency of the \ptmiss requirement is independent from the dilepton trigger efficiencies, the bias introduced by the \ptmiss requirement is negligible. The efficiencies are close to unity in both data and simulation. An uncertainty arising from the modelling of the trigger efficiencies in simulation is estimated by two variations of the scale factors. First, the scale factors are varied within their uncertainties coherently for all leptons. Second, to account for potential differential effects not covered by the coherent variations, simulated events are divided into categories according to the $\eta$ of the leptons, and the scale factors are varied in opposite directions for each category. A final trigger uncertainty is derived by taking the maximal deviation produced by the two variations in each bin.

The uncertainties from modelling of the lepton identification and isolation efficiencies are determined using the tag-and-probe method with \Zjets event samples~\cite{bib:tp,bib:TOP-15-003_paper}. The differences between lepton identification and isolation efficiencies in data and simulation in bins of $\eta$ and \pt are generally less than 10\% for electrons, while differences for muons are negligible. The lepton identification uncertainty is estimated by varying the scale factors within their uncertainties.

The uncertainty arising from the jet energy scale (JES) is determined by varying the 19 sources of uncertainty in the JES in bins of \pt and $\eta$ and taking the quadrature sum of the effects~\cite{Khachatryan:2016kdb}. The JES variations are also propagated to the uncertainties in \ptmiss. The uncertainty from the jet energy resolution (JER) is determined by the variation of the JER in simulation by $\pm$~1 standard deviation in different $\eta$ regions~\cite{Khachatryan:2016kdb}. An additional uncertainty from the calculation of \ptmiss\ is estimated by varying the energies of the reconstructed particles not clustered into jets within their respective resolutions and recalculating the \ptmiss.

The uncertainty from the modelling of the number of pileup events is obtained by changing the inelastic proton-proton cross section assumed in simulation by ${\pm}4.6\%$, corresponding to the uncertainty in the measurement of this cross section presented in~Ref.~\cite{Aaboud:2016mmw}.

The uncertainty due to imperfect modelling of the \cPqb\ tagging efficiency is determined by varying the measured scale factor for \cPqb\ tagging efficiencies within its uncertainties. An additional shape uncertainty is determined by dividing the {\cPqb} jet distributions in \pt and $\eta$ at their medians to form two bins in each variable.
The {\cPqb} tagging scale factors in the first bin are scaled up according to their uncertainties, while those in the second bin are scaled down and vice versa. The variations are performed separately for the \pt and $\eta$ distributions, and independently for heavy-flavour ({\cPqb} and {\cPqc}) and light-flavour ({\cPqu}, {\cPqd}, {\cPqs}, and gluon) jets.

The efficiency of the kinematic reconstruction of the top quarks is found to be consistent between data and simulation within around 0.2\%. An associated uncertainty is derived by varying the scale factor that describes the ratio of the kinematic reconstruction efficiency in data and simulation by ${\pm}0.2\%$.

The uncertainty in the integrated luminosity of the 2016 data sample recorded by CMS is $2.5\%$~\cite{bib:CMS-PAS-LUM-17-001} and is applied coherently to the normalisation of all simulated distributions.

\subsection{Theoretical sources of uncertainty}
The uncertainty arising from the missing higher-order terms in the simulation of the signal process at the ME level is assessed by varying the renormalisation and factorisation scales in the \POWHEG\ simulation up and down by factors of two with respect to their nominal values. In the \POWHEG\ simulation, the nominal scales are defined as $m^2_{\cPqt} + p^2_{\mathrm{T,\cPqt}}$, where $p_{\mathrm{T},\cPqt}$ denotes the \pt of the top quark in the \ttbar rest frame. In total, three variations are applied: one with the factorisation scale fixed, one with the renormalisation scale fixed, and one with both scales varied coherently together. The final uncertainty is taken as the maximum deviation from the nominal prediction from each of the three variations. In the parton-shower simulation, the corresponding uncertainty is estimated by varying the scale of initial- and final-state radiation separately up and down by factors of 2 and $\sqrt{2}$, respectively, as suggested in Ref.~\cite{Skands:2014pea}.

The effect of the uncertainty from the choice of PDF is assessed by reweighting the signal simulation according to the prescription provided for the NNPDF3.0 PDF set~\cite{bib:NNPDF}. An additional uncertainty is independently derived by varying the \alpS value within its uncertainty in the PDF set. The dependence of the measurement on the assumed \mt\ value is estimated by varying the chosen \mt\ in the default setup by $\pm1\GeV$ with respect to the default value of $172.5\GeV$.

The uncertainty originating from the scheme used to match the ME-level calculation to the parton-shower simulation is derived by varying the $h_\mathrm{damp}$ parameter in \POWHEG\ by factors of 1.42 and 0.63, according to the results of a tuning of this parameter from Ref.~\cite{bib:CMS:2016kle}.

The uncertainty related to the modelling of the underlying event is estimated by varying the parameters used to derive the CUETP8M2T4 tune in the default setup. The default setup in \PYTHIA\ includes a model of colour reconnection based on multiple-particle interactions (MPI) with early resonance decays switched off. To estimate an uncertainty from this choice of model, the analysis is repeated with three other models of colour reconnection within \PYTHIA: the MPI-based scheme with early resonance decays switched on, a gluon-move scheme~\cite{Argyropoulos:2014zoa}, and a QCD-inspired scheme~\cite{Christiansen:2015yqa}. The total uncertainty from colour reconnection modelling is estimated by taking the maximum deviation from the nominal result.

The uncertainty from imperfect knowledge of the {\cPqb} quark fragmentation function is assessed by varying the Bowler--Lund function within its uncertainties \cite{Bowler:1981sb}. In addition, the analysis is repeated with the Peterson model for {\cPqb} quark fragmentation~\cite{PhysRevD.27.105}. An uncertainty from the semileptonic branching fraction of \cPqb\ hadrons is estimated by correcting the \ttbar simulation to match the branching fraction in Ref.~\cite{Patrignani:2016xqp}. Since \ttbar events containing electrons or muons that originate from $\tau$ decays are considered as backgrounds, the measured differential cross sections are sensitive to the value of the $\tau$ semileptonic branching fraction used in the simulation. Hence, an uncertainty is derived by varying the branching fractions by $1.5\%$~\cite{Patrignani:2016xqp}.
Since the \cPqb\ tagging efficiency depends on many simulation parameters, it is recalculated for each variation of the sources of theoretical uncertainty, with the exception of the PDFs, the semileptonic branching fraction of \cPqb\ hadrons, the JES, and the JER. Finally, the normalisations of all backgrounds except \ttbar other are varied up and down by ${\pm} 30\%$~\cite{bib:TOP-15-003_paper}.

The total uncertainty in each bin of each measurement is determined by summing the experimental and theoretical uncertainties in quadrature and ranges from 4--25\%, depending on the observable and the bin. In Section \ref{sec:results}, figures showing the contribution of each systematic uncertainty, the statistical uncertainty, and the total uncertainty in each bin for selected normalised parton-level differential cross sections as a function of top-quark-related kinematic observables are provided. For most bins in a majority of these distributions, the JES is the dominant systematic uncertainty. In the first three bins of the \pttt distribution, the dominant uncertainty arises from the measurement of the energies of reconstructed particles not clustered into jets.

\section{Results}
\label{sec:results}
In this section, the results of all differential cross section measurements are presented.
\subsection{Measured observables}
The differential cross sections are measured as functions of numerous kinematic observables of the top quarks and their decay products, the \ttbar\ system, and the total number of jets in the event.These observables are listed below. For one group of observables both parton- and particle-level measurements are presented, while for a second group only particle-level measurements are given.

\begin{itemize}
\item [Observables measured at parton and particle levels:]
\item \pt of the top quark (\pttop)
\item \pt of the top antiquark (\ptantitop)
\item \pt of the top quark or top antiquark with largest \pt (\pttop (leading))
\item \pt of the top quark or top antiquark with second-largest \pt (\pttop (trailing))
\item \pt of the top quark in the rest frame of the \ttbar system (\pttop (\ttbar RF))
\item rapidity of the top quark (\ytop)
\item rapidity of the top antiquark (\yantitop)
\item rapidity of the top quark or top antiquark with largest \pt (\ytop (leading))
\item rapidity of the top quark or top antiquark with second-largest \pt (\ytop (trailing))
\item difference in absolute rapidity between the top quark and antiquark (\delytt)
\item absolute difference in azimuthal angle between the top quark and antiquark (\delphitt)
\item \pt of the \ttbar system (\pttt)
\item rapidity of the \ttbar system (\ytt)
\item invariant mass of the \ttbar system (\mtt)
\end{itemize}

\begin{itemize}
\item [Observables measured at particle level only:]
\item \pt of the lepton (\ptlep)
\item \pt of the antilepton (\ptalep)
\item \pt of the lepton or antilepton with largest \pt (\ptlep (leading))
\item \pt of the lepton or antilepton with second-largest \pt (\ptlep (trailing))
\item pseudorapidity of the lepton (\etalep)
\item pseudorapidity of the antilepton (\etaalep)
\item pseudorapidity of the lepton or antilepton with largest \pt (\etalep (leading))
\item pseudorapidity of the lepton or antilepton with second-largest \pt (\etalep (trailing))
\item \pt of the dilepton system (\ptll)
\item invariant mass of the dilepton system (\mll)
\item absolute difference in azimuthal angle between the lepton and antilepton (\delphill)
\item difference in absolute pseudorapidity between the lepton and antilepton (\deletall)
\item \pt of the {\cPqb} jet with largest \pt (\ptb (leading))
\item \pt of the {\cPqb} jet with second-largest \pt (\ptb (trailing))
\item pseudorapidity of the {\cPqb} jet with largest \pt (\etab (leading))
\item pseudorapidity of the {\cPqb} jet with second-largest \pt (\etab (trailing))
\item \pt of the \bbbar\ system (\ptbb)
\item invariant mass of the \bbbar\ system (\mbb)
\item multiplicity of jets with $\pt^{\text{jet}} > 30\GeV$ (\Nj)
\end{itemize}

The measurements of top quark \pt\ are sensitive to higher-order QCD and electroweak corrections in the SM, \mt, PDFs, and potential BSM physics signals.
In order to probe the modelling of the top quark \pt\ as thoroughly as possible, various differential cross sections related to the \pt of top quarks are measured. These include: the separate \pt of the top quarks and antiquarks in the laboratory frame and, in order to suppress the effects of initial- and final-state radiation (ISR and FSR), in the \ttbar rest frame (RF), and the largest (leading) and second-largest \pt\ (trailing) top quark or antiquark in an event. Similarly, the rapidity distributions are determined separately for top quarks and antiquarks, as well as the rapidity of the leading and trailing top quark or antiquark in an event. The differential cross sections as a function of the differences in absolute rapidities between the top quark and antiquark and in absolute pseudorapidities between the lepton and antilepton are measured to allow the extraction of the \ttbar and leptonic charge asymmetries described in Section~\ref{sec:ca}. The \pt of the \ttbar system and \Nj\ distributions are measured since they are especially sensitive to the higher-order terms in the perturbative calculations. The rapidity and invariant mass distributions of the \ttbar system are measured because of their potential to reduce gluon PDF uncertainties at large fractions of the proton longitudinal momentum carried by the gluon. In addition, for small values of \mtt, the \mtt distribution is sensitive to \mt, while for large values of \mtt, it is sensitive to BSM scenarios in which heavy states decay to \ttbar pairs. The measurements of the lepton kinematic observables test the modelling of the top quark decays and spin correlations in the \ttbar pair. Measuring the {\cPqb} jet kinematic observables further tests the modelling of the top quark decays, while also testing the parton shower and hadronisation models.

\subsection{Theoretical predictions}
{\tolerance=800 All data are compared to predictions from \pwhgpy, \pwhghpp, and \mgamcpy. Where possible, parton-level measurements are also compared to predictions based on the following calculations at beyond-NLO precision:\par}

\begin{itemize}
\item A calculation with full NNLO precision in QCD and including electroweak corrections of order $\alpS^{2}\alphaew$, $\alpS\alphaew^{2}$, and $\alphaew^{3}$ (\nnloew)~\cite{Czakon:2017wor}. The dynamic renormalisation and factorisation scales are set to \mT/2 for \pttop and \ptantitop and \HT/4 for \ytop, \yantitop, \pttt, \ytt, \mtt, and \delytt, where $\mT = \sqrt{\smash[b]{\mt^{2} + ({\pttop})^{2}}}$ and \HT is the sum of the top quark and antiquark \mT values. Predictions are provided for both the LUXQED17~\cite{Manohar:2016nzj} and NNPDF3.1\_qed PDF~\cite{Bertone:2017bme} sets with \mt = 173.3\GeV. In order to probe the sensitivity of the results to the value of \mt, an additional prediction for the LUXQED17 PDF set with \mt = 172.5\GeV is provided.

\item A prediction~\cite{Czakon:2018nun} that combines the NNLO QCD calculations with the double resummation of soft and small-mass logarithms to NNLL' accuracy, matched with both the standard soft-gluon resummation at NNLL accuracy and the fixed-order calculation at NNLO accuracy (\nnlonnllprime). These corrections are expected to affect the high-energy tails of the \ttbar differential distributions. The calculation is performed using the NNPDF3.1 PDF set~\cite{Ball:2017nwa}, and dynamic renormalisation and factorisation scales (\mt/2 for \pttop and \HT/4 for \mtt). Predictions are provided for \mt values of 173.3 and 172.5\GeV.

\item An approximate next-to-NNLO calculation~\cite{bib:kidonakis_13TeV} (\annnlo) based on the resummation of soft-gluon contributions in the double-differential cross section at NNLL accuracy in the moment-space approach. The NNPDF3.0 PDF set is used and \mt\ is set to 172.5\GeV. The renormalisation and factorisation scales are set to \mT for the \pttop distribution and \mt for the \ytop distribution.

\item An approximate NNLO calculation~\cite{bib:difftop} (\annlo), based on QCD threshold expansions beyond the leading-logarithmic approximation using the CT14nnlo~\cite{bib:Dulat:2015mca} PDF set. The top quark mass and dynamic factorisation and renormalisation scales are set to \mt = 172.5\GeV.
\end{itemize}

The \nnloew predictions include uncertainties from variations of the renormalisation and factorisation scales and from the PDFs. The \nnlonnllprime and \annnlo predictions include uncertainties from scale variations only. The \annlo prediction includes uncertainties from the PDFs only.

For the \nnloew calculations, predictions for the \pttop, \ptantitop, \ytop, \yantitop, \pttt, \mtt, and \delytt distributions are provided. For the \nnlonnllprime calculation, predictions for the average of the \pttop and \ptantitop distributions and for the \mtt distribution are provided. For the \annnlo calculation, predictions for the \pttop and \ytop distributions are provided. For the \annlo calculation, predictions for the \pttop distribution and the average of the \ytop and \yantitop distributions are provided. Since the differences between the averaged predictions and the corresponding separate predictions for top quark and antiquark are expected to be small, the averaged predictions are compared to the top quark distributions in data.

\subsection{Commentary on results}

\begin{figure*}[!phtb]
\centering
\includegraphics[width=0.49\textwidth]{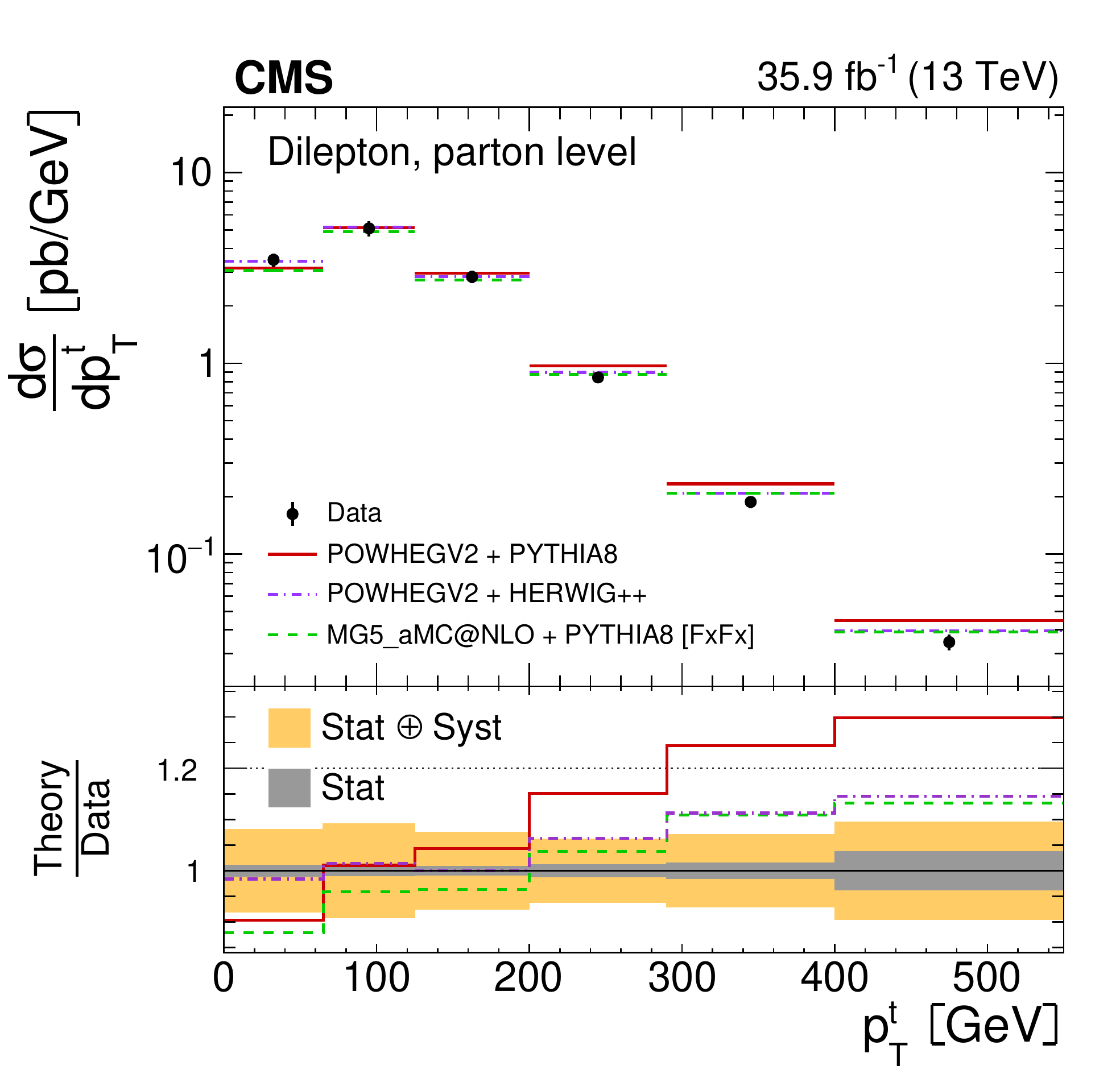}
\includegraphics[width=0.49\textwidth]{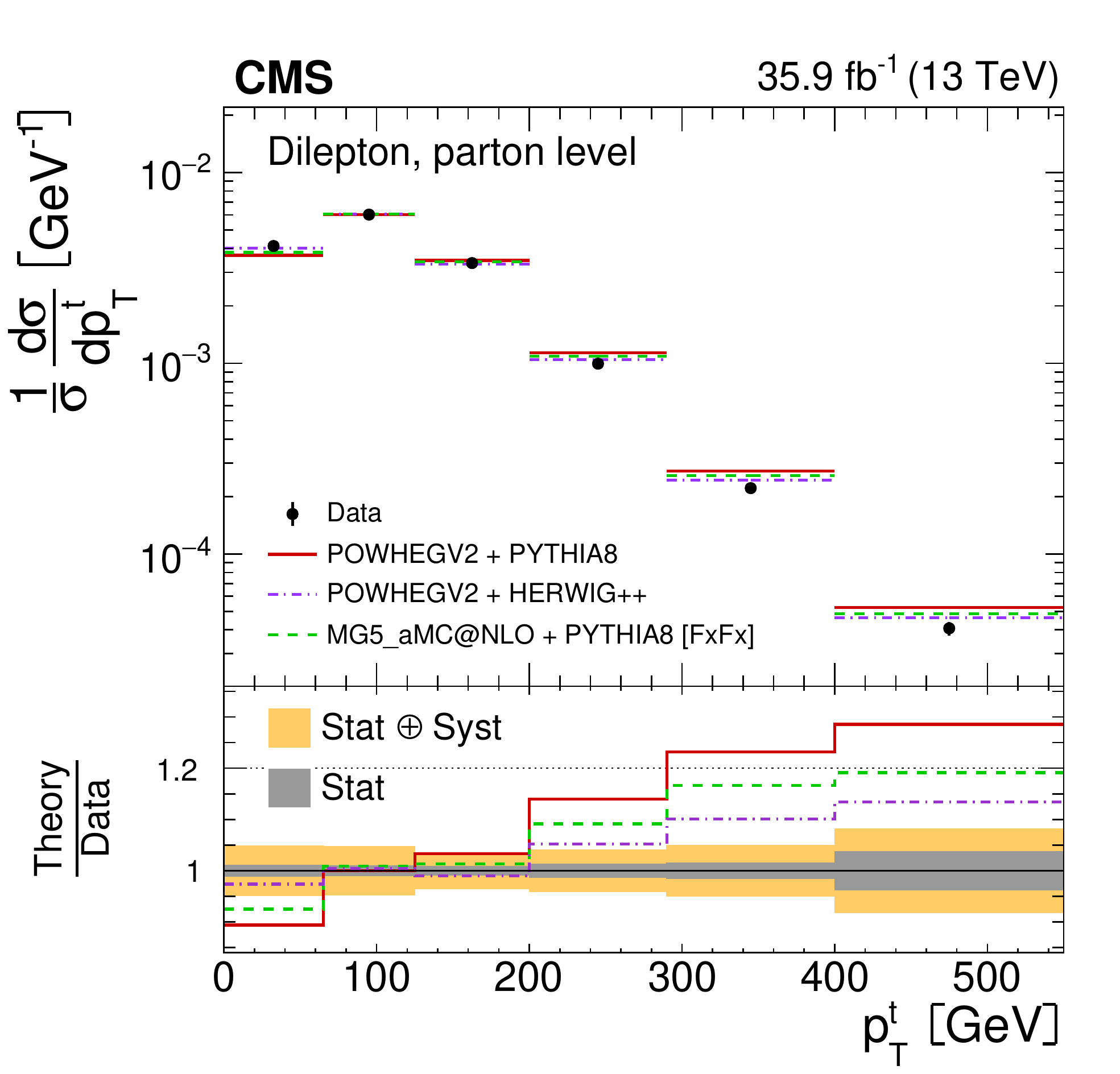} \\
\includegraphics[width=0.49\textwidth]{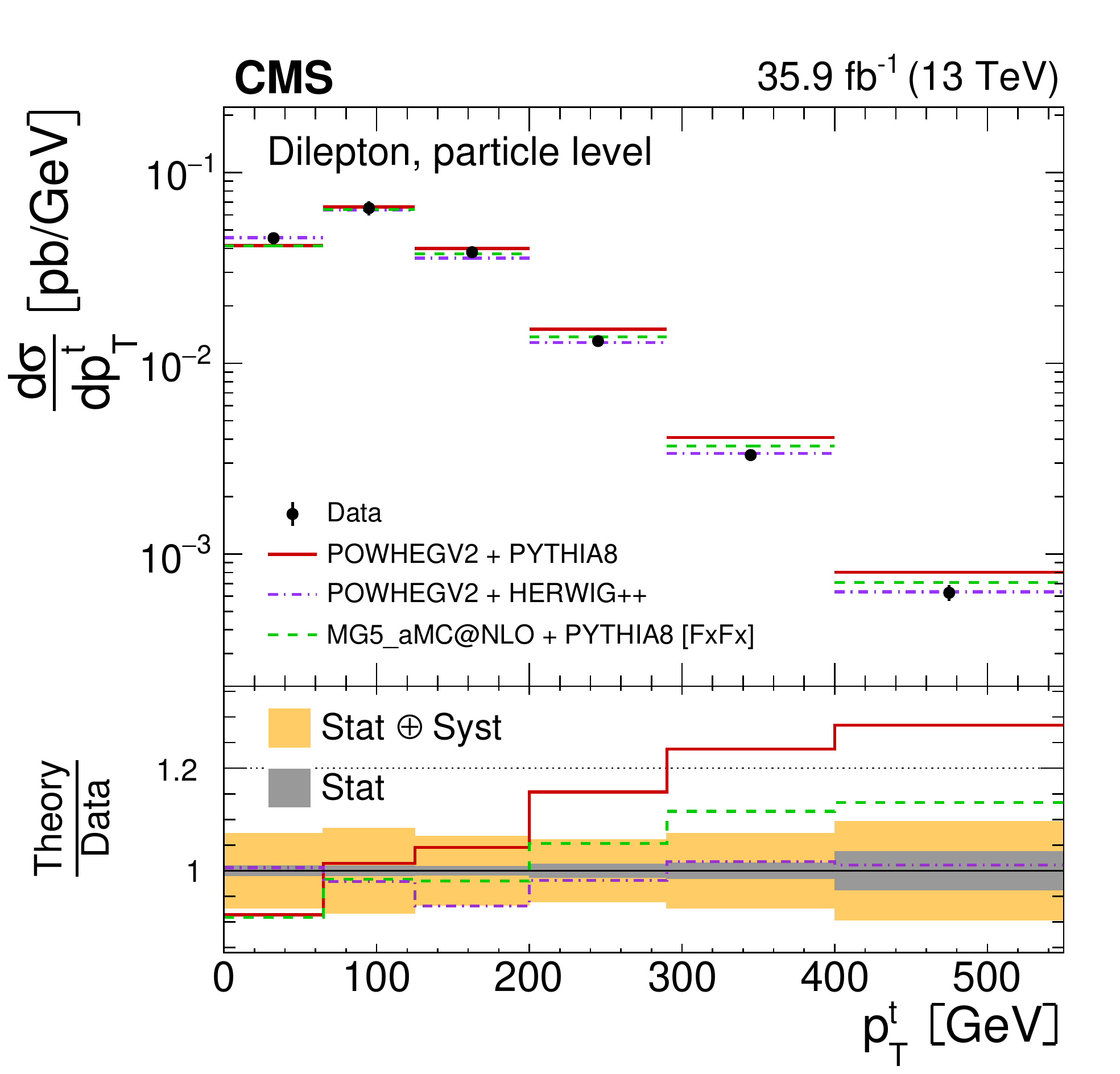}
\includegraphics[width=0.49\textwidth]{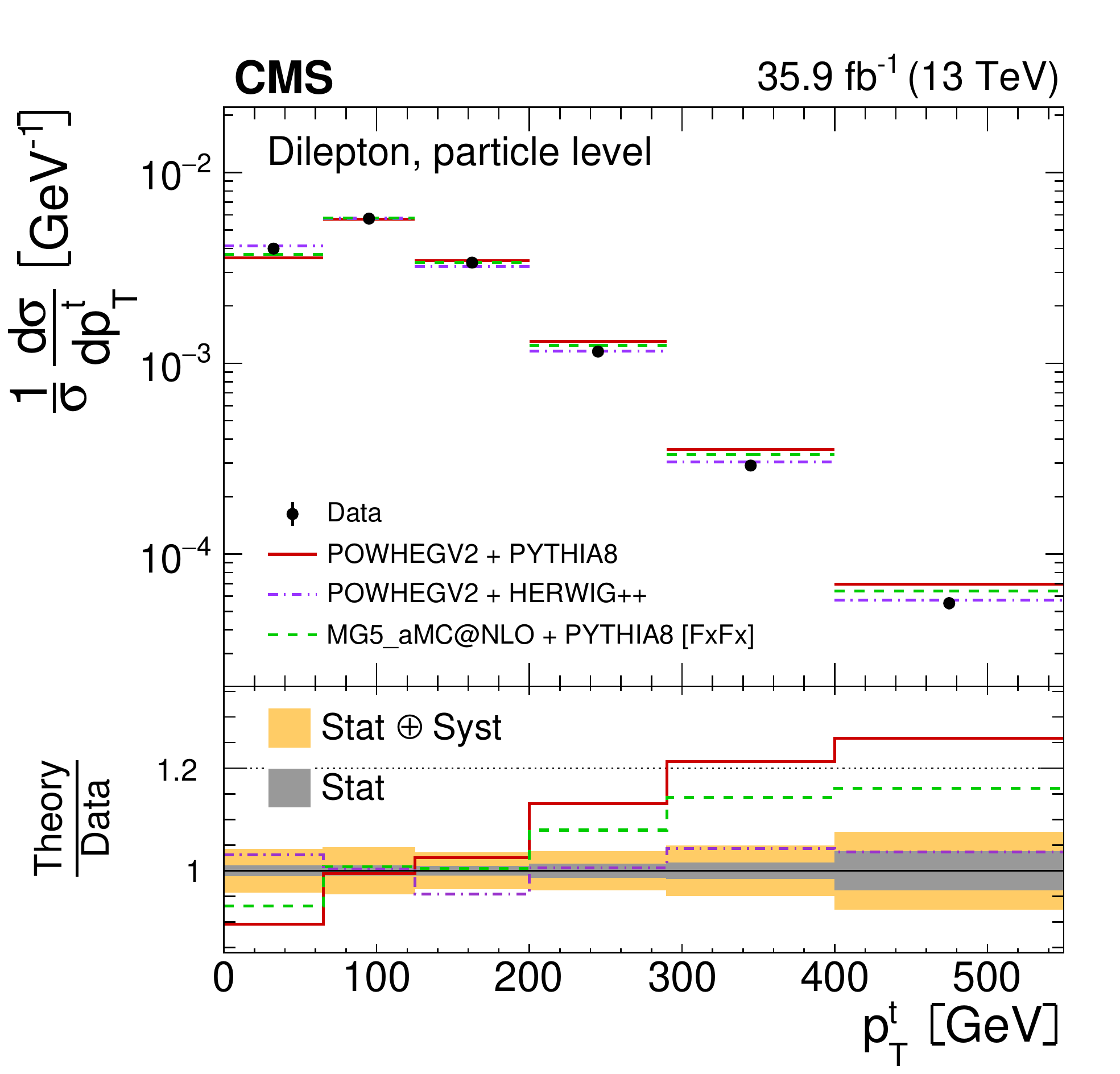}
\caption{The differential \ttbar production cross sections as a function of \pttop are shown for the data (points) and the MC predictions (lines). The vertical lines on the points indicate the total uncertainty in the data. The left and right columns correspond to absolute and normalised measurements, respectively. The upper row corresponds to measurements at the parton level in the full phase space and the lower row to the particle level in a fiducial phase space. The lower panel in each plot shows the ratios of the theoretical predictions to the data. The dark and light bands show the relative statistical and total uncertainties in the data, respectively.}
\label{fig:diffxsec:res_toppt}
\end{figure*}

\begin{figure*}[!phtb]
\centering
\includegraphics[width=0.49\textwidth]{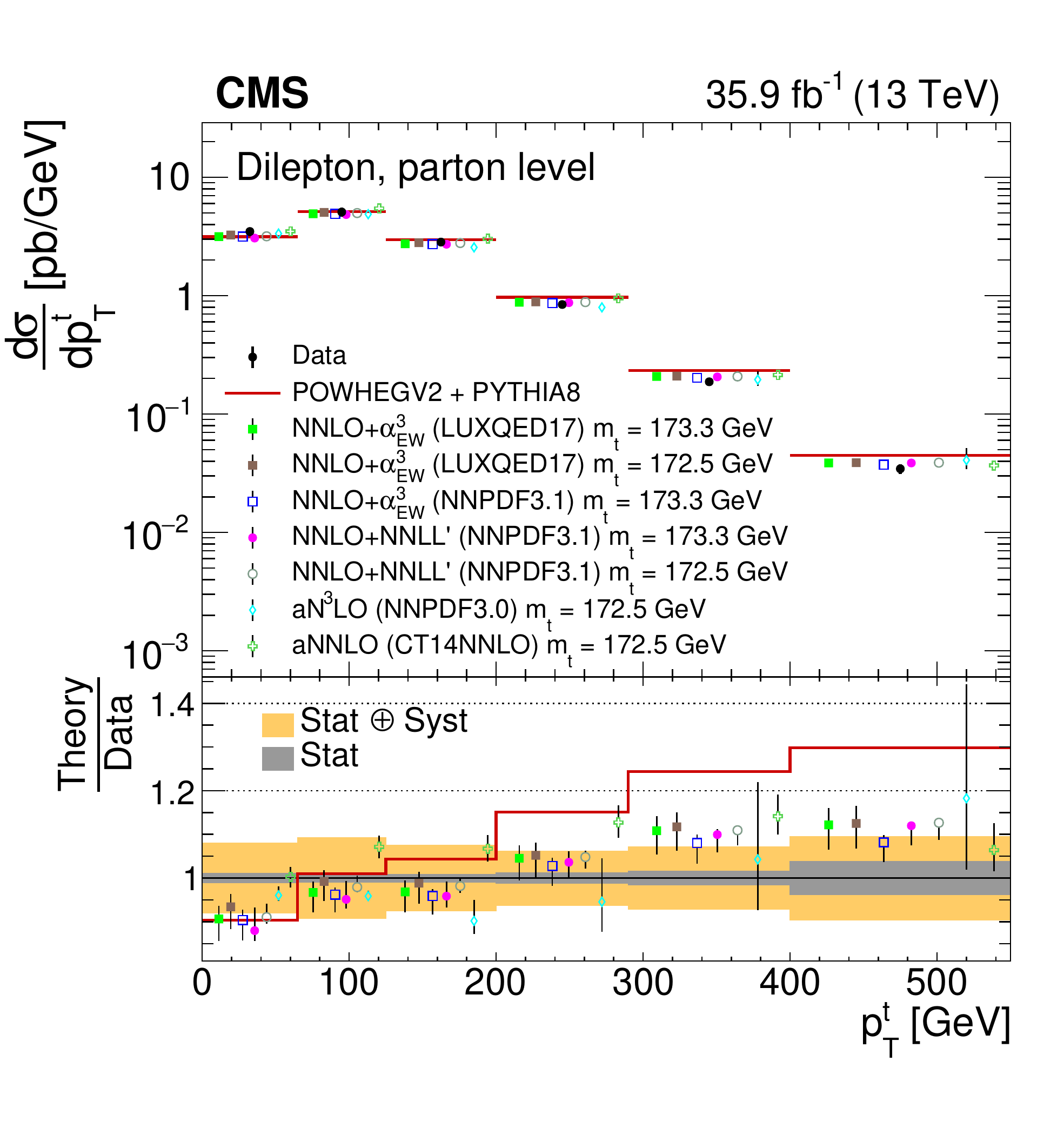}
\includegraphics[width=0.49\textwidth]{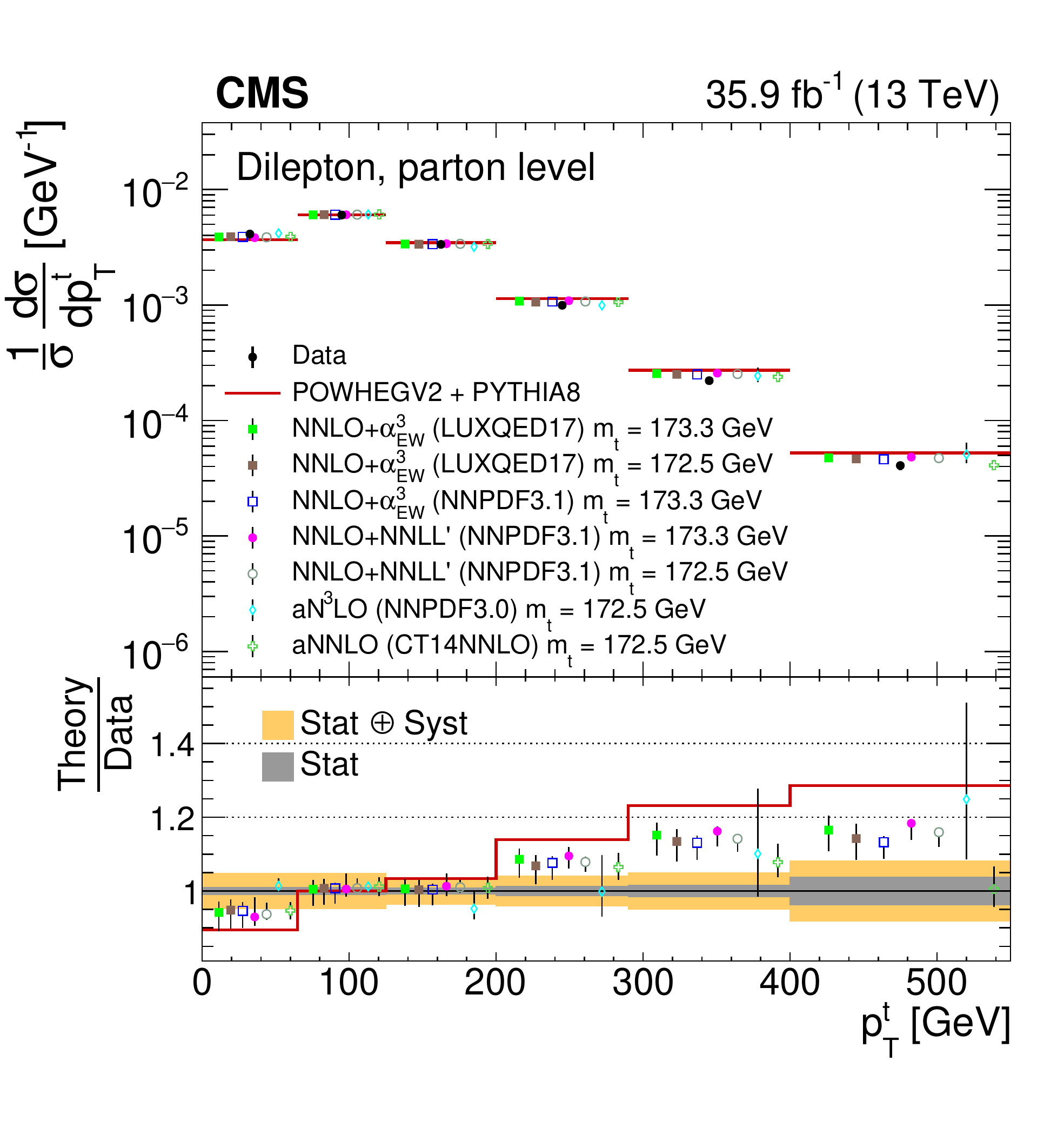} \\
\caption{The differential \ttbar production cross sections at the parton level in the full phase space as a function of \pttop are shown for the data (filled circles), the theoretical predictions with beyond-NLO precision (other points) and the prediction from \pwhgpy (solid line). The vertical lines on the filled circles and other points indicate the total uncertainty in the data and theoretical predictions, respectively. The left and right plots correspond to absolute and normalised measurements, respectively. The lower panel in each plot shows the ratios of the theoretical predictions to the data. The dark and light bands show the relative statistical and total uncertainties in the data, respectively.}
\label{fig:diffxsec:res_toppt_bnlo}
\includegraphics[width=0.75\textwidth]{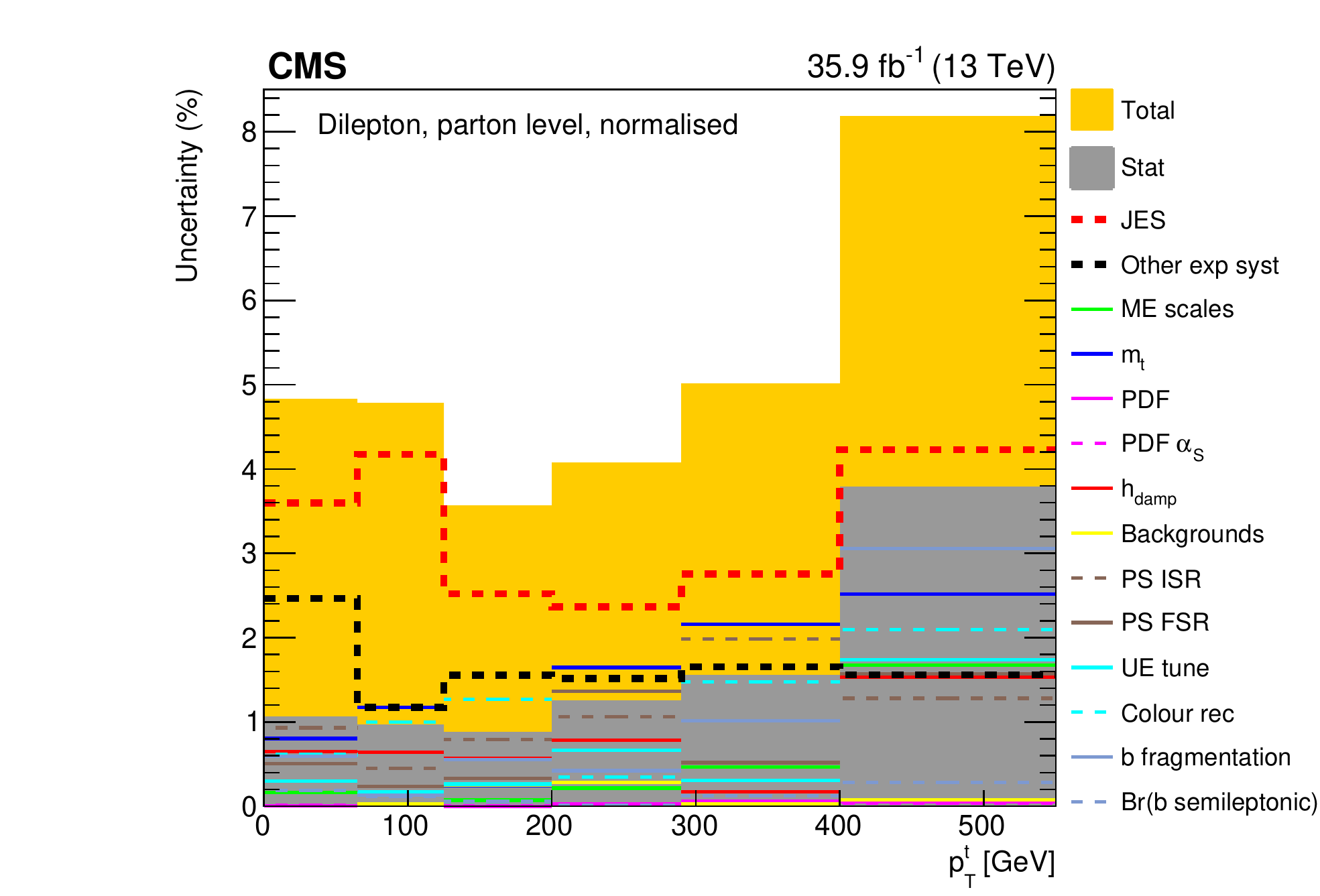}
\caption{The contributions of each source of systematic uncertainty in each bin is shown for the measurement of the normalised \ttbar production cross sections as a function of \pttop. The sources affecting the JES are added in quadrature and shown as a single component. Additional experimental systematic uncertainties are also added in quadrature and shown as a single component. Contributions from theoretical uncertainties are shown separately. The statistical and total uncertainties, corresponding to the quadrature addition of statistical and systematic uncertainties, are shown by the dark and light filled histograms, respectively.}
\label{fig:diffxsec:unc_breakdown_toppt}
\end{figure*}

\begin{figure*}[!phtb]
\centering
\includegraphics[width=0.49\textwidth]{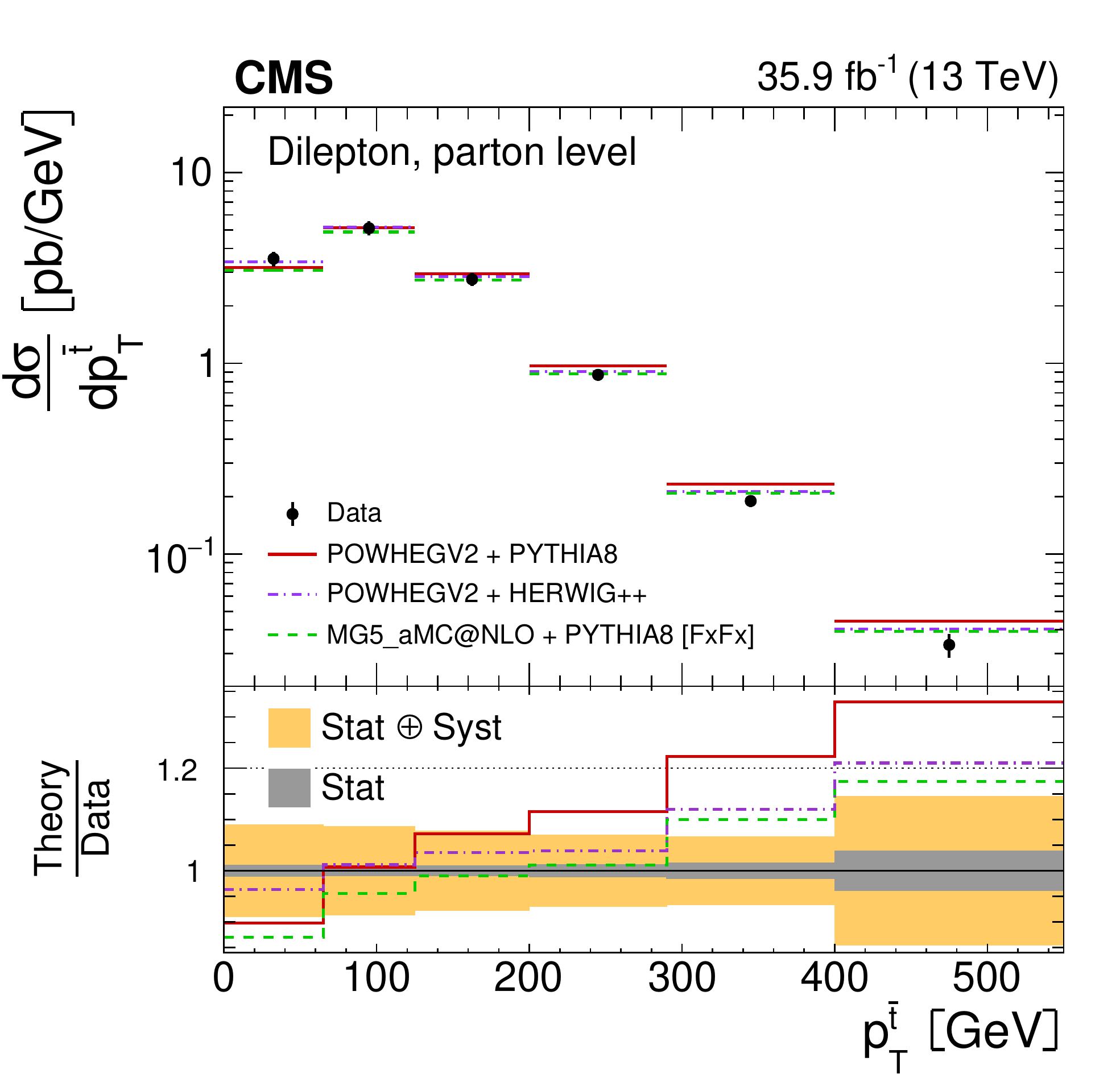}
\includegraphics[width=0.49\textwidth]{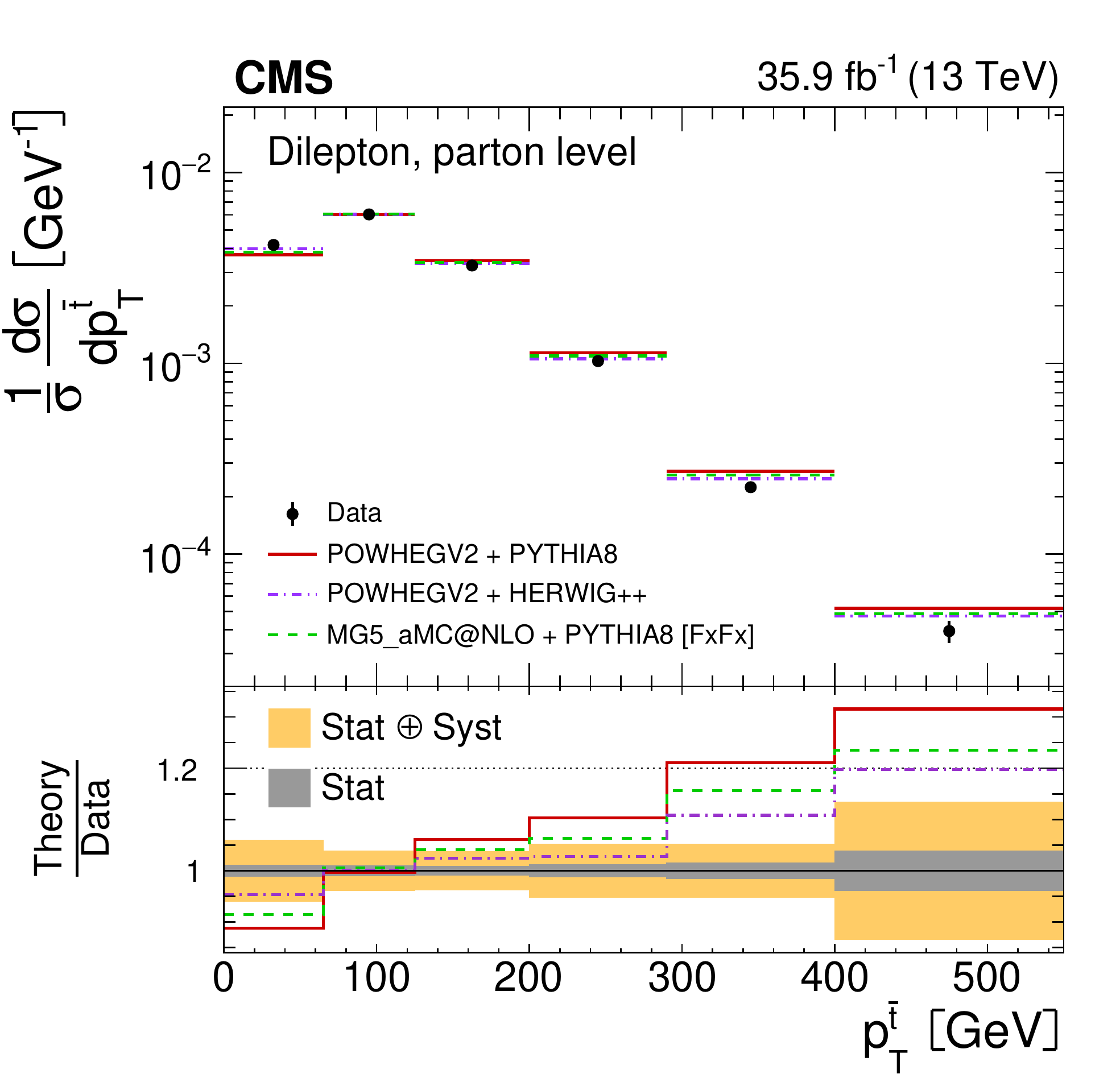} \\
\includegraphics[width=0.49\textwidth]{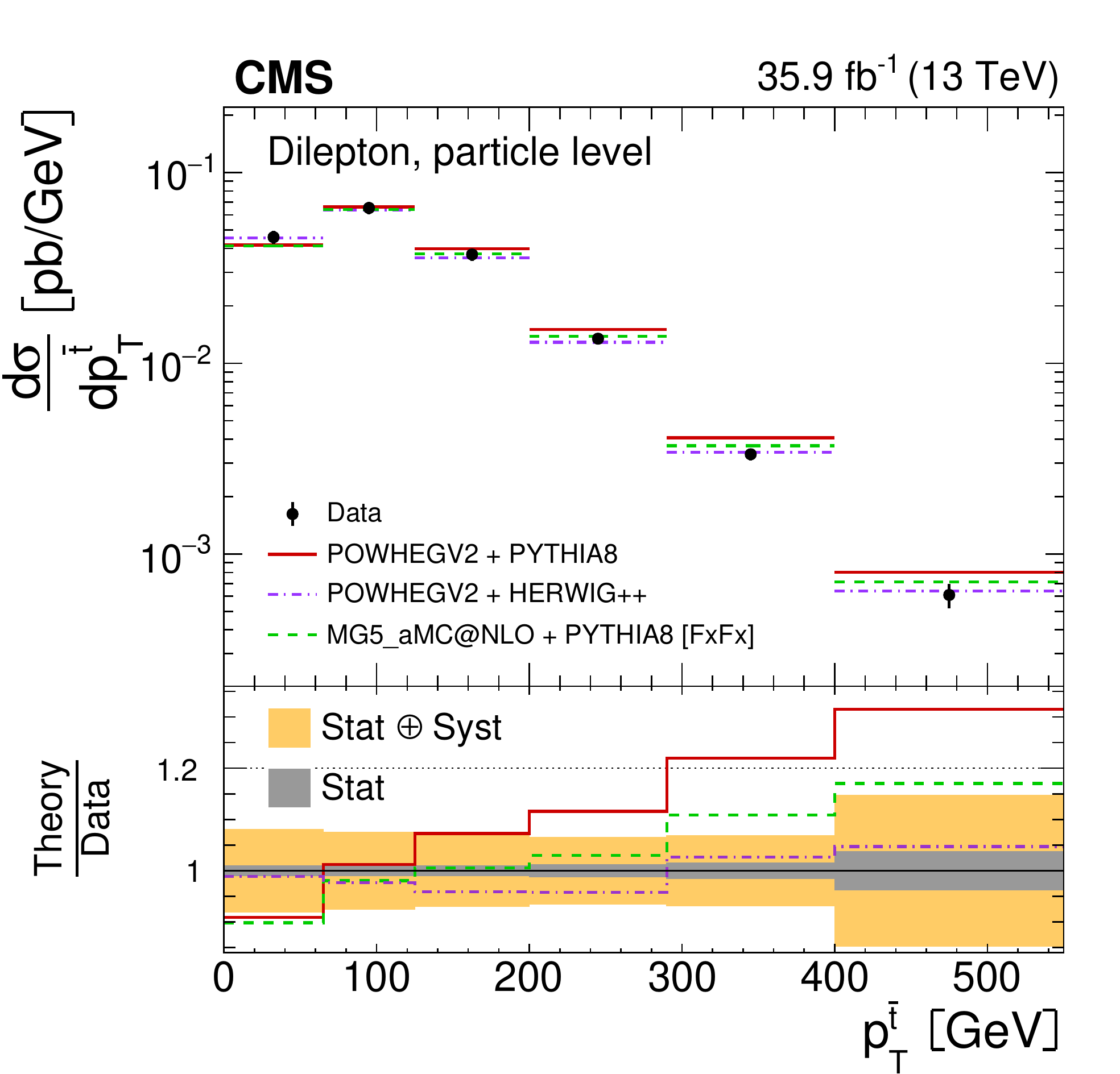}
\includegraphics[width=0.49\textwidth]{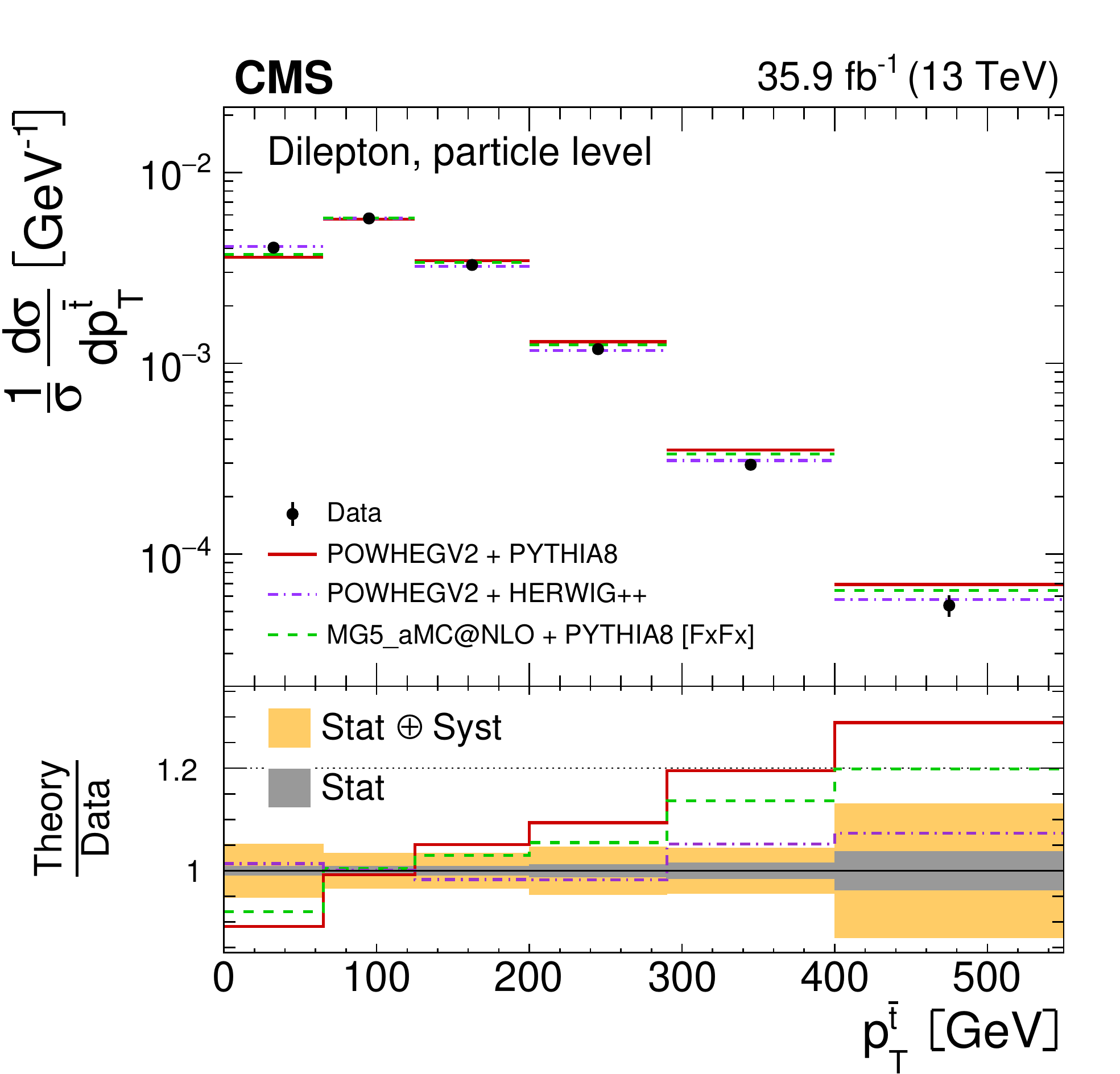}
\caption{The differential \ttbar production cross sections as a function of \ptantitop are shown for the data (points) and the MC predictions (lines). The vertical lines on the points indicate the total uncertainty in the data. The left and right columns correspond to absolute and normalised measurements, respectively. The upper row corresponds to measurements at the parton level in the full phase space and the lower row to the particle level in a fiducial phase space. The lower panel in each plot shows the ratios of the theoretical predictions to the data. The dark and light bands show the relative statistical and total uncertainties in the data, respectively.}
\label{fig:diffxsec:res_antitoppt}
\end{figure*}

\begin{figure*}[!phtb]
\centering
\includegraphics[width=0.49\textwidth]{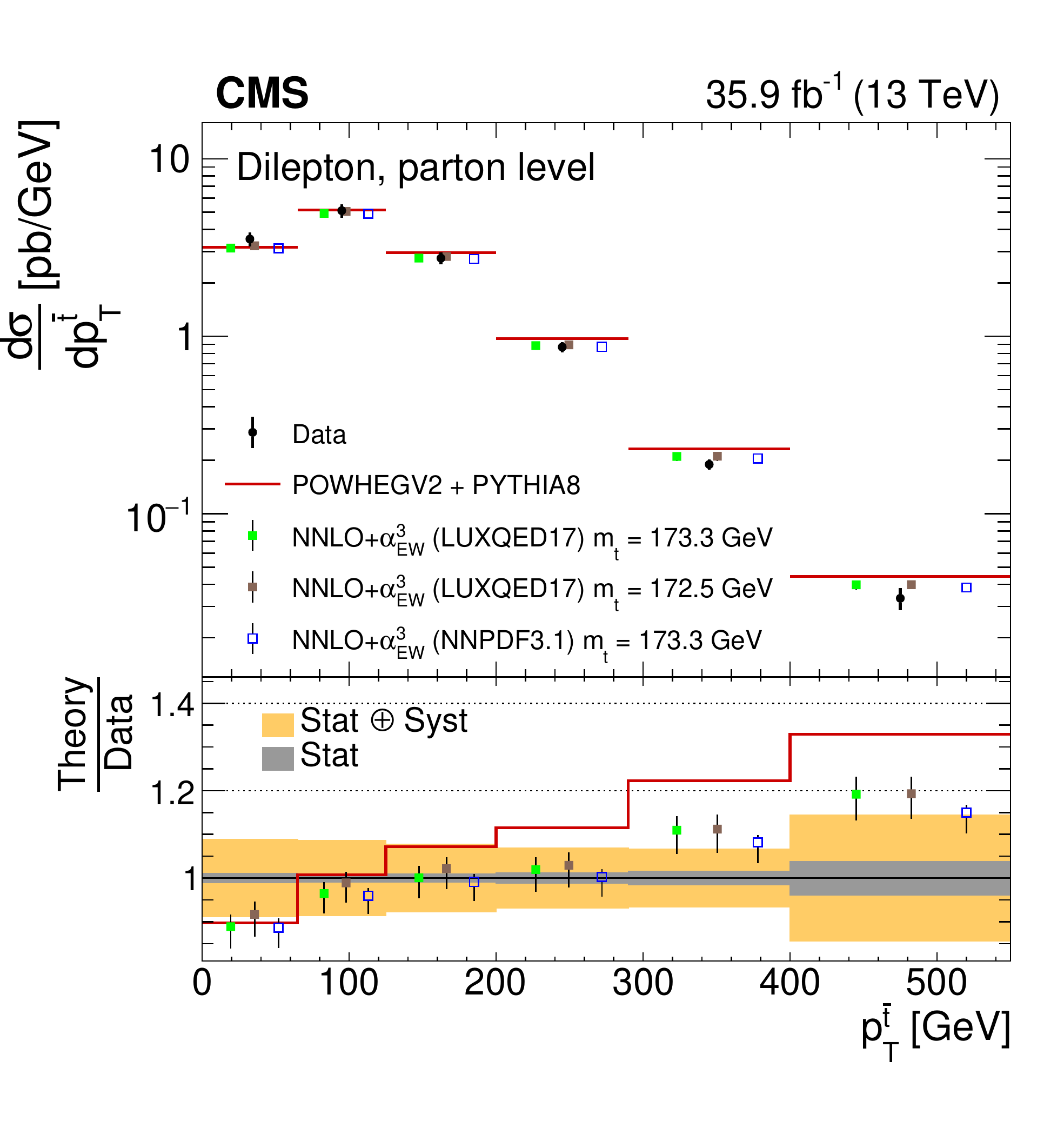}
\includegraphics[width=0.49\textwidth]{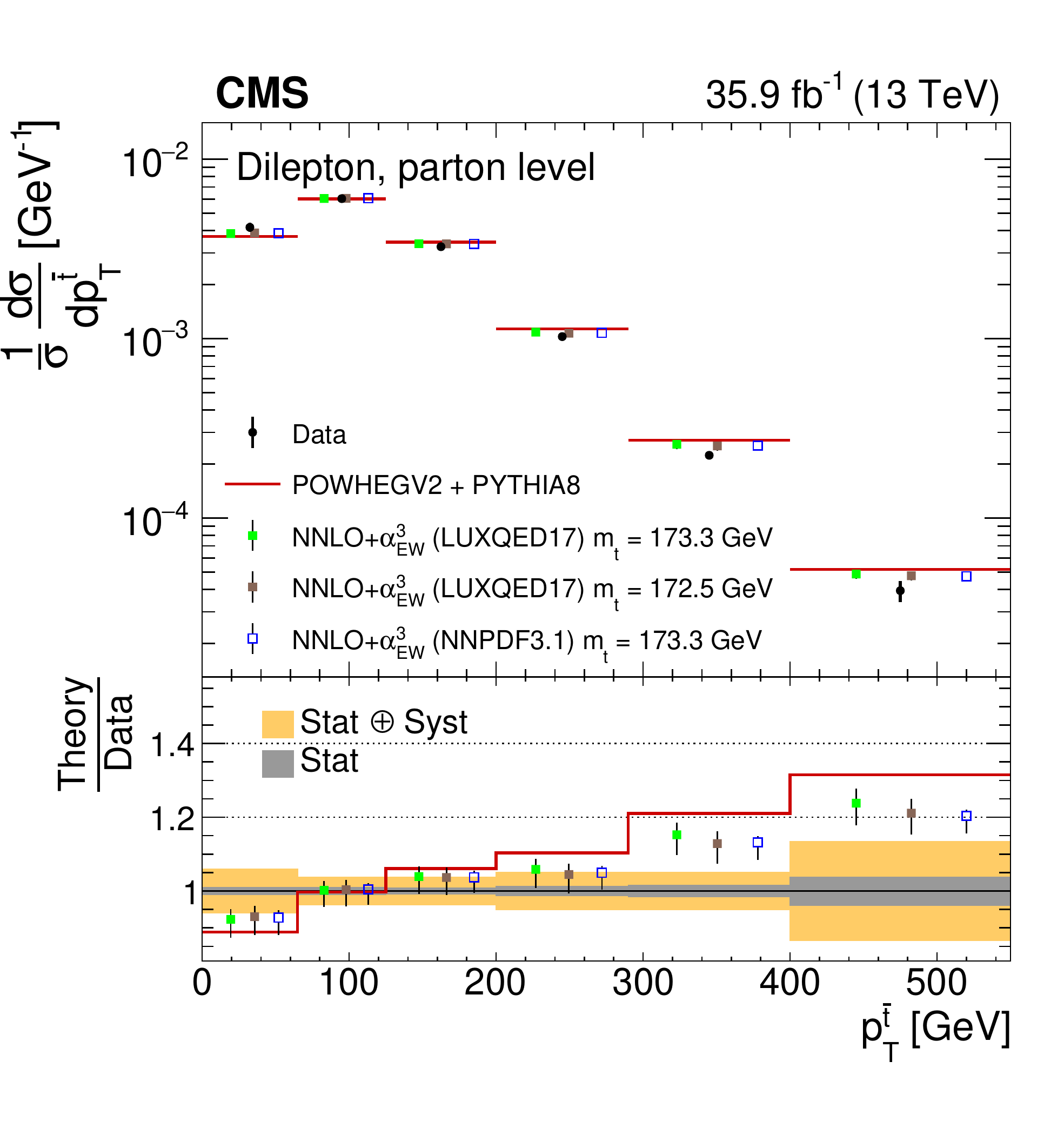}
\caption{The differential \ttbar production cross sections at the parton level in the full phase space as a function of \ptantitop are shown for the data (filled circles), the theoretical predictions with beyond-NLO precision (other points) and the prediction from \pwhgpy (solid line). The vertical lines on the filled circles and other points indicate the total uncertainty in the data and theoretical predictions, respectively. The left and right plots correspond to absolute and normalised measurements, respectively. The lower panel in each plot shows the ratios of the theoretical predictions to the data. The dark and light bands show the relative statistical and total uncertainties in the data, respectively.}
\label{fig:diffxsec:res_antitoppt_bnlo}
\includegraphics[width=0.75\textwidth]{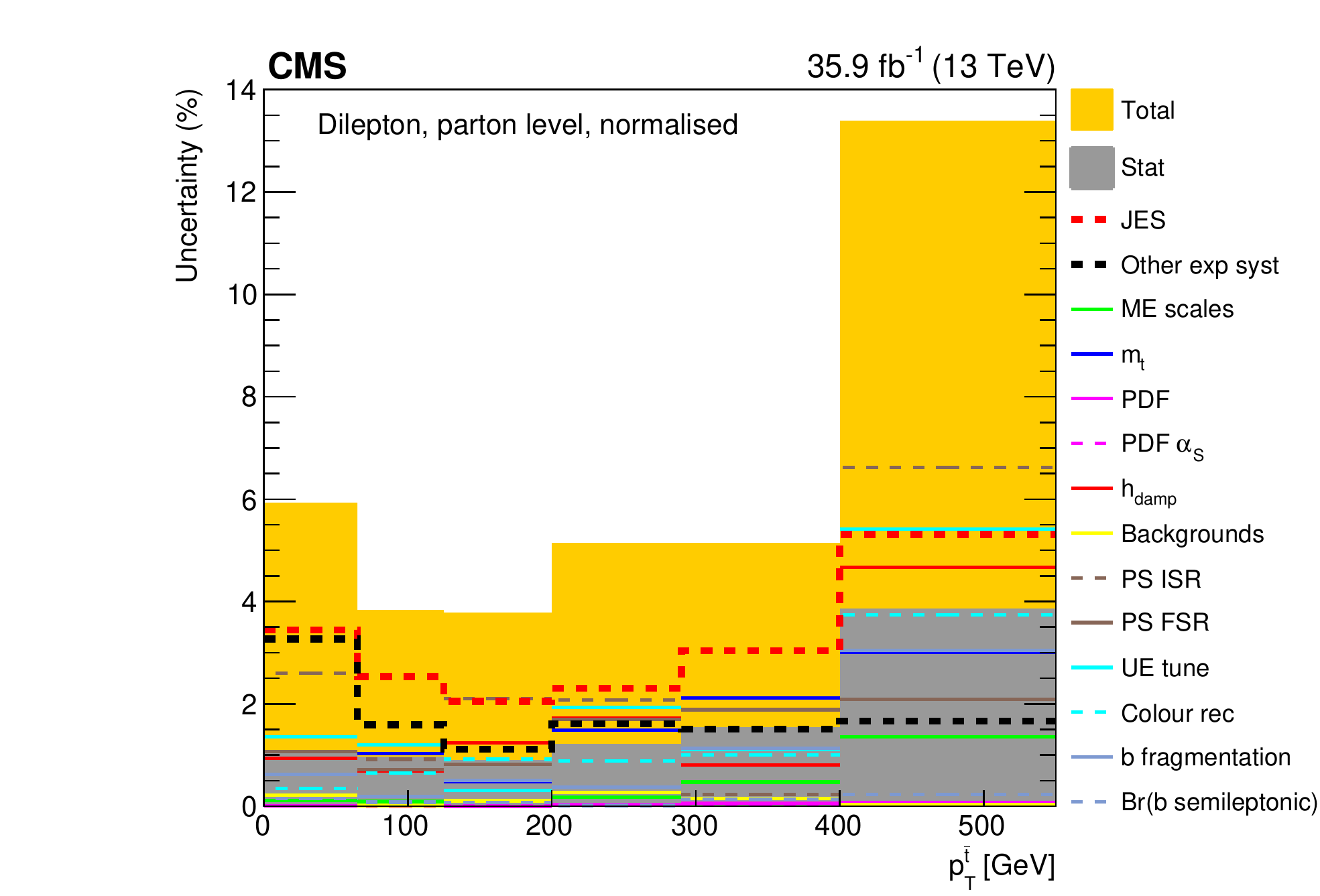}
\caption{The contributions of each source of systematic uncertainty to the total systematic uncertainty in each bin is shown for the measurement of the normalised \ttbar production cross sections  as a function of \ptantitop. The sources affecting the JES are added in quadrature and shown as a single component. Additional experimental systematic uncertainties are also added in quadrature and shown as a single component. Contributions from theoretical uncertainties are shown separately. The statistical and total uncertainties, corresponding to the quadrature addition of statistical and systematic uncertainties, are shown by the dark and light filled histograms, respectively.}
\label{fig:diffxsec:unc_breakdown_antitoppt}
\end{figure*}

\begin{figure*}[!phtb]
\centering
\includegraphics[width=0.49\textwidth]{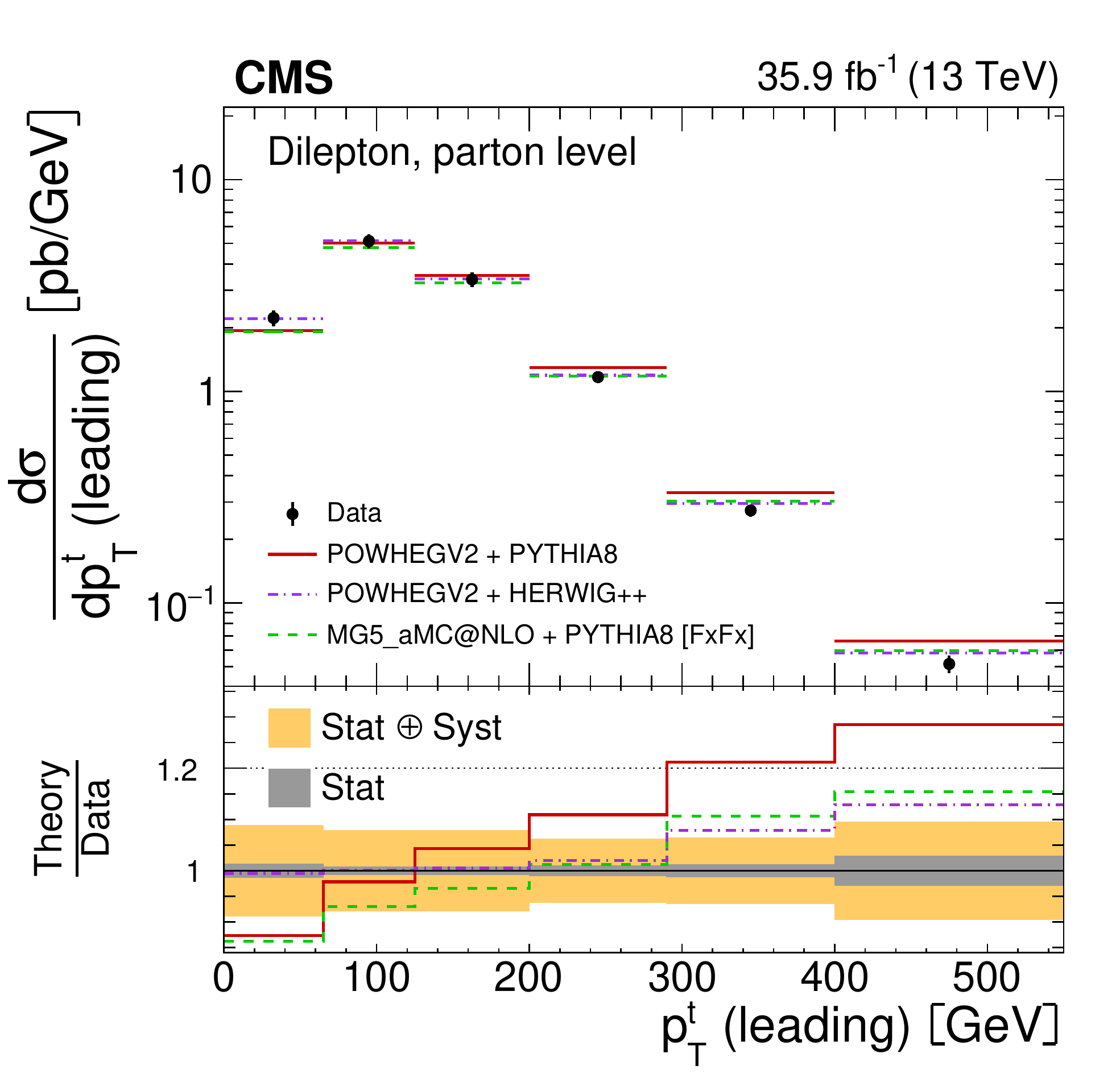}
\includegraphics[width=0.49\textwidth]{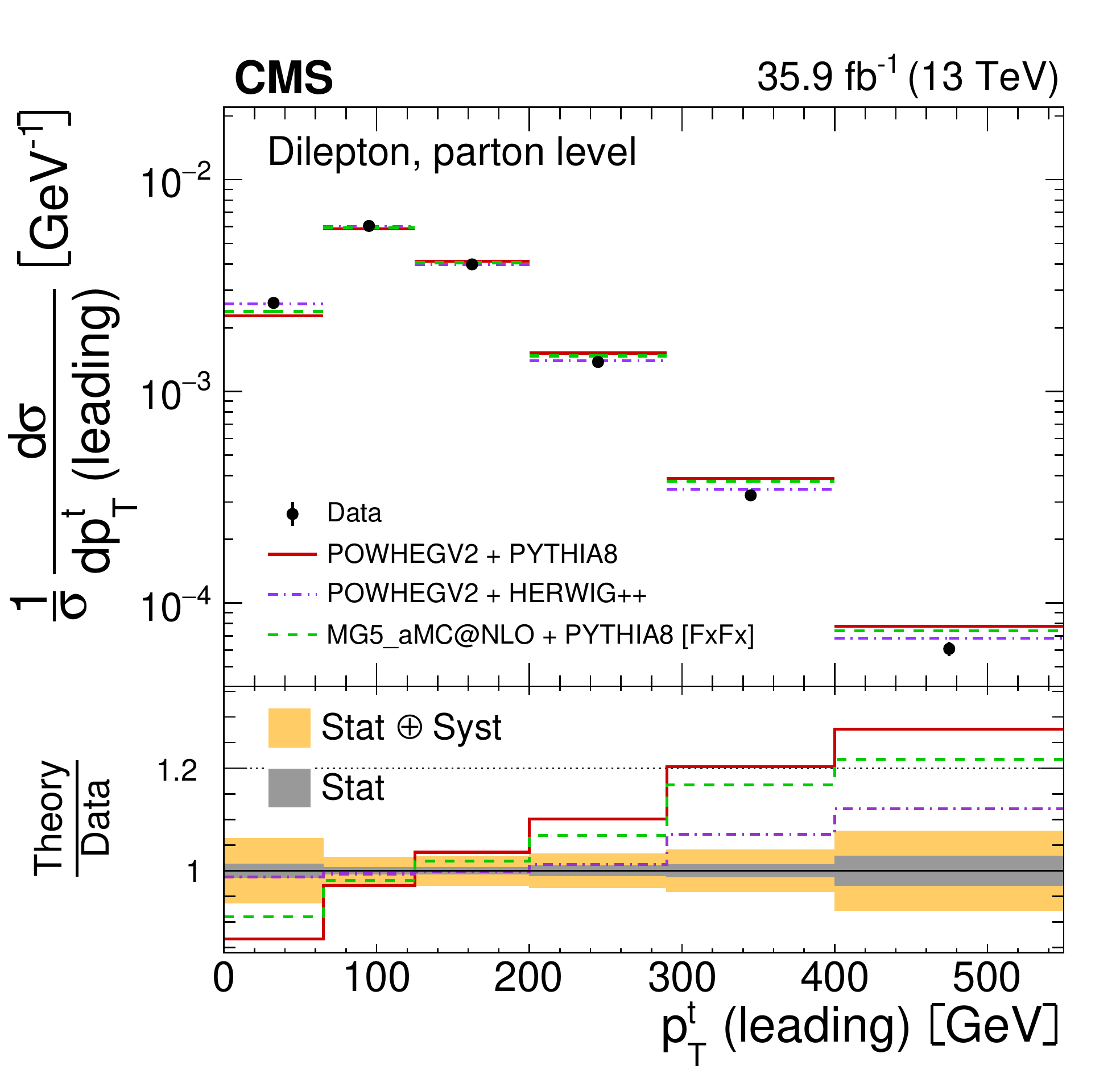} \\
\includegraphics[width=0.49\textwidth]{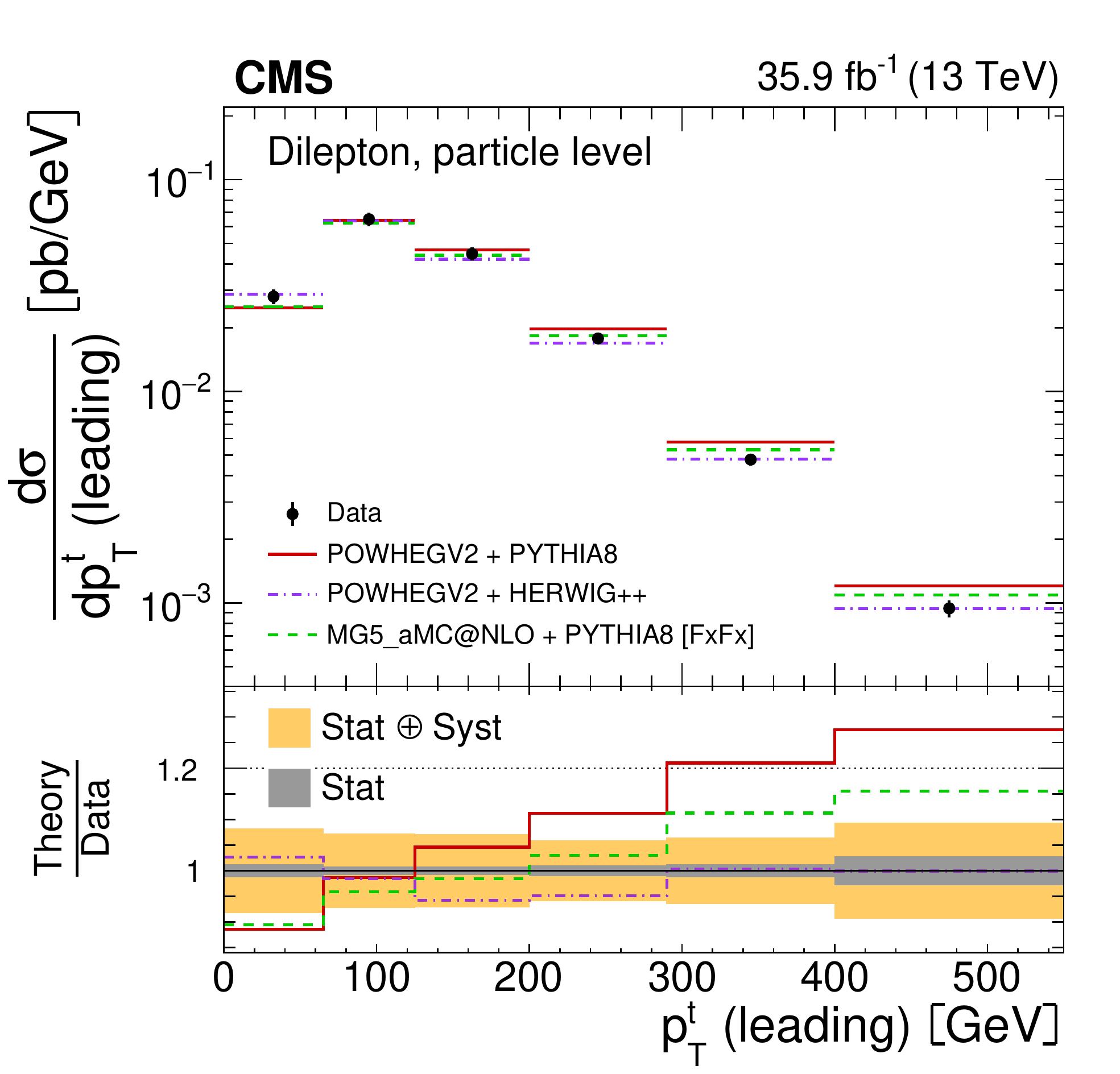}
\includegraphics[width=0.49\textwidth]{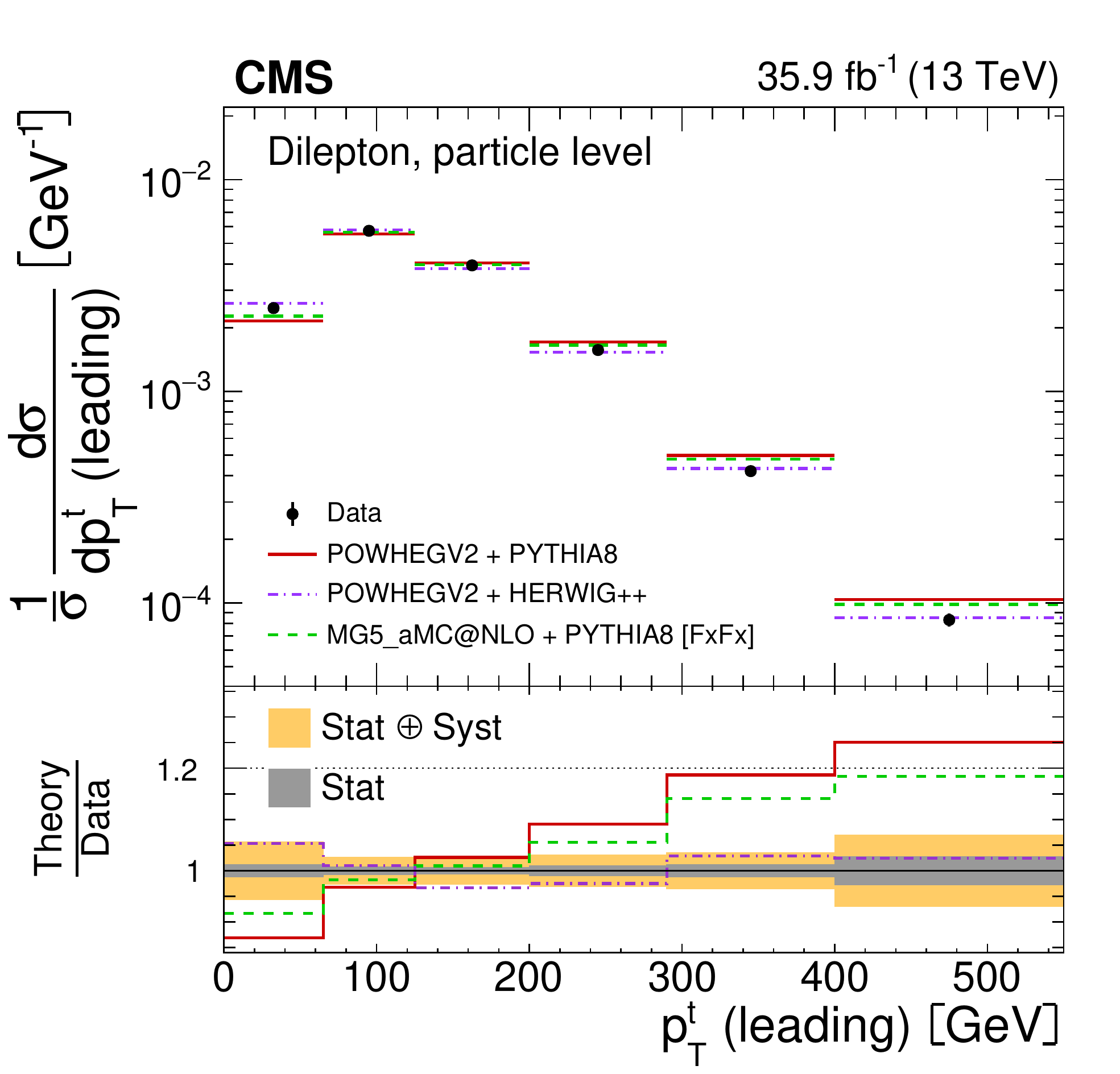}
\caption{The differential \ttbar production cross sections as a function of \pttop (leading) are shown for the data (points) and the MC predictions (lines). The vertical lines on the points indicate the total uncertainty in the data. The left and right columns correspond to absolute and normalised measurements, respectively. The upper row corresponds to measurements at the parton level in the full phase space and the lower row to the particle level in a fiducial phase space. The lower panel in each plot shows the ratios of the theoretical predictions to the data. The dark and light bands show the relative statistical and total uncertainties in the data, respectively.}
\label{fig:diffxsec:res_leadingtoppt}
\end{figure*}

\begin{figure*}[!phtb]
\centering
\includegraphics[width=0.49\textwidth]{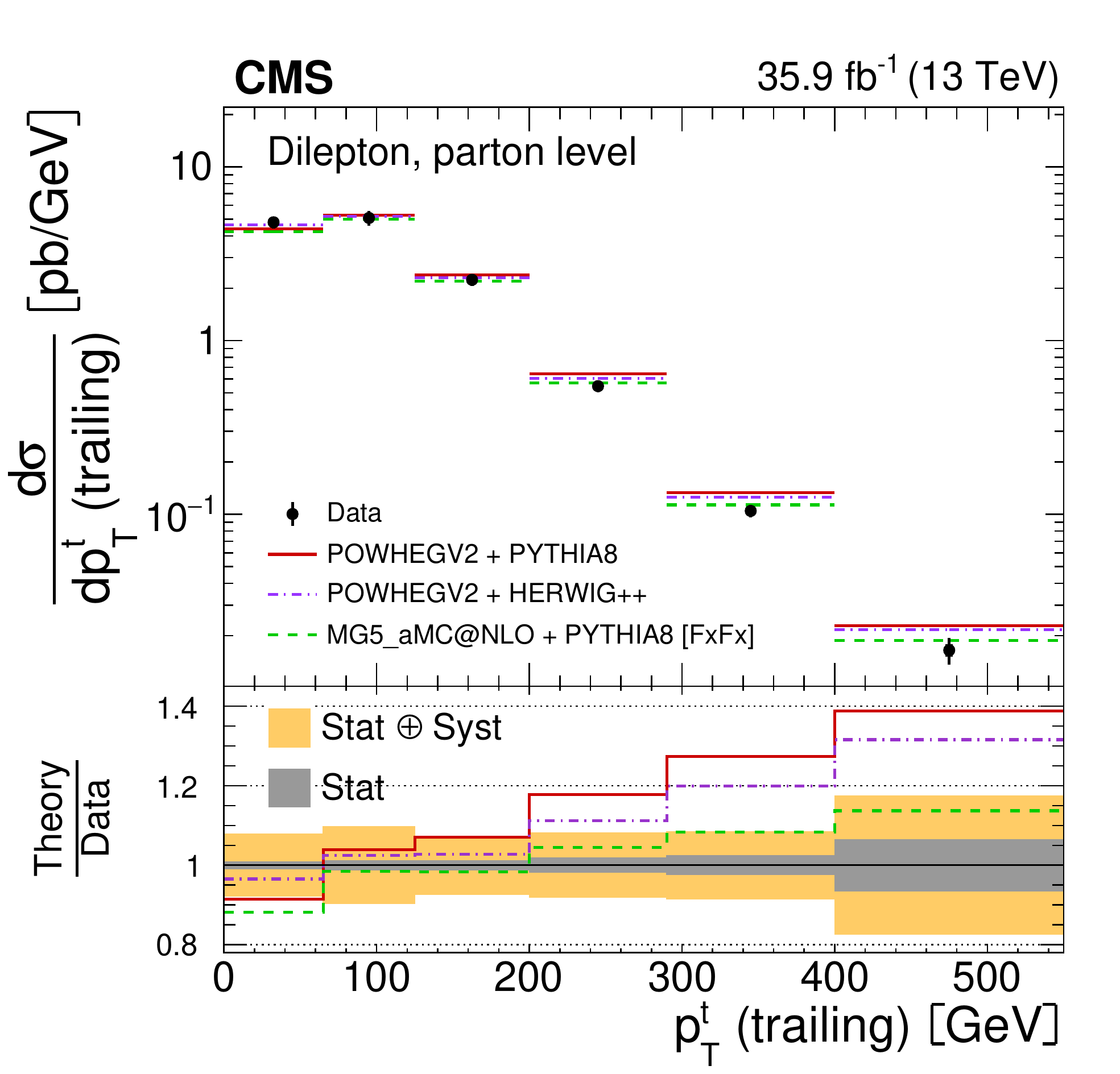}
\includegraphics[width=0.49\textwidth]{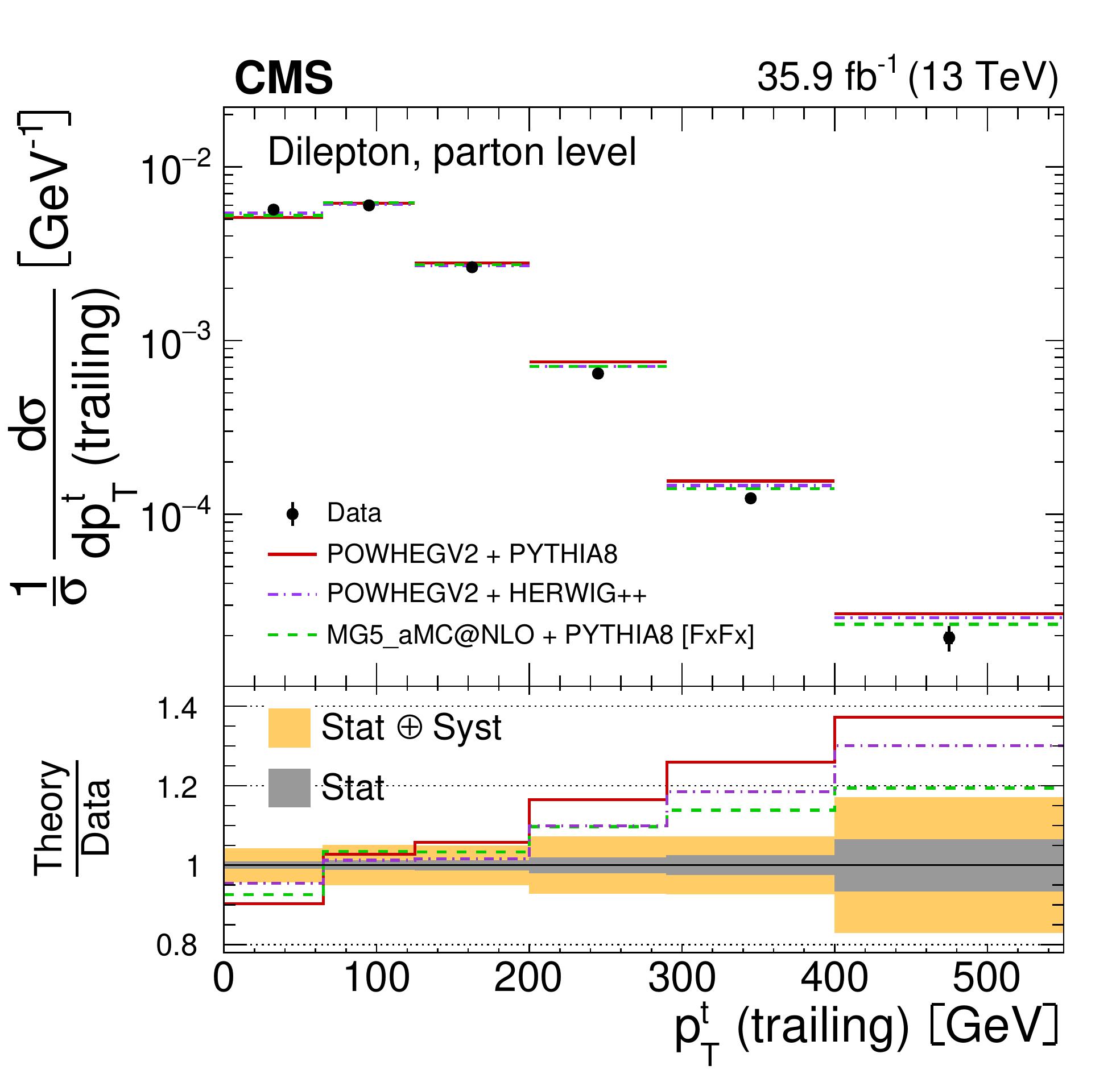} \\
\includegraphics[width=0.49\textwidth]{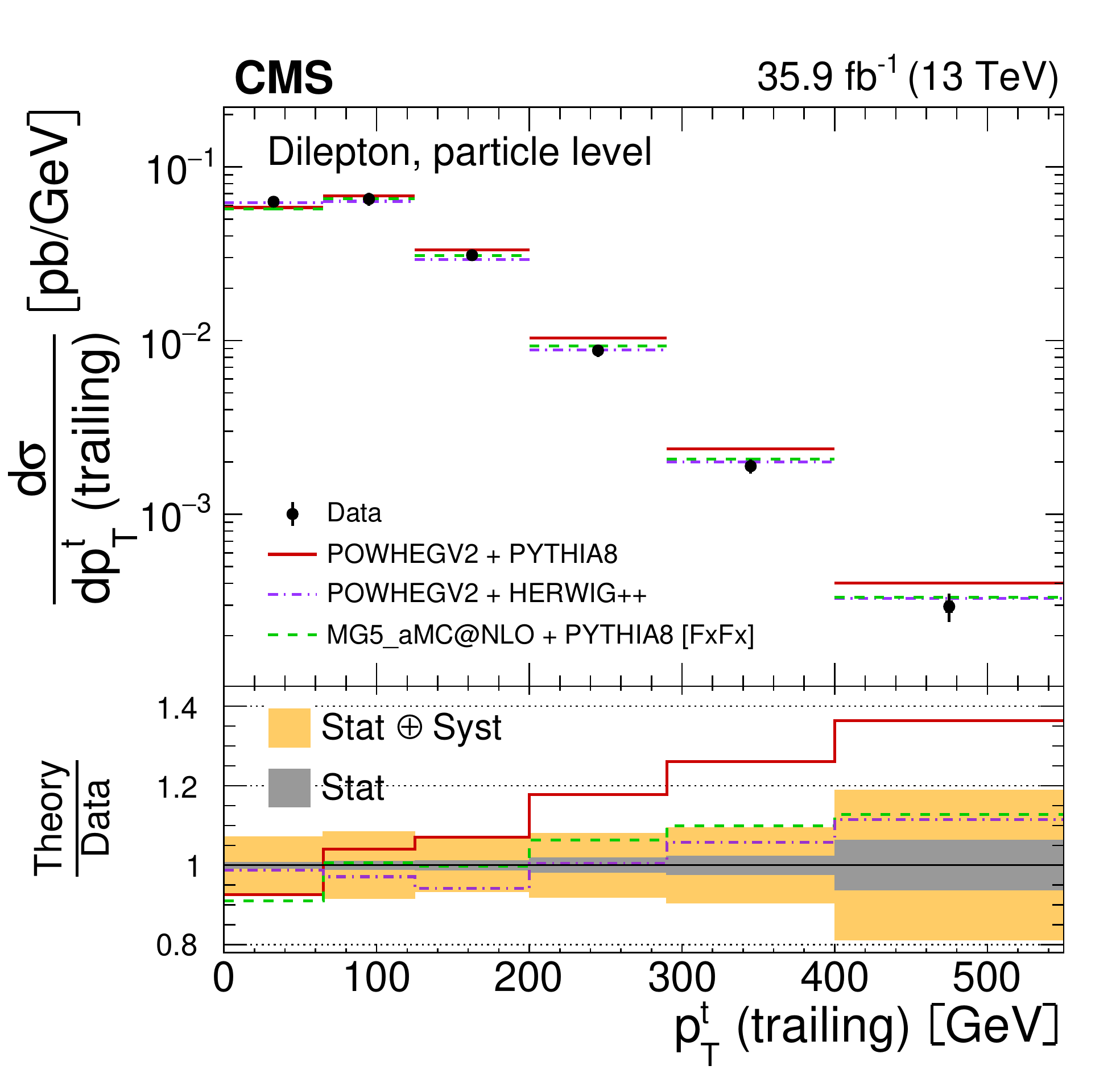}
\includegraphics[width=0.49\textwidth]{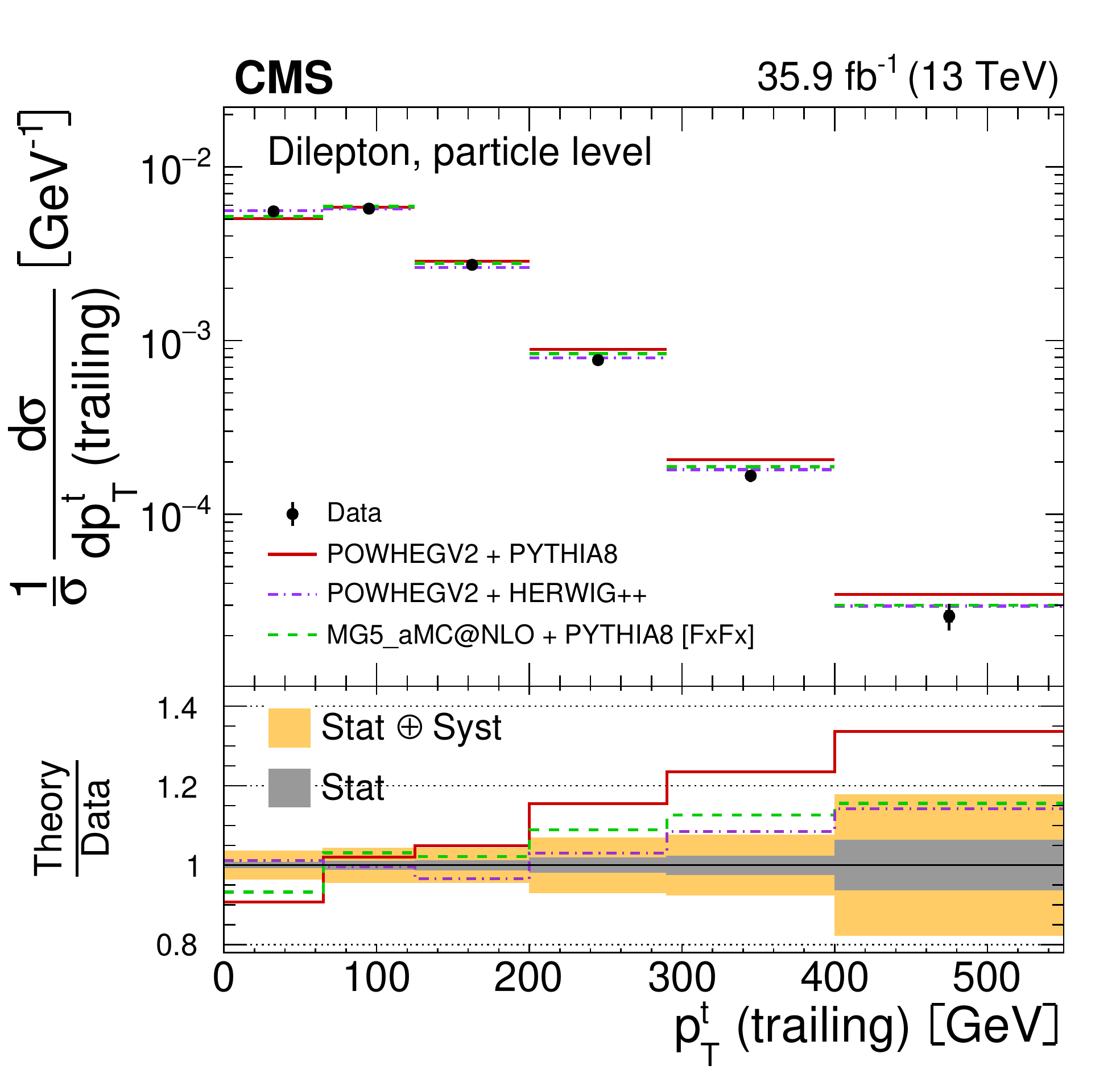}
\caption{The differential \ttbar production cross sections as a function of \pttop (trailing) are shown for the data (points) and the MC predictions (lines). The vertical lines on the points indicate the total uncertainty in the data. The left and right columns correspond to absolute and normalised measurements, respectively. The upper row corresponds to measurements at the parton level in the full phase space and the lower row to the particle level in a fiducial phase space. The lower panel in each plot shows the ratios of the theoretical predictions to the data. The dark and light bands show the relative statistical and total uncertainties in the data, respectively.}
\label{fig:diffxsec:res_subleadingtoppt}
\end{figure*}

\begin{figure*}[!phtb]
\centering
\includegraphics[width=0.49\textwidth]{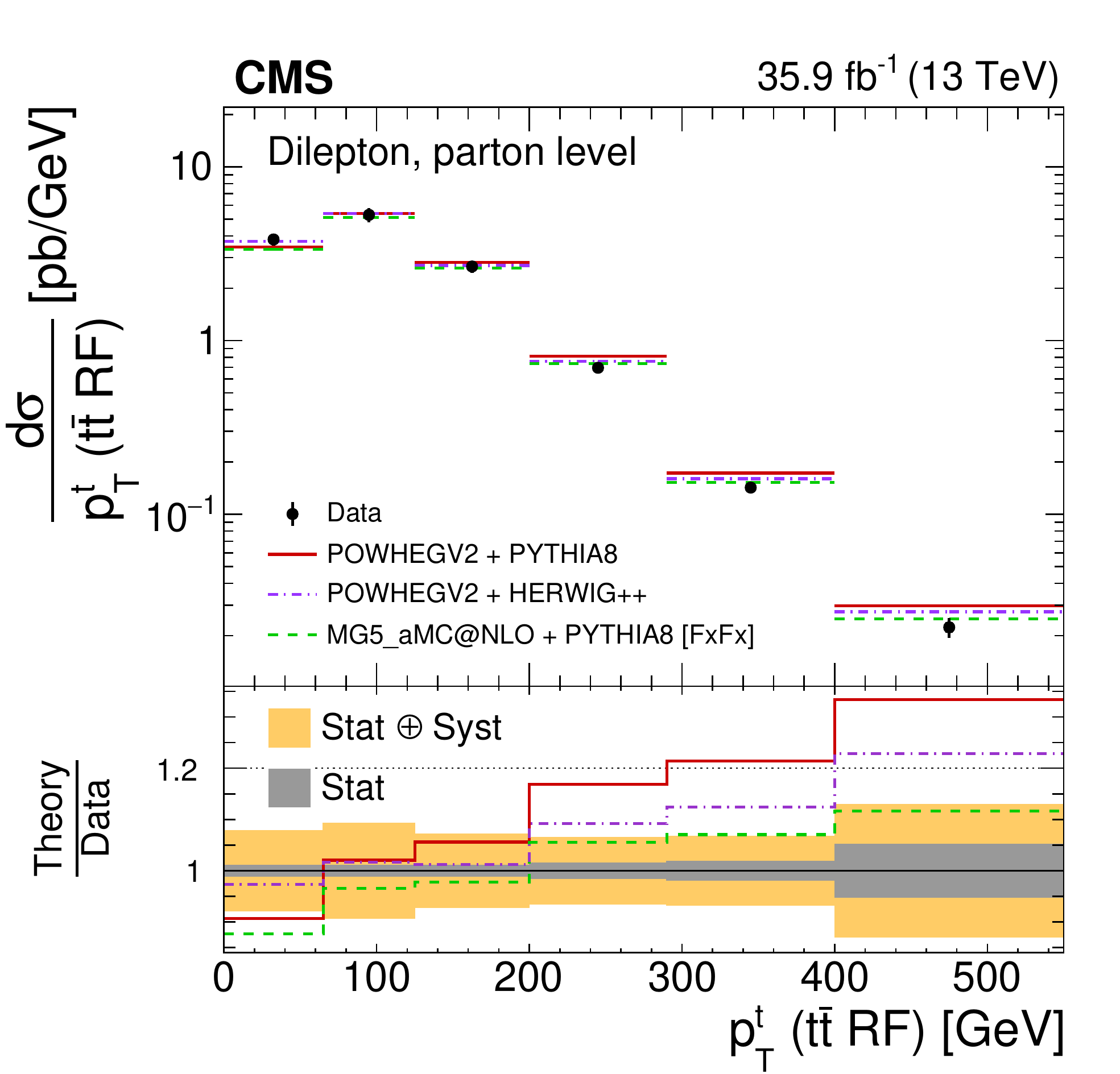}
\includegraphics[width=0.49\textwidth]{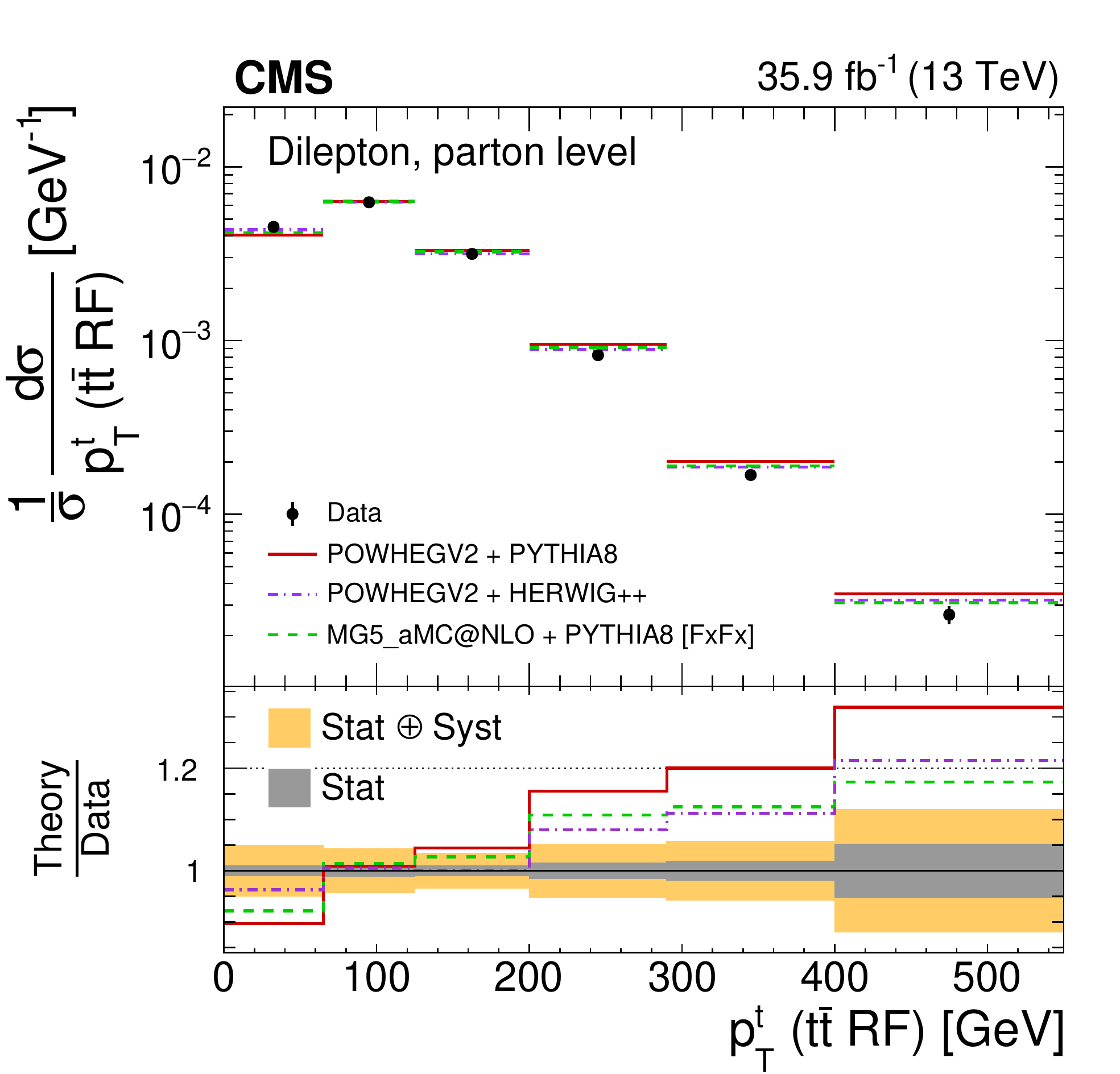} \\
\includegraphics[width=0.49\textwidth]{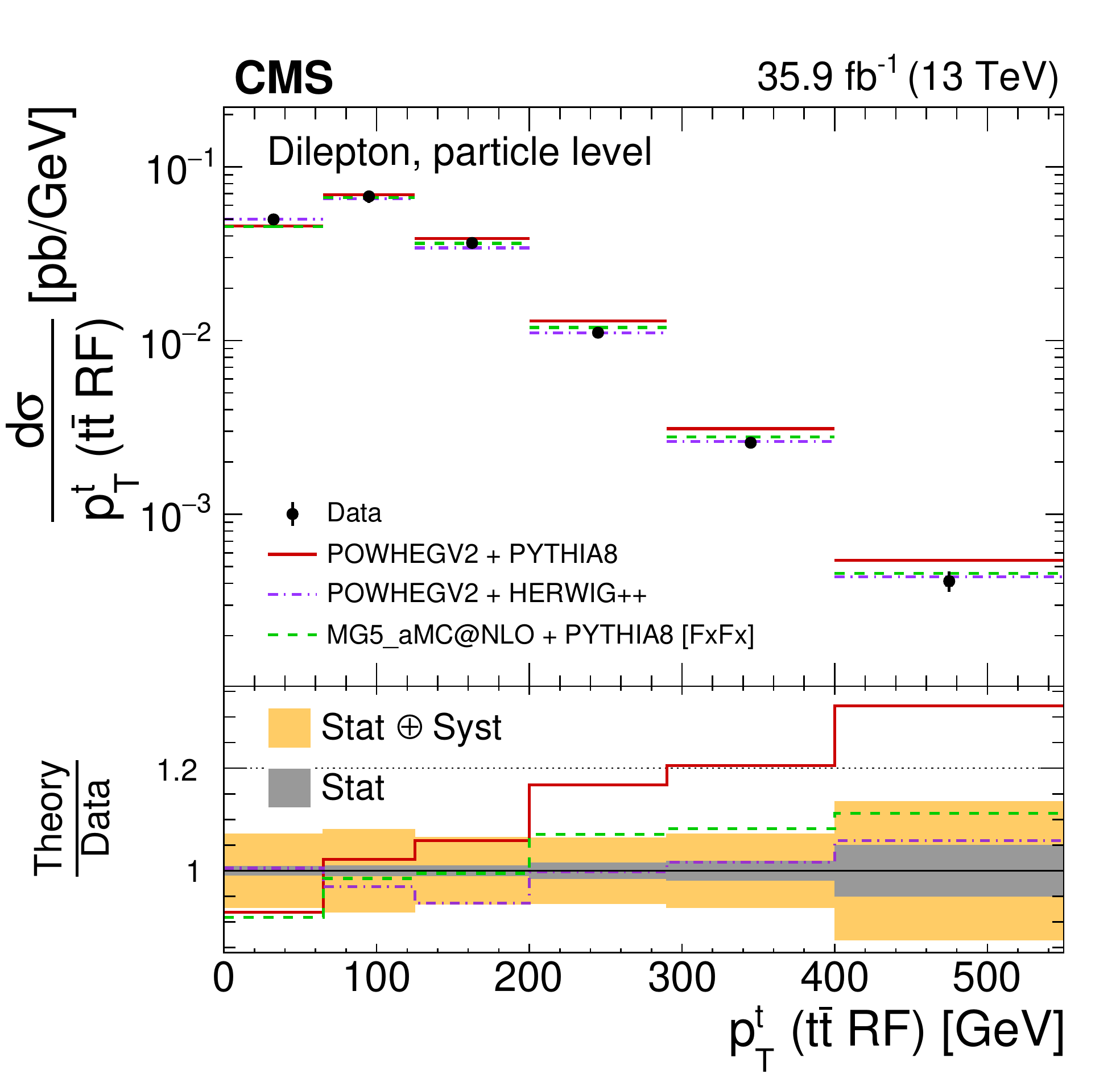}
\includegraphics[width=0.49\textwidth]{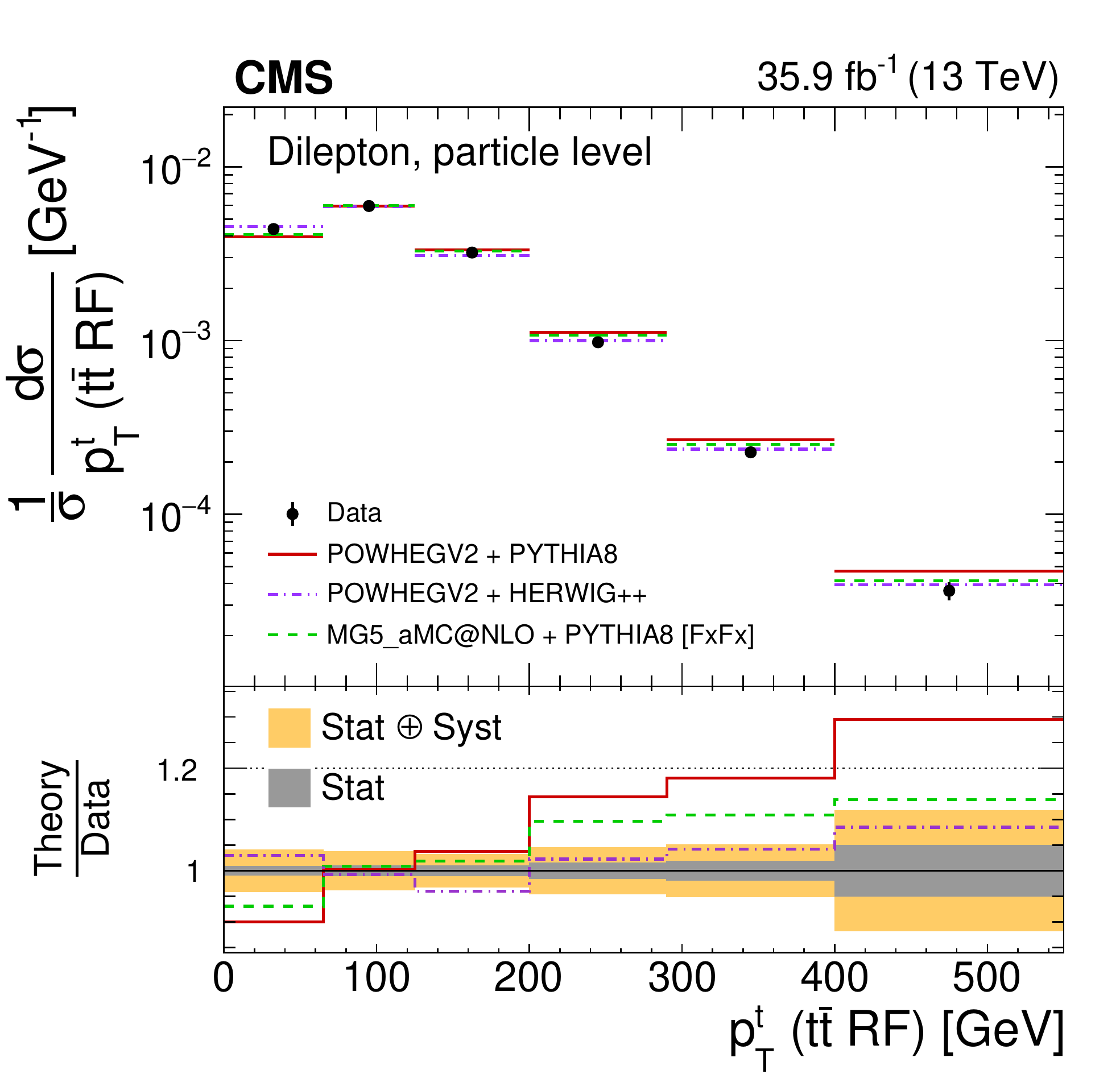}
\caption{The differential \ttbar production cross sections as a function of \pttop (\ttbar RF) are shown for the data (points) and the MC predictions (lines). The vertical lines on the points indicate the total uncertainty in the data. The left and right columns correspond to absolute and normalised measurements, respectively. The upper row corresponds to measurements at the parton level in the full phase space and the lower row to the particle level in a fiducial phase space. The lower panel in each plot shows the ratios of the theoretical predictions to the data. The dark and light bands show the relative statistical and total uncertainties in the data, respectively.}
\label{fig:diffxsec:res_toppt_ttrestframe}
\end{figure*}

\begin{figure*}[!phtb]
\centering
\includegraphics[width=0.49\textwidth]{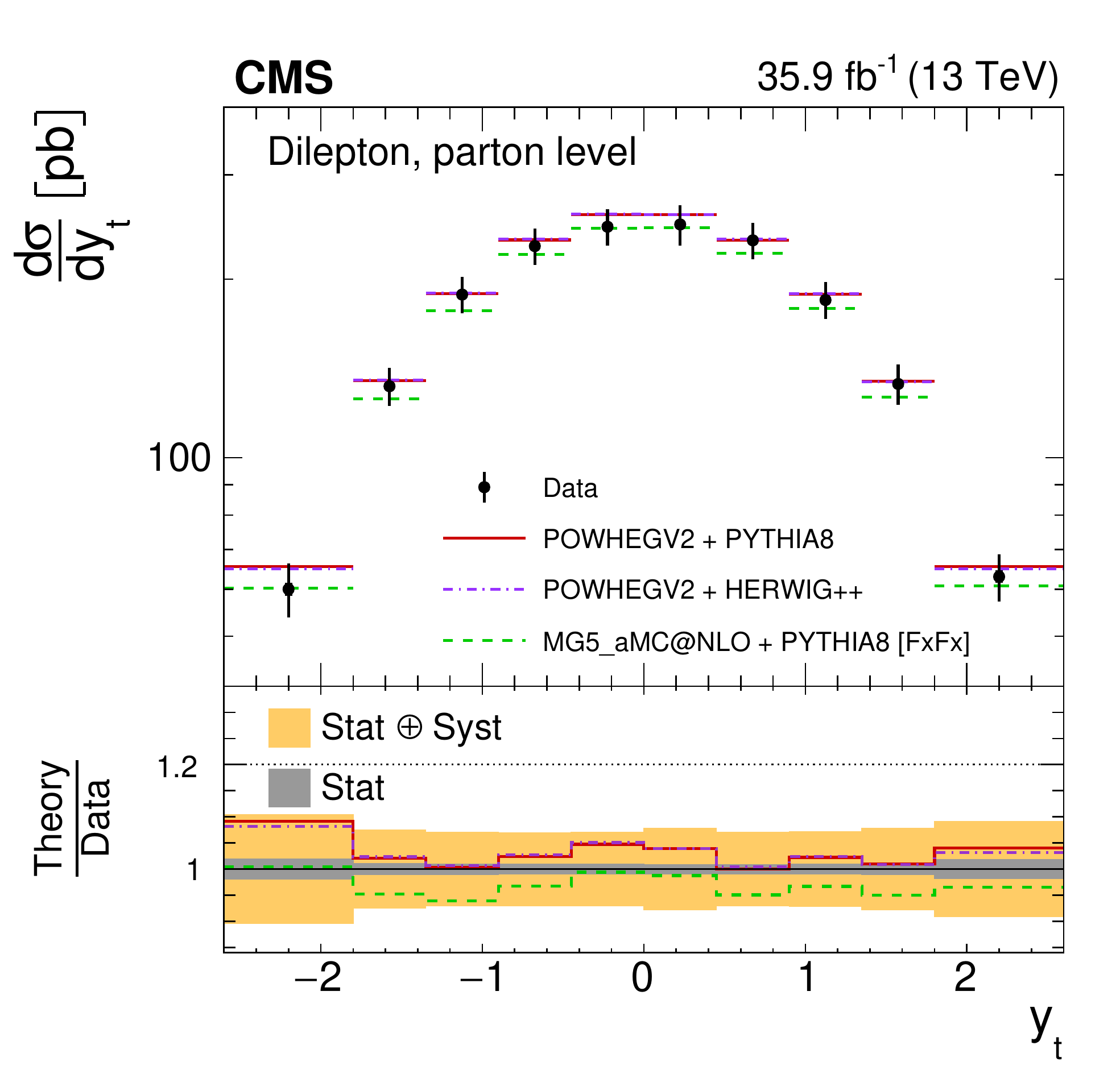}
\includegraphics[width=0.49\textwidth]{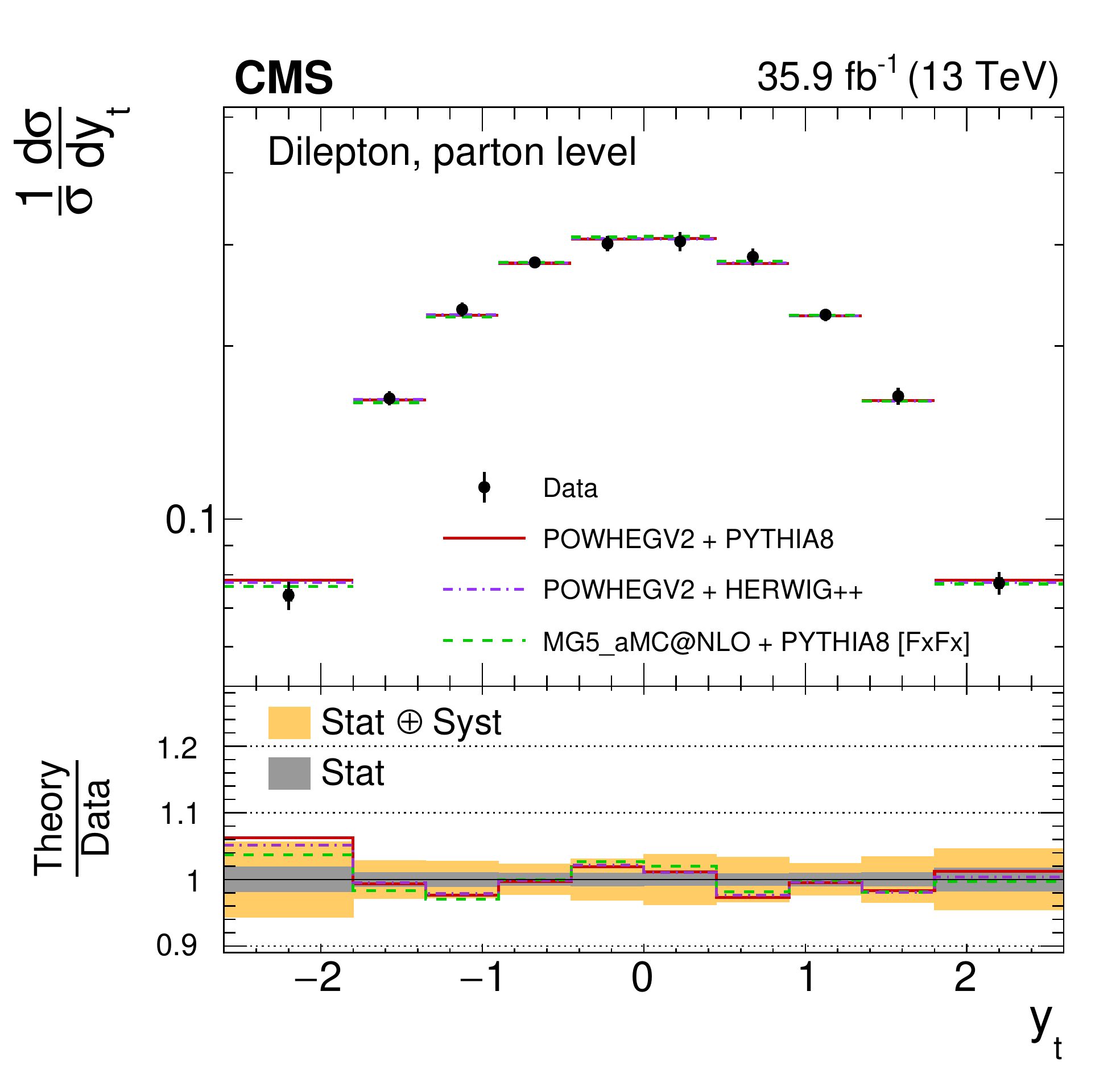} \\
\includegraphics[width=0.49\textwidth]{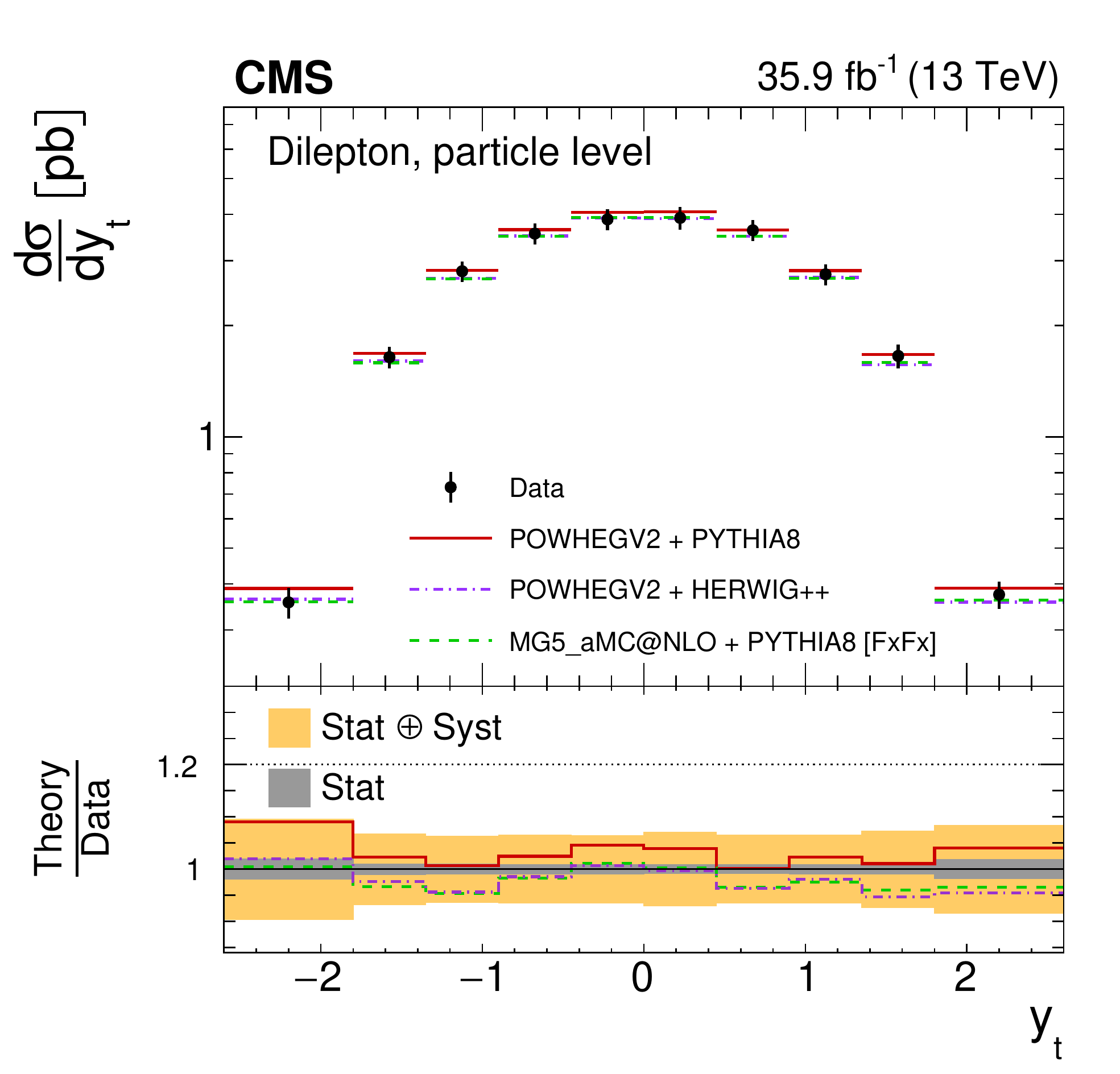}
\includegraphics[width=0.49\textwidth]{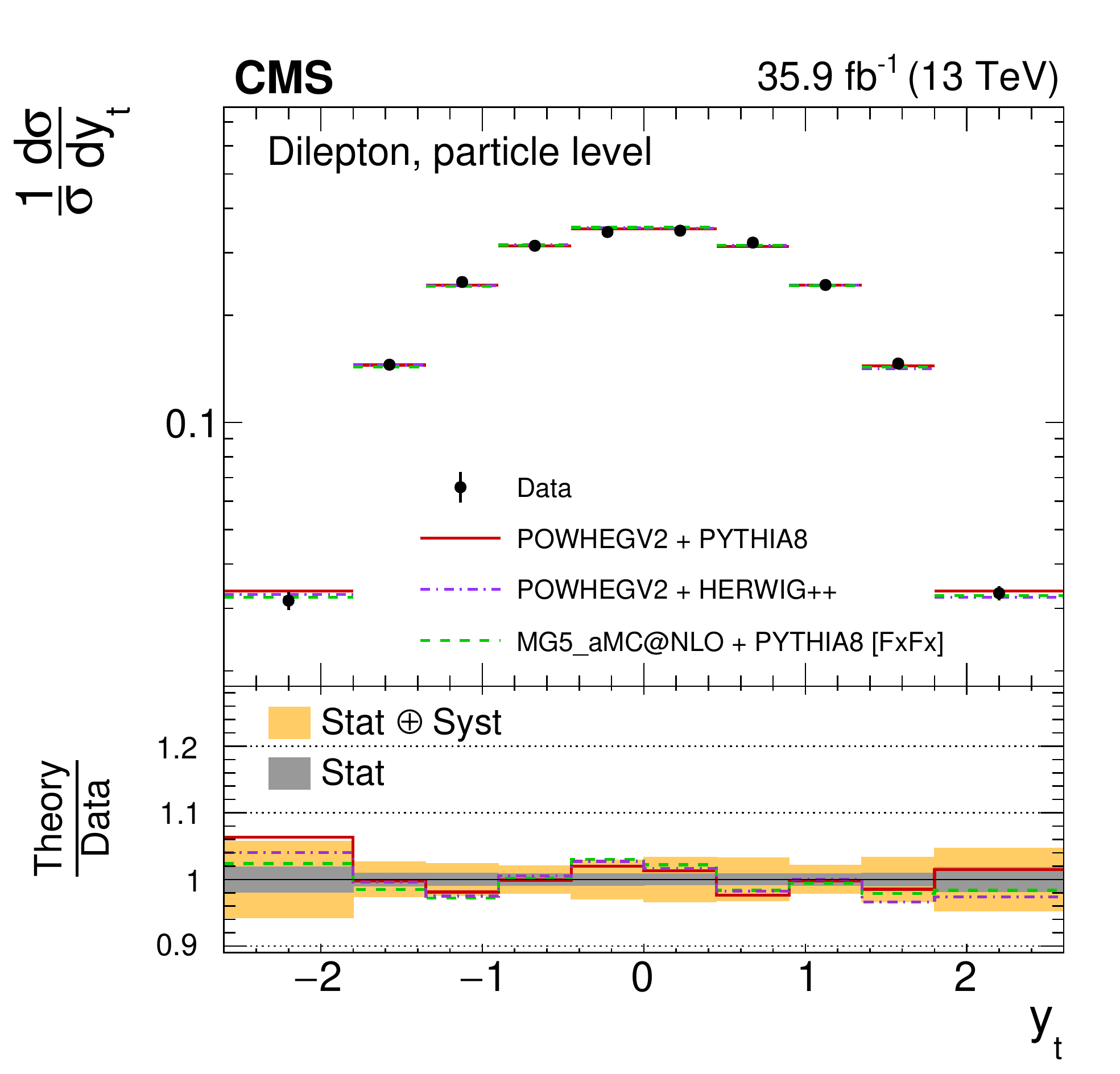}
\caption{The differential \ttbar production cross sections as a function of \ytop are shown for the data (points) and the MC predictions (lines). The vertical lines on the points indicate the total uncertainty in the data. The left and right columns correspond to absolute and normalised measurements, respectively. The upper row corresponds to measurements at the parton level in the full phase space and the lower row to the particle level in a fiducial phase space. The lower panel in each plot shows the ratios of the theoretical predictions to the data. The dark and light bands show the relative statistical and total uncertainties in the data, respectively.}
\label{fig:diffxsec:res_topy}
\end{figure*}

\begin{figure*}[!phtb]
\centering
\includegraphics[width=0.49\textwidth]{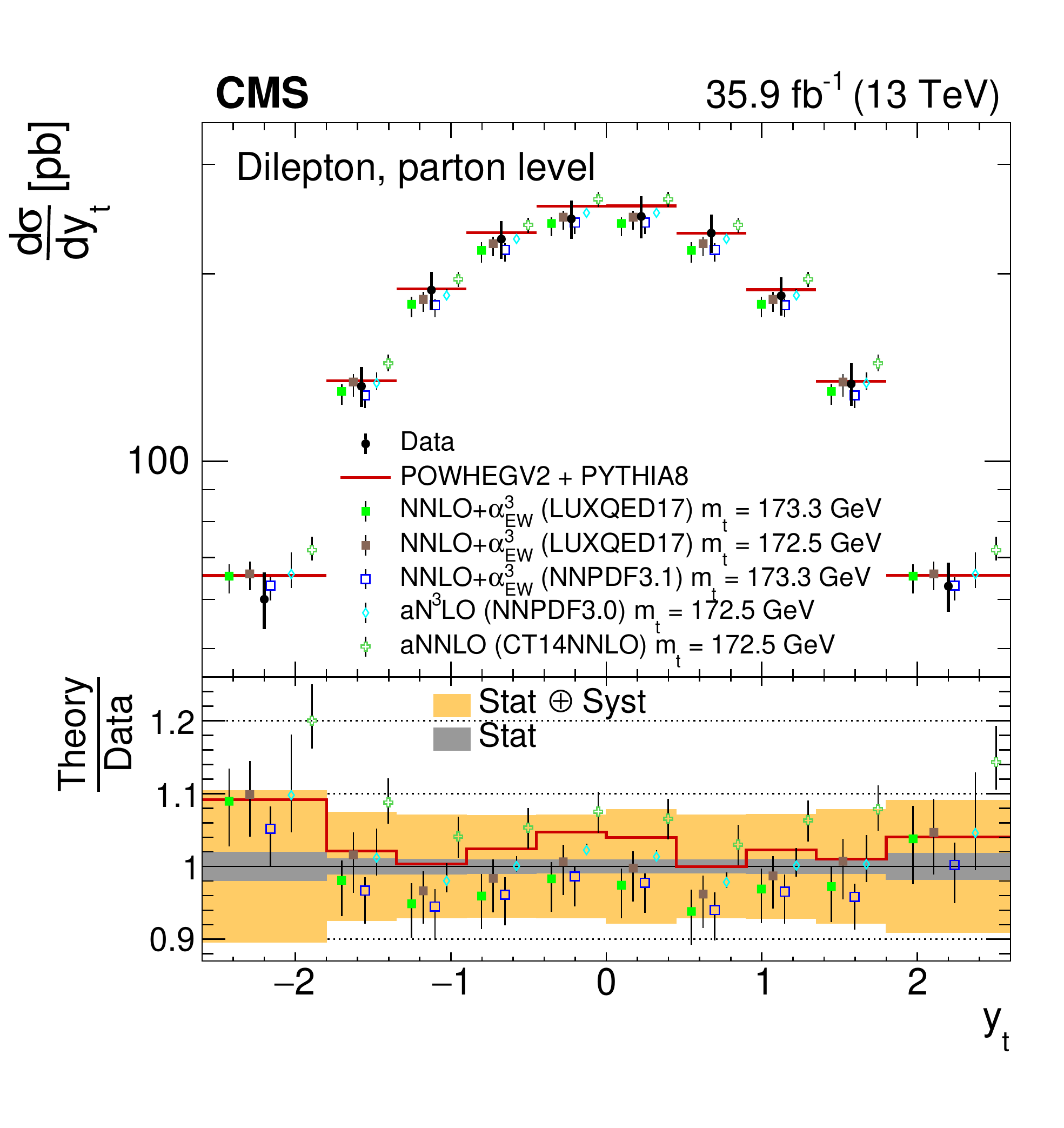}
\includegraphics[width=0.49\textwidth]{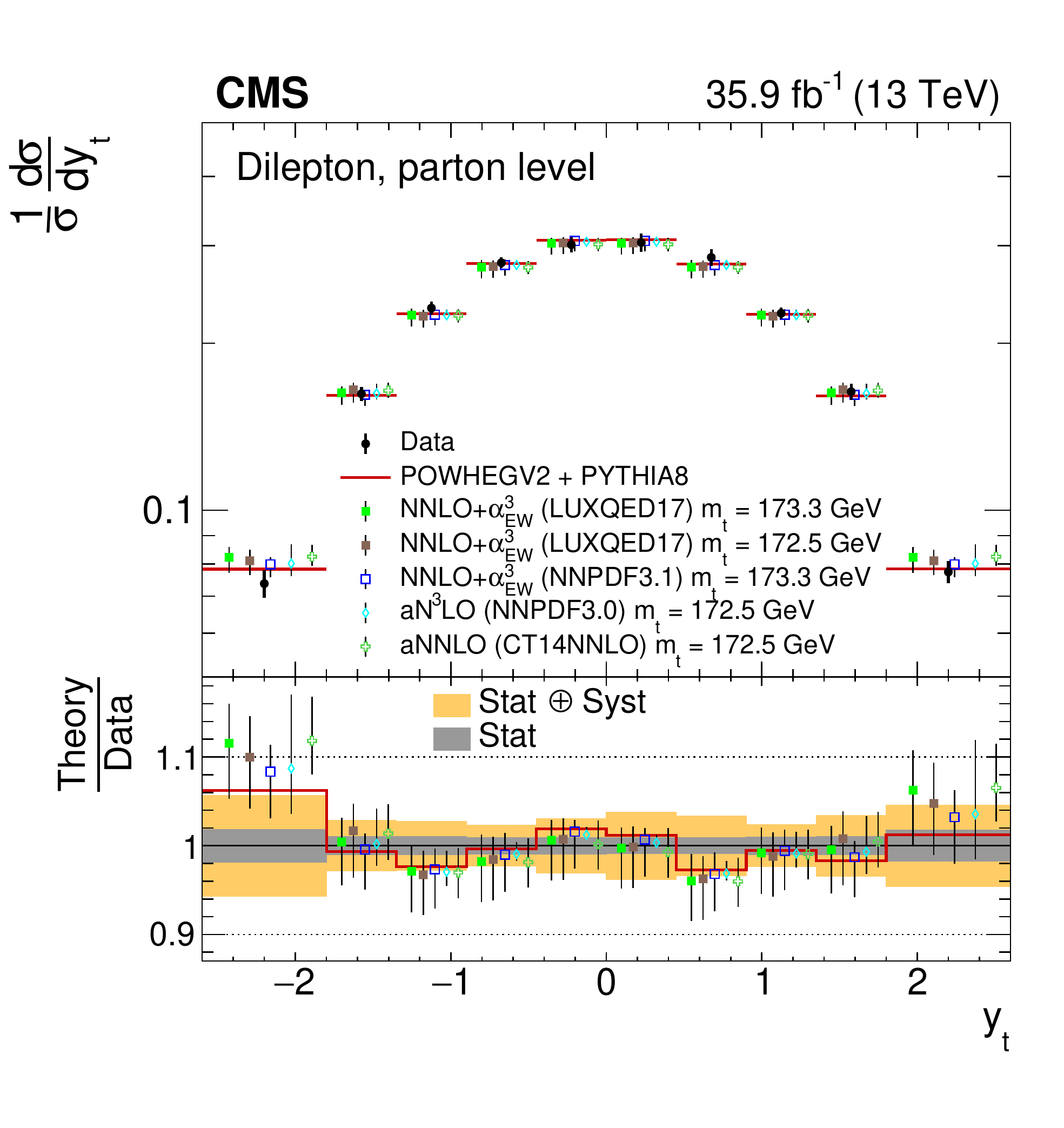}
\caption{The differential \ttbar production cross sections at the parton level in the full phase space as a function of \ytop  are shown for the data (filled circles), the theoretical predictions with beyond-NLO precision (other points) and the prediction from \pwhgpy (solid line). The vertical lines on the filled circles and other points indicate the total uncertainty in the data and theoretical predictions, respectively. The left and right plots correspond to absolute and normalised measurements, respectively. The lower panel in each plot shows the ratios of the theoretical predictions to the data. The dark and light bands show the relative statistical and total uncertainties in the data, respectively.}
\label{fig:diffxsec:res_topy_bnlo}
\includegraphics[width=0.75\textwidth]{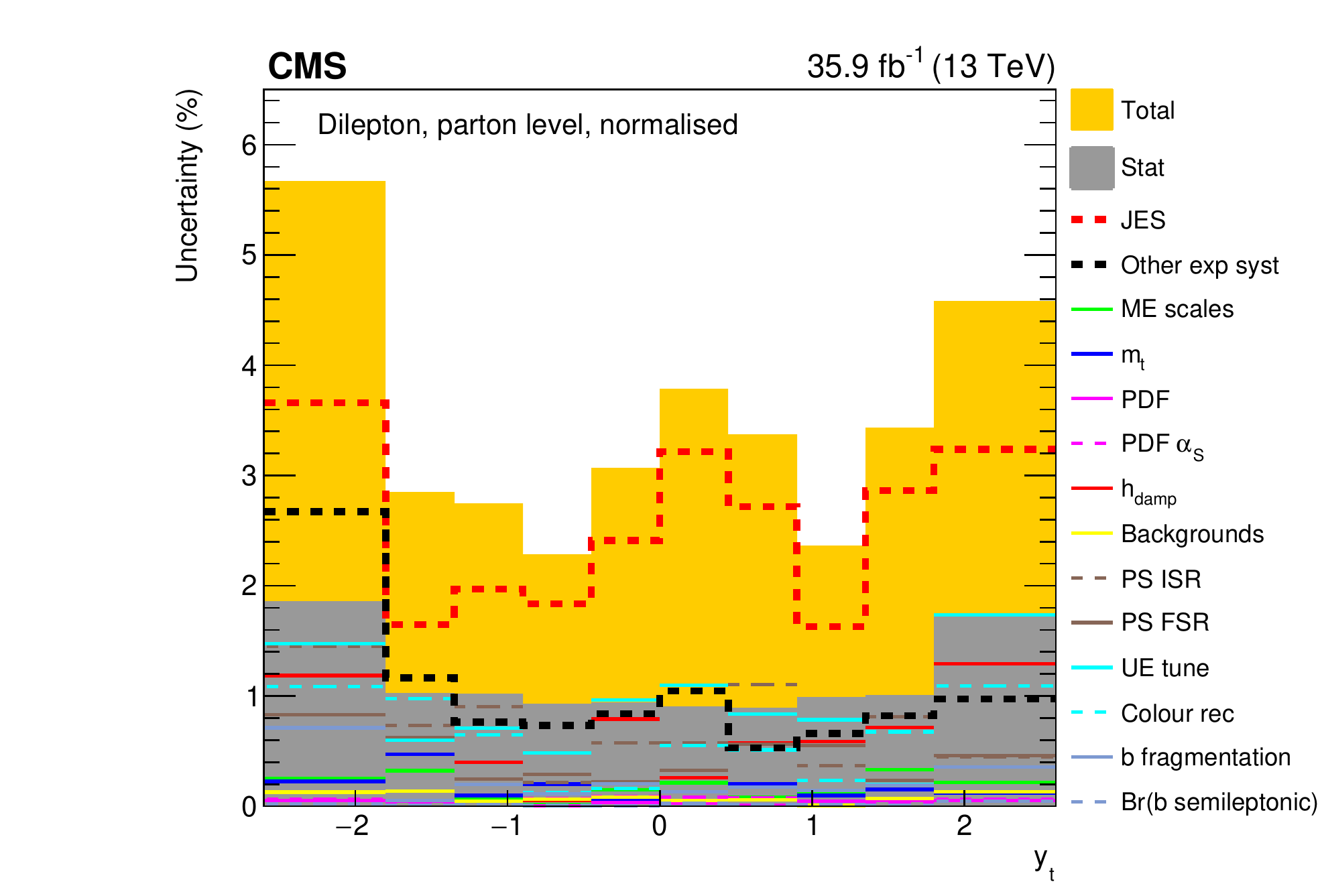}
\caption{The contributions of each source of systematic uncertainty to the total systematic uncertainty in each bin is shown for the measurement of the normalised \ttbar production cross sections  as a function of \ytop. The sources affecting the JES are added in quadrature and shown as a single component. Additional experimental systematic uncertainties are also added in quadrature and shown as a single component. Contributions from theoretical uncertainties are shown separately. The statistical and total uncertainties, corresponding to the quadrature addition of statistical and systematic uncertainties, are shown by the dark and light filled histograms, respectively.}
\label{fig:diffxsec:unc_breakdown_topy}
\end{figure*}

\begin{figure*}[!phtb]
\centering
\includegraphics[width=0.49\textwidth]{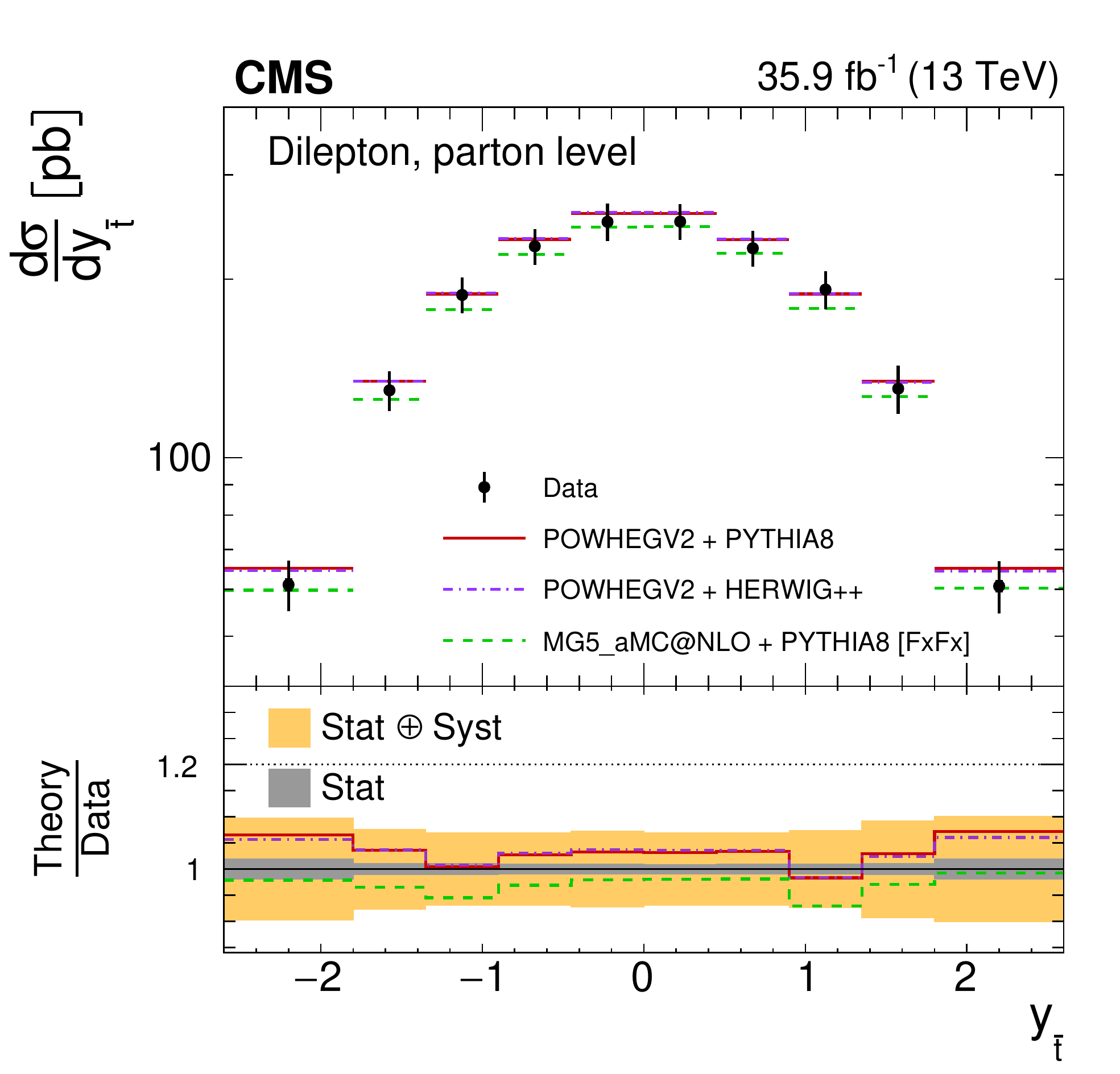}
\includegraphics[width=0.49\textwidth]{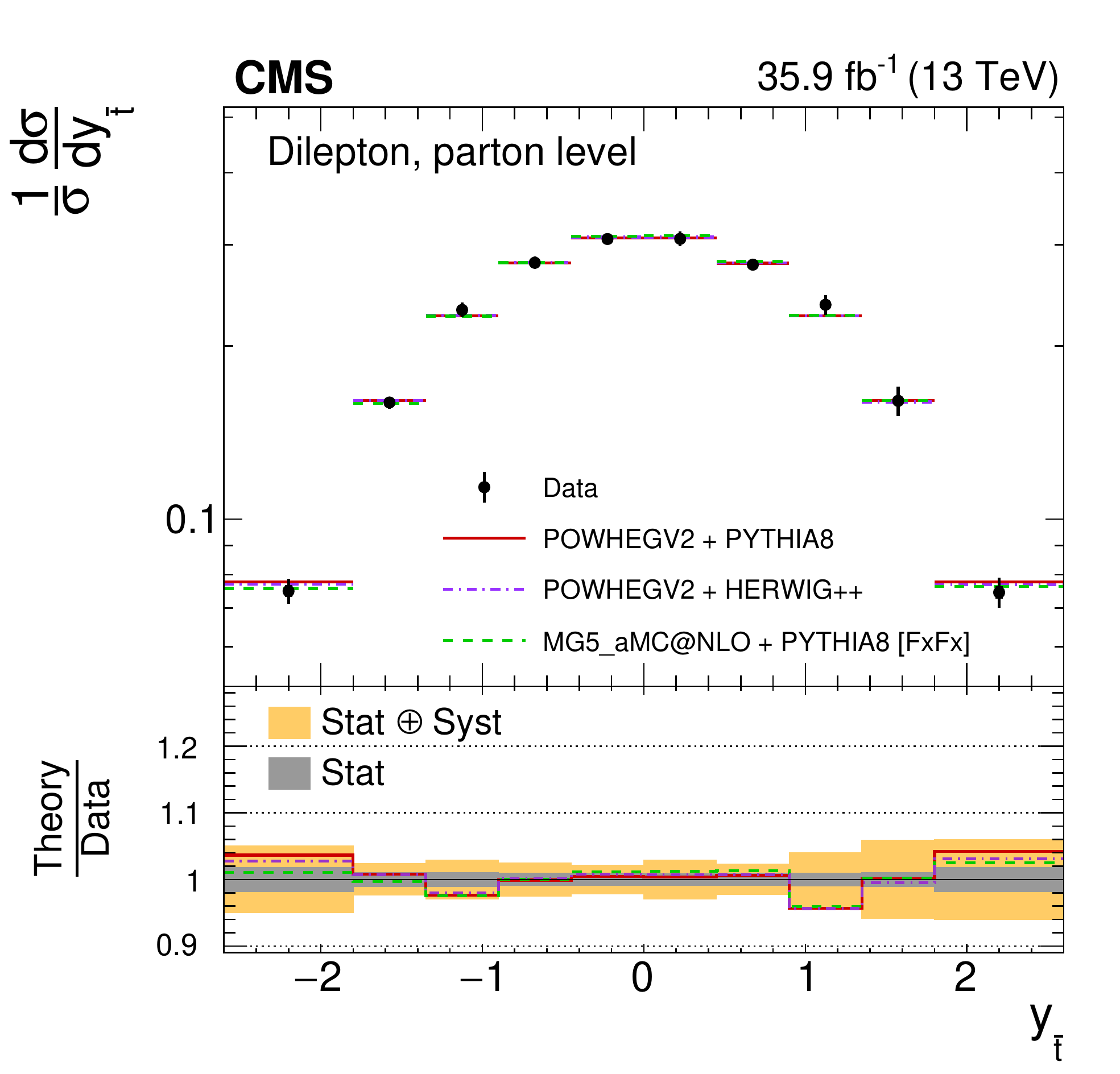} \\
\includegraphics[width=0.49\textwidth]{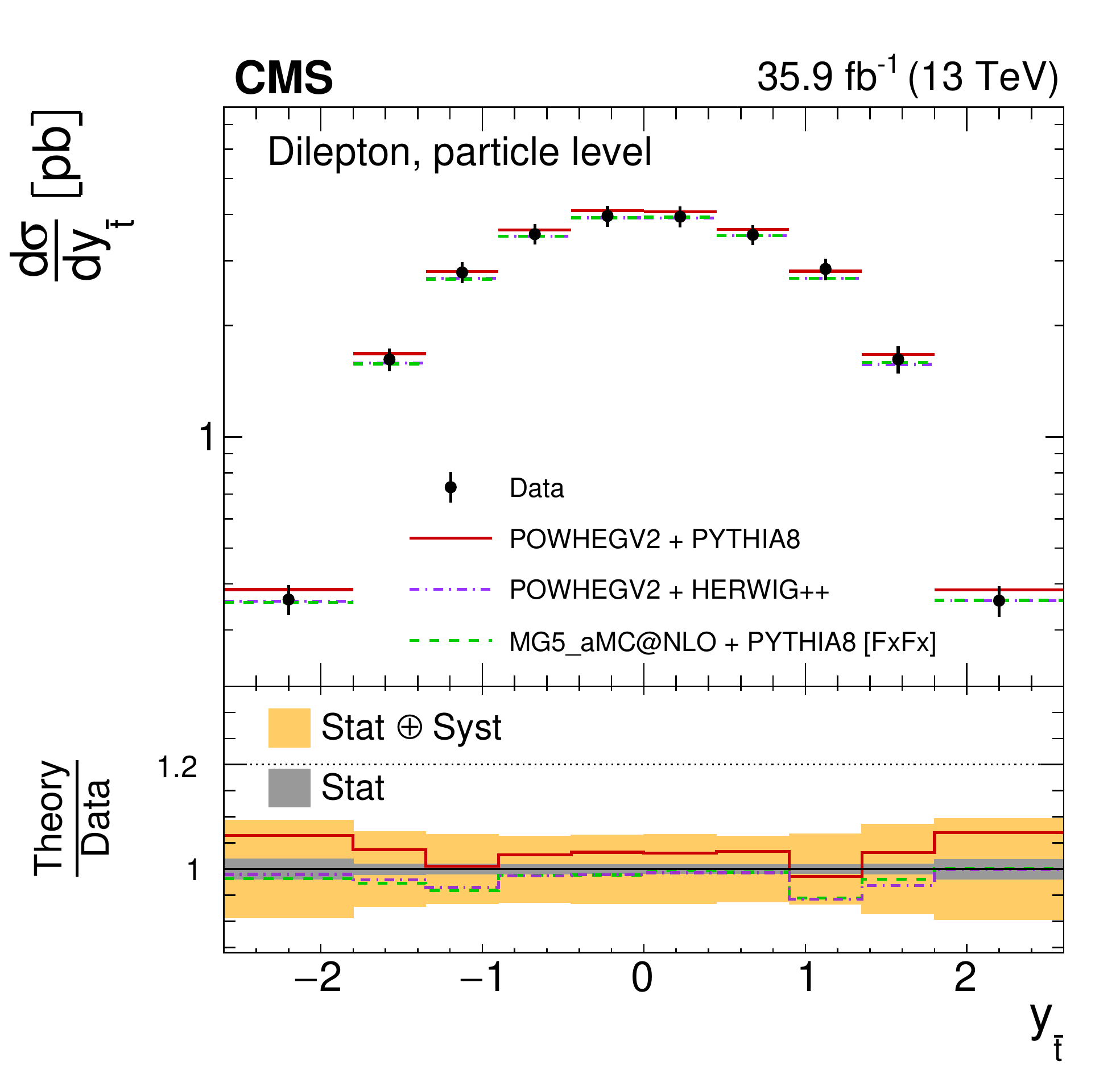}
\includegraphics[width=0.49\textwidth]{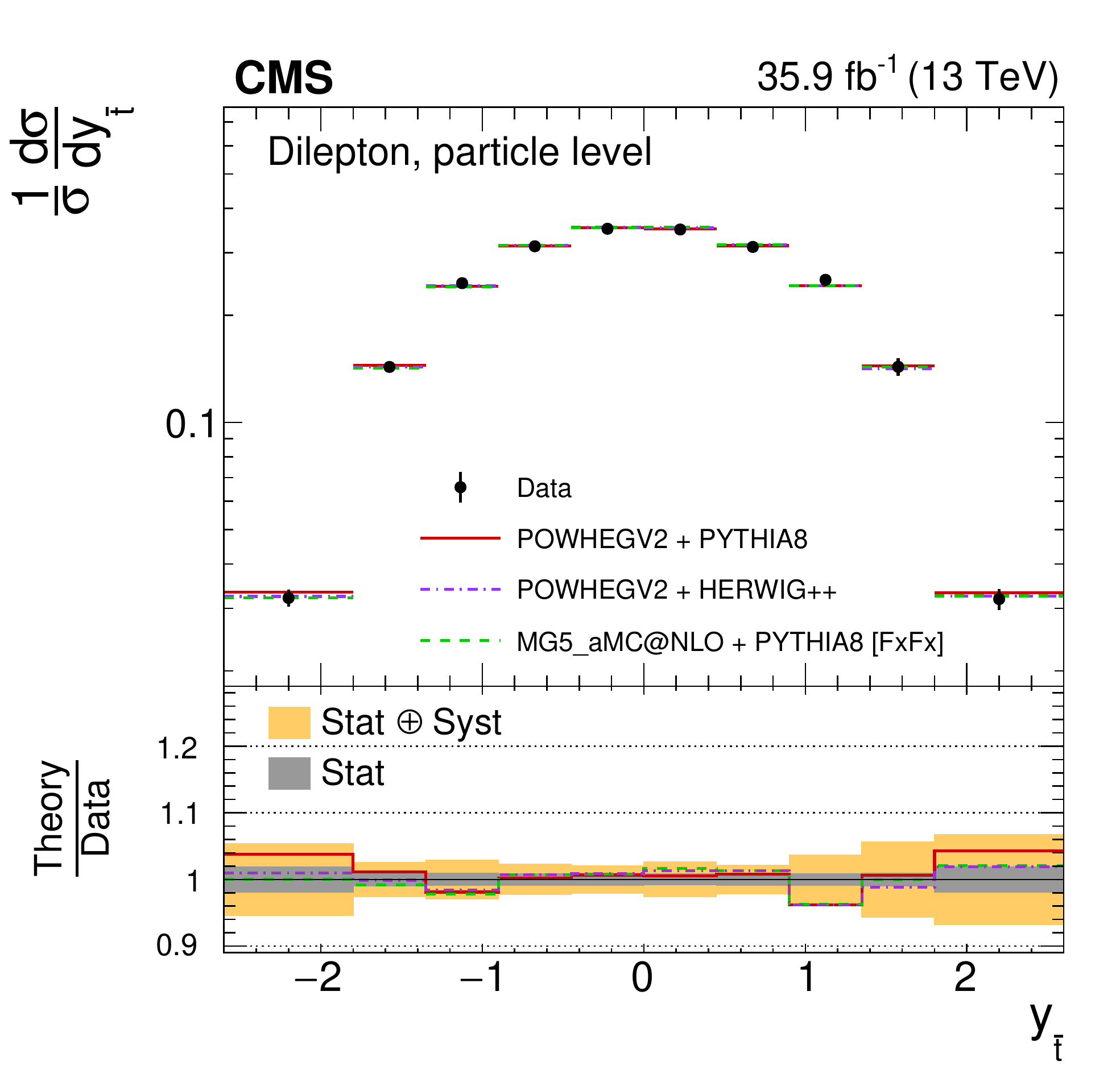}
\caption{The differential \ttbar production cross sections as a function of \yantitop are shown for the data (points) and the MC predictions (lines). The vertical lines on the points indicate the total uncertainty in the data. The left and right columns correspond to absolute and normalised measurements, respectively. The upper row corresponds to measurements at the parton level in the full phase space and the lower row to the particle level in a fiducial phase space. The lower panel in each plot shows the ratios of the theoretical predictions to the data. The dark and light bands show the relative statistical and total uncertainties in the data, respectively.}
\label{fig:diffxsec:res_antitopy}
\end{figure*}

\begin{figure*}[!phtb]
\centering
\includegraphics[width=0.49\textwidth]{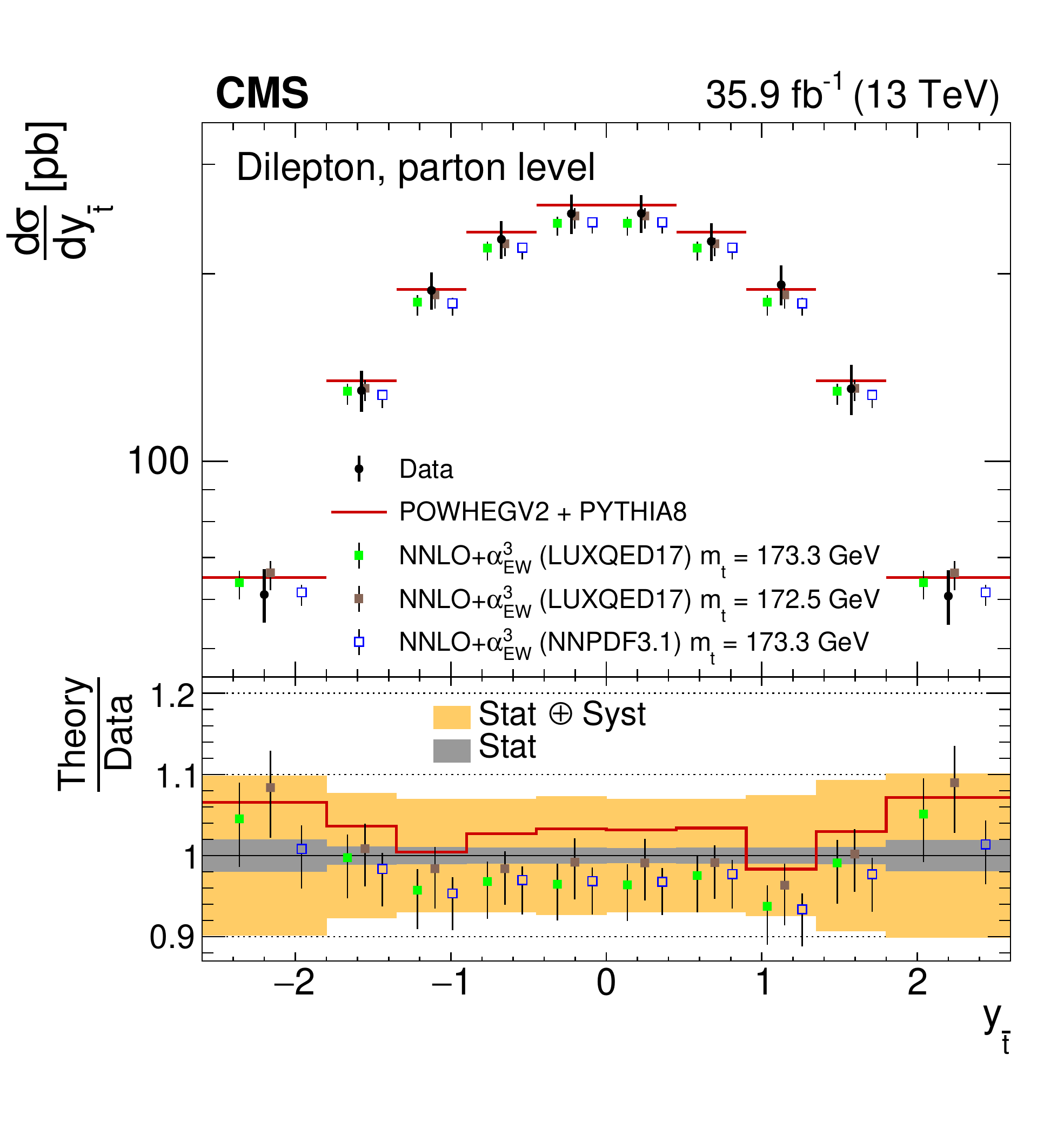}
\includegraphics[width=0.49\textwidth]{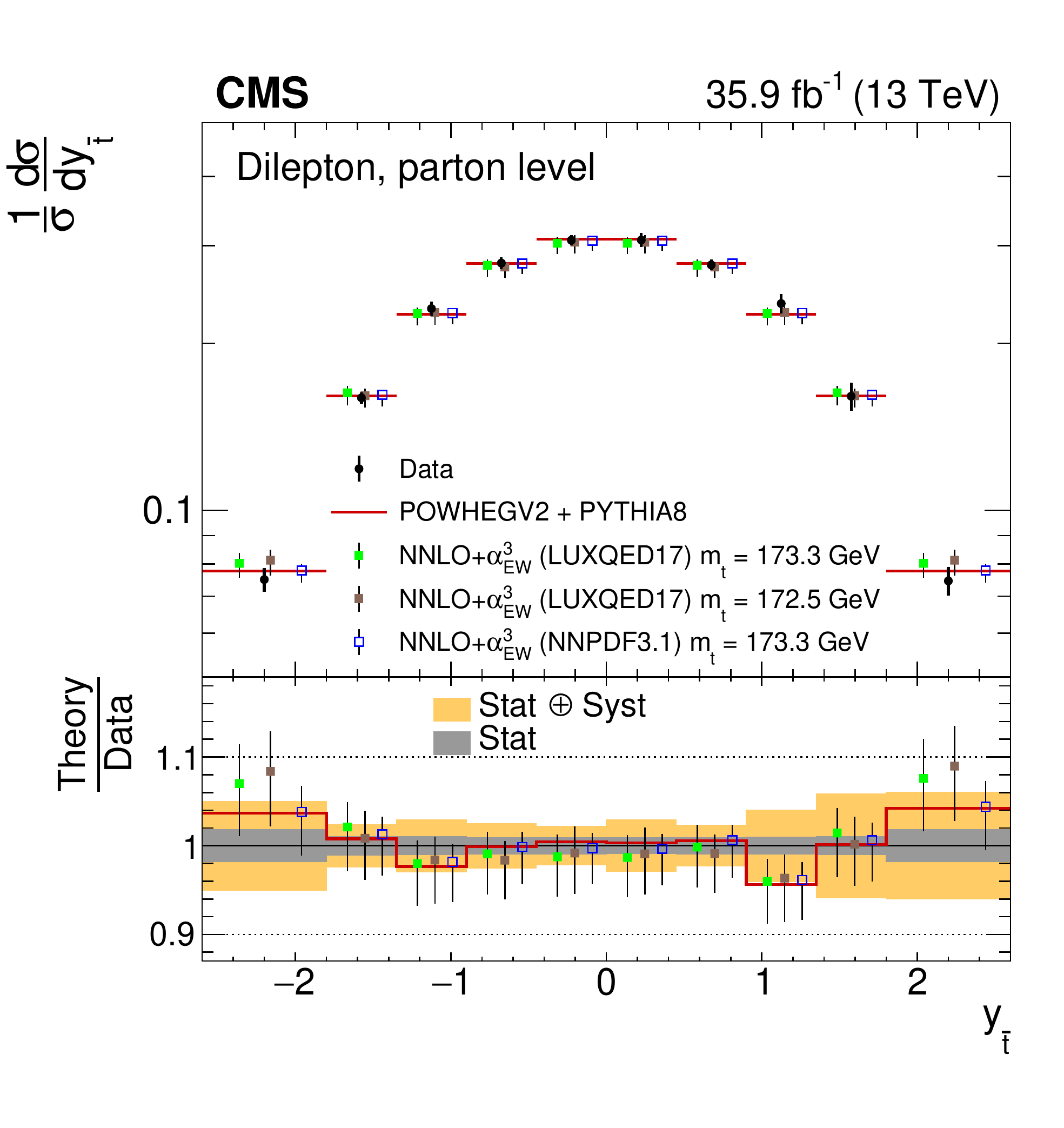}
\caption{The differential \ttbar production cross sections at the parton level in the full phase space as a function of \yantitop are shown for the data (filled circles), the theoretical predictions with beyond-NLO precision (other points) and the prediction from \pwhgpy (solid line). The vertical lines on the filled circles and other points indicate the total uncertainty in the data and theoretical predictions, respectively.  The left and right plots correspond to absolute and normalised measurements, respectively. The lower panel in each plot shows the ratios of the theoretical predictions to the data. The dark and light bands show the relative statistical and total uncertainties in the data, respectively.}
\label{fig:diffxsec:res_antitopy_bnlo}
\includegraphics[width=0.75\textwidth]{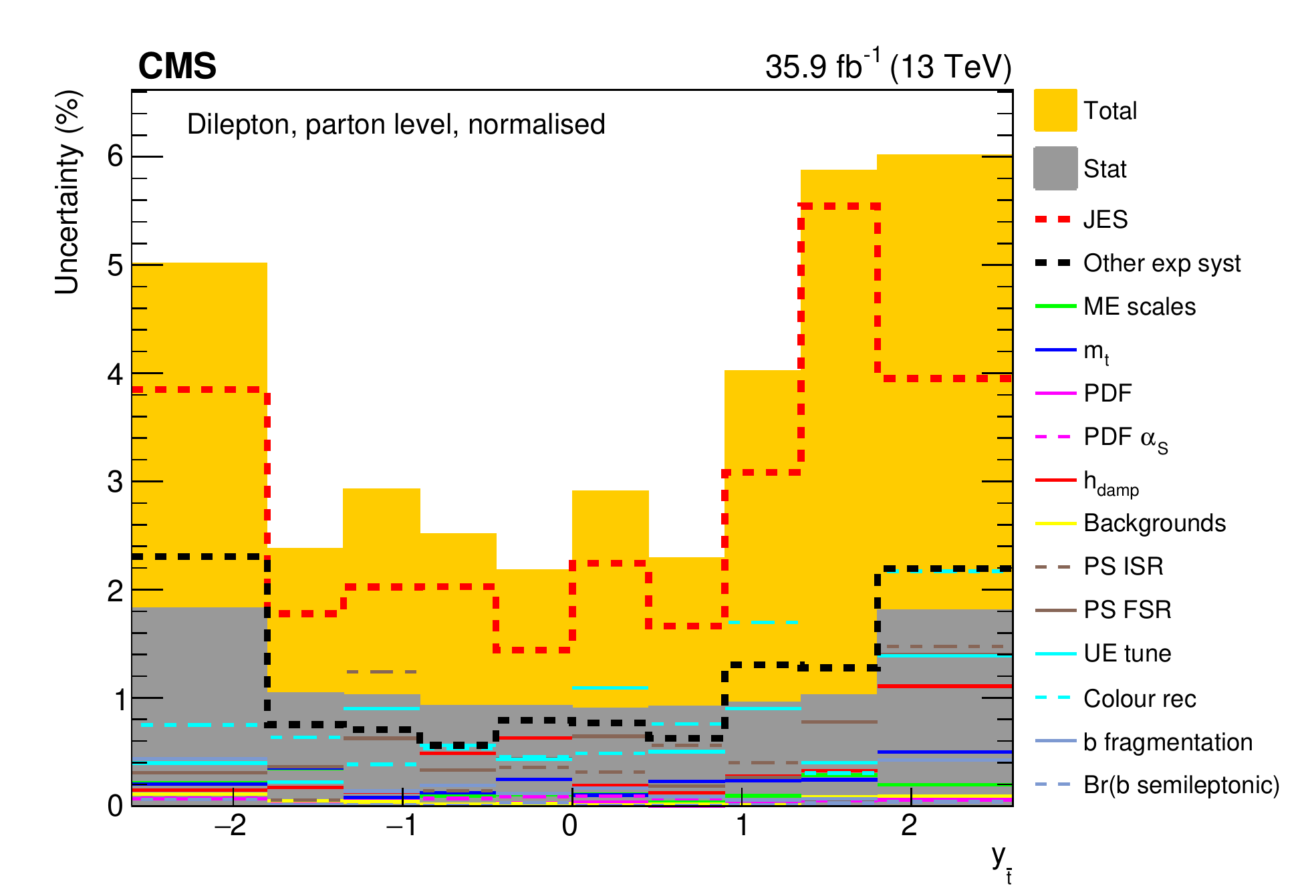}
\caption{The contributions of each source of systematic uncertainty to the total systematic uncertainty in each bin is shown for the measurement of the normalised \ttbar production cross sections  as a function of \yantitop. The sources affecting the JES are added in quadrature and shown as a single component. Additional experimental systematic uncertainties are also added in quadrature and shown as a single component. Contributions from theoretical uncertainties are shown separately. The statistical and total uncertainties, corresponding to the quadrature addition of statistical and systematic uncertainties, are shown by the dark and light filled histograms, respectively.}
\label{fig:diffxsec:unc_breakdown_antitopy}
\end{figure*}

\begin{figure*}[!phtb]
\centering
\includegraphics[width=0.49\textwidth]{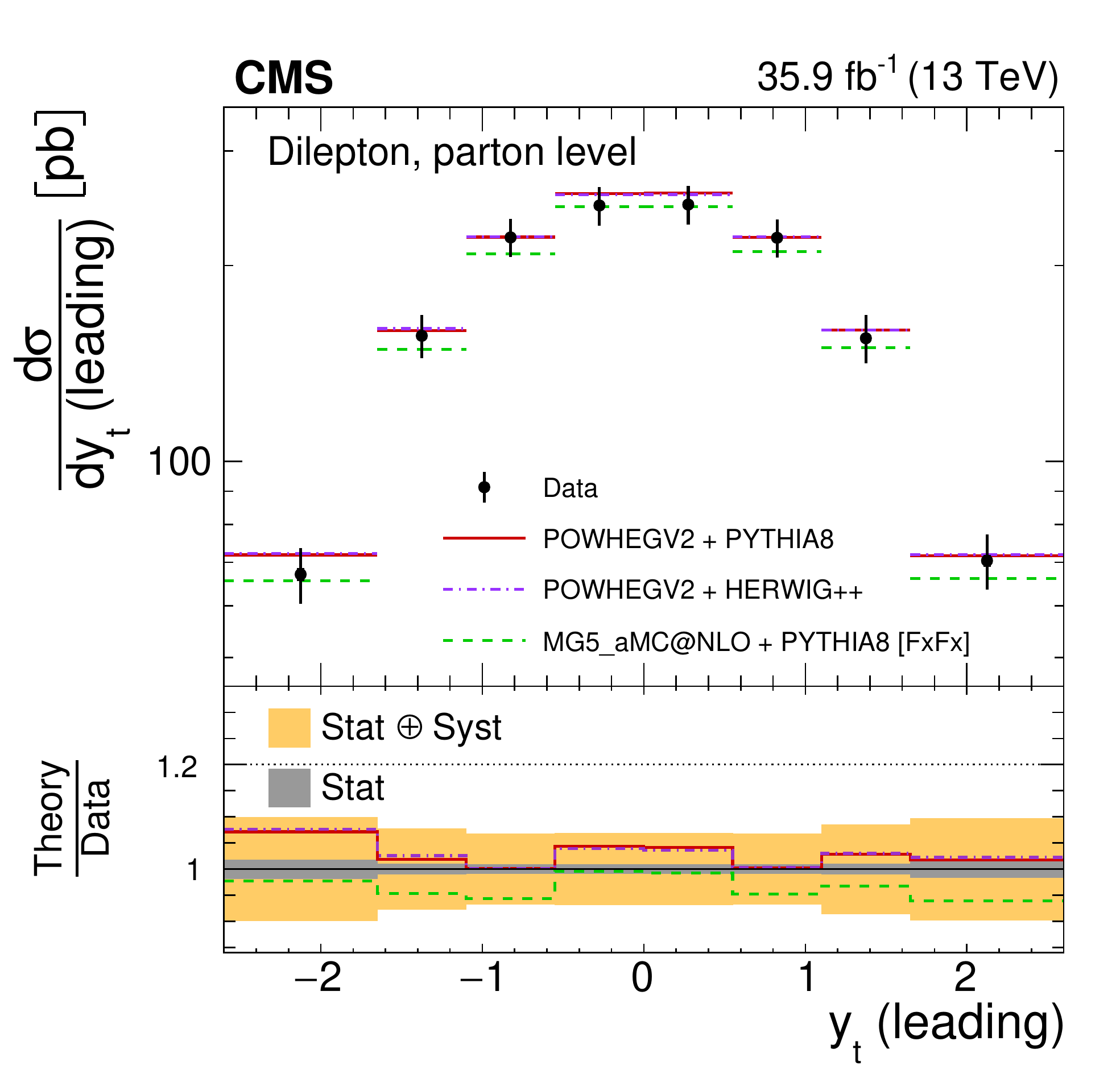}
\includegraphics[width=0.49\textwidth]{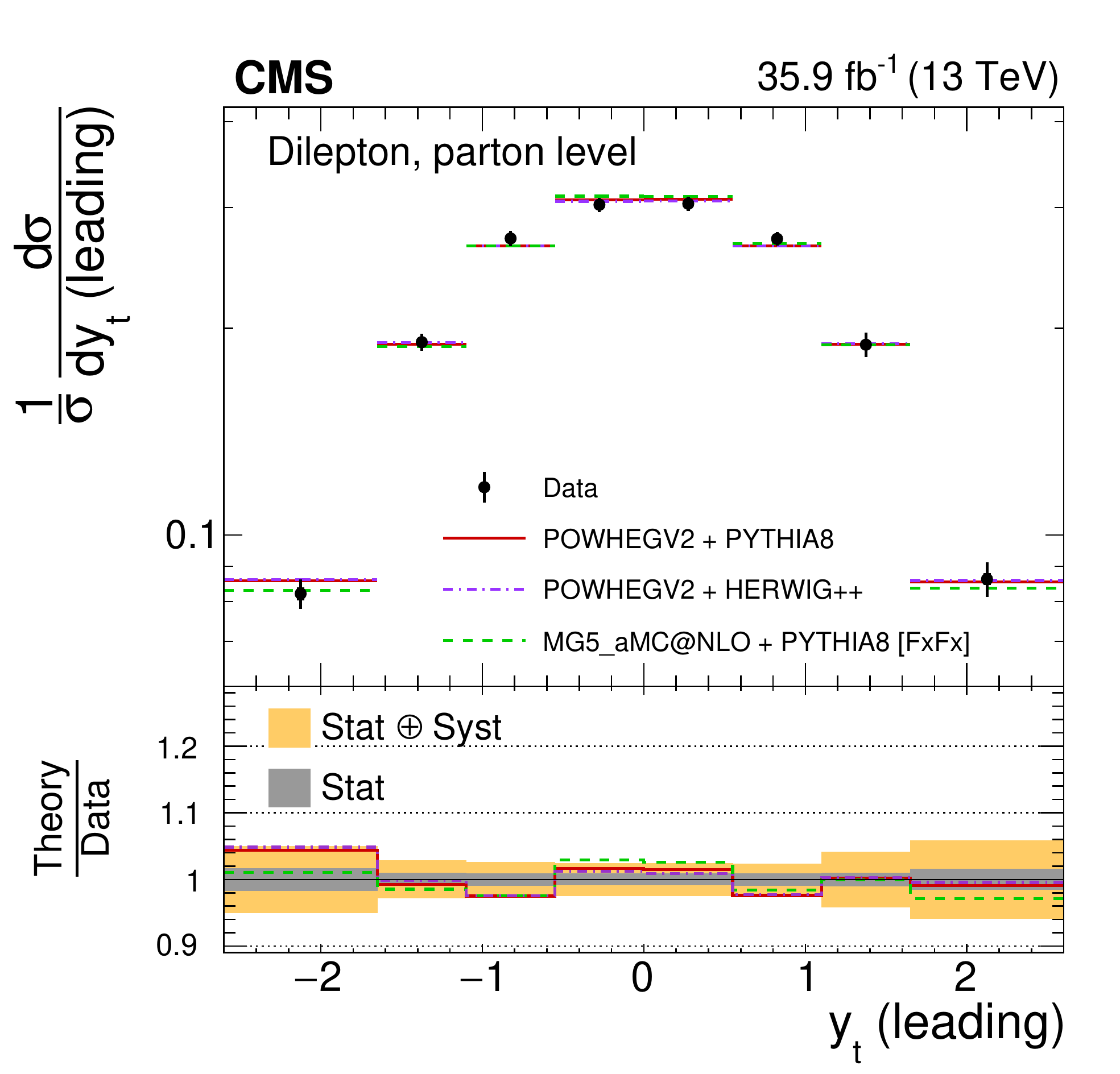} \\
\includegraphics[width=0.49\textwidth]{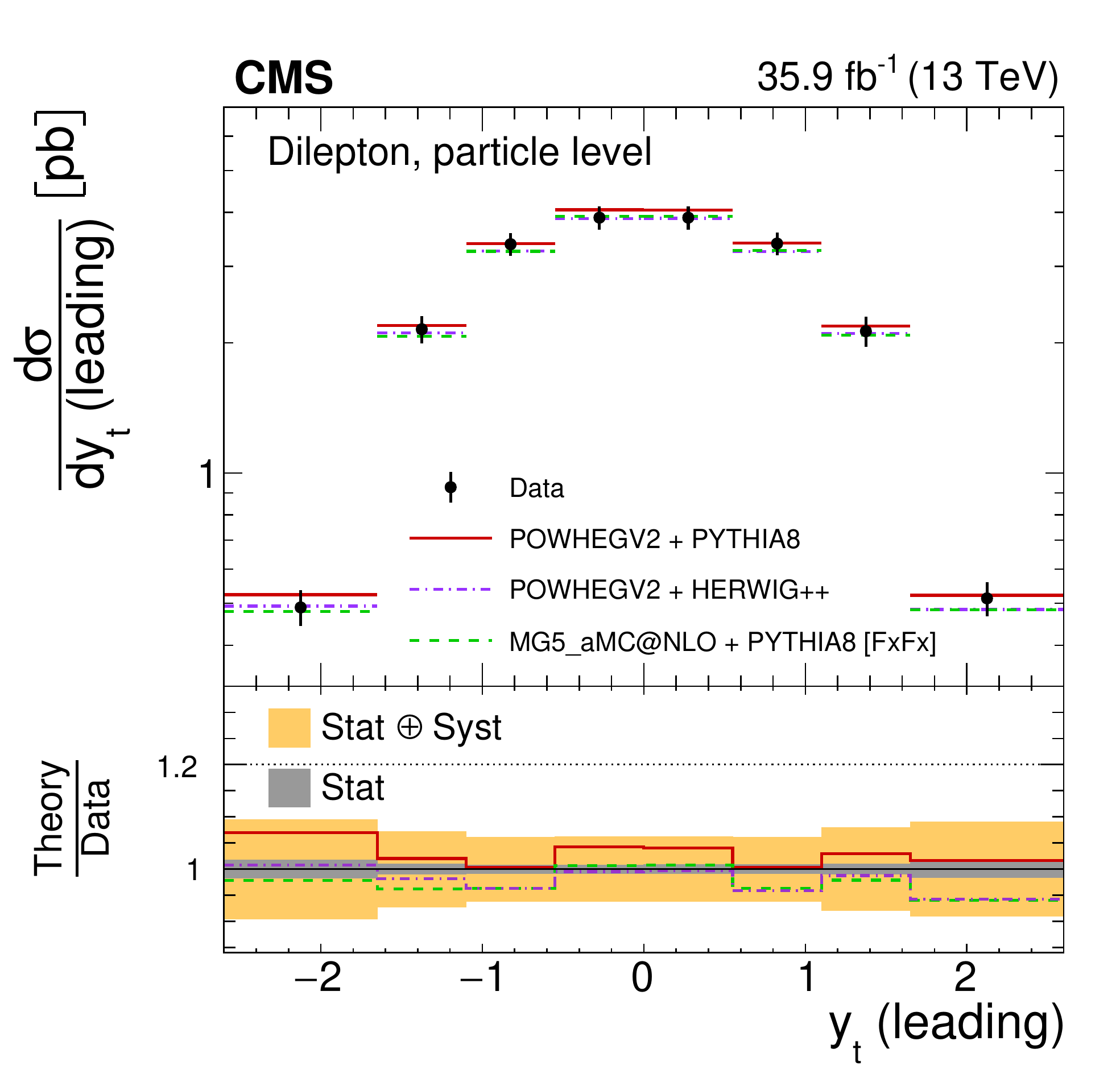}
\includegraphics[width=0.49\textwidth]{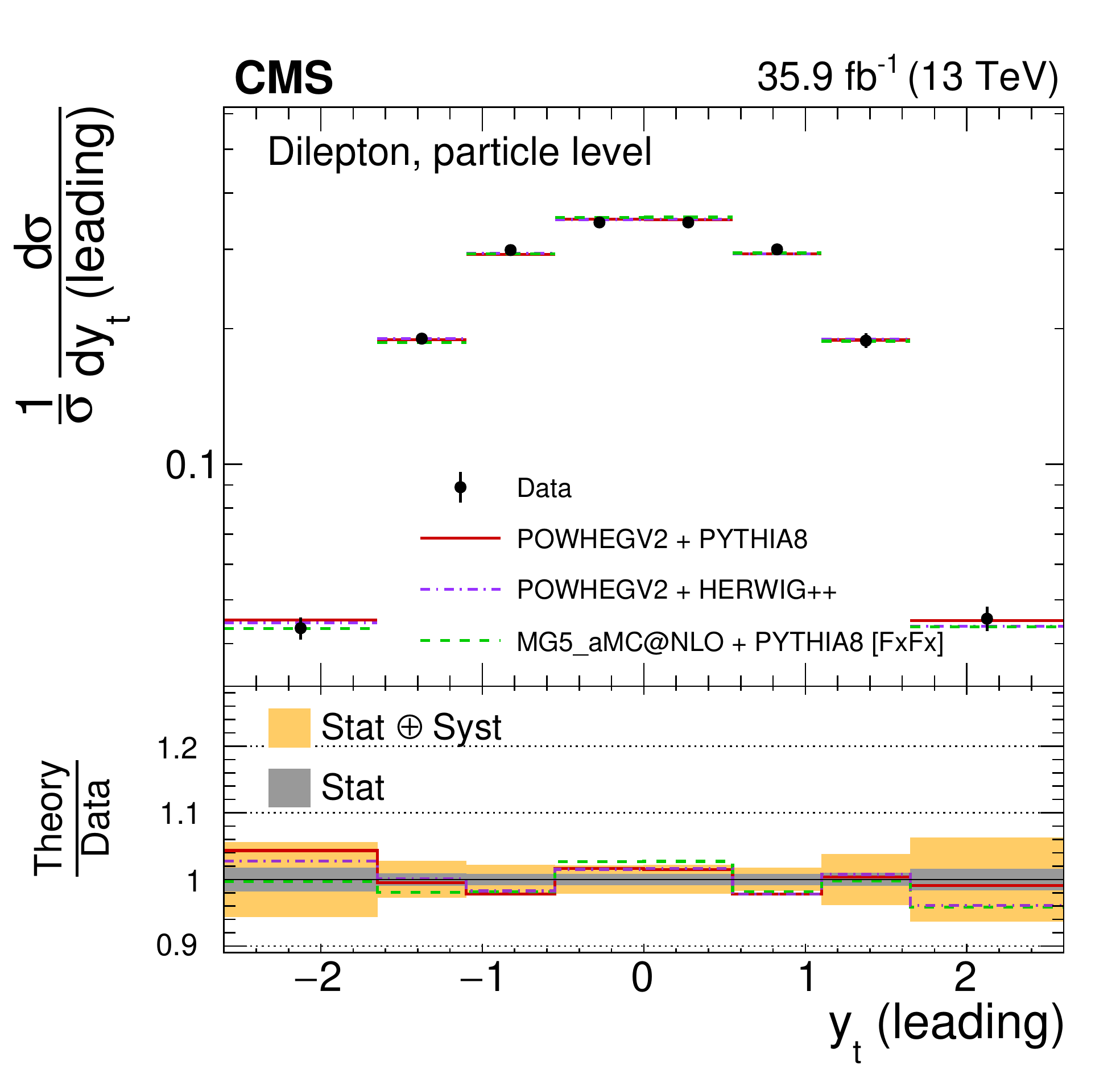}
\caption{The differential \ttbar production cross sections as a function of \ytop (leading) are shown for the data (points) and the MC predictions (lines). The vertical lines on the points indicate the total uncertainty in the data. The left and right columns correspond to absolute and normalised measurements, respectively. The upper row corresponds to measurements at the parton level in the full phase space and the lower row to the particle level in a fiducial phase space. The lower panel in each plot shows the ratios of the theoretical predictions to the data. The dark and light bands show the relative statistical and total uncertainties in the data, respectively.}
\label{fig:diffxsec:res_topy_leading}
\end{figure*}

\begin{figure*}[!phtb]
\centering
\includegraphics[width=0.49\textwidth]{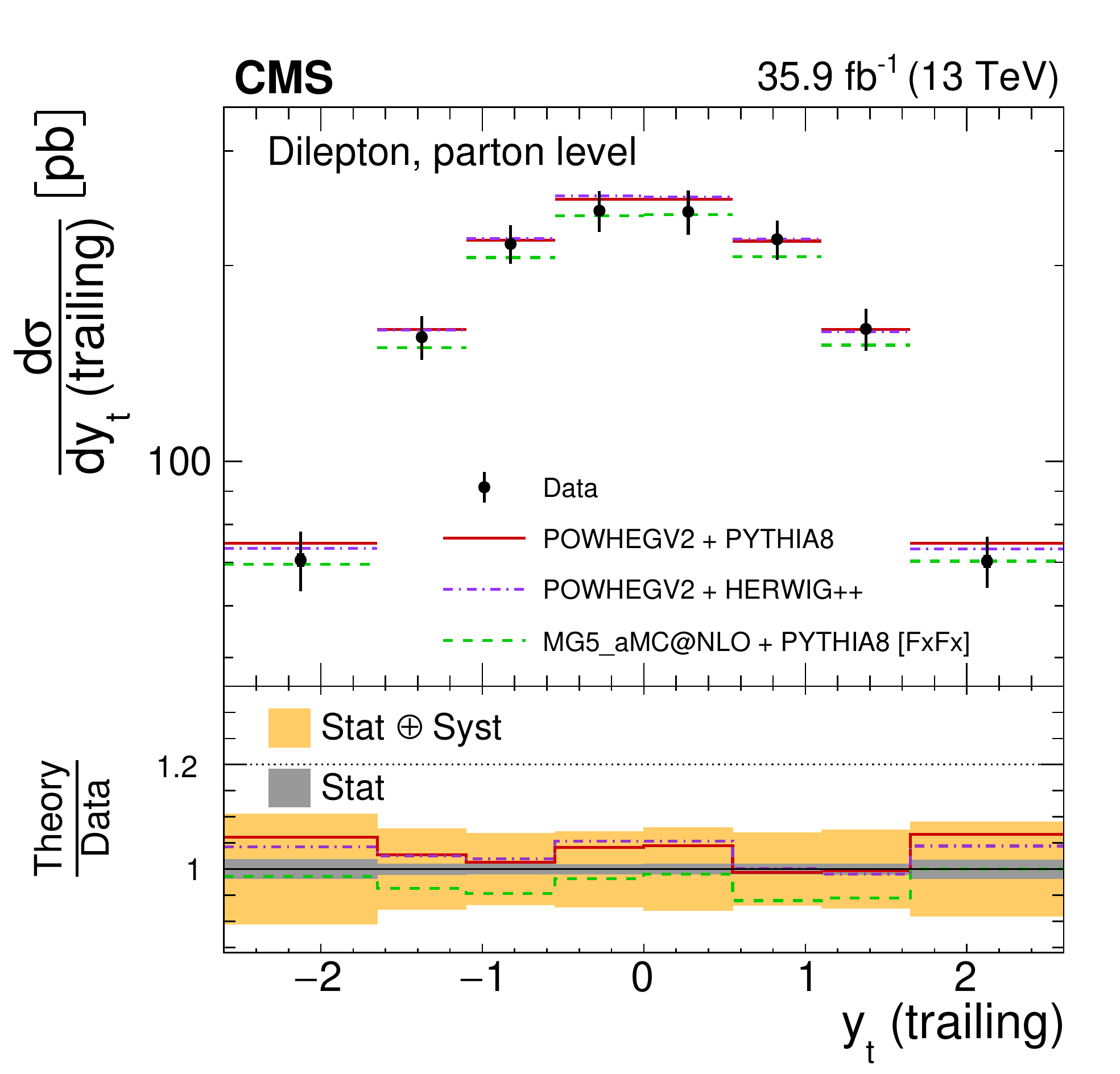}
\includegraphics[width=0.49\textwidth]{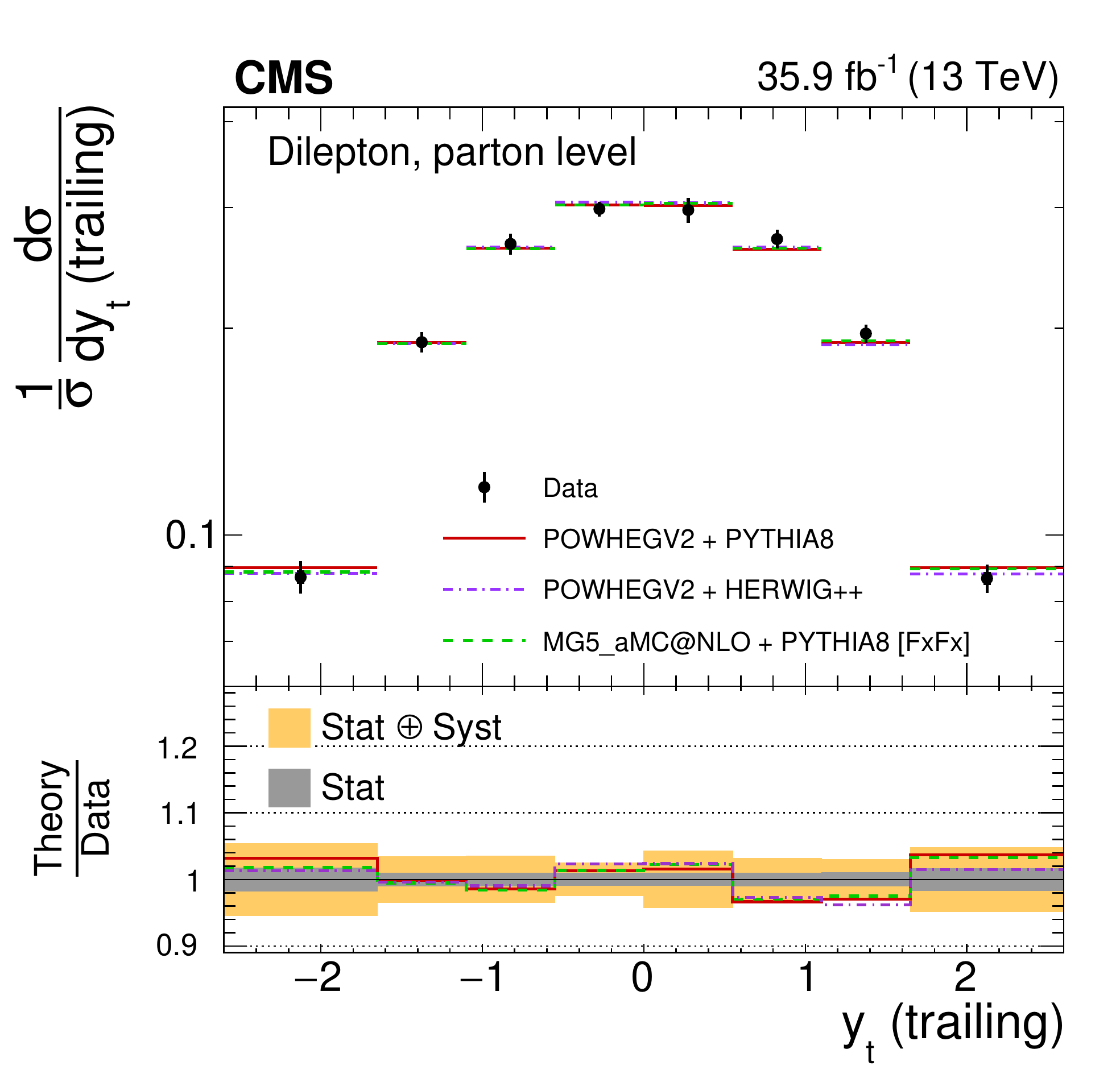} \\
\includegraphics[width=0.49\textwidth]{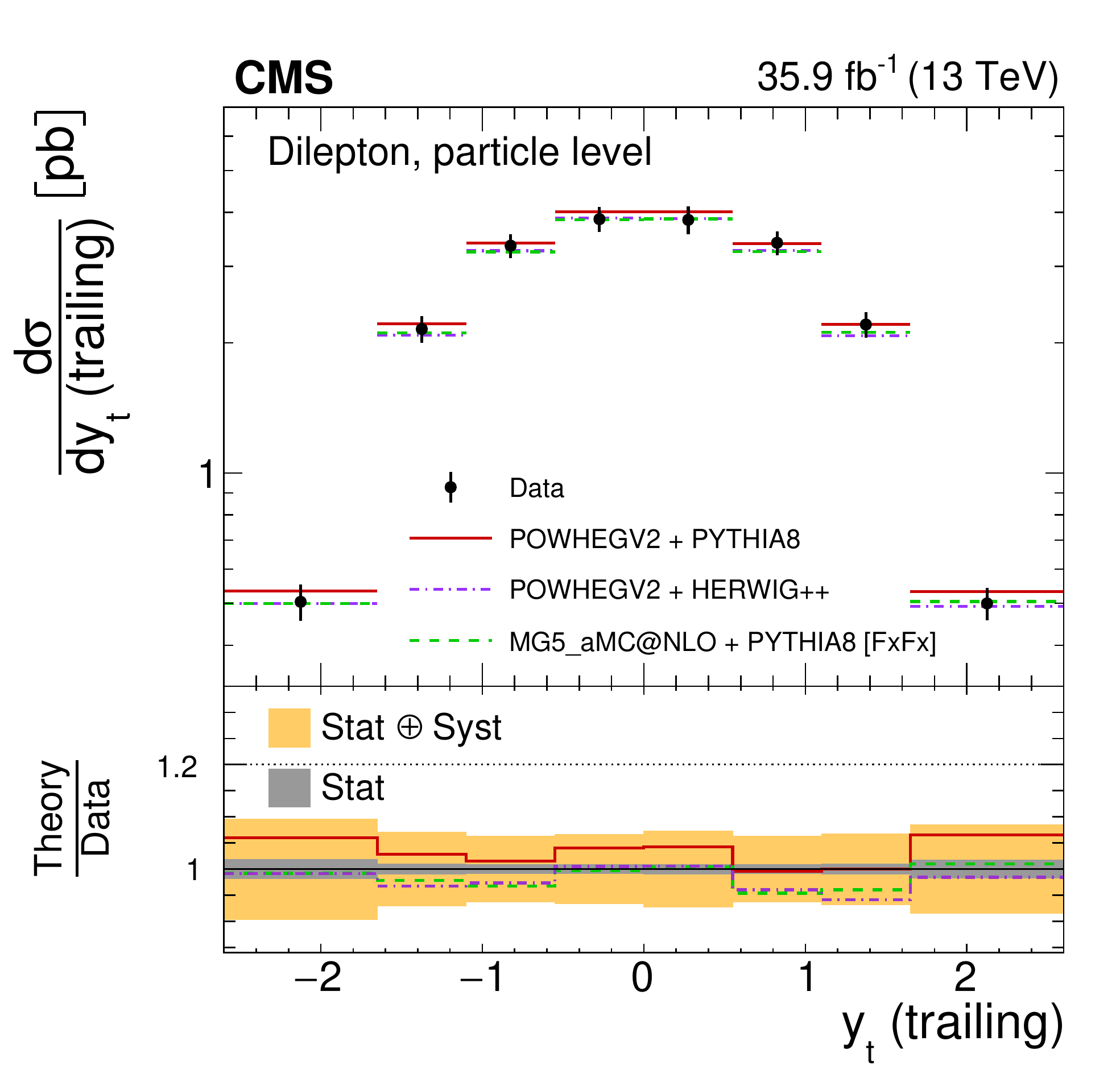}
\includegraphics[width=0.49\textwidth]{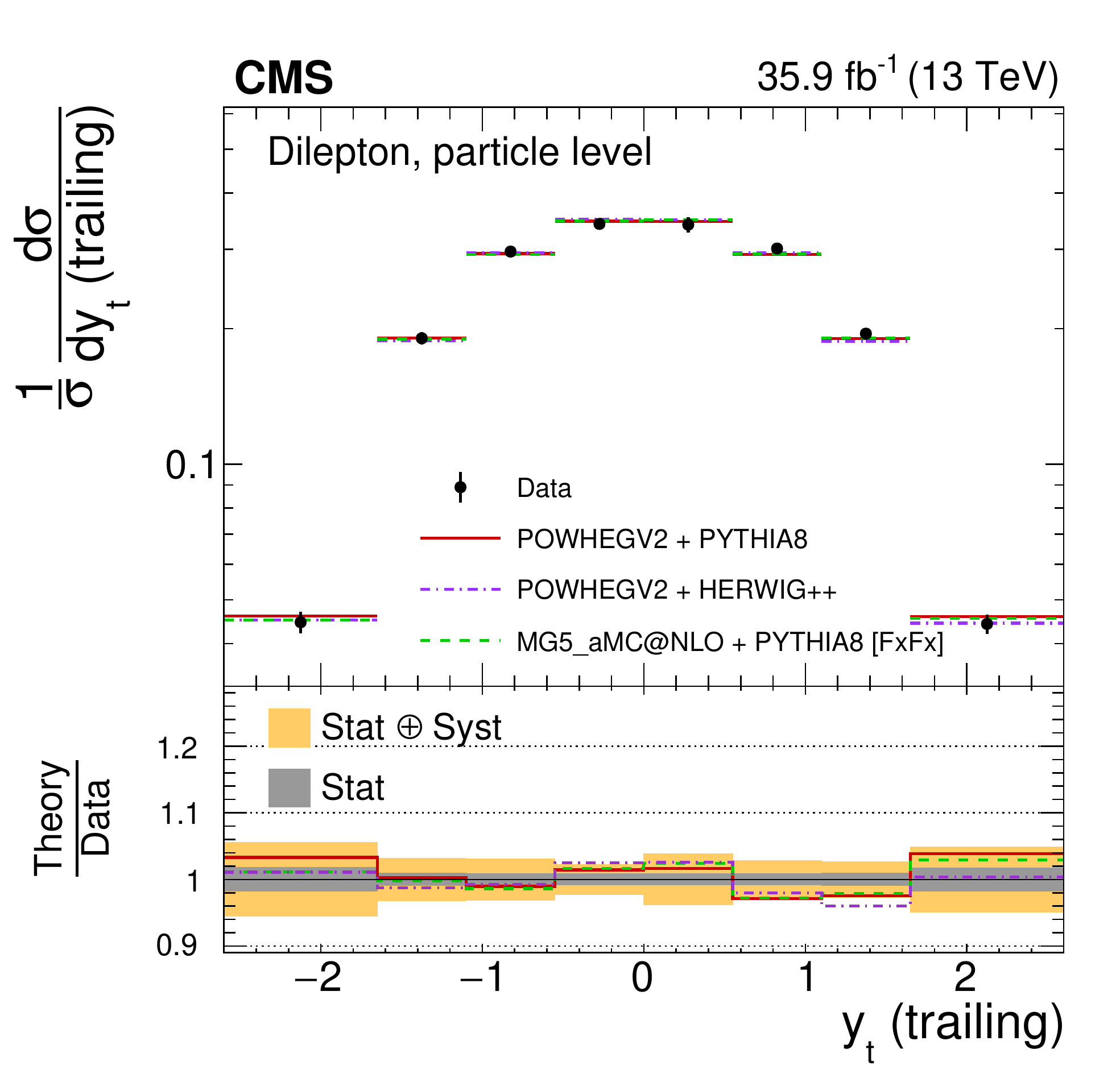}
\caption{The differential \ttbar production cross sections  as a function of \ytop (trailing) are shown for the data (points) and the MC predictions (lines). The vertical lines on the points indicate the total uncertainty in the data. The left and right columns correspond to absolute and normalised measurements, respectively. The upper row corresponds to measurements at the parton level in the full phase space and the lower row to the particle level in a fiducial phase space. The lower panel in each plot shows the ratios of the theoretical predictions to the data. The dark and light bands show the relative statistical and total uncertainties in the data, respectively.}
\label{fig:diffxsec:res_topy_trailing}
\end{figure*}

\begin{figure*}[!phtb]
\centering
\includegraphics[width=0.49\textwidth]{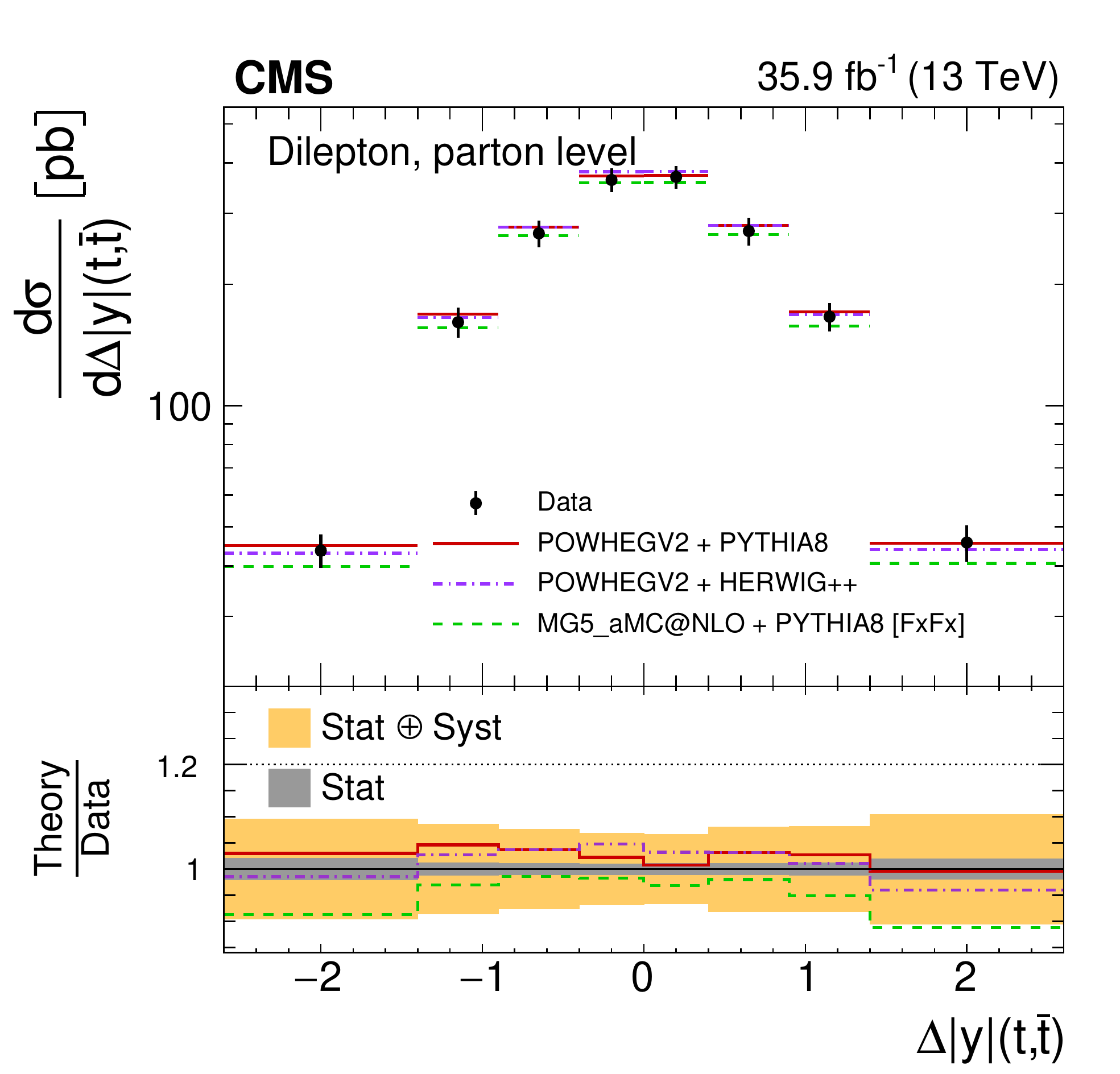}
\includegraphics[width=0.49\textwidth]{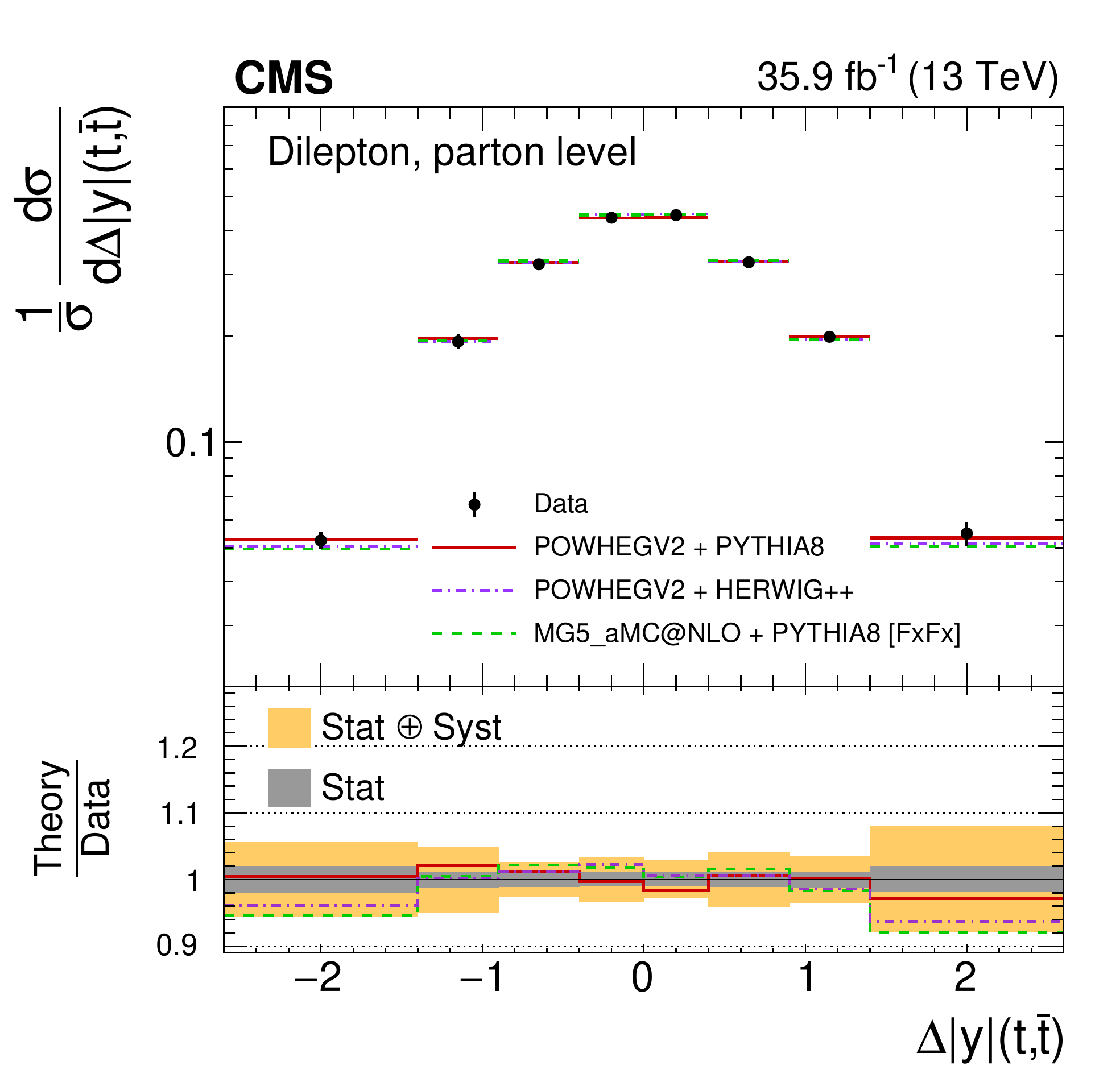} \\
\includegraphics[width=0.49\textwidth]{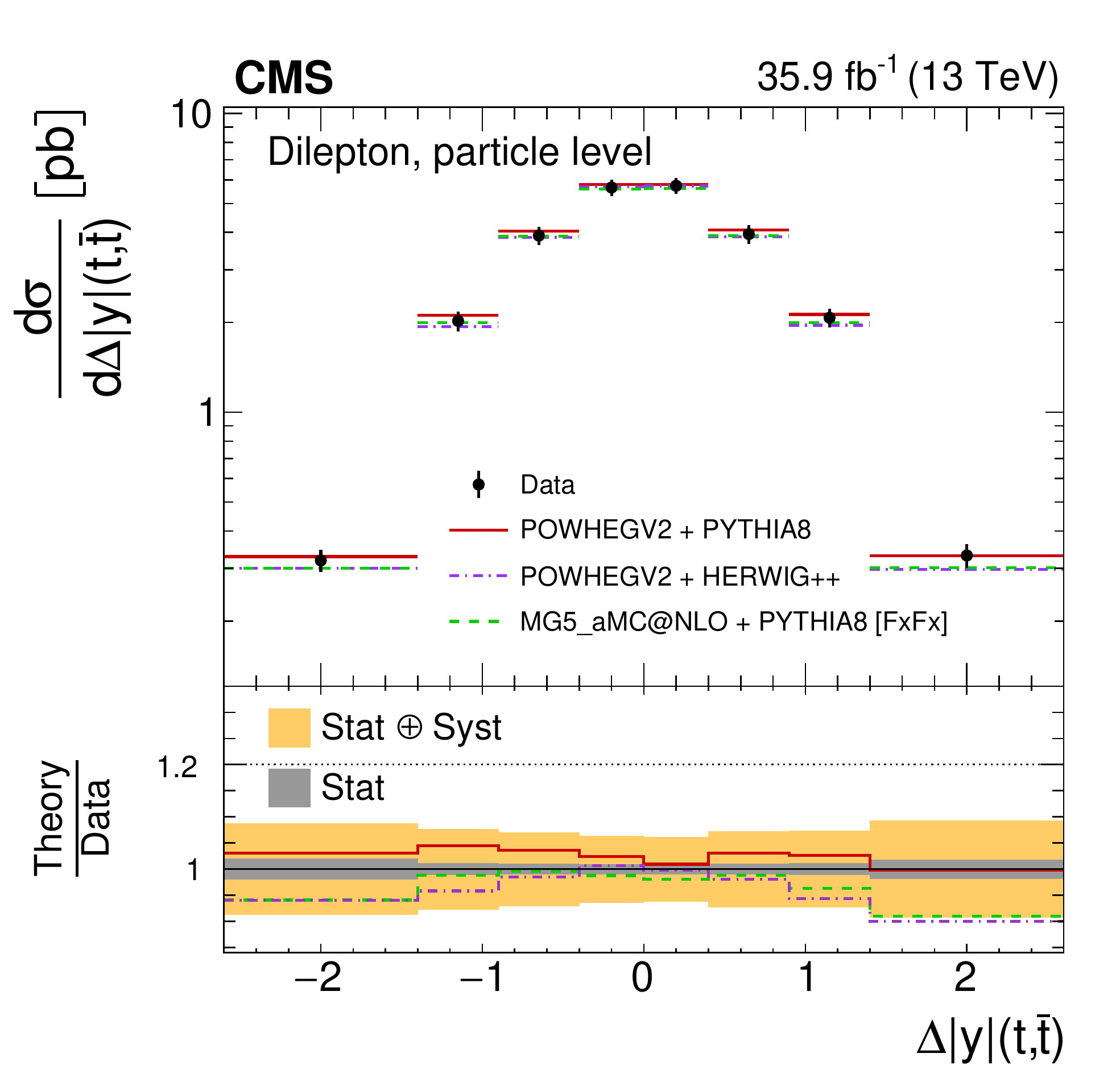}
\includegraphics[width=0.49\textwidth]{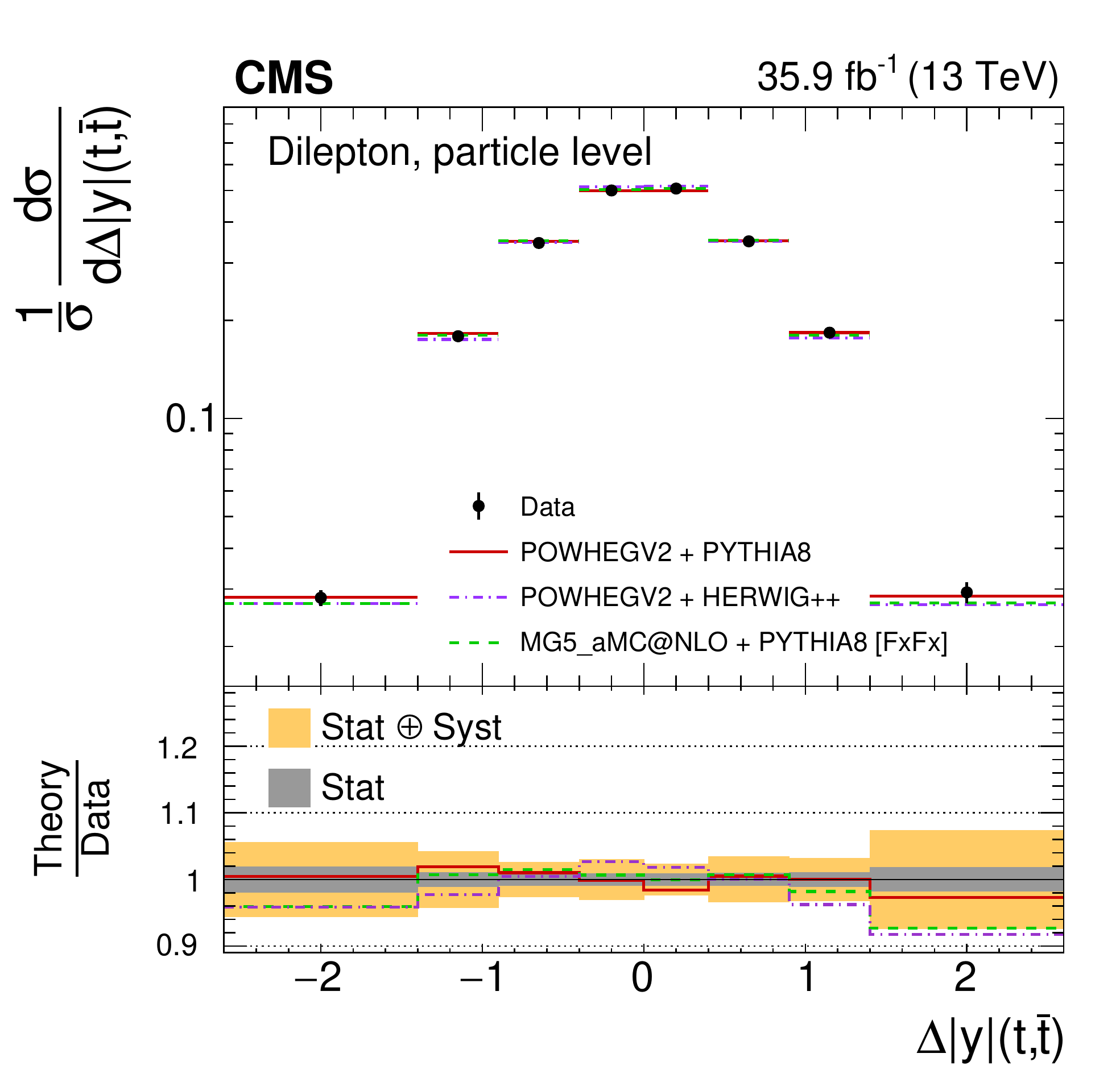}
\caption{The differential \ttbar production cross sections as a function of \delytt are shown for the data (points) and the MC predictions (lines). The vertical lines on the points indicate the total uncertainty in the data. The left and right columns correspond to absolute and normalised measurements, respectively. The upper row corresponds to measurements at the parton level in the full phase space and the lower row to the particle level in a fiducial phase space. The lower panel in each plot shows the ratios of the theoretical predictions to the data. The dark and light bands show the relative statistical and total uncertainties in the data, respectively.}
\label{fig:diffxsec:res_deltay}
\end{figure*}

\begin{figure*}[!phtb]
\centering
\includegraphics[width=0.49\textwidth]{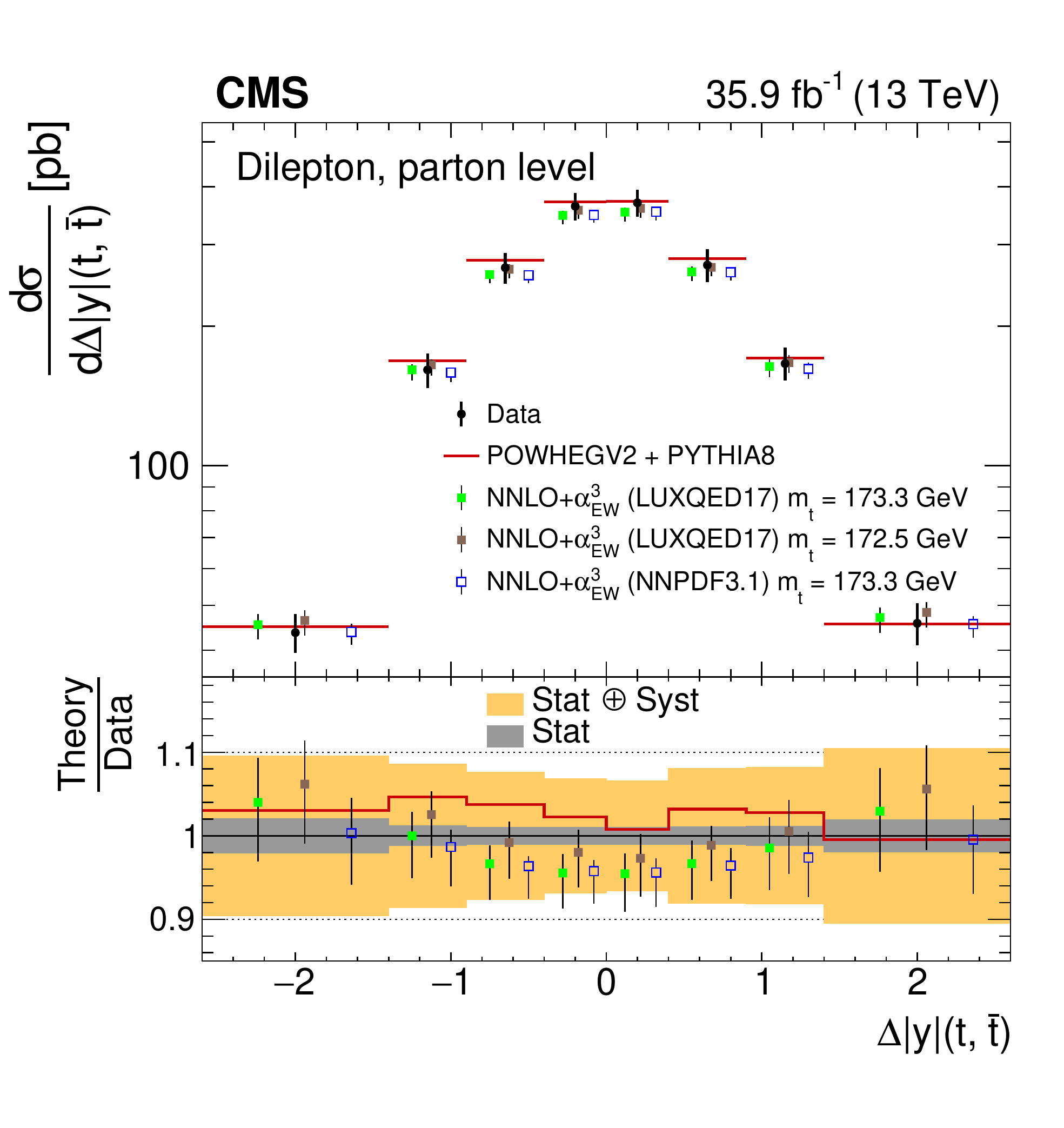}
\includegraphics[width=0.49\textwidth]{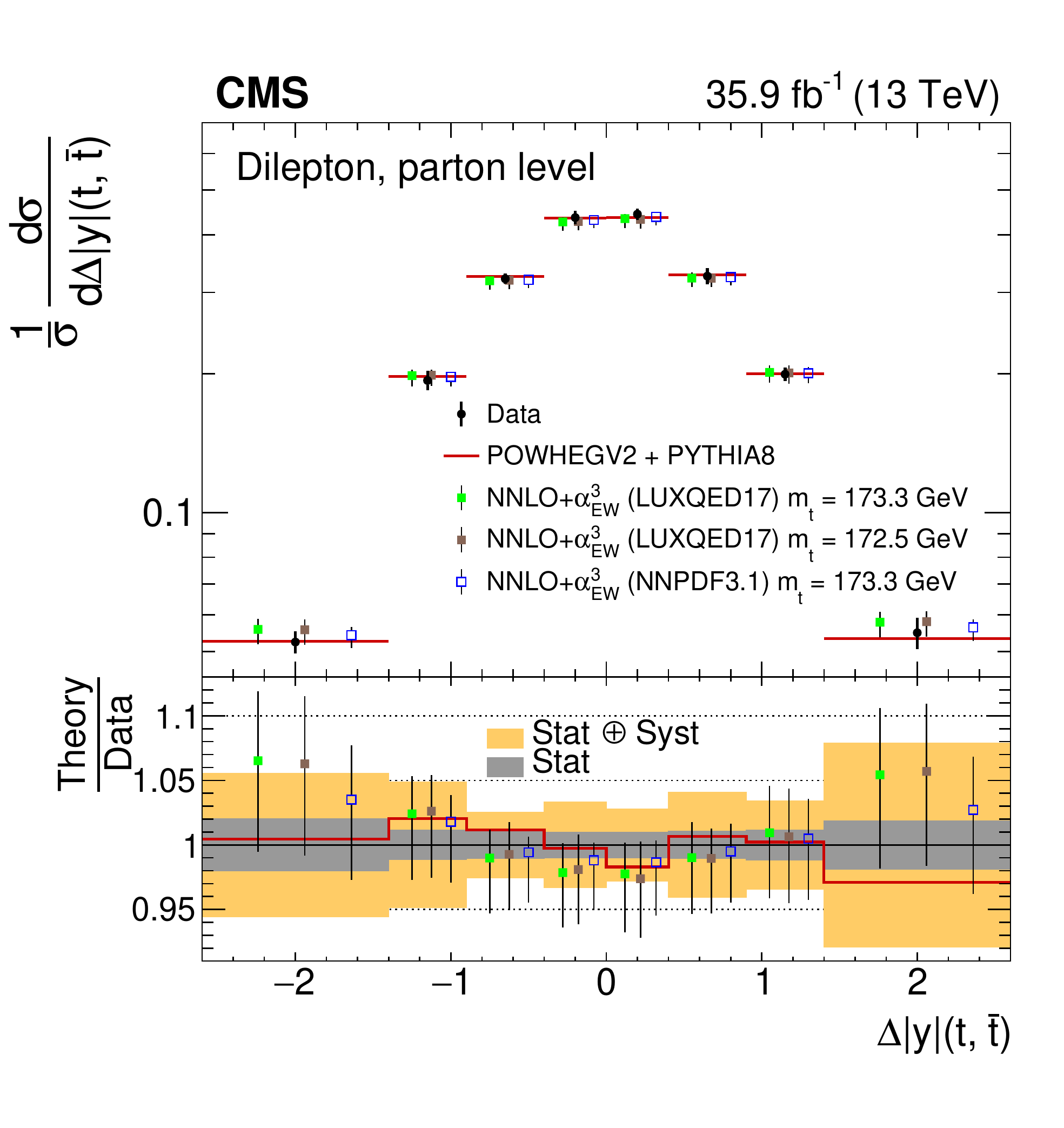}
\caption{The differential \ttbar production cross sections at the parton level in the full phase space as a function of \delytt are shown for the data (filled circles), the theoretical predictions with beyond-NLO precision (other points) and the prediction from \pwhgpy (solid line). The vertical lines on the filled circles and other points indicate the total uncertainty in the data and theoretical predictions, respectively.  The left and right plots correspond to absolute and normalised measurements, respectively. The lower panel in each plot shows the ratios of the theoretical predictions to the data. The dark and light bands show the relative statistical and total uncertainties in the data, respectively.}
\label{fig:diffxsec:res_deltay_bnlo}
\includegraphics[width=0.75\textwidth]{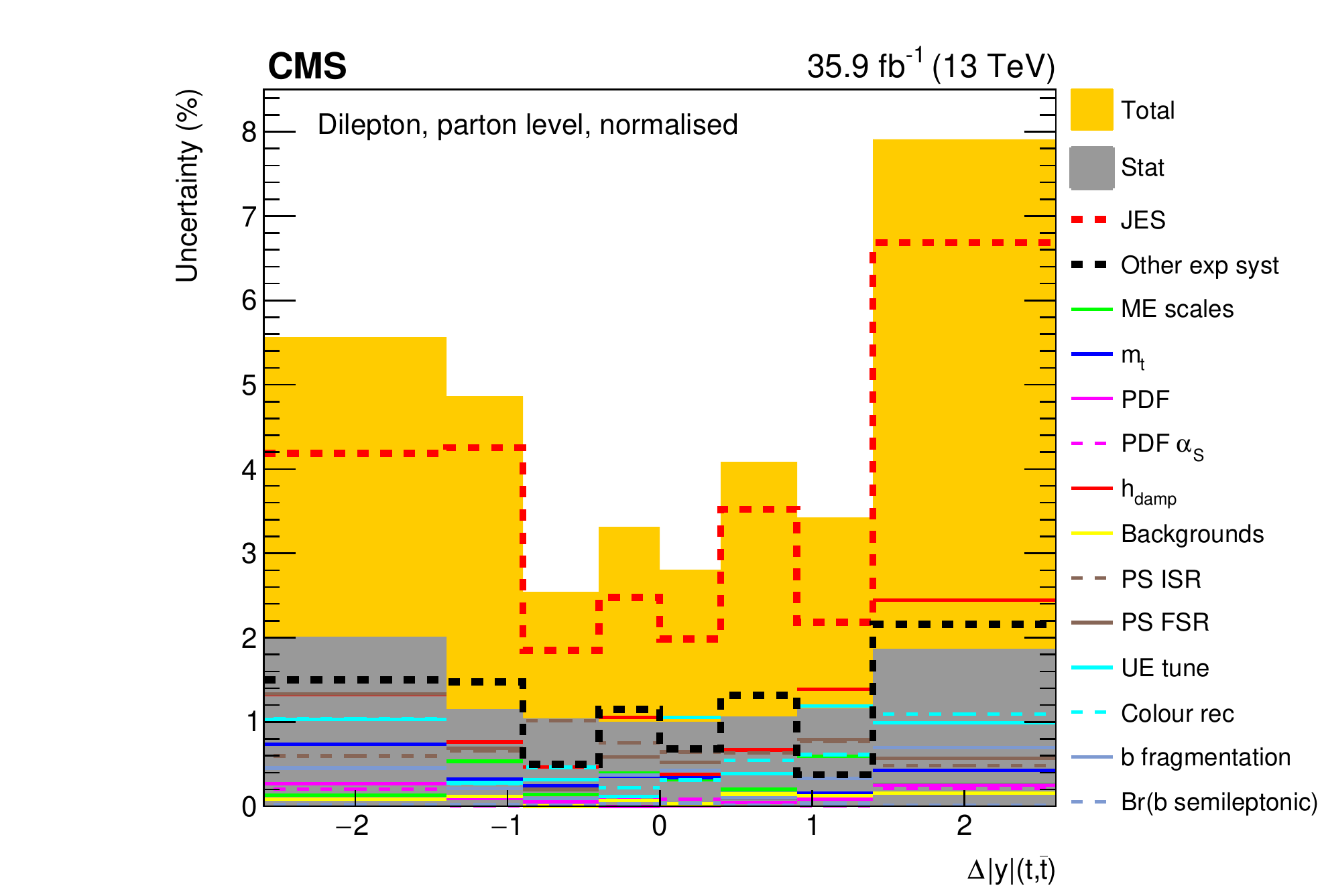}
\caption{The contributions of each source of systematic uncertainty to the total systematic uncertainty in each bin is shown for the measurement of the normalised \ttbar production cross sections  as a function of \delytt. The sources affecting the JES are added in quadrature and shown as a single component. Additional experimental systematic uncertainties are also added in quadrature and shown as a single component. Contributions from theoretical uncertainties are shown separately. The statistical and total uncertainties, corresponding to the quadrature addition of statistical and systematic uncertainties, are shown by the dark and light filled histograms, respectively.}
\label{fig:diffxsec:unc_breakdown_res_ttdeltay}
\end{figure*}

\begin{figure*}[!phtb]
\centering
\includegraphics[width=0.49\textwidth]{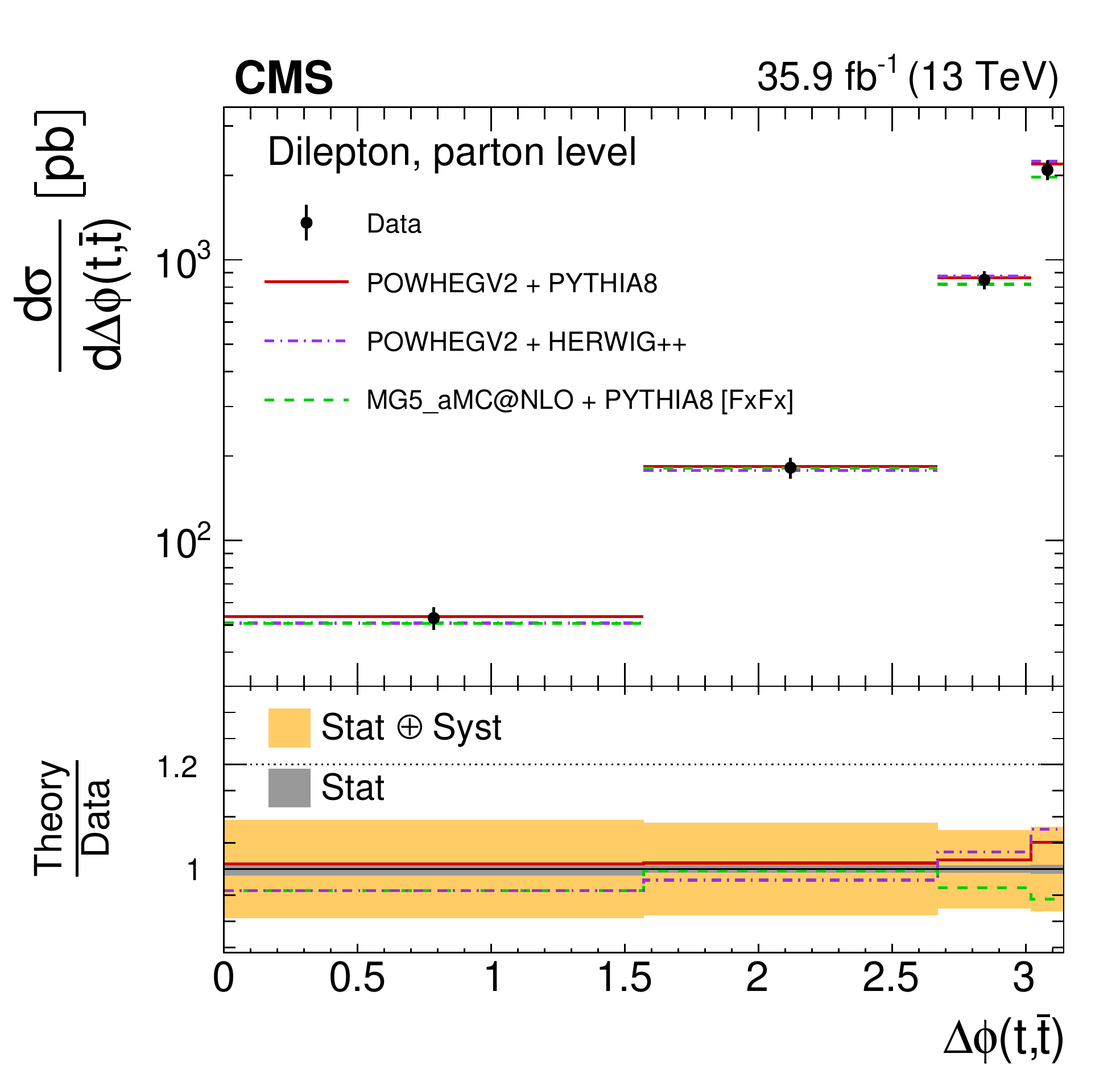}
\includegraphics[width=0.49\textwidth]{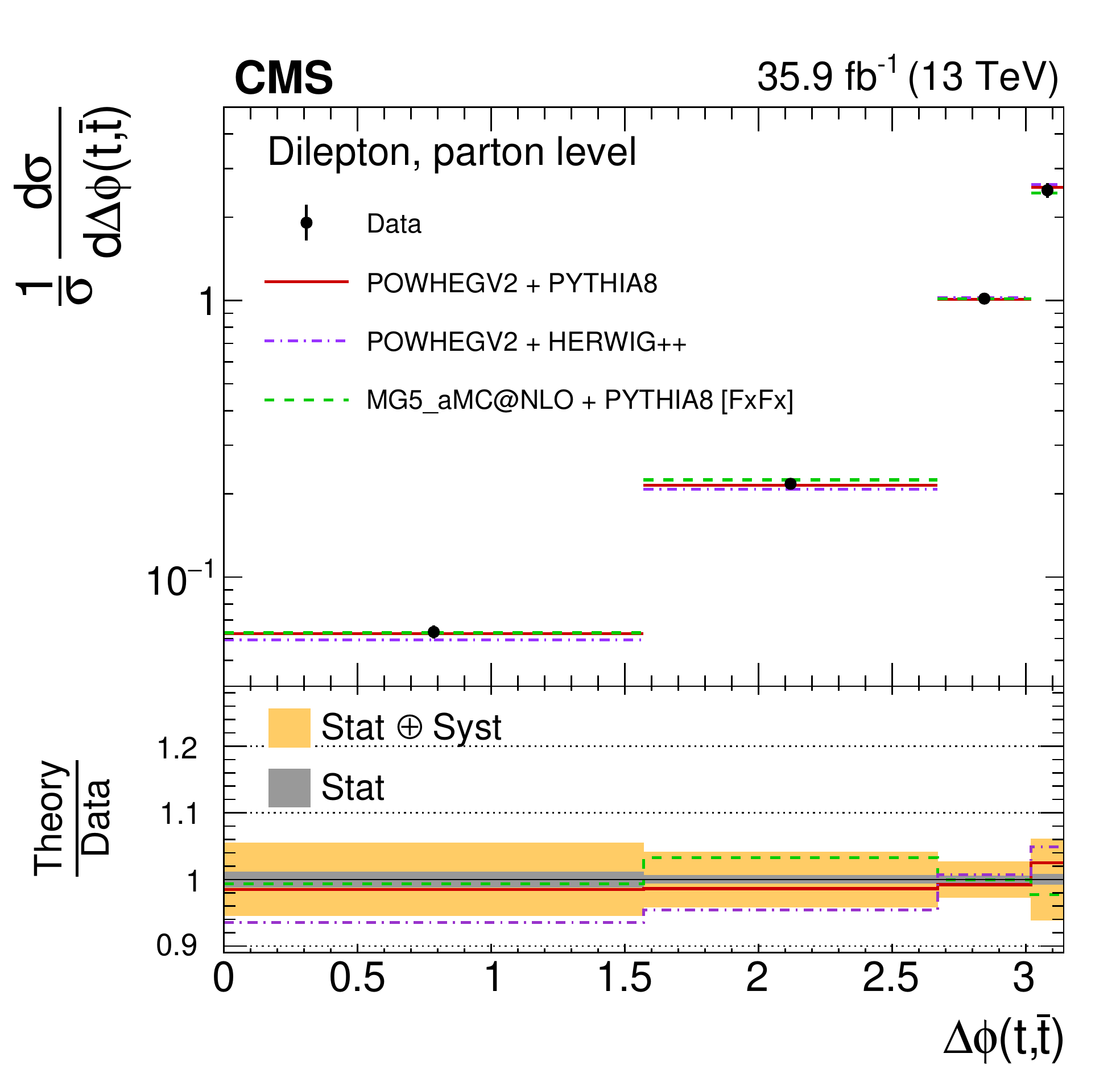} \\
\includegraphics[width=0.49\textwidth]{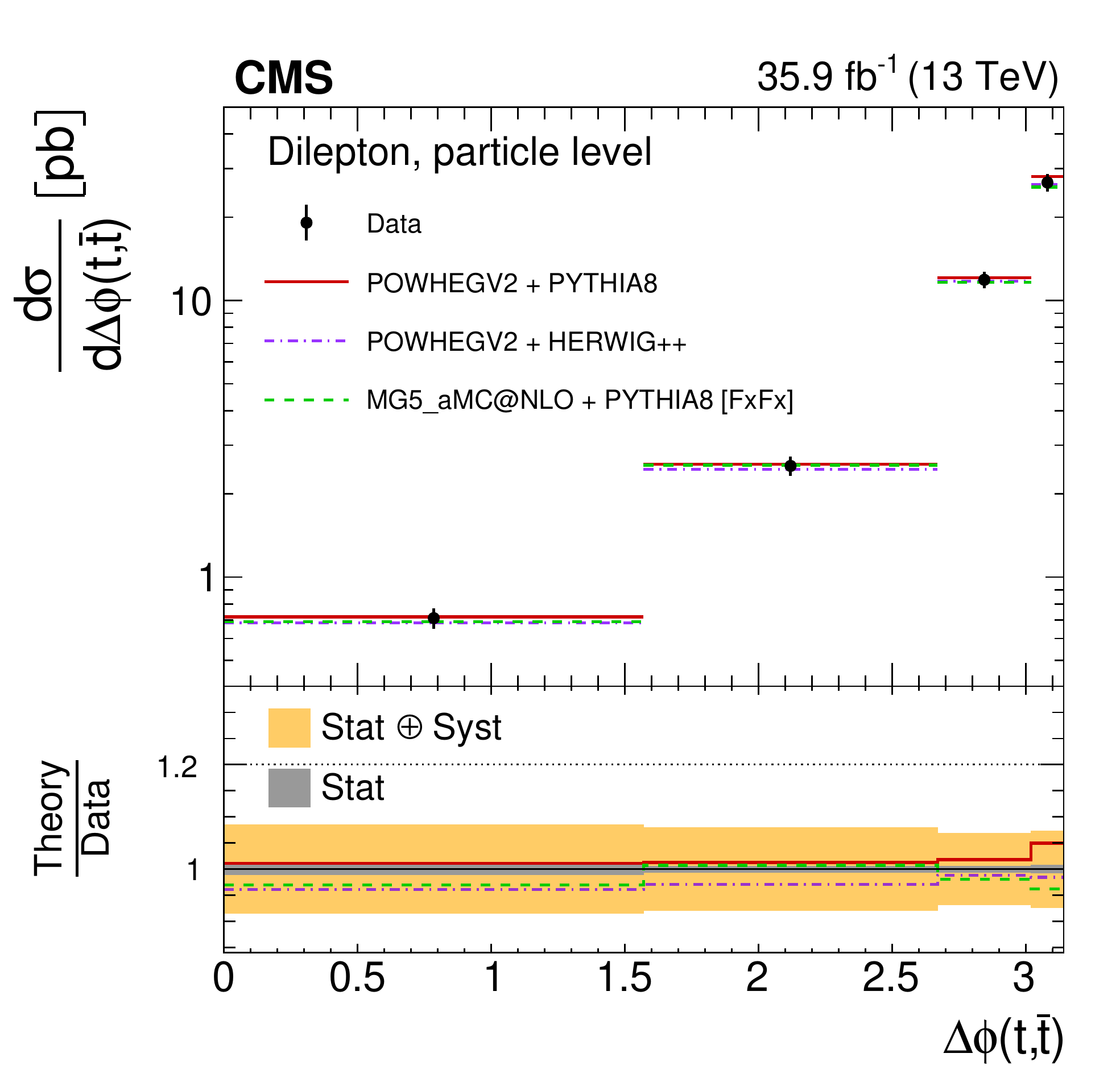}
\includegraphics[width=0.49\textwidth]{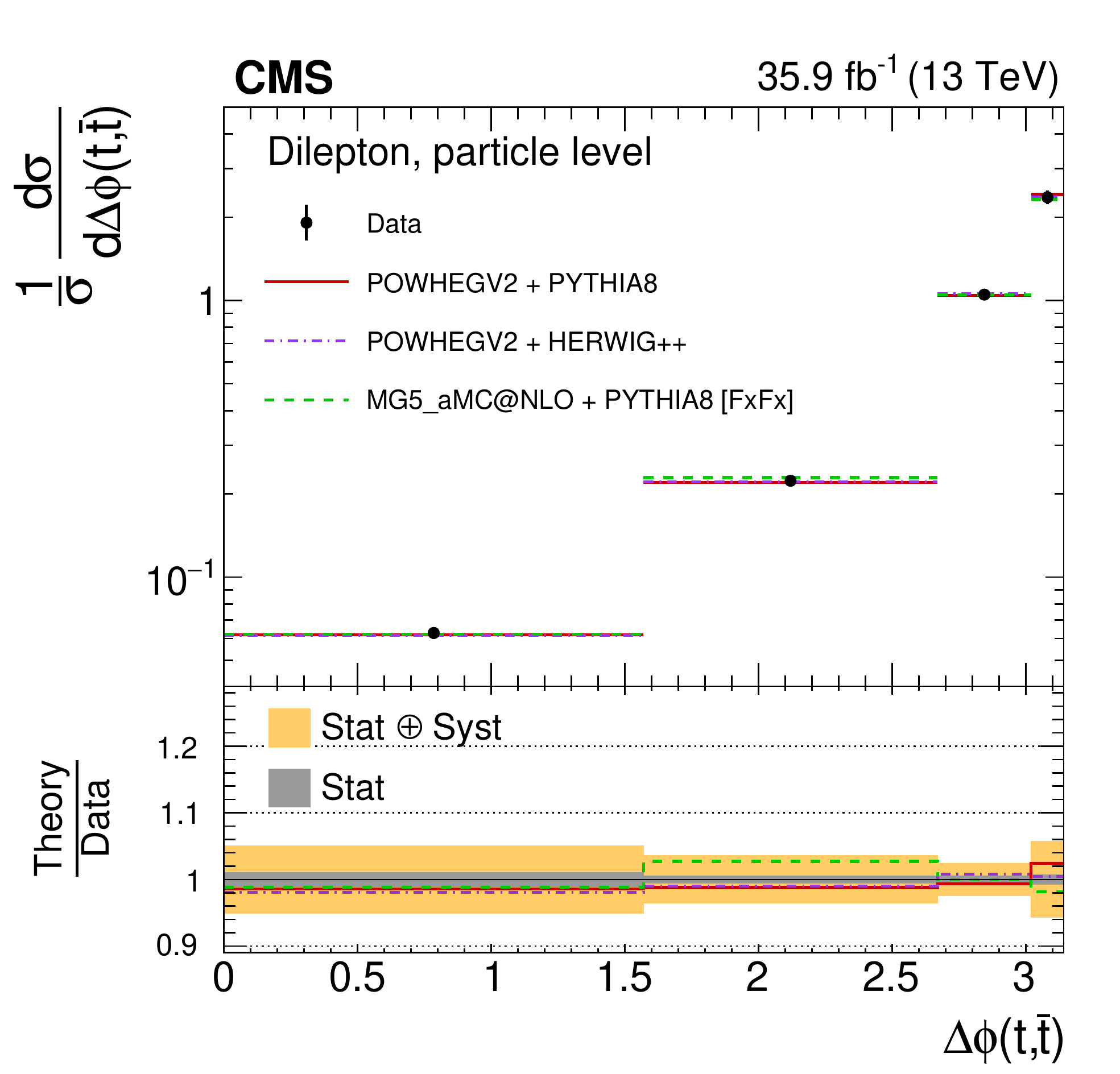}
\caption{The differential \ttbar production cross sections as a function of \delphitt are shown for the data (points) and the MC predictions (lines). The vertical lines on the points indicate the total uncertainty in the data. The left and right columns correspond to absolute and normalised measurements, respectively. The upper row corresponds to measurements at the parton level in the full phase space and the lower row to the particle level in a fiducial phase space. The lower panel in each plot shows the ratios of the theoretical predictions to the data. The dark and light bands show the relative statistical and total uncertainties in the data, respectively.}
\label{fig:diffxsec:res_ttdely}
\end{figure*}

\begin{figure*}[!phtb]
\centering
\includegraphics[width=0.49\textwidth]{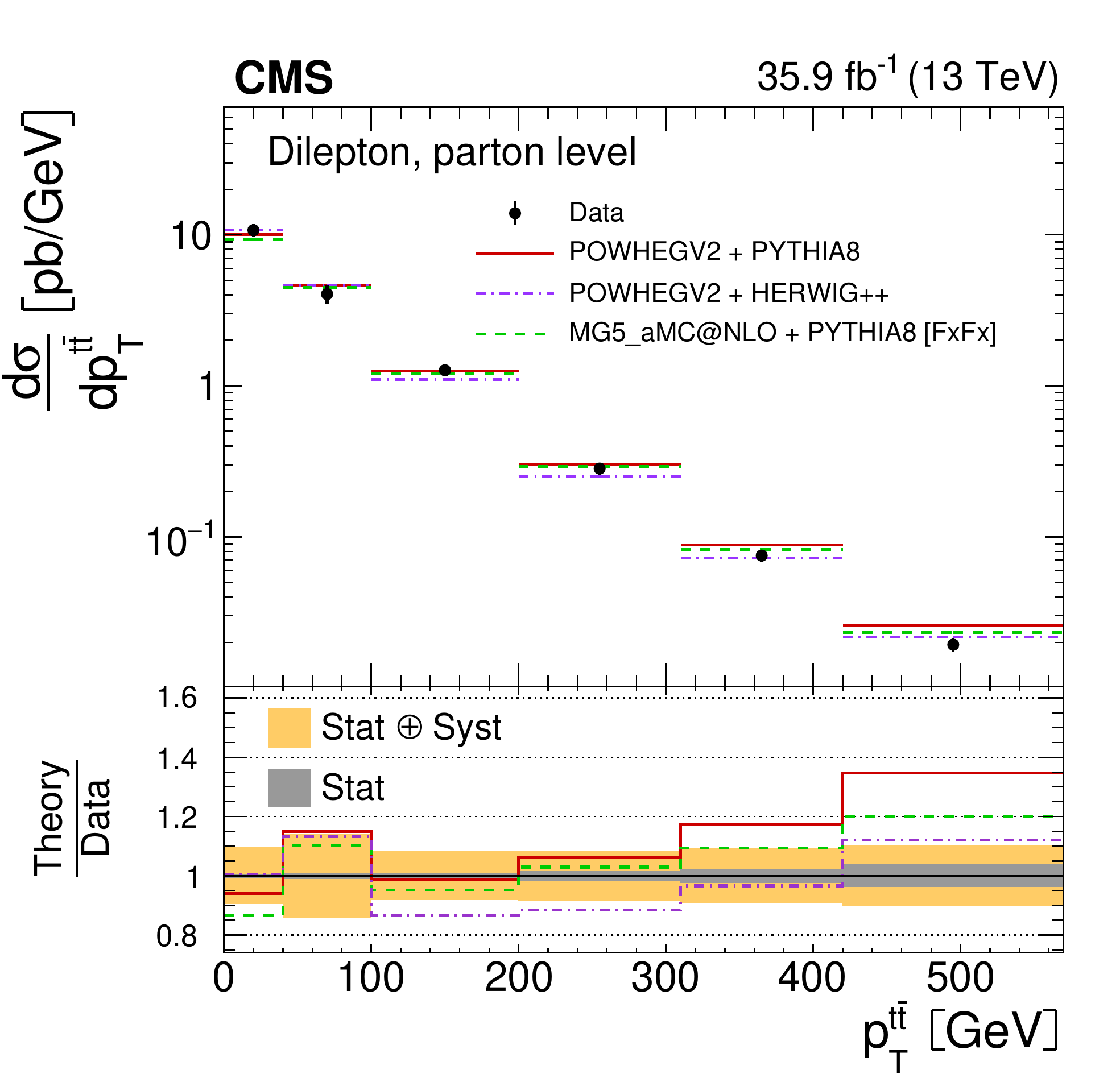}
\includegraphics[width=0.49\textwidth]{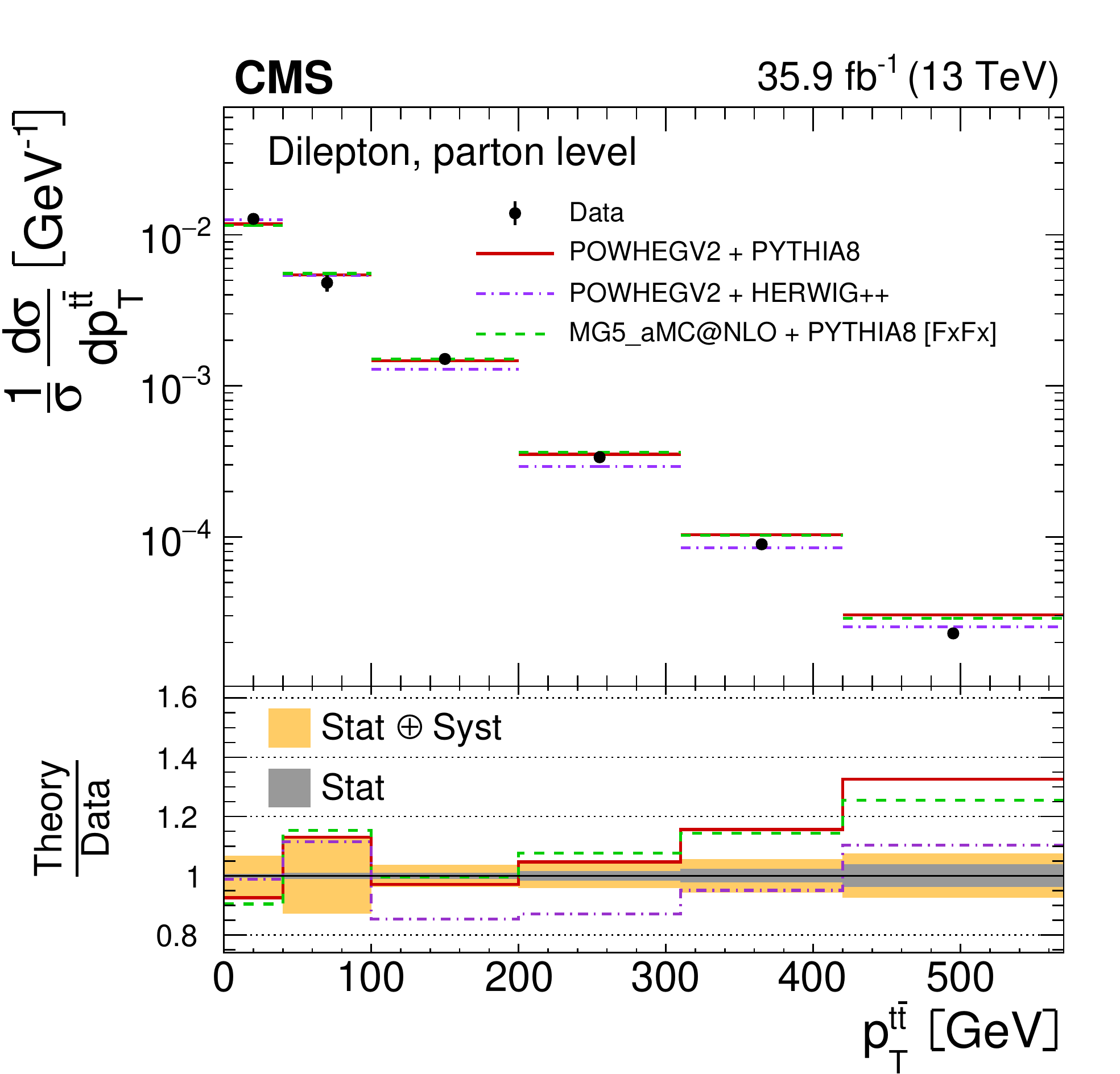} \\
\includegraphics[width=0.49\textwidth]{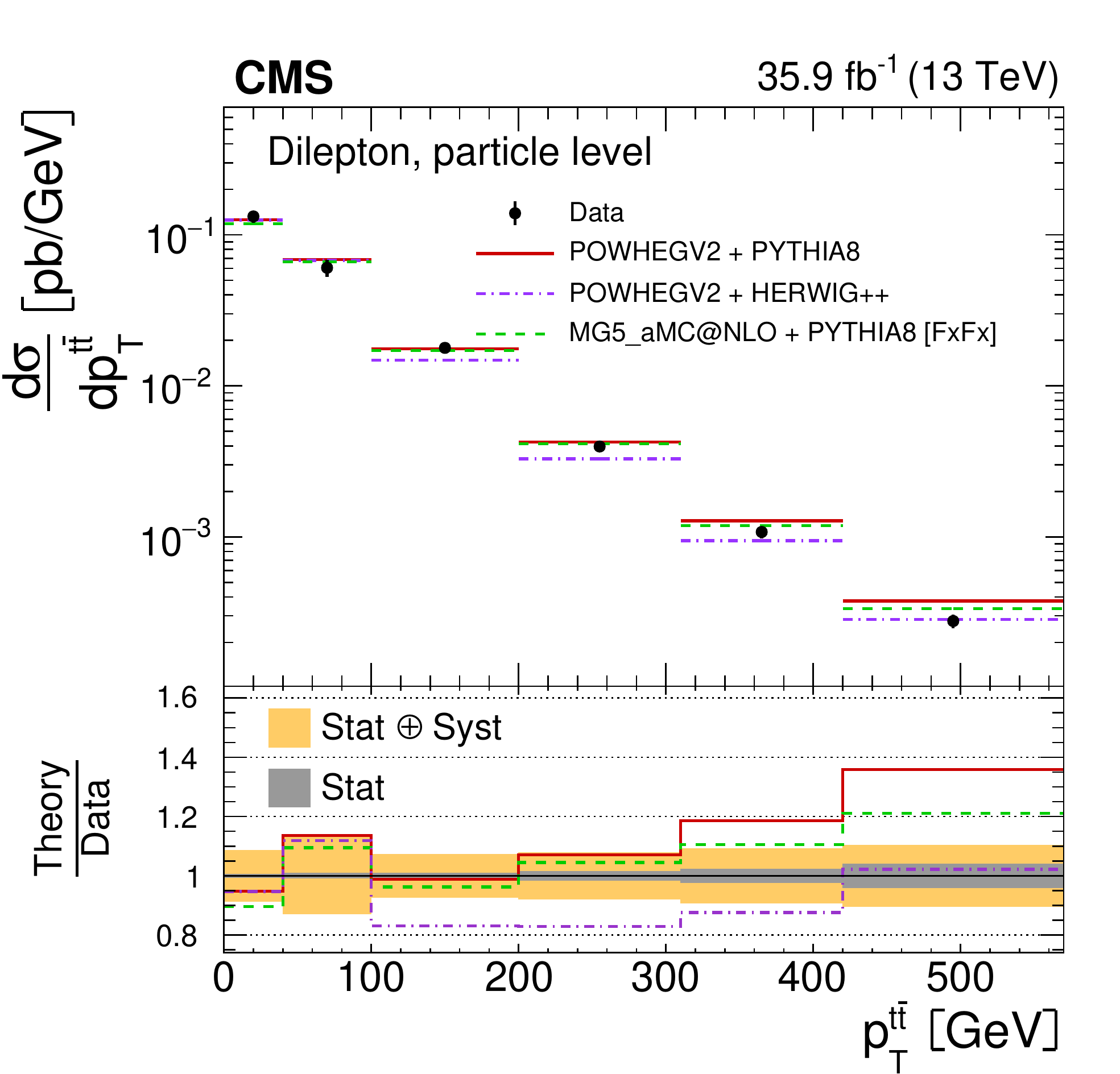}
\includegraphics[width=0.49\textwidth]{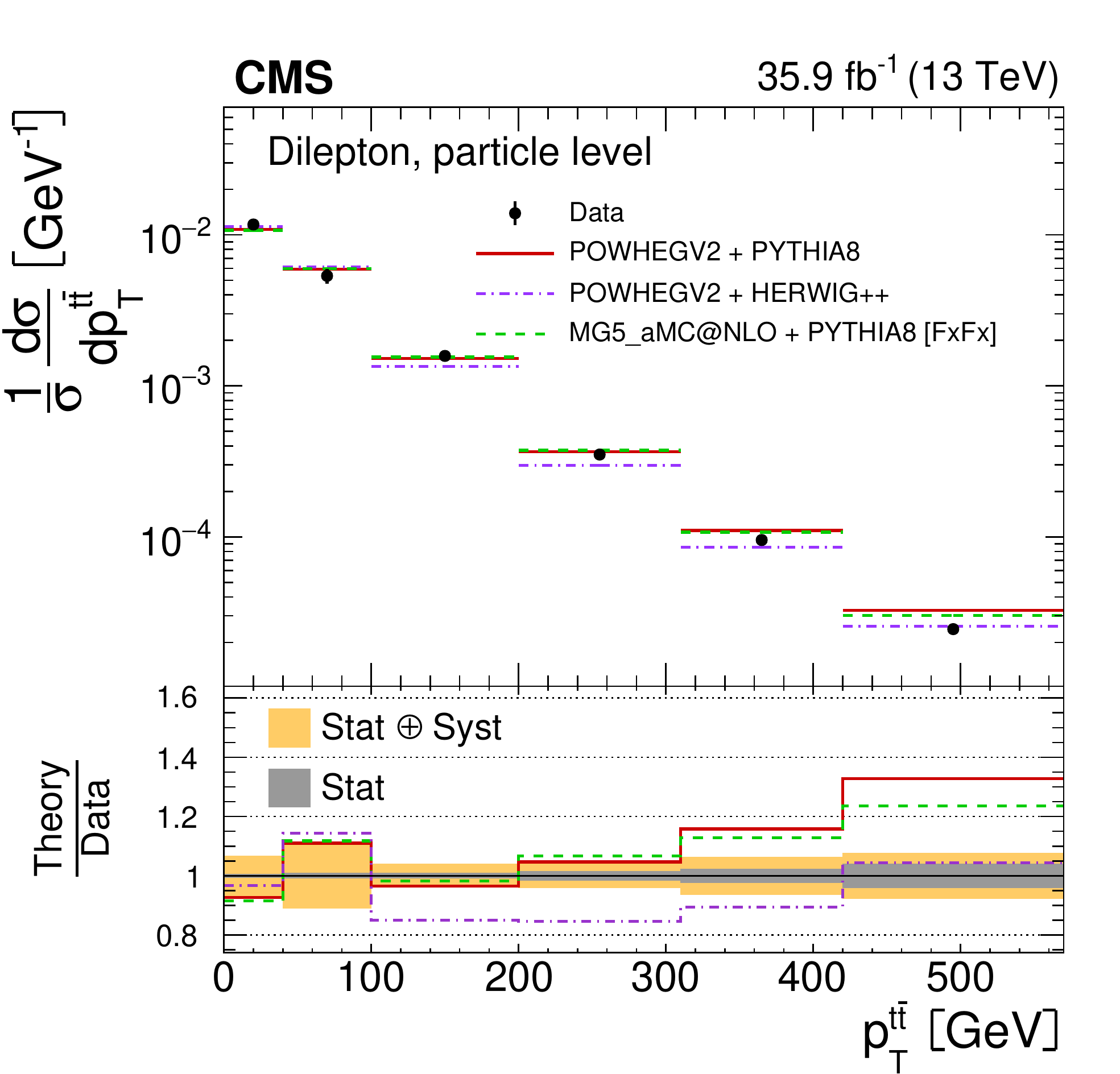}
\caption{The differential \ttbar production cross sections as a function of \pttt are shown for the data (points) and the MC predictions (lines). The vertical lines on the points indicate the total uncertainty in the data. The left and right columns correspond to absolute and normalised measurements, respectively. The upper row corresponds to measurements at the parton level in the full phase space and the lower row to the particle level in a fiducial phase space. The lower panel in each plot shows the ratios of the theoretical predictions to the data. The dark and light bands show the relative statistical and total uncertainties in the data, respectively.}
\label{fig:diffxsec:res_ttpt}
\end{figure*}

\begin{figure*}[!phtb]
\centering
\includegraphics[width=0.49\textwidth]{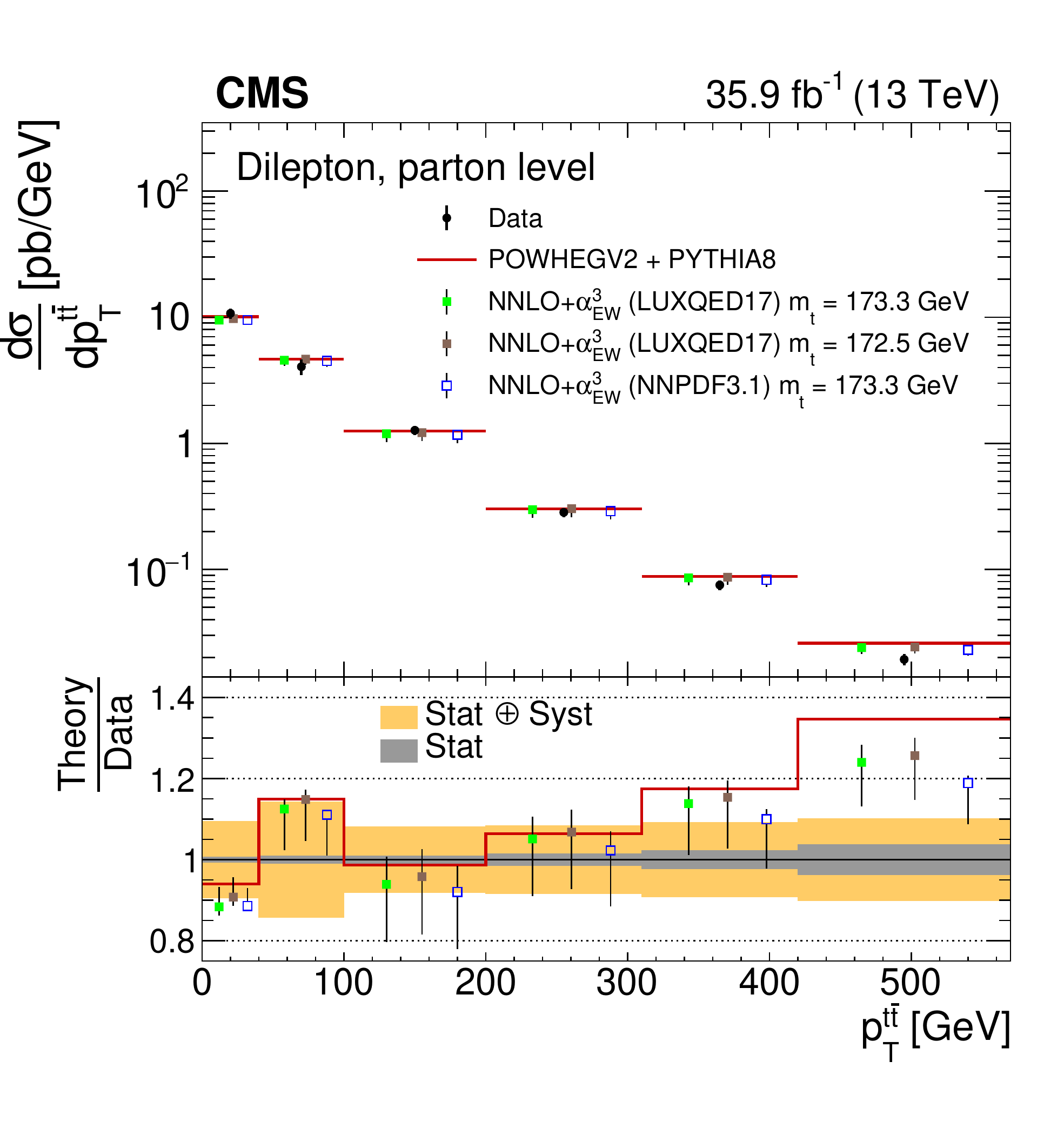}
\includegraphics[width=0.49\textwidth]{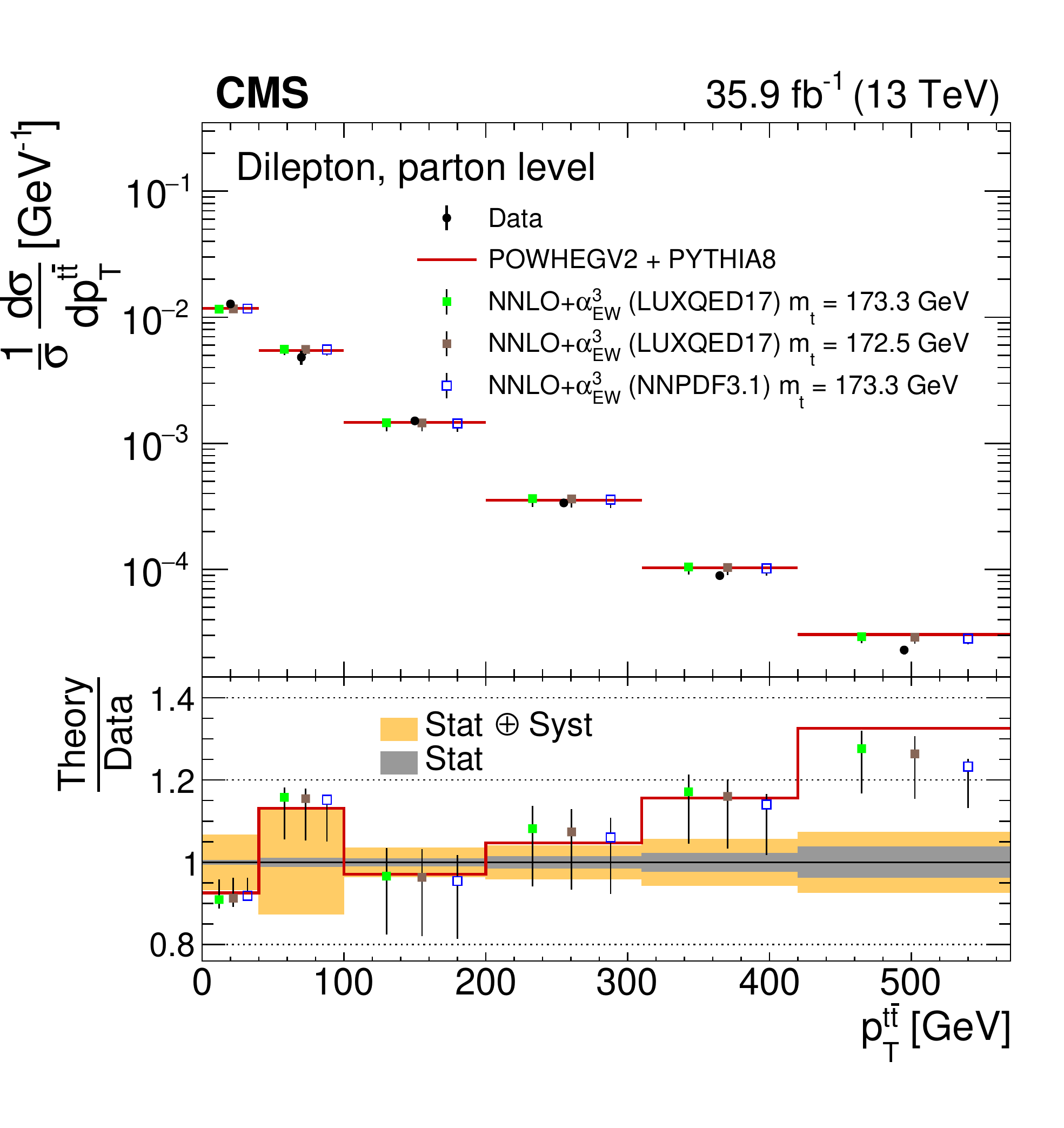}
\caption{The differential \ttbar production cross sections at the parton level in the full phase space as a function of \pttt are shown for the data (filled circles), the theoretical predictions with beyond-NLO precision (other points) and the prediction from \pwhgpy (solid line). The vertical lines on the filled circles and other points indicate the total uncertainty in the data and theoretical predictions, respectively.  The left and right plots correspond to absolute and normalised measurements, respectively. The lower panel in each plot shows the ratios of the theoretical predictions to the data. The dark and light bands show the relative statistical and total uncertainties in the data, respectively.}
\label{fig:diffxsec:res_ttpt_bnlo}
\includegraphics[width=0.75\textwidth]{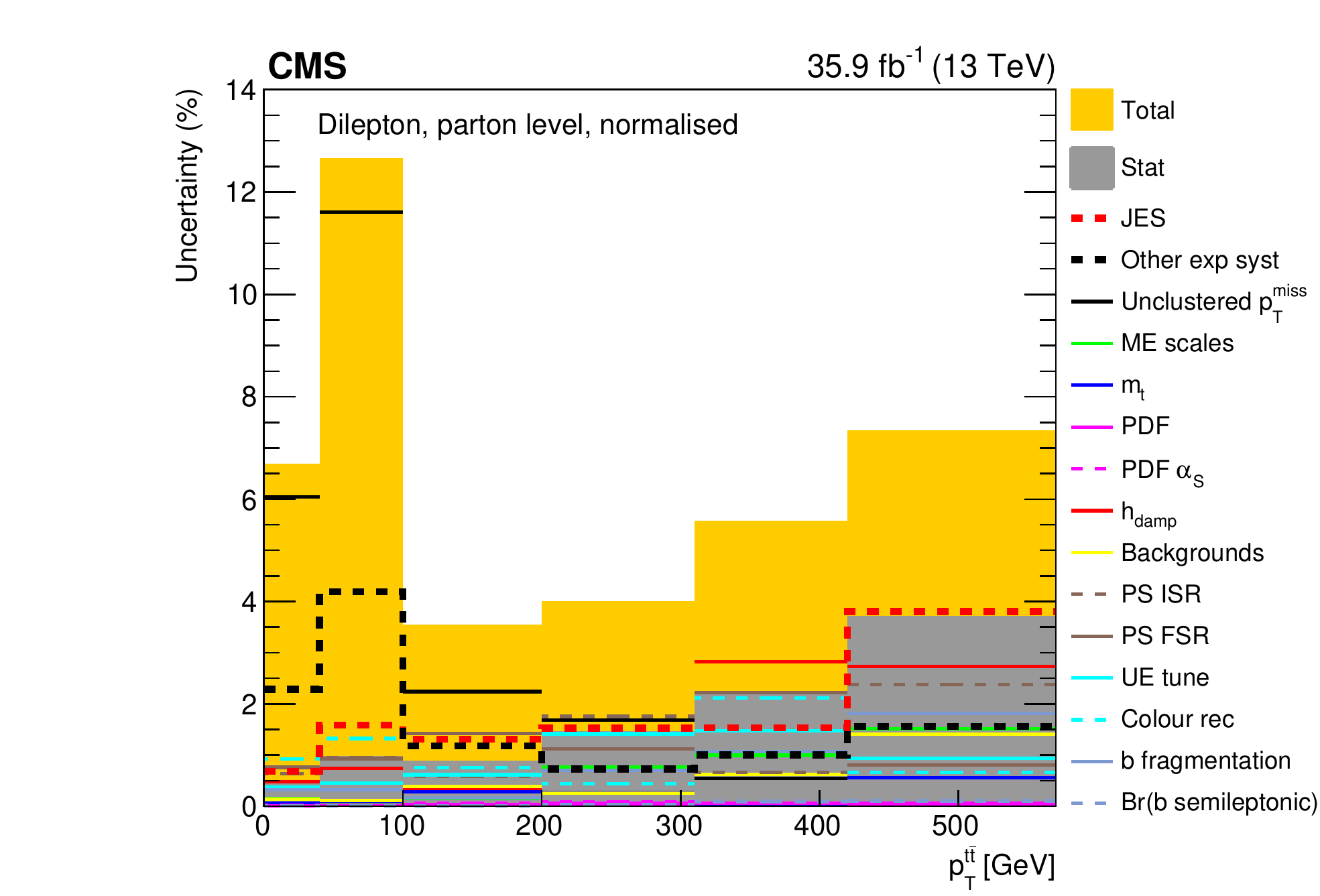}
\caption{The contributions of each source of systematic uncertainty to the total systematic uncertainty in each bin is shown for the measurement of the normalised \ttbar production cross sections  as a function of \pttt. The sources affecting the JES are added in quadrature and shown as a single component. As the contribution from unclustered \ptmiss\ is dominant in lower bins, it is shown separately. Additional experimental systematic uncertainties are added in quadrature and shown as a single component. Contributions from theoretical uncertainties are shown separately. Related theoretical uncertainties are grouped by colour.}
\label{fig:diffxsec:unc_breakdown_res_ttpt}
\end{figure*}

\begin{figure*}[!phtb]
\centering
\includegraphics[width=0.49\textwidth]{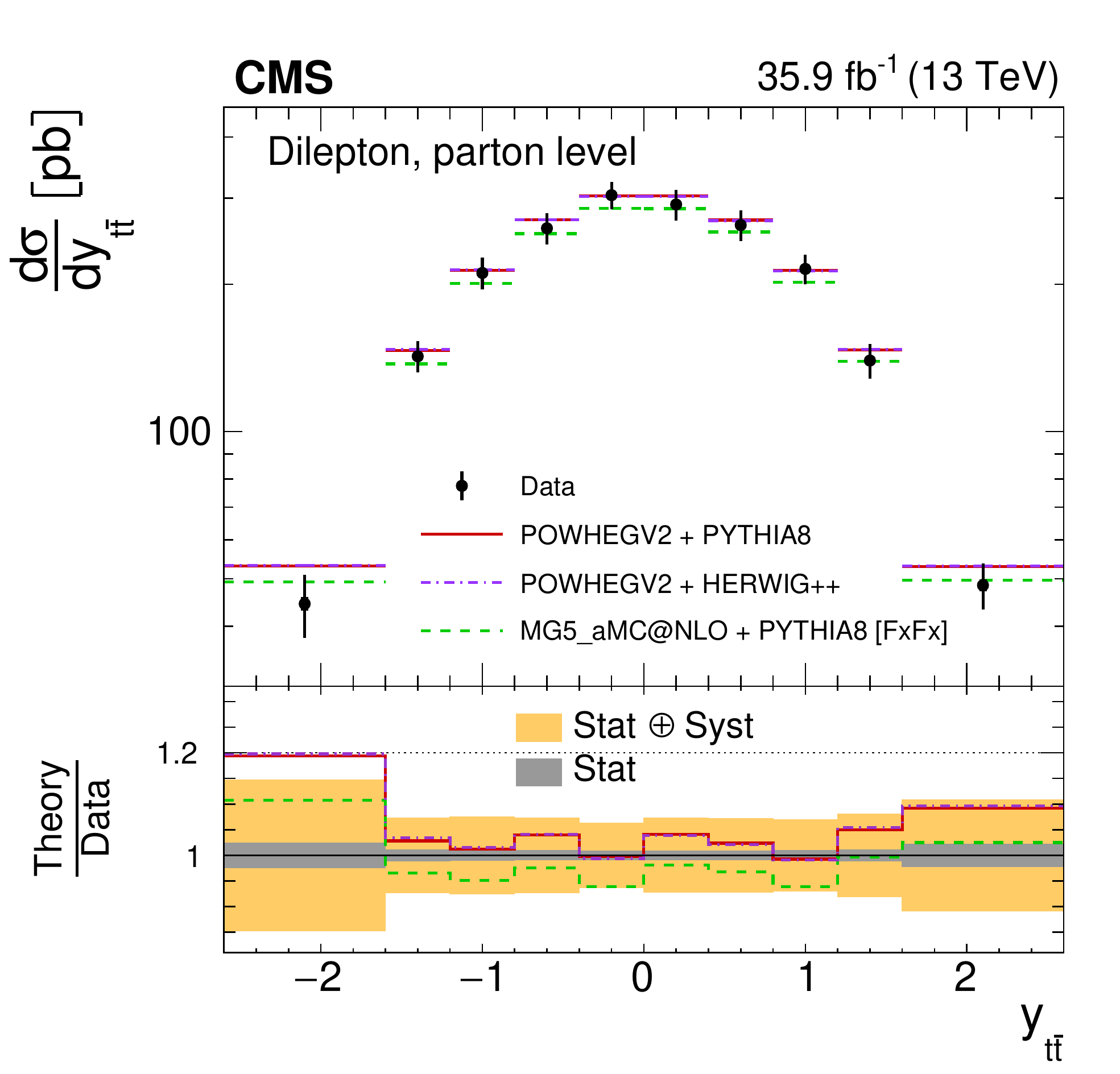}
\includegraphics[width=0.49\textwidth]{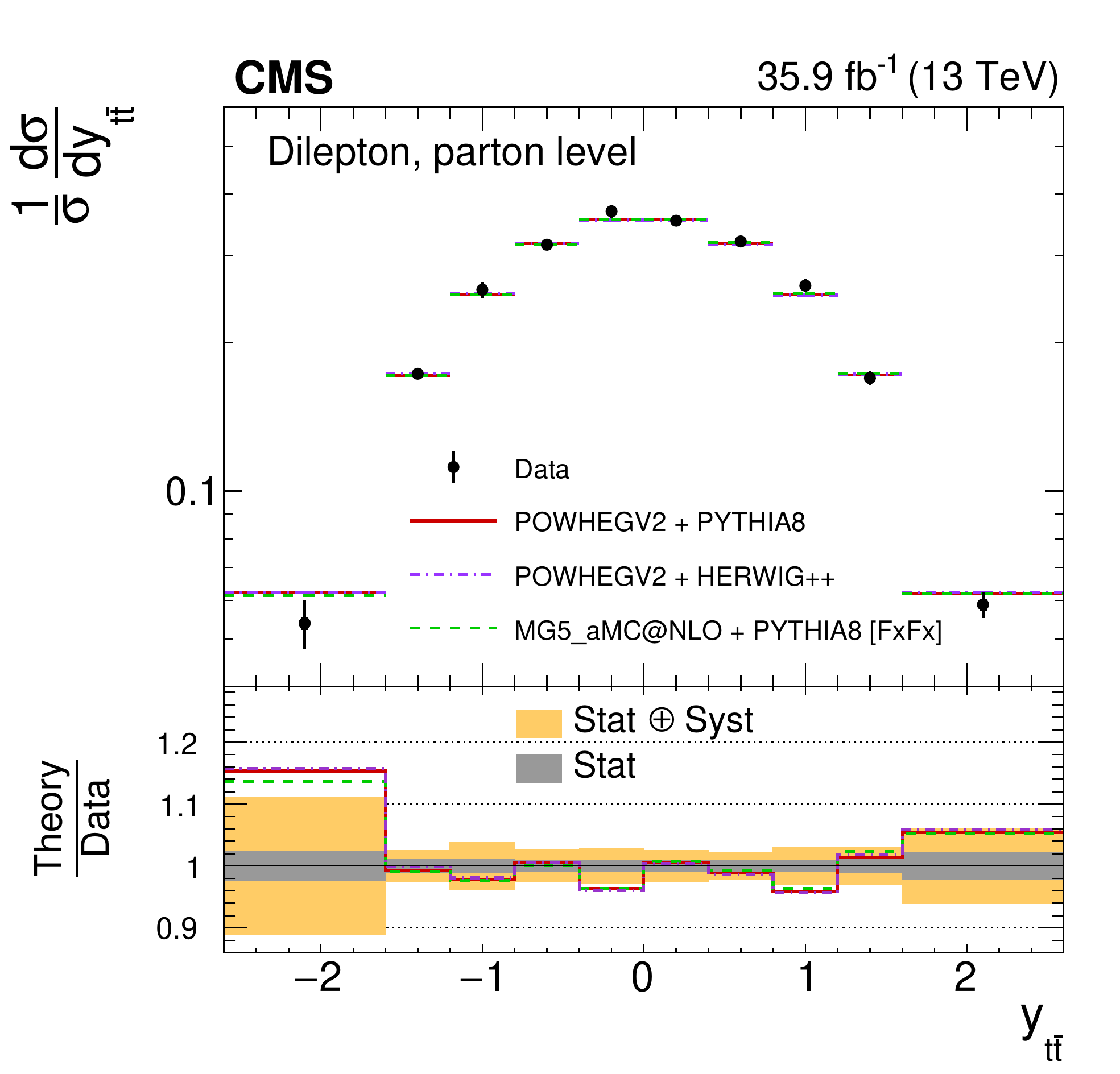} \\
\includegraphics[width=0.49\textwidth]{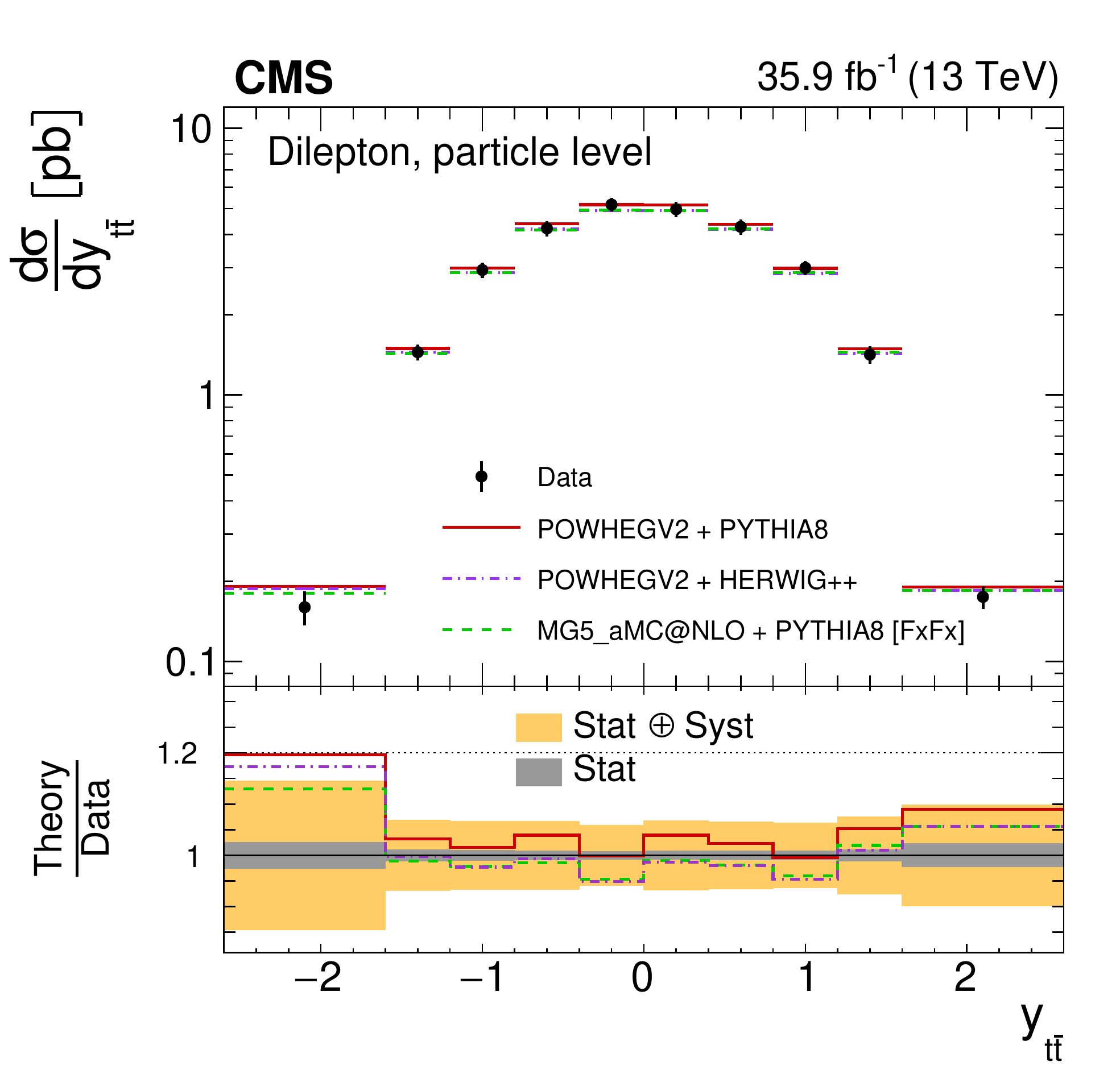}
\includegraphics[width=0.49\textwidth]{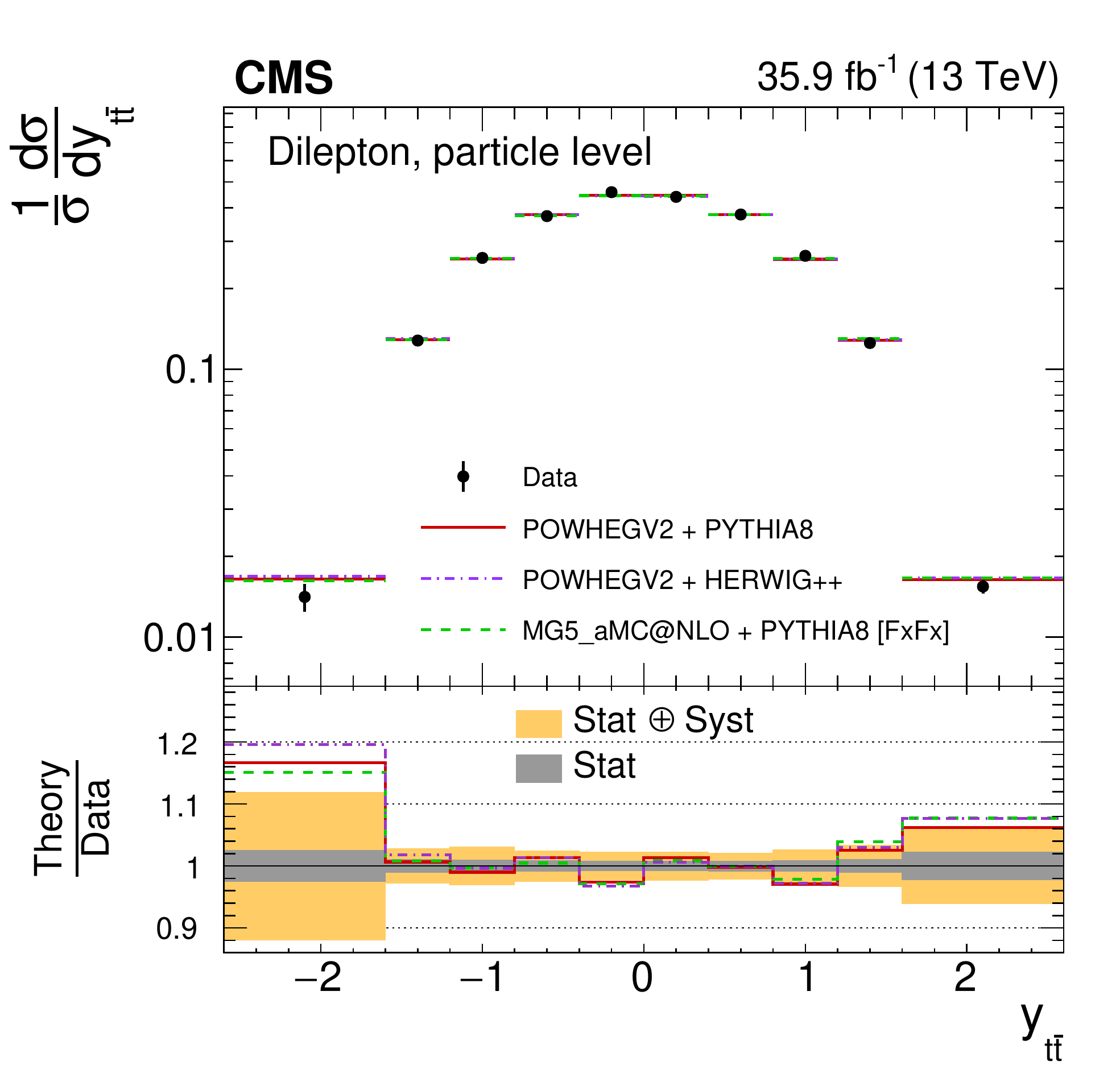}
\caption{The differential \ttbar production cross sections as a function of \ytt are shown for the data (points) and the MC predictions (lines). The vertical lines on the points indicate the total uncertainty in the data. The left and right columns correspond to absolute and normalised measurements, respectively. The upper row corresponds to measurements at the parton level in the full phase space and the lower row to the particle level in a fiducial phase space. The lower panel in each plot shows the ratios of the theoretical predictions to the data. The dark and light bands show the relative statistical and total uncertainties in the data, respectively.}
\label{fig:diffxsec:res_tty}
\end{figure*}

\begin{figure*}[!phtb]
\centering
\includegraphics[width=0.49\textwidth]{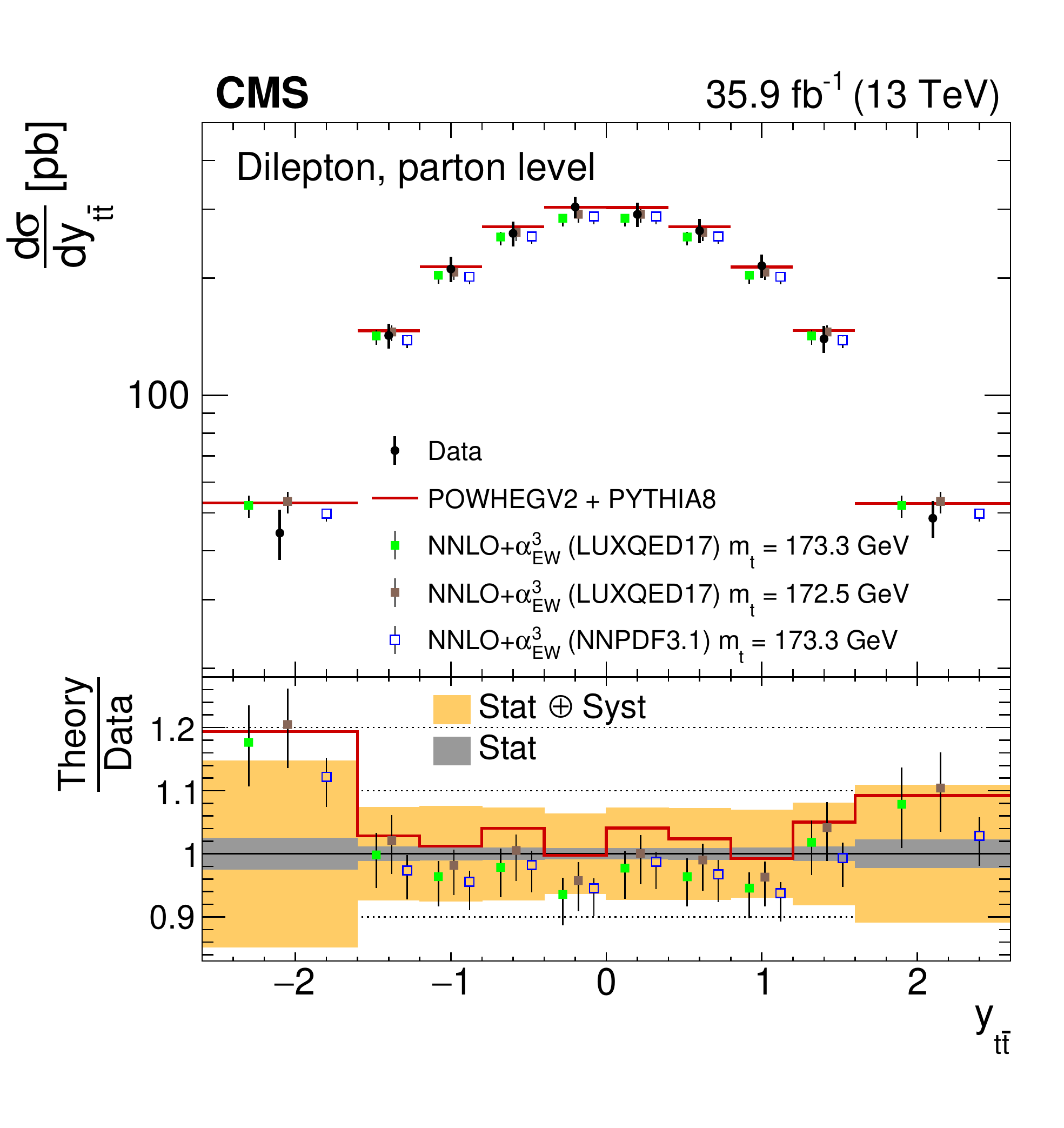}
\includegraphics[width=0.49\textwidth]{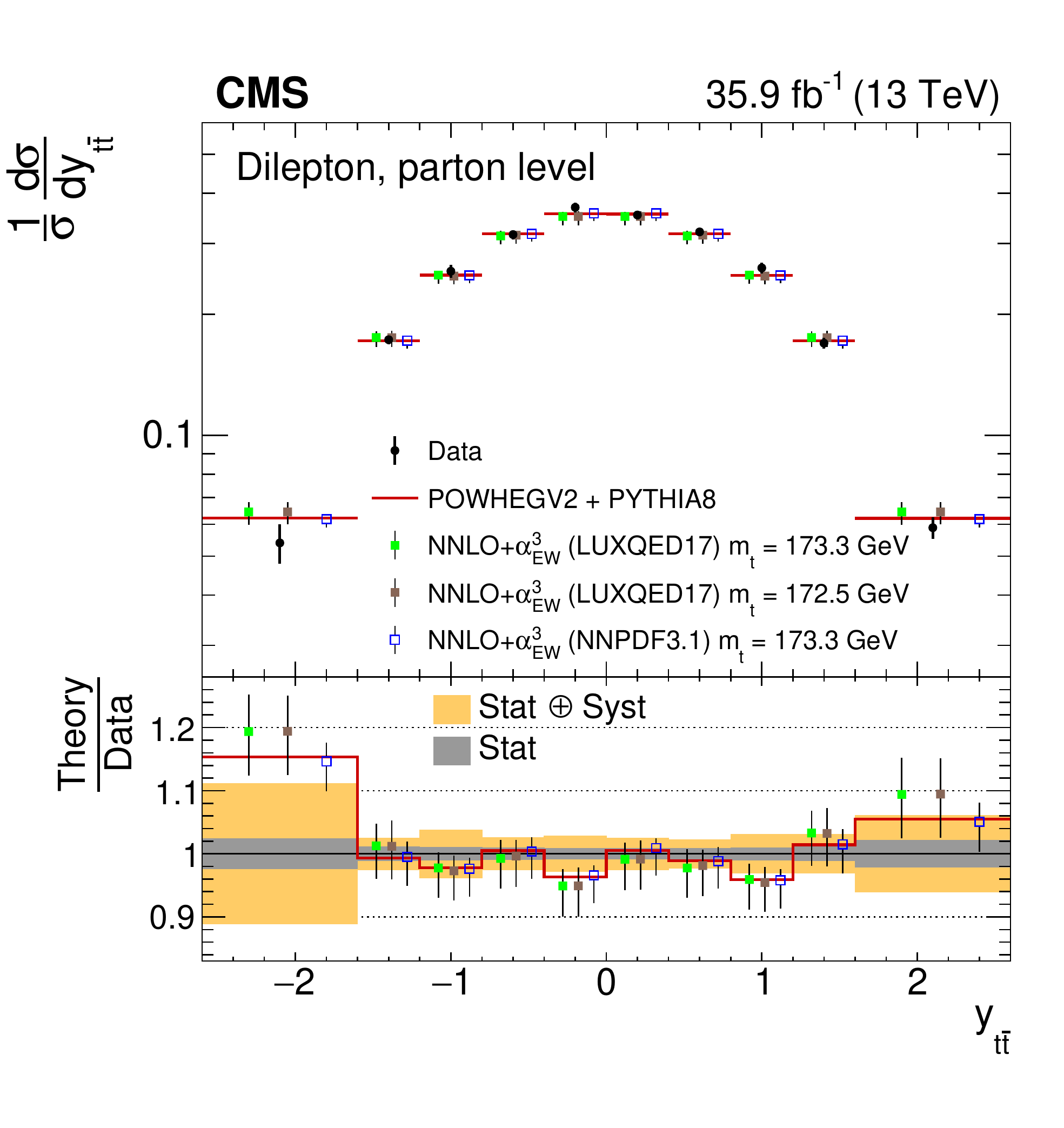}
\caption{The differential \ttbar production cross sections at the parton level in the full phase space as a function of \ytt are shown for the data (filled circles), the theoretical predictions with beyond-NLO precision (other points) and the prediction from \pwhgpy (solid line). The vertical lines on the filled circles and other points indicate the total uncertainty in the data and theoretical predictions, respectively. The left and right plots correspond to absolute and normalised measurements, respectively. The lower panel in each plot shows the ratios of the theoretical predictions to the data. The dark and light bands show the relative statistical and total uncertainties in the data, respectively.}
\label{fig:diffxsec:res_tty_bnlo}
\includegraphics[width=0.75\textwidth]{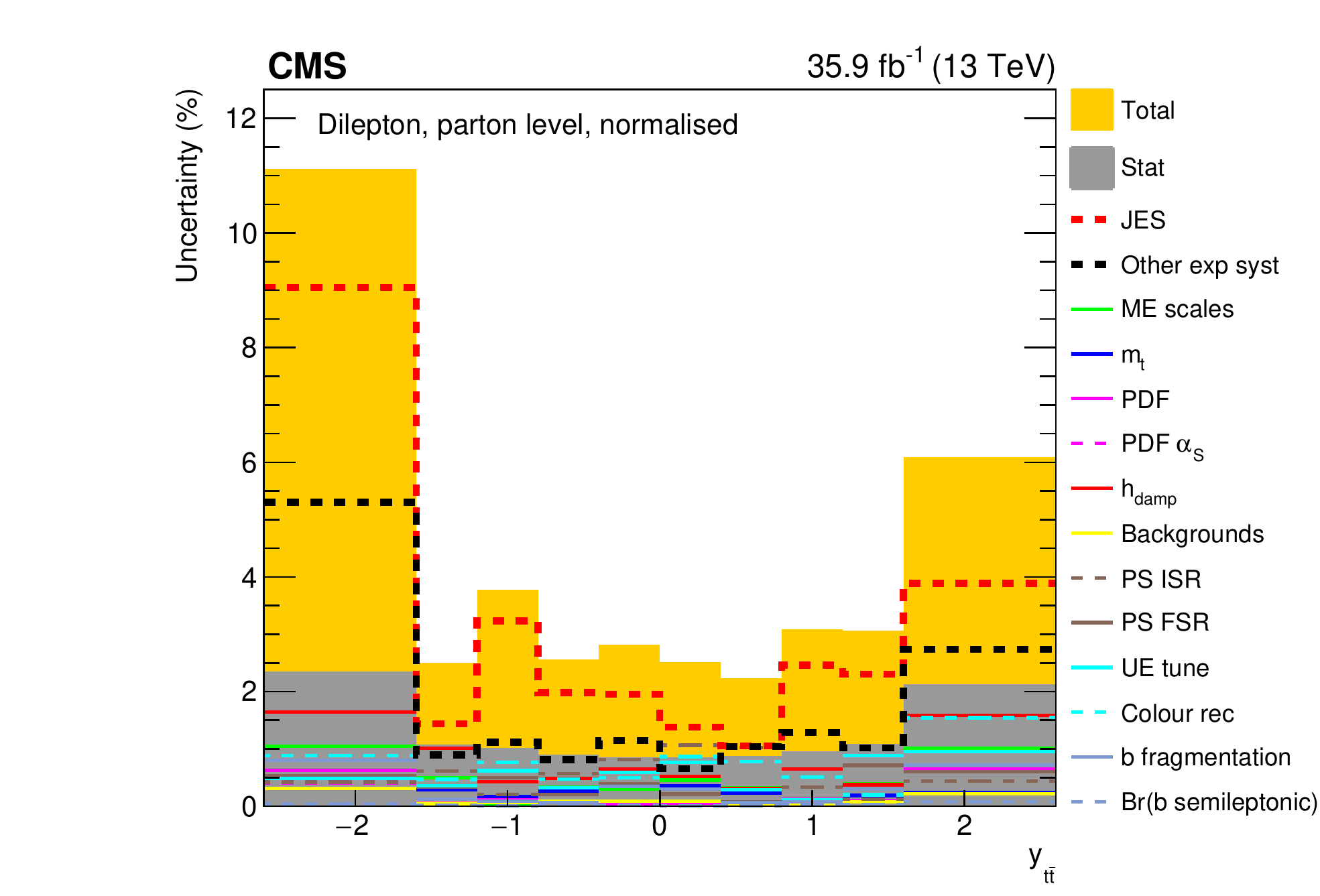}
\caption{The contributions of each source of systematic uncertainty to the total systematic uncertainty in each bin is shown for the measurement of the normalised \ttbar production cross sections  as a function of \ytt. The sources affecting the JES are added in quadrature and shown as a single component. Additional experimental systematic uncertainties are also added in quadrature and shown as a single component. Contributions from theoretical uncertainties are shown separately. The statistical and total uncertainties, corresponding to the quadrature addition of statistical and systematic uncertainties, are shown by the dark and light filled histograms, respectively.}
\label{fig:diffxsec:unc_breakdown_res_tty}
\end{figure*}

\begin{figure*}[!phtb]
\centering
\includegraphics[width=0.49\textwidth]{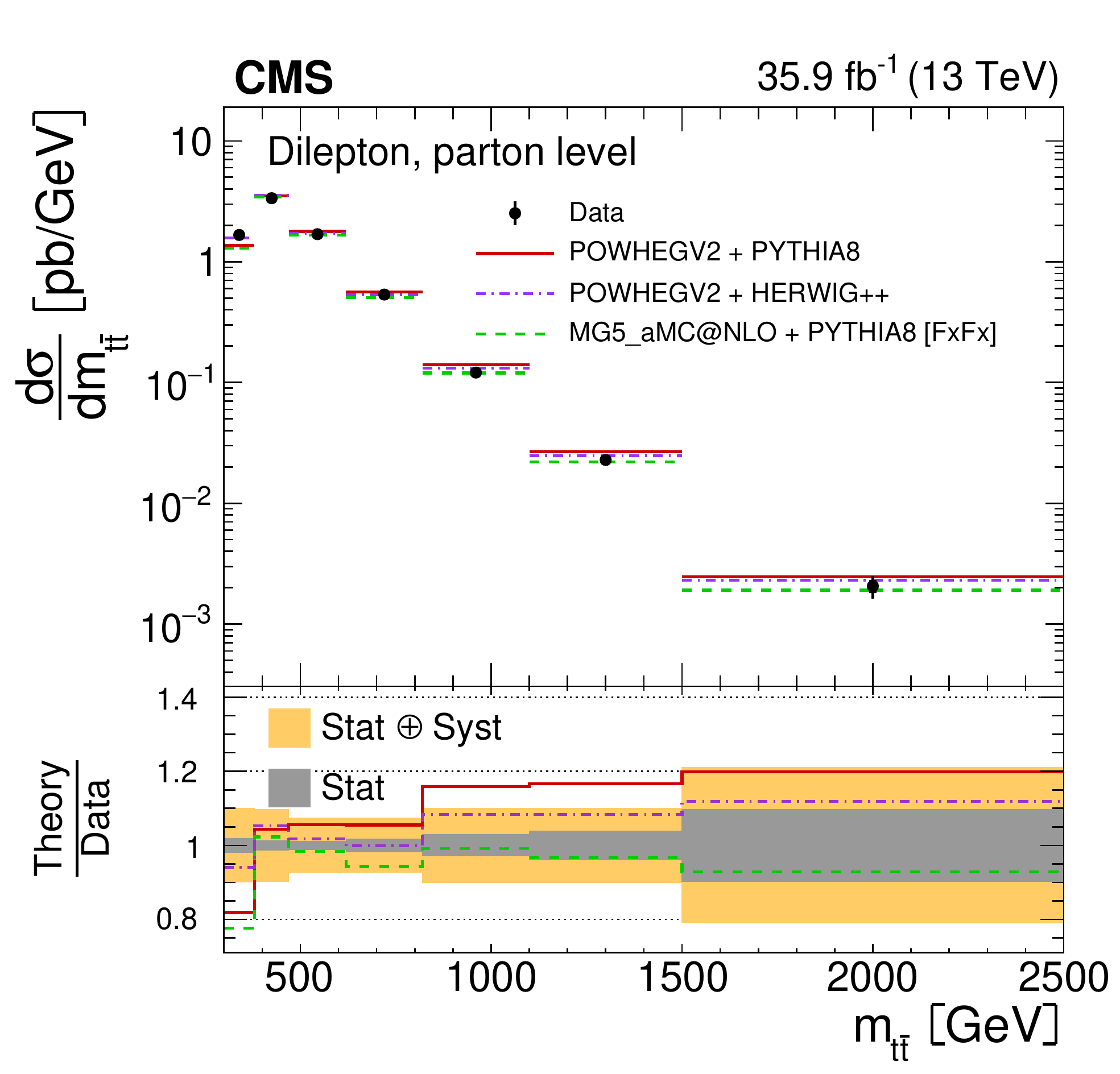}
\includegraphics[width=0.49\textwidth]{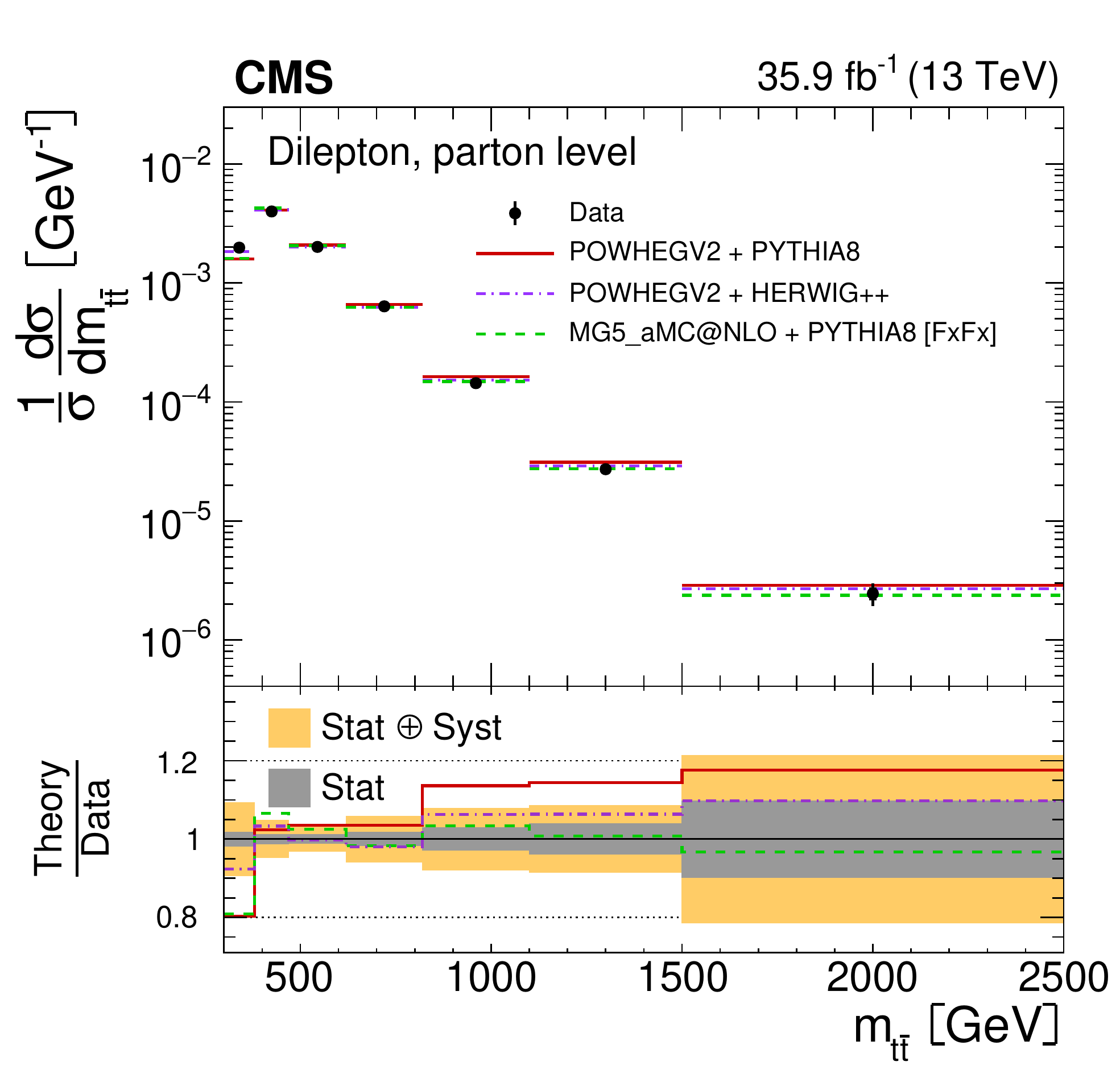} \\
\includegraphics[width=0.49\textwidth]{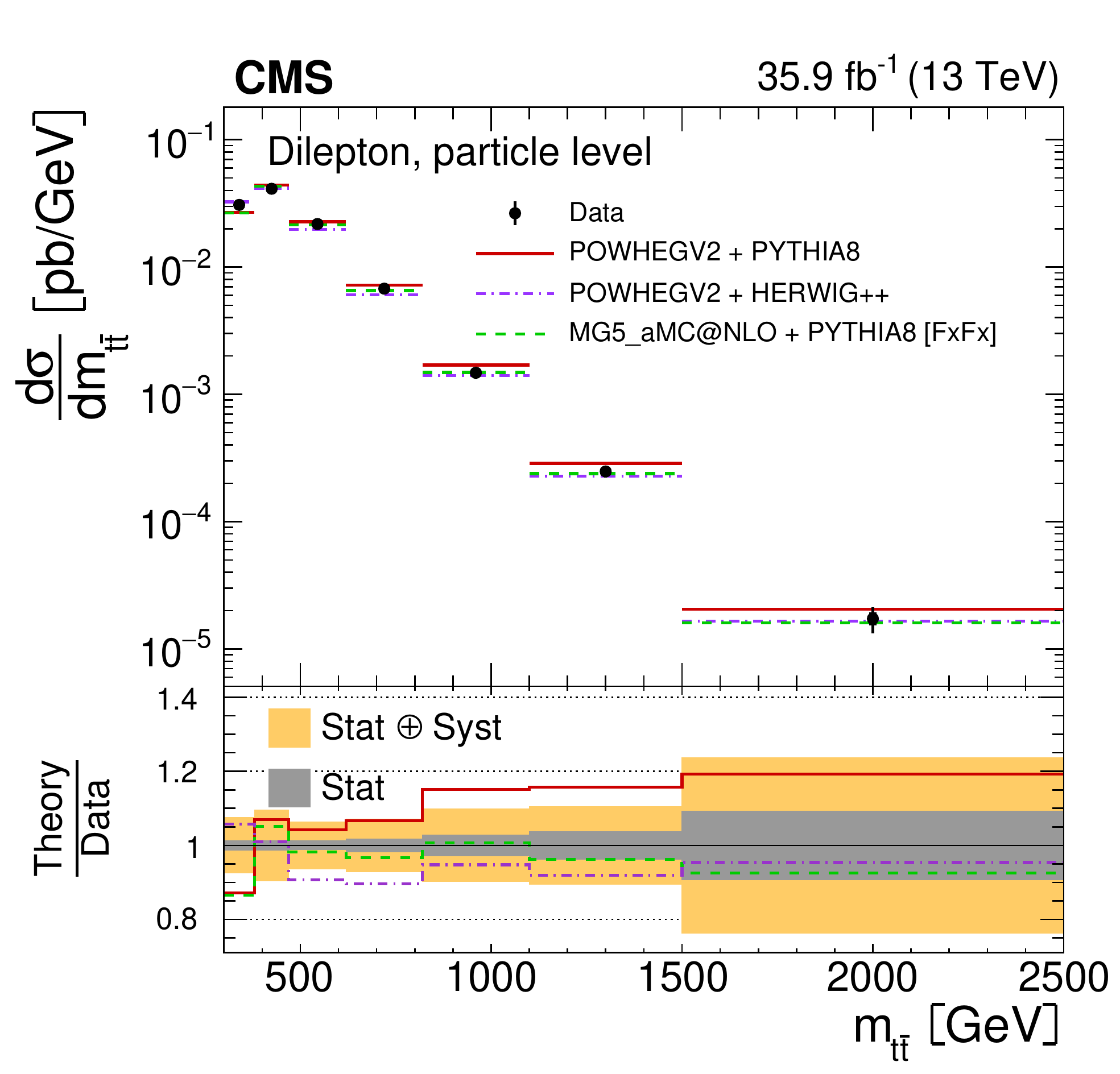}
\includegraphics[width=0.49\textwidth]{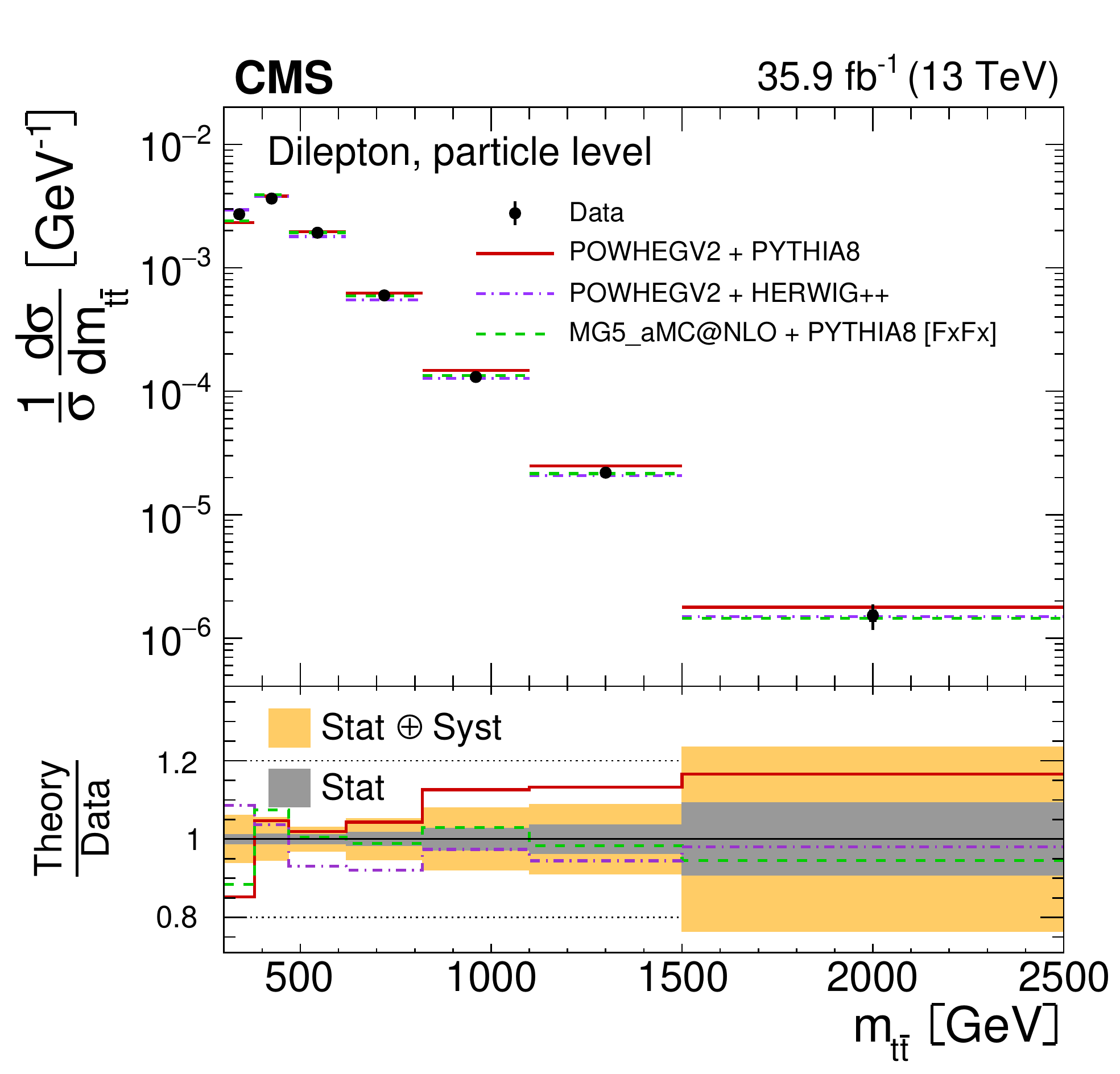}
\caption{The differential \ttbar production cross sections as a function of \mtt are shown for the data (points) and the MC predictions (lines). The vertical lines on the points indicate the total uncertainty in the data. The left and right columns correspond to absolute and normalised measurements, respectively. The upper row corresponds to measurements at the parton level in the full phase space and the lower row to the particle level in a fiducial phase space. The lower panel in each plot shows the ratios of the theoretical predictions to the data. The dark and light bands show the relative statistical and total uncertainties in the data, respectively.}
\label{fig:diffxsec:res_mtt}
\end{figure*}

\begin{figure*}[!phtb]
\centering
\includegraphics[width=0.49\textwidth]{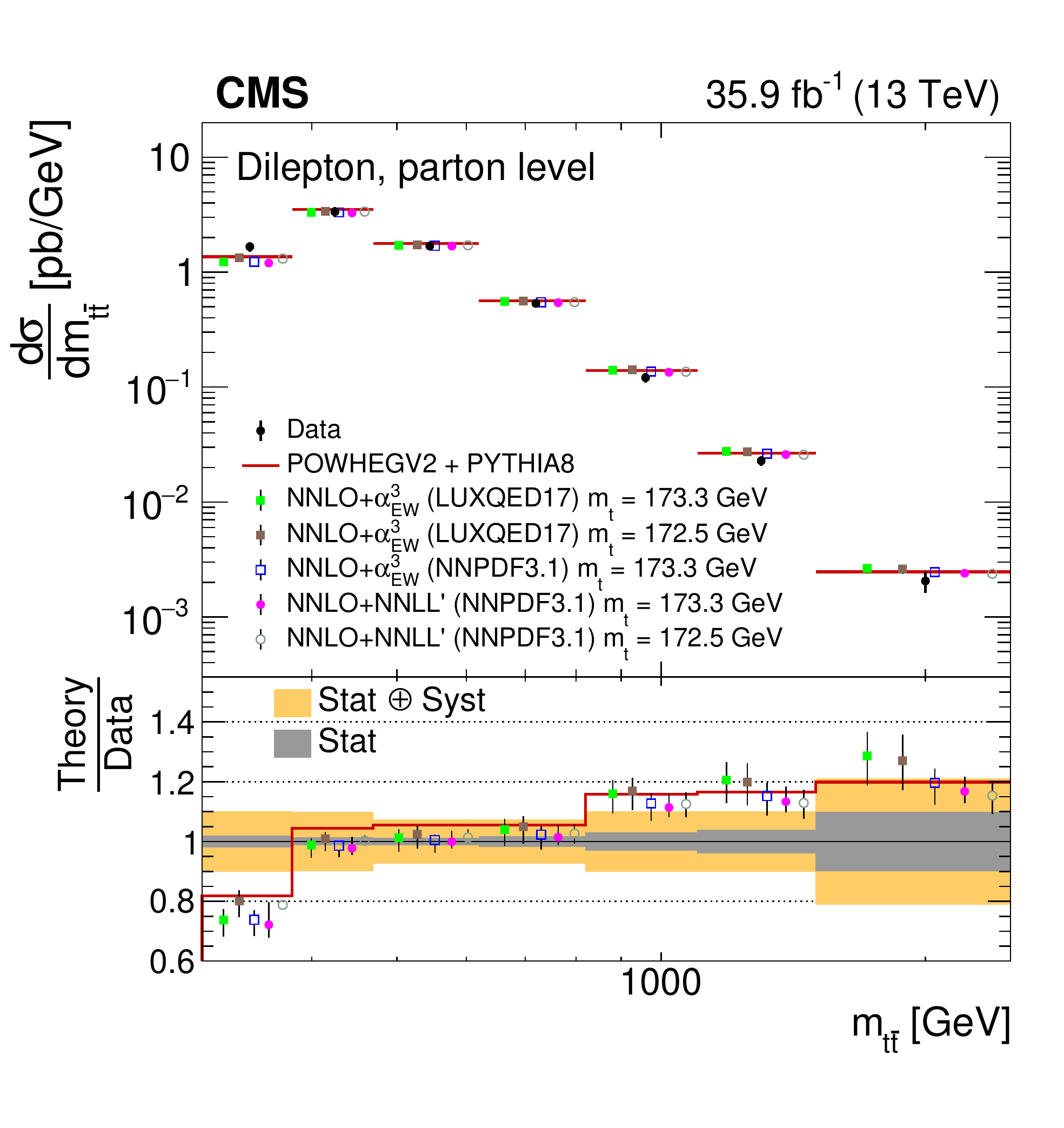}
\includegraphics[width=0.49\textwidth]{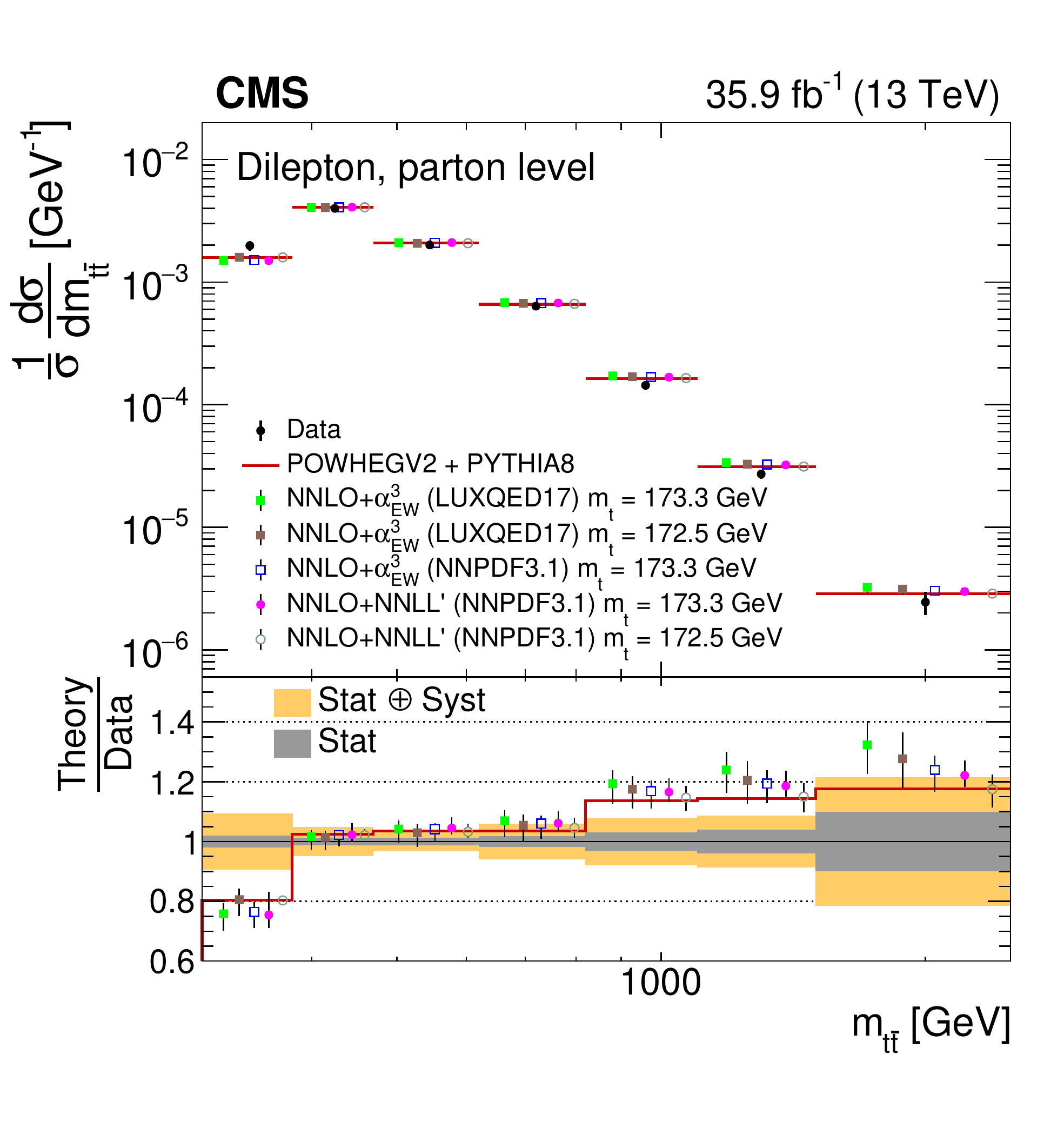}
\caption{The differential \ttbar production cross sections at the parton level in the full phase space as a function of \mtt are shown for the data (filled circles), the theoretical predictions with beyond-NLO precision (other points) and the prediction from \pwhgpy (solid line). The vertical lines on the filled circles and other points indicate the total uncertainty in the data and theoretical predictions, respectively.  The left and right plots correspond to absolute and normalised measurements, respectively. The lower panel in each plot shows the ratios of the theoretical predictions to the data. The dark and light bands show the relative statistical and total uncertainties in the data, respectively.}
\label{fig:diffxsec:res_mtt_bnlo}
\includegraphics[width=0.75\textwidth]{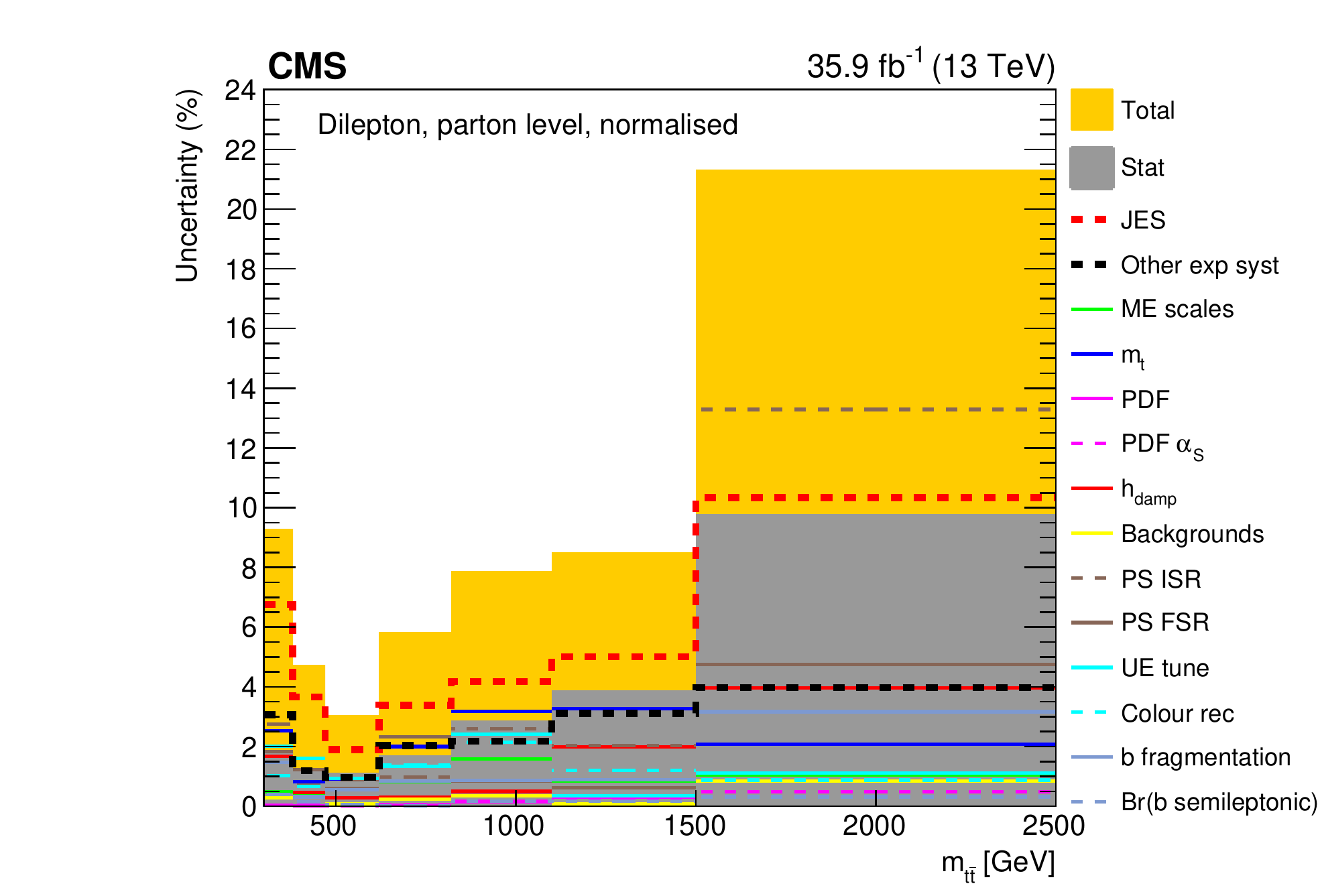}
\caption{The contributions of each source of systematic uncertainty to the total systematic uncertainty in each bin is shown for the measurement of the normalised \ttbar production cross sections  as a function of \mtt. The sources affecting the JES are added in quadrature and shown as a single component. Additional experimental systematic uncertainties are also added in quadrature and shown as a single component. Contributions from theoretical uncertainties are shown separately. The statistical and total uncertainties, corresponding to the quadrature addition of statistical and systematic uncertainties, are shown by the dark and light filled histograms, respectively.}
\label{fig:diffxsec:unc_breakdown_res_mtt}
\end{figure*}

All measured differential cross sections, along with figures giving the contribution of each source of uncertainty to the total uncertainty for selected normalised parton-level measurements, are shown in Figs.~\ref{fig:diffxsec:res_toppt}--\ref{fig:diffxsec:unc_breakdown_res_mtt}. Absolute and normalised results at the particle and parton levels for a given observable are grouped together in each figure. Within the figure, the upper row corresponds to the parton-level measurement in the full phase space, and the lower row to the particle-level measurement in the fiducial phase space. The left column corresponds to the absolute measurement and the right column to the normalised measurement. In each plot the top panel shows the measured differential cross section with the predictions overlayed and the bottom panel shows the ratios of the predictions to the measured distribution and the statistical and total uncertainties in the measured distribution. When predictions with beyond-NLO precision are available, additional figures with comparisons of these predictions to data are included. In addition, the numerical values of the measured differential cross sections in each bin and associated uncertainties for all observables are tabulated in Tables \ref{tab:norm_parton0}--\ref{tab:norm_parton13} in Appendix \ref{app:tables} for parton level and in Tables \ref{tab:norm_particle0}--\ref{tab:norm_particle32} in Appendix \ref{app:tables_particle} for particle level.

\begin{figure}[!phtb]
\centering
\includegraphics[width=0.49\textwidth]{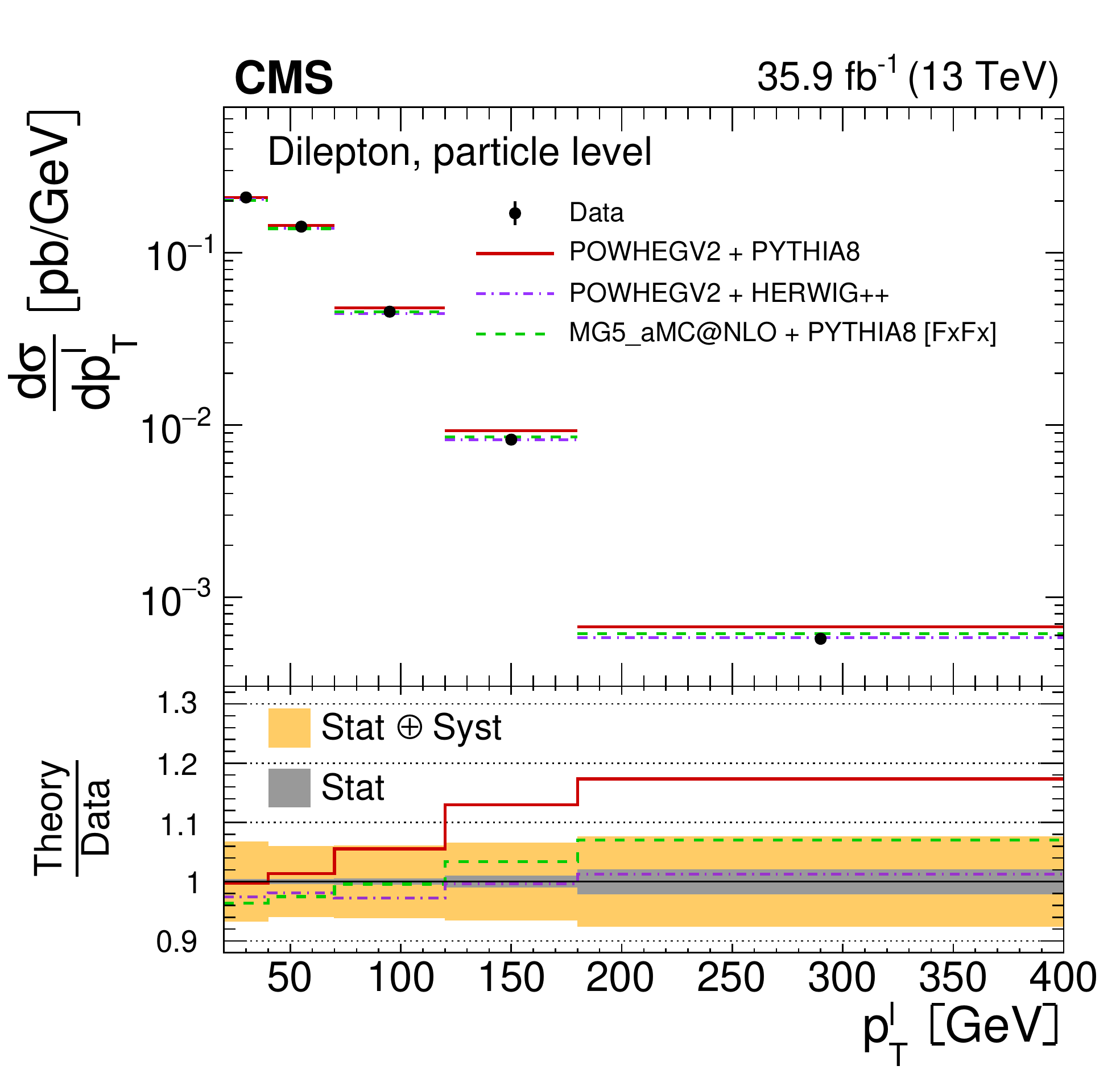}
\includegraphics[width=0.49\textwidth]{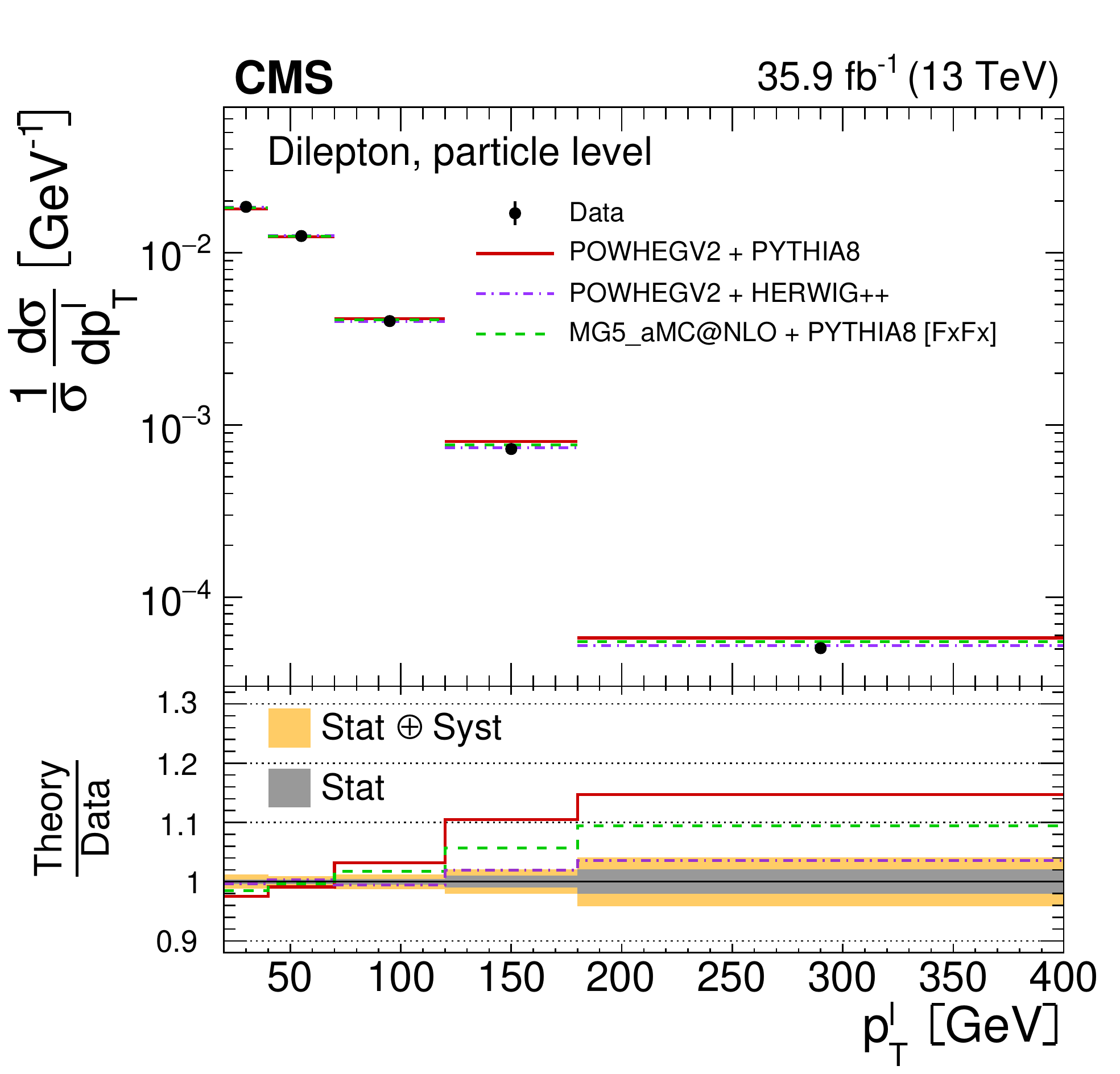}
\caption{The differential \ttbar production cross sections as a function of \ptlep in a fiducial phase space at the particle level are shown for the data (points) and the MC predictions (lines). The vertical lines on the points indicate the total uncertainty in the data. The left and right plots correspond to absolute and normalised measurements, respectively. The lower panel in each plot shows the ratios of the theoretical predictions to the data. The dark and light bands show the relative statistical and total uncertainties in the data, respectively.}
\label{fig:diffxsec:res_ptlep}
\end{figure}

\begin{figure}[!phtb]
\centering
\includegraphics[width=0.49\textwidth]{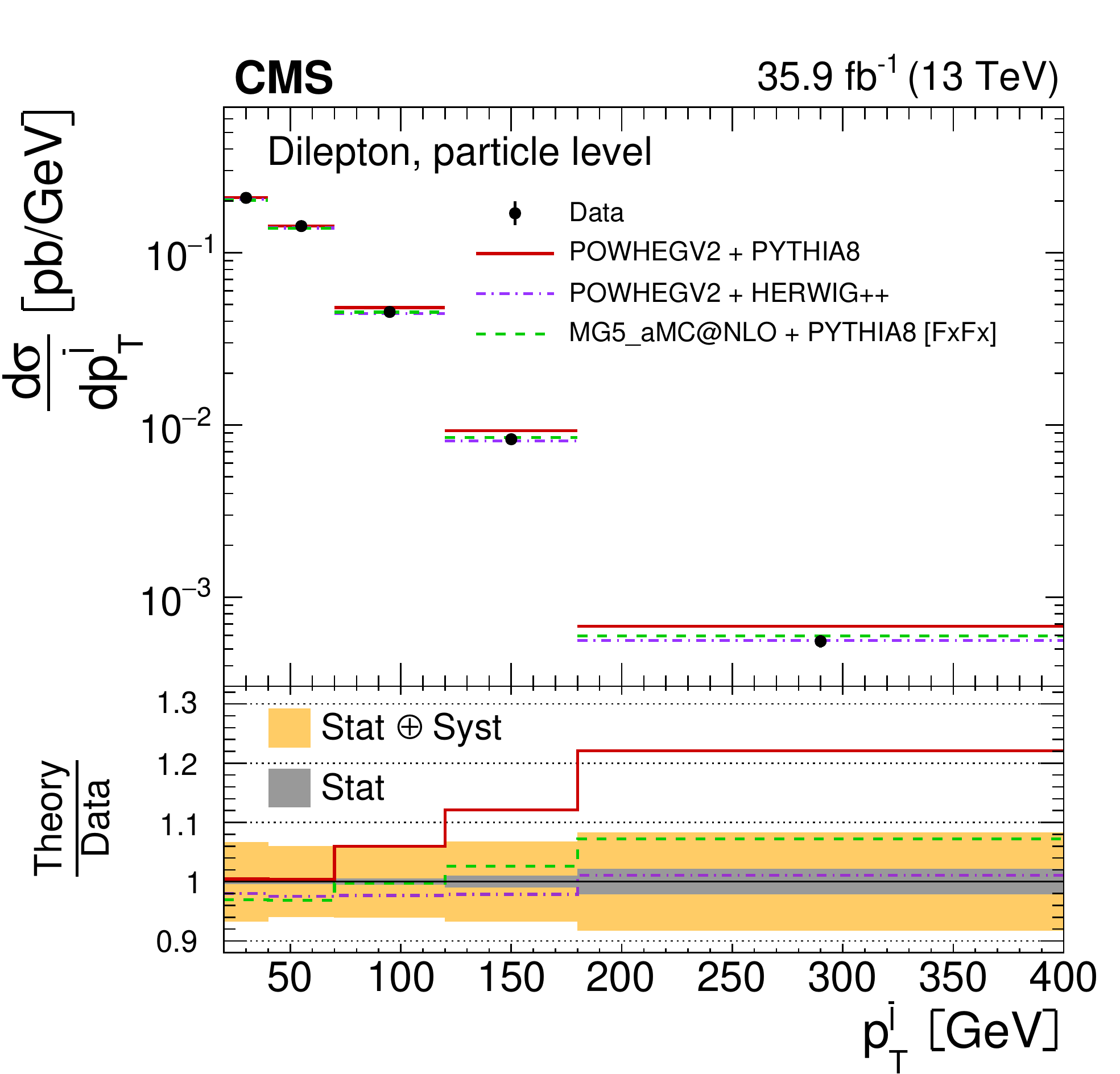}
\includegraphics[width=0.49\textwidth]{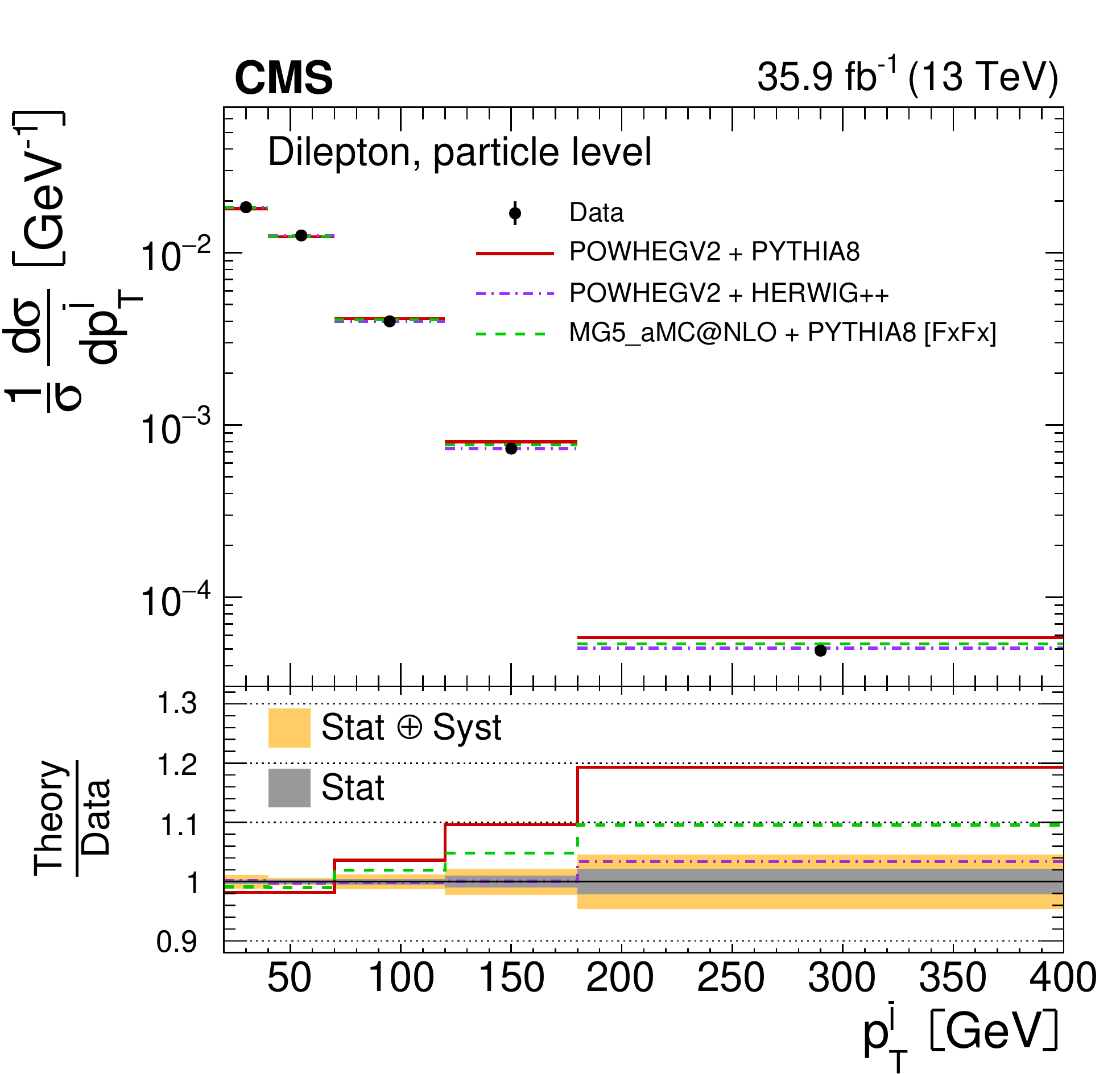}
\caption{The differential \ttbar production cross sections as a function of \ptalep in a fiducial phase space at the particle level are shown for the data (points) and the MC predictions (lines). The vertical lines on the points indicate the total uncertainty in the data. The left and right plots correspond to absolute and normalised measurements, respectively. The lower panel in each plot shows the ratios of the theoretical predictions to the data. The dark and light bands show the relative statistical and total uncertainties in the data, respectively.}
\label{fig:diffxsec:res_ptantilep}
\end{figure}

\begin{figure}[!phtb]
\centering
\includegraphics[width=0.49\textwidth]{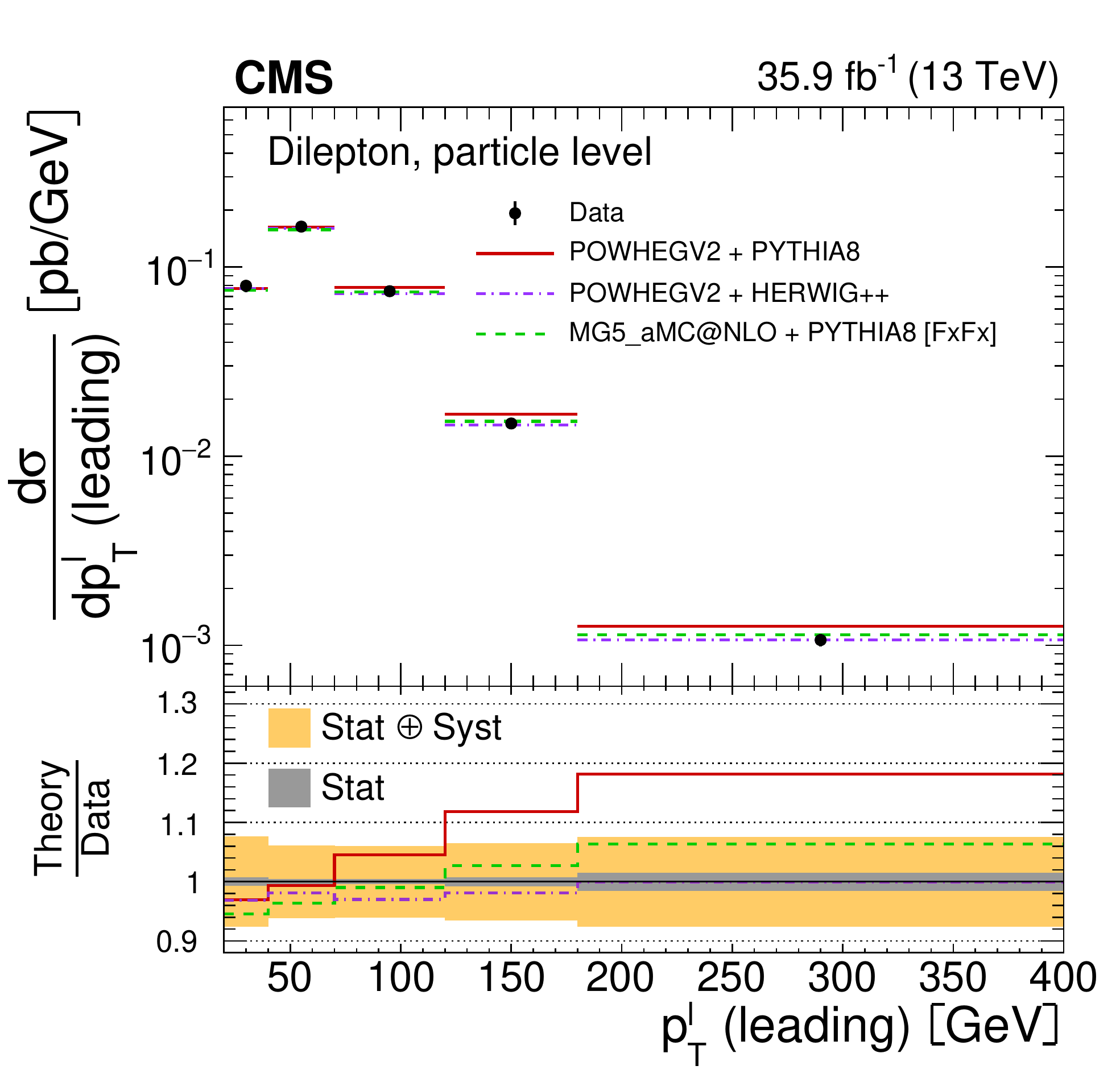}
\includegraphics[width=0.49\textwidth]{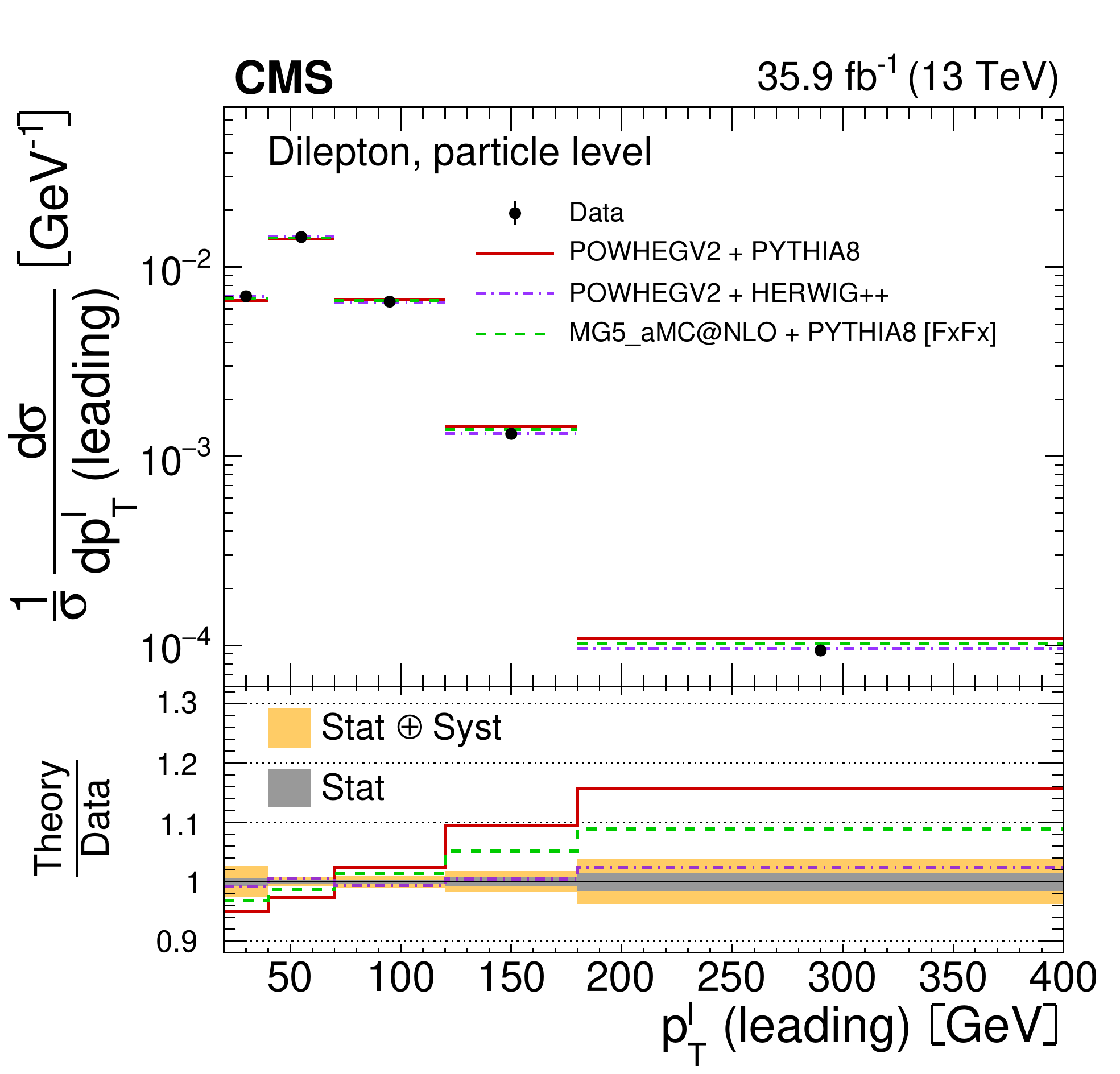}
\caption{The differential \ttbar production cross sections as a function of \ptlep (leading) in a fiducial phase space at the particle level are shown for the data (points) and the MC predictions (lines). The vertical lines on the points indicate the total uncertainty in the data. The left and right plots correspond to absolute and normalised measurements, respectively. The lower panel in each plot shows the ratios of the theoretical predictions to the data. The dark and light bands show the relative statistical and total uncertainties in the data, respectively.}
\label{fig:diffxsec:res_ptlepleading}
\end{figure}

\begin{figure}[!phtb]
\centering
\includegraphics[width=0.49\textwidth]{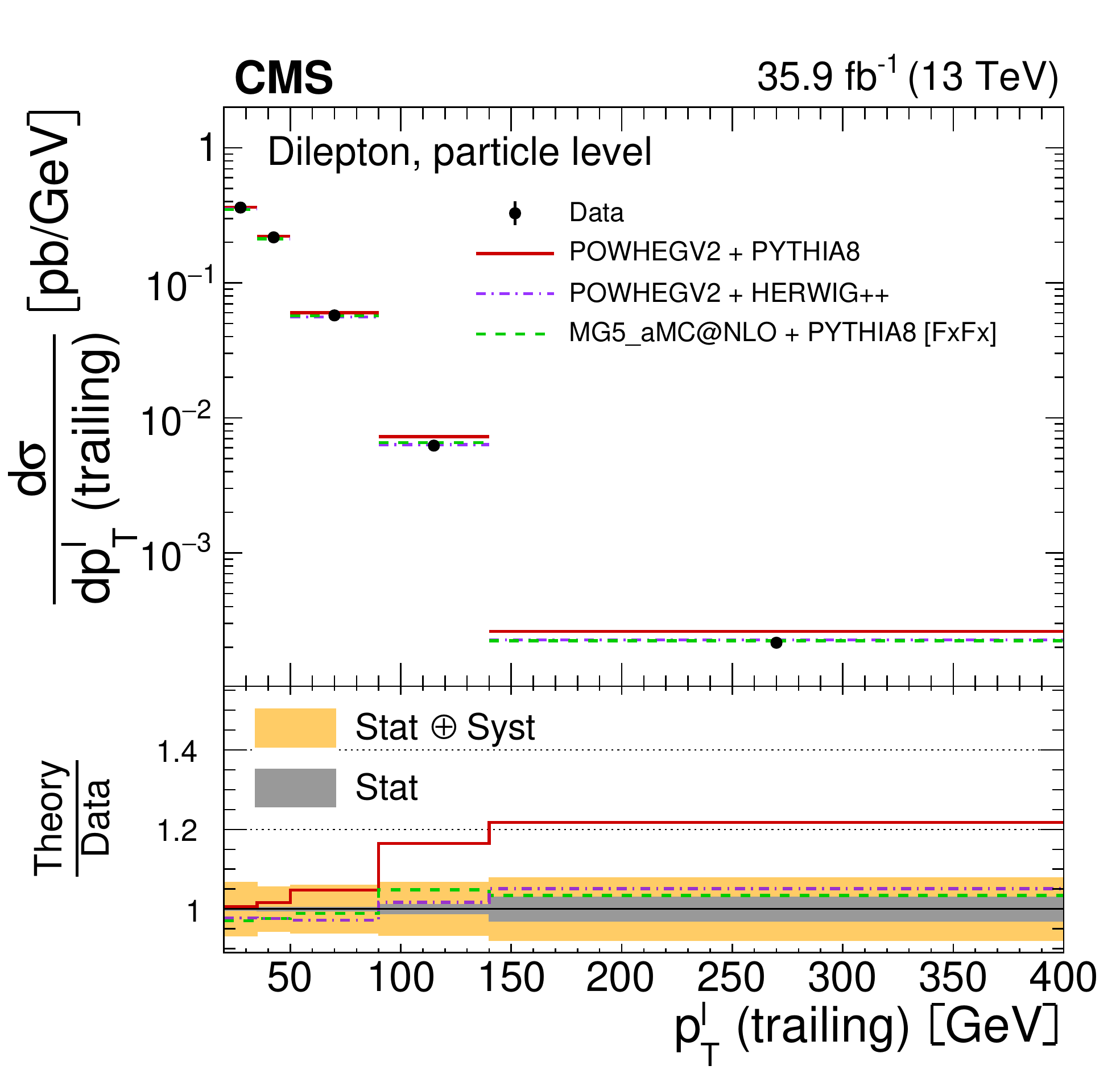}
\includegraphics[width=0.49\textwidth]{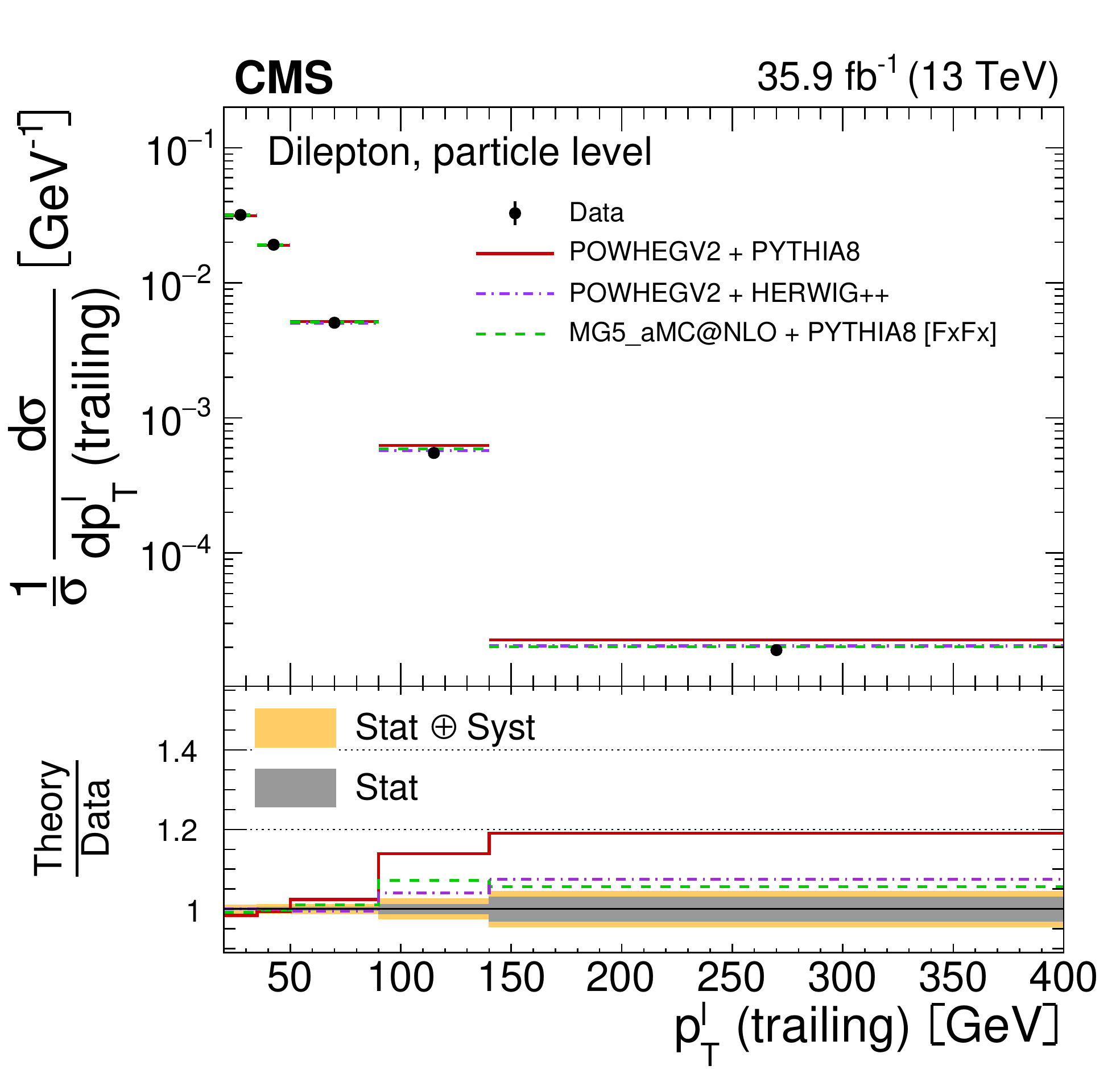}
\caption{The differential \ttbar production cross sections as a function of \ptlep (trailing) in a fiducial phase space at the particle level are shown for the data (points) and the MC predictions (lines). The vertical lines on the points indicate the total uncertainty in the data. The left and right plots correspond to absolute and normalised measurements, respectively. The lower panel in each plot shows the ratios of the theoretical predictions to the data. The dark and light bands show the relative statistical and total uncertainties in the data, respectively.}
\label{fig:diffxsec:res_ptleptsubleading}
\end{figure}

\begin{figure}[!phtb]
\centering
\includegraphics[width=0.49\textwidth]{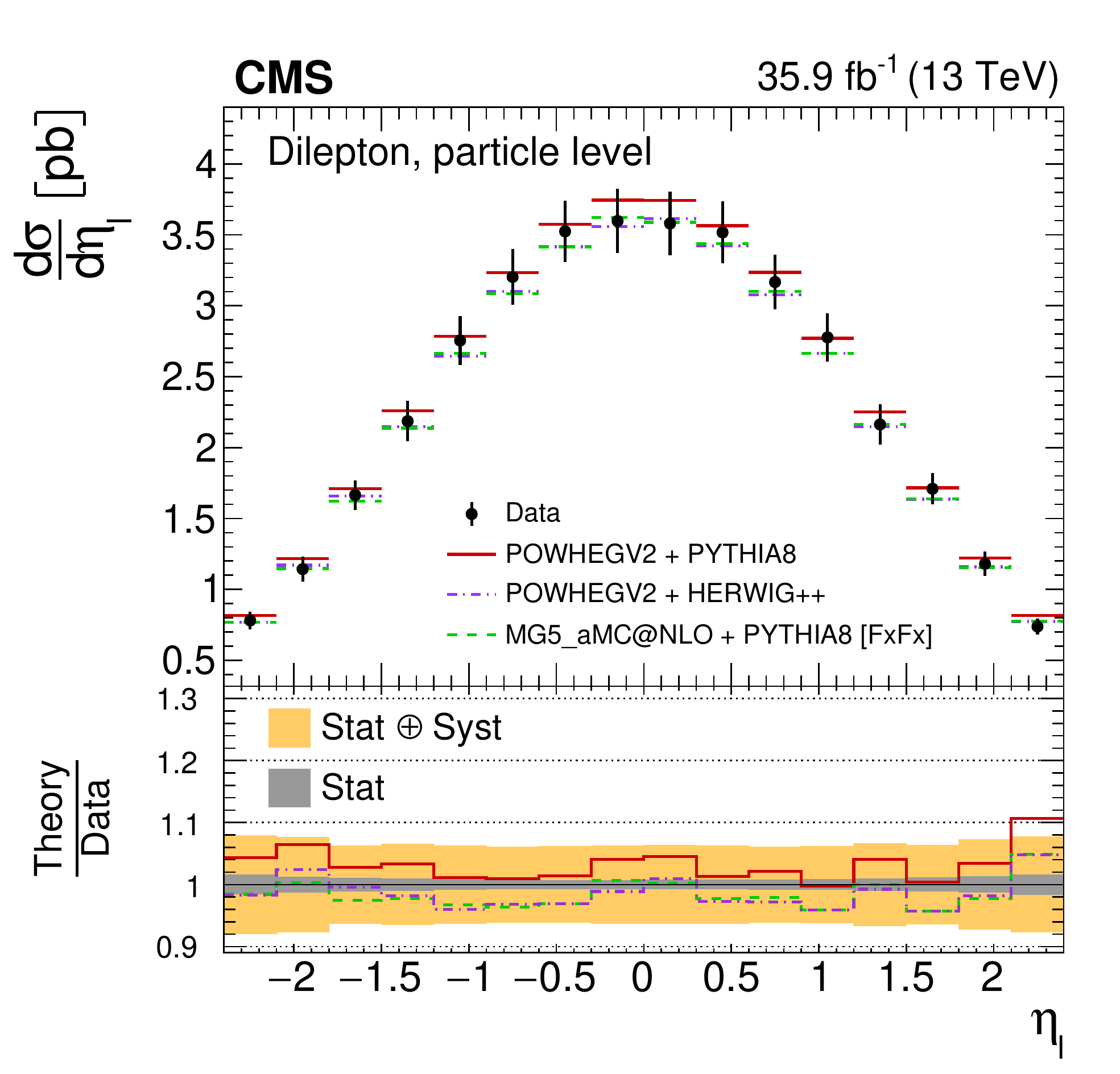}
\includegraphics[width=0.49\textwidth]{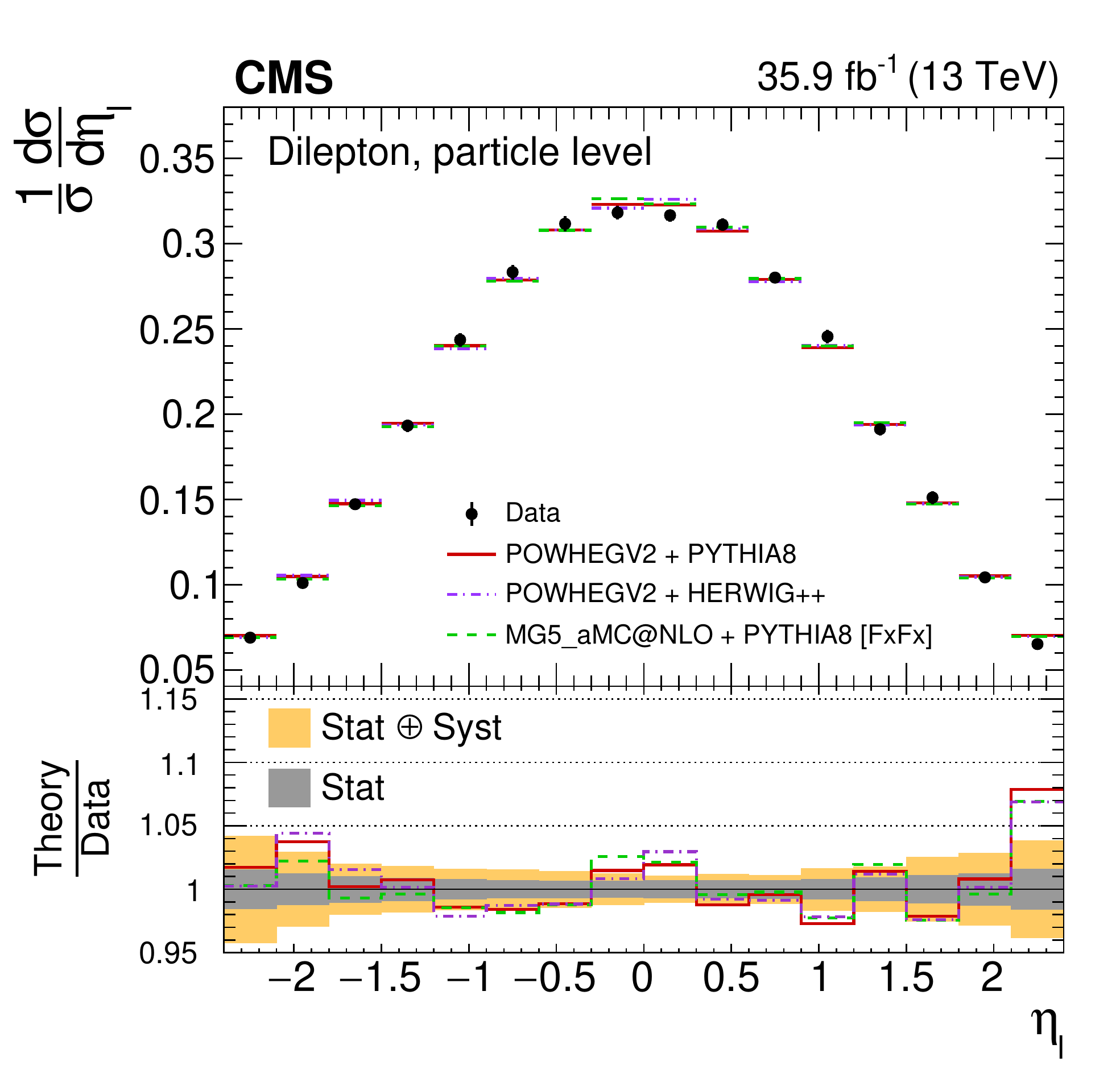}
\caption{The differential \ttbar production cross sections as a function of \etalep in a fiducial phase space at the particle level are shown for the data (points) and the MC predictions (lines). The vertical lines on the points indicate the total uncertainty in the data. The left and right plots correspond to absolute and normalised measurements, respectively. The lower panel in each plot shows the ratios of the theoretical predictions to the data. The dark and light bands show the relative statistical and total uncertainties in the data, respectively.}
\label{fig:diffxsec:res_etalep}
\end{figure}

\begin{figure}[!phtb]
\centering
\includegraphics[width=0.49\textwidth]{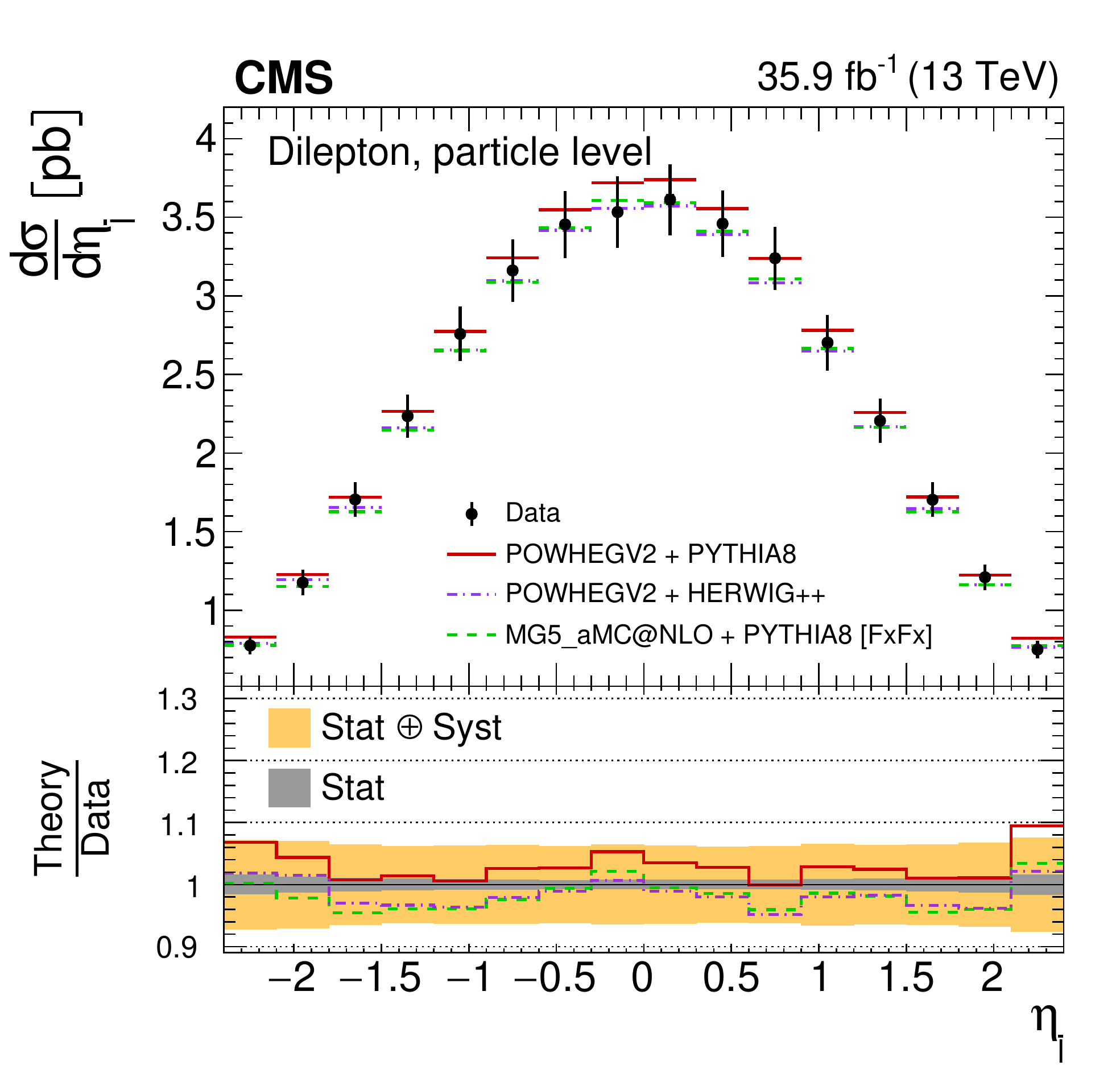}
\includegraphics[width=0.49\textwidth]{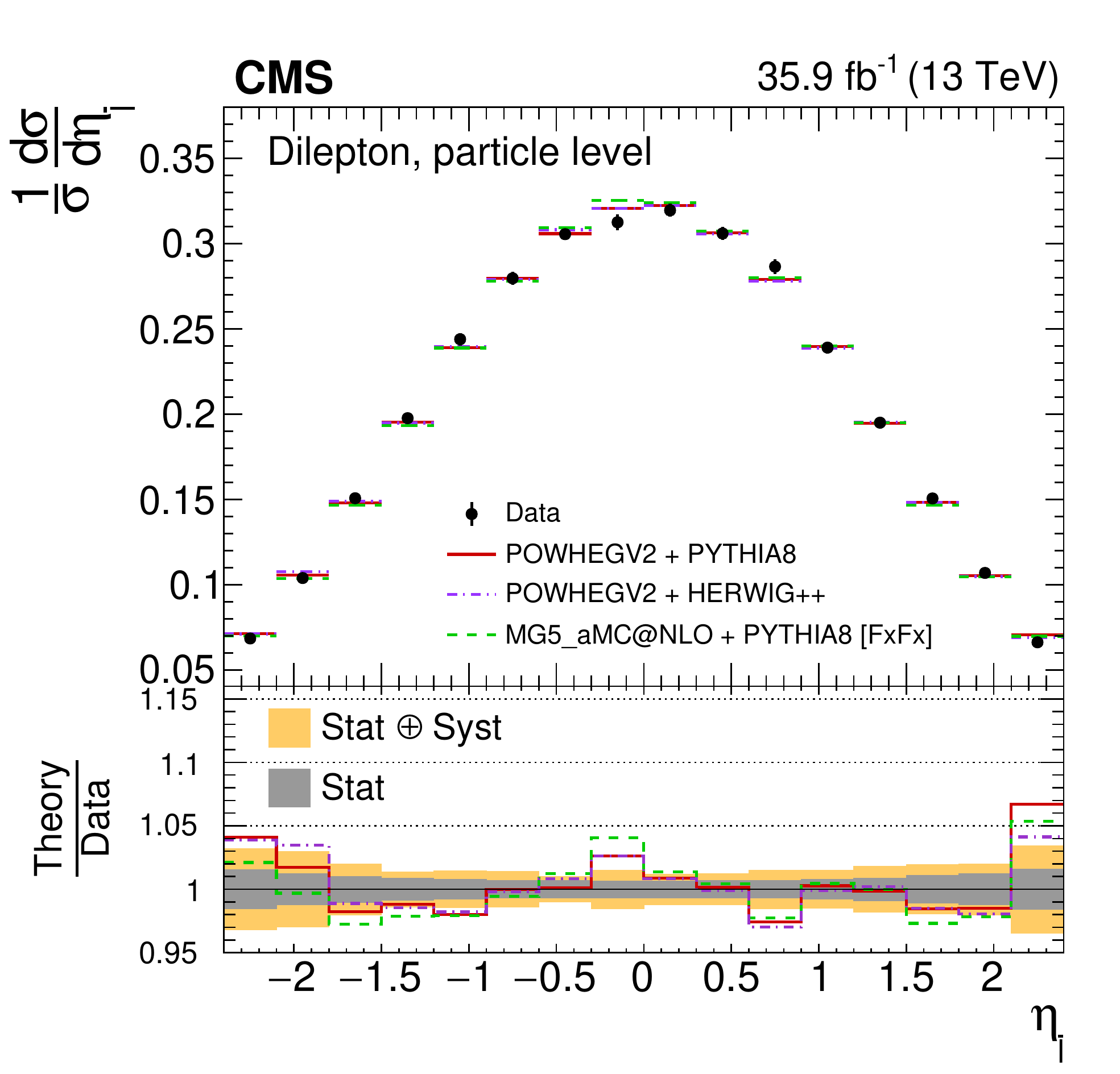}
\caption{The differential \ttbar production cross sections as a function of \etaalep in a fiducial phase space at the particle level are shown for the data (points) and the MC predictions (lines). The vertical lines on the points indicate the total uncertainty in the data. The left and right plots correspond to absolute and normalised measurements, respectively. The lower panel in each plot shows the ratios of the theoretical predictions to the data. The dark and light bands show the relative statistical and total uncertainties in the data, respectively.}
\label{fig:diffxsec:res_etaantilep}
\end{figure}

\begin{figure}[!phtb]
\centering
\includegraphics[width=0.49\textwidth]{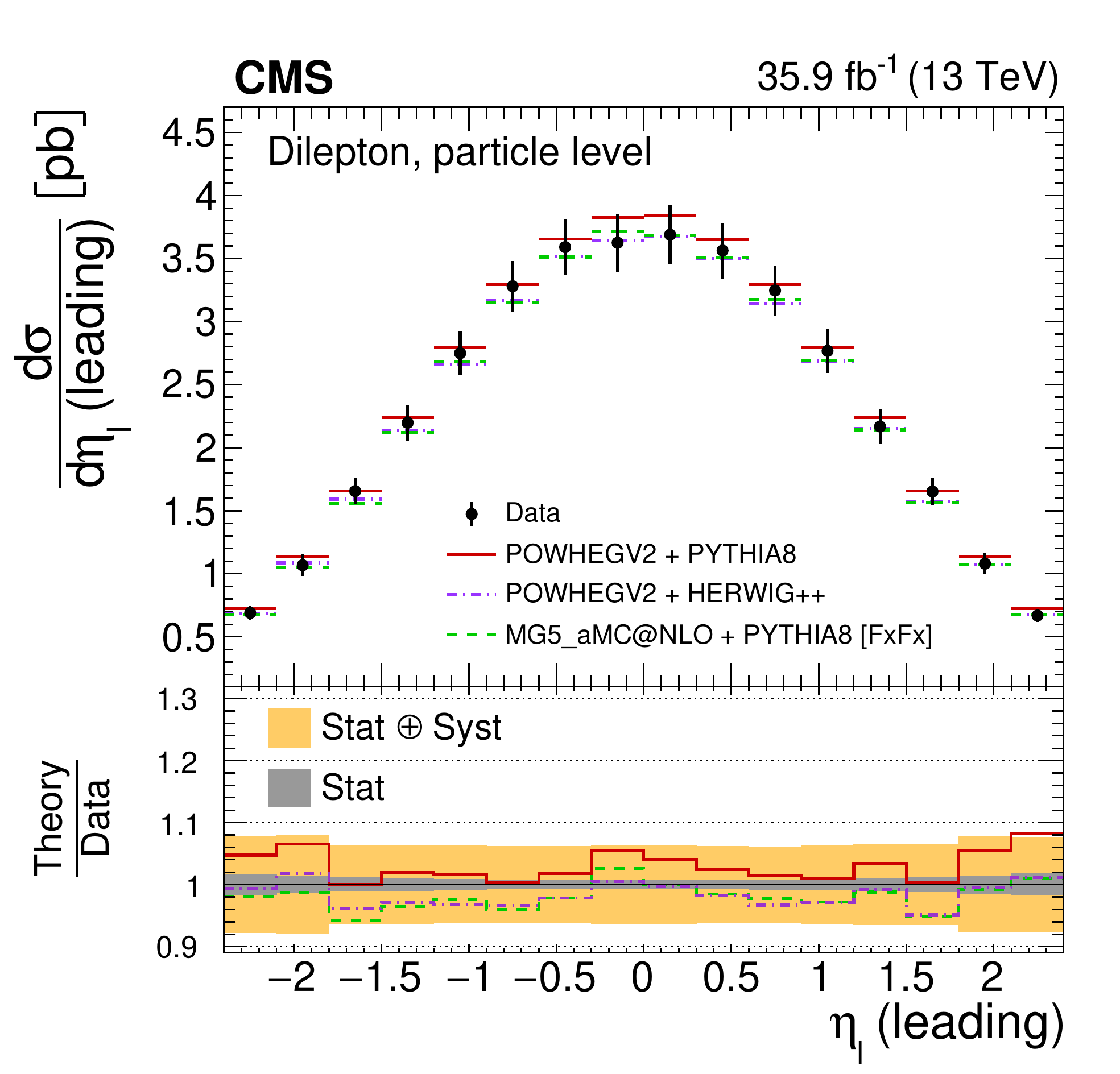}
\includegraphics[width=0.49\textwidth]{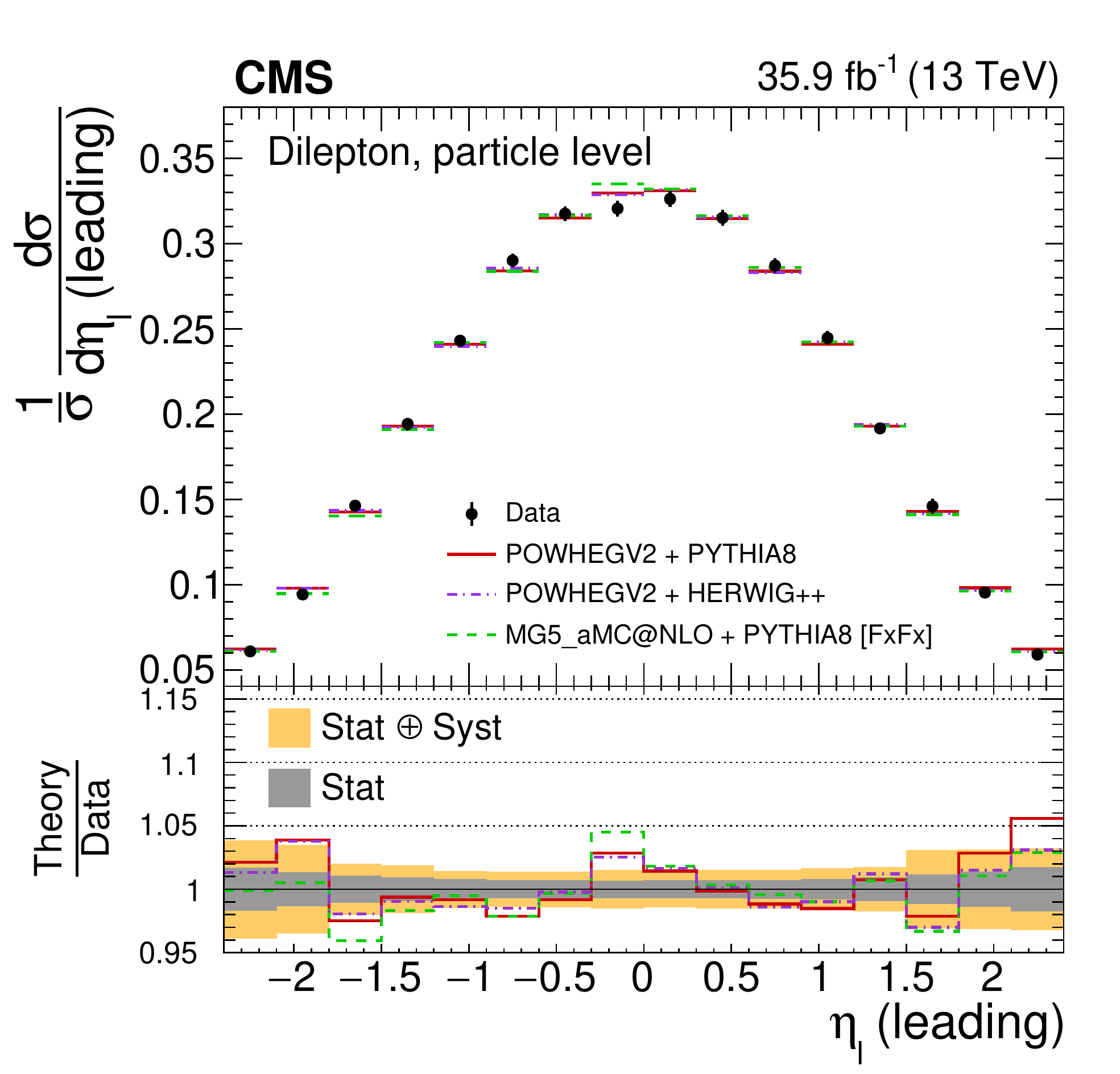}
\caption{The differential \ttbar production cross sections as a function of \etalep (leading) in a fiducial phase space at the particle level are shown for the data (points) and the MC predictions (lines). The vertical lines on the points indicate the total uncertainty in the data. The left and right plots correspond to absolute and normalised measurements, respectively. The lower panel in each plot shows the ratios of the theoretical predictions to the data. The dark and light bands show the relative statistical and total uncertainties in the data, respectively.}
\label{fig:diffxsec:res_etalepleading}
\end{figure}

\begin{figure}[!phtb]
\centering
\includegraphics[width=0.49\textwidth]{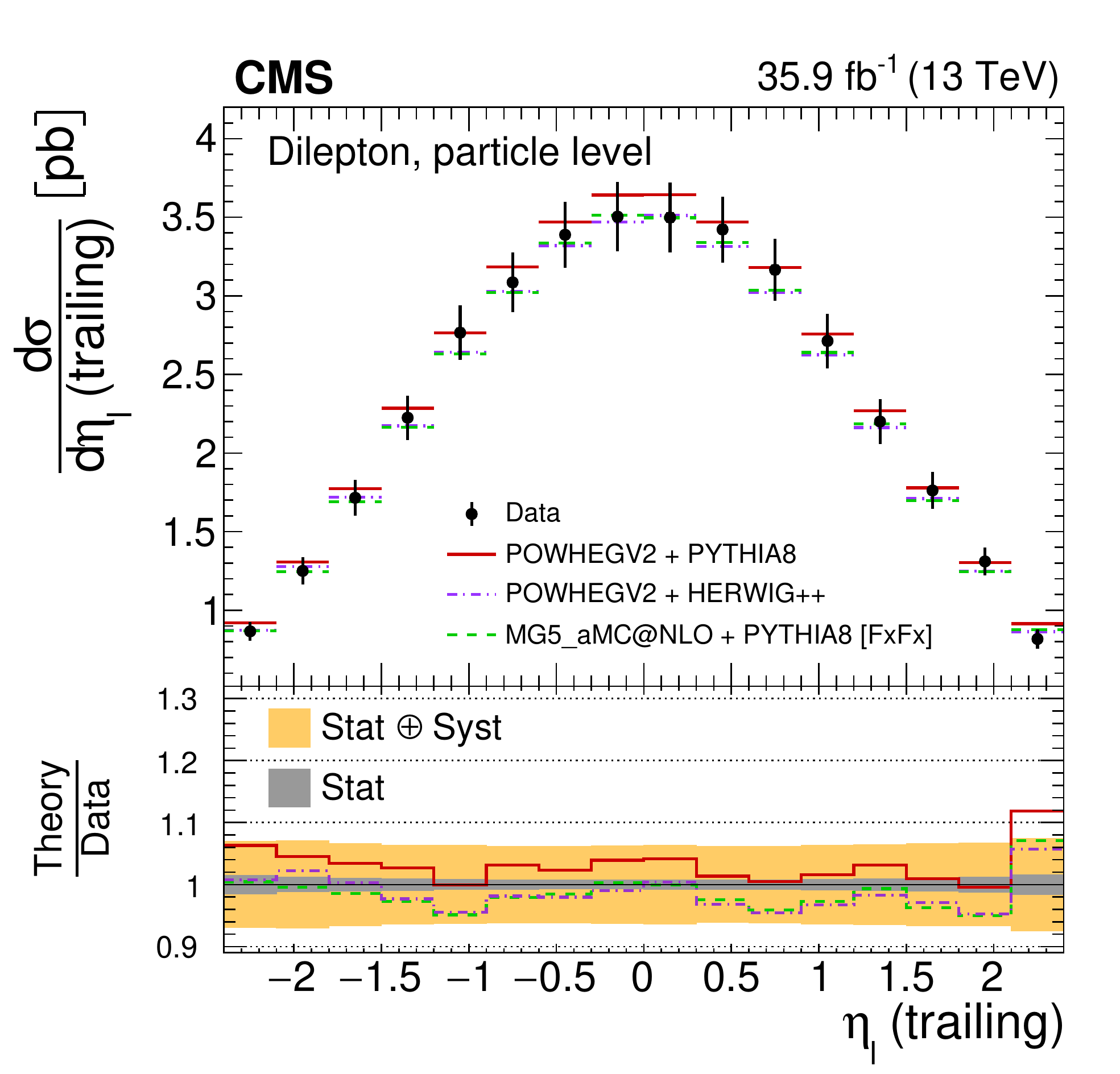}
\includegraphics[width=0.49\textwidth]{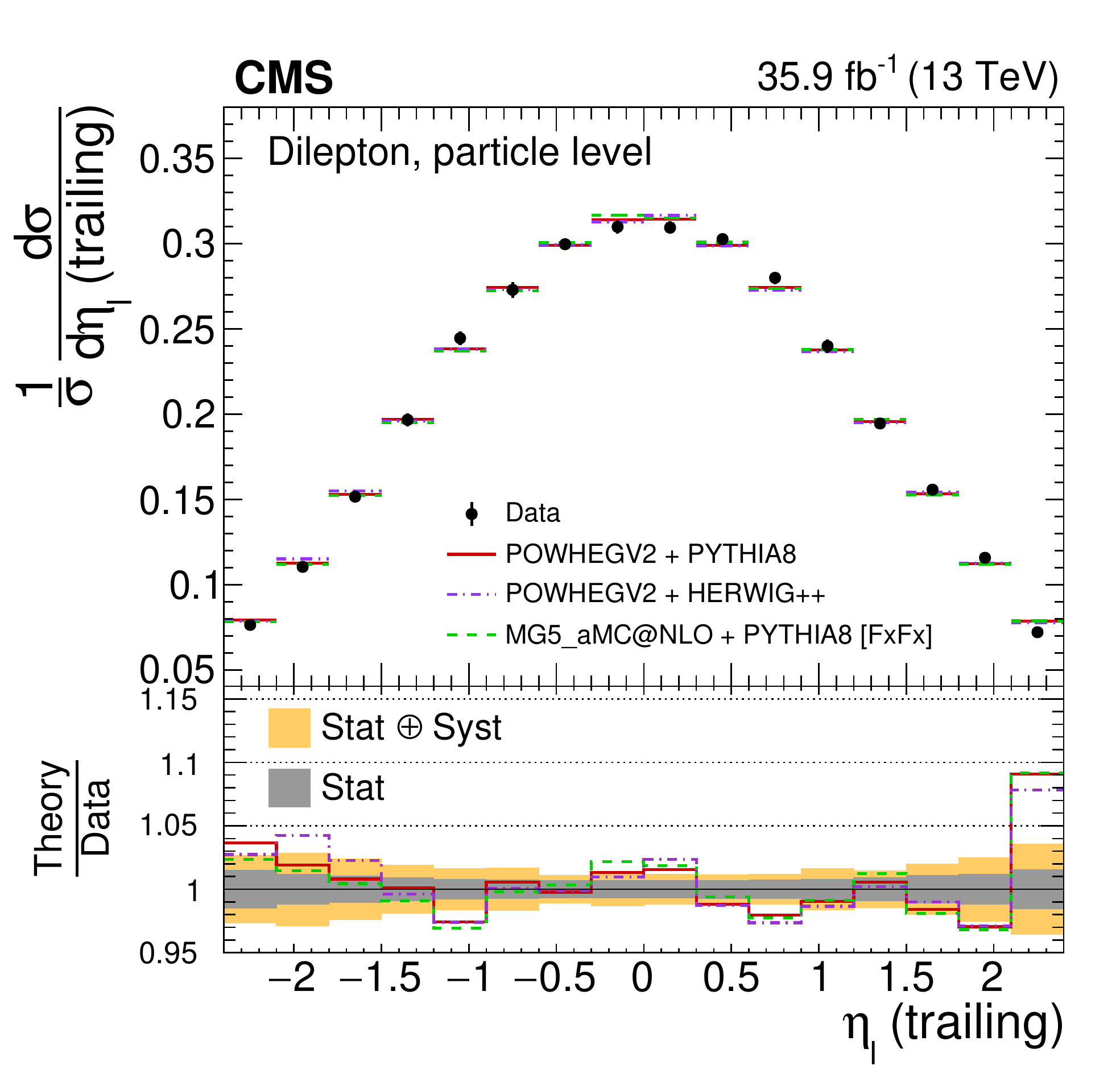}
\caption{The differential \ttbar production cross sections as a function of \etalep (trailing) in a fiducial phase space at the particle level are shown for the data (points) and the MC predictions (lines). The vertical lines on the points indicate the total uncertainty in the data. The left and right plots correspond to absolute and normalised measurements, respectively. The lower panel in each plot shows the ratios of the theoretical predictions to the data. The dark and light bands show the relative statistical and total uncertainties in the data, respectively.}
\label{fig:diffxsec:res_etalepsubleading}
\end{figure}

\begin{figure*}[!phtb]
\centering
\includegraphics[width=0.49\textwidth]{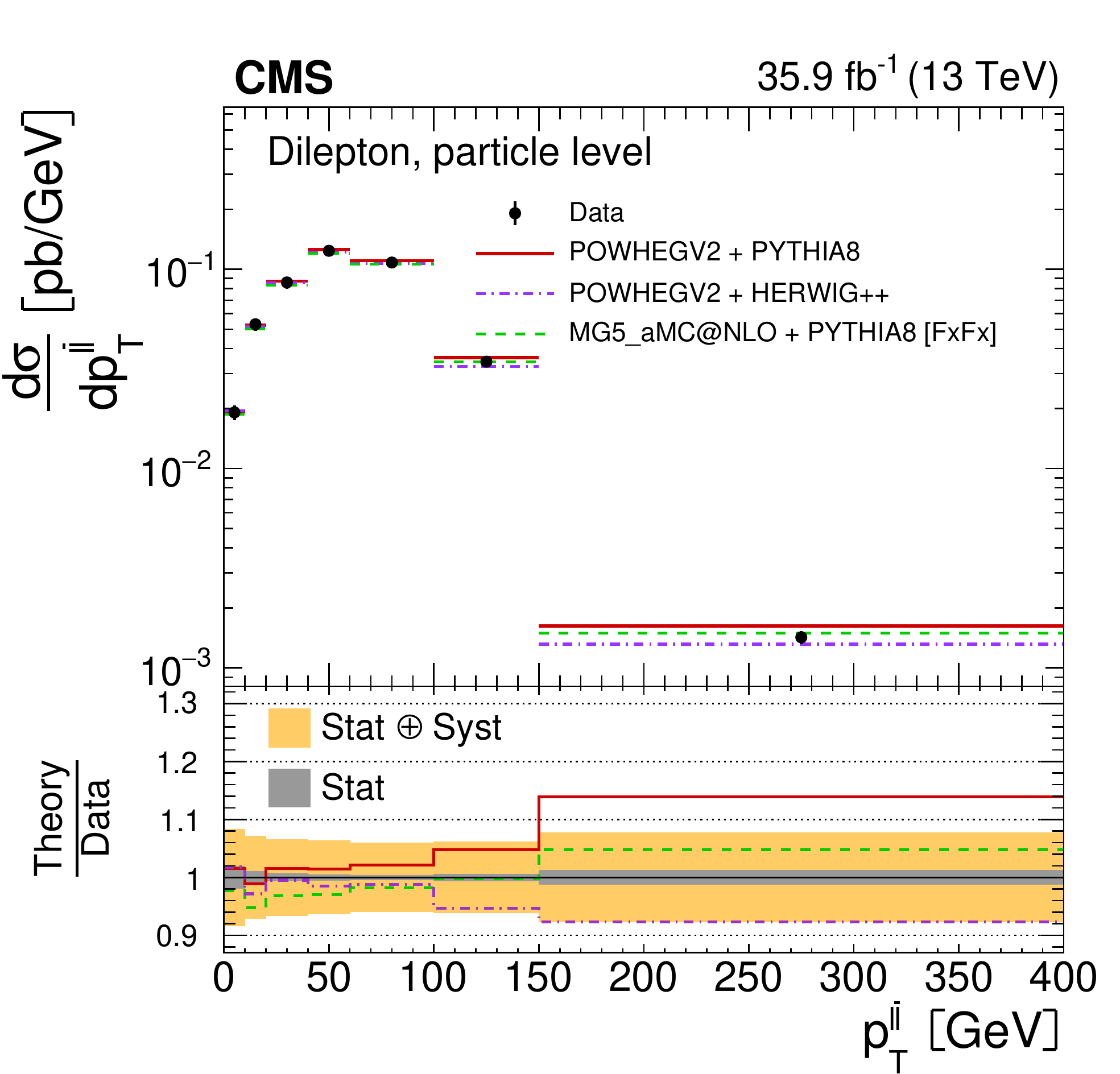}
\includegraphics[width=0.49\textwidth]{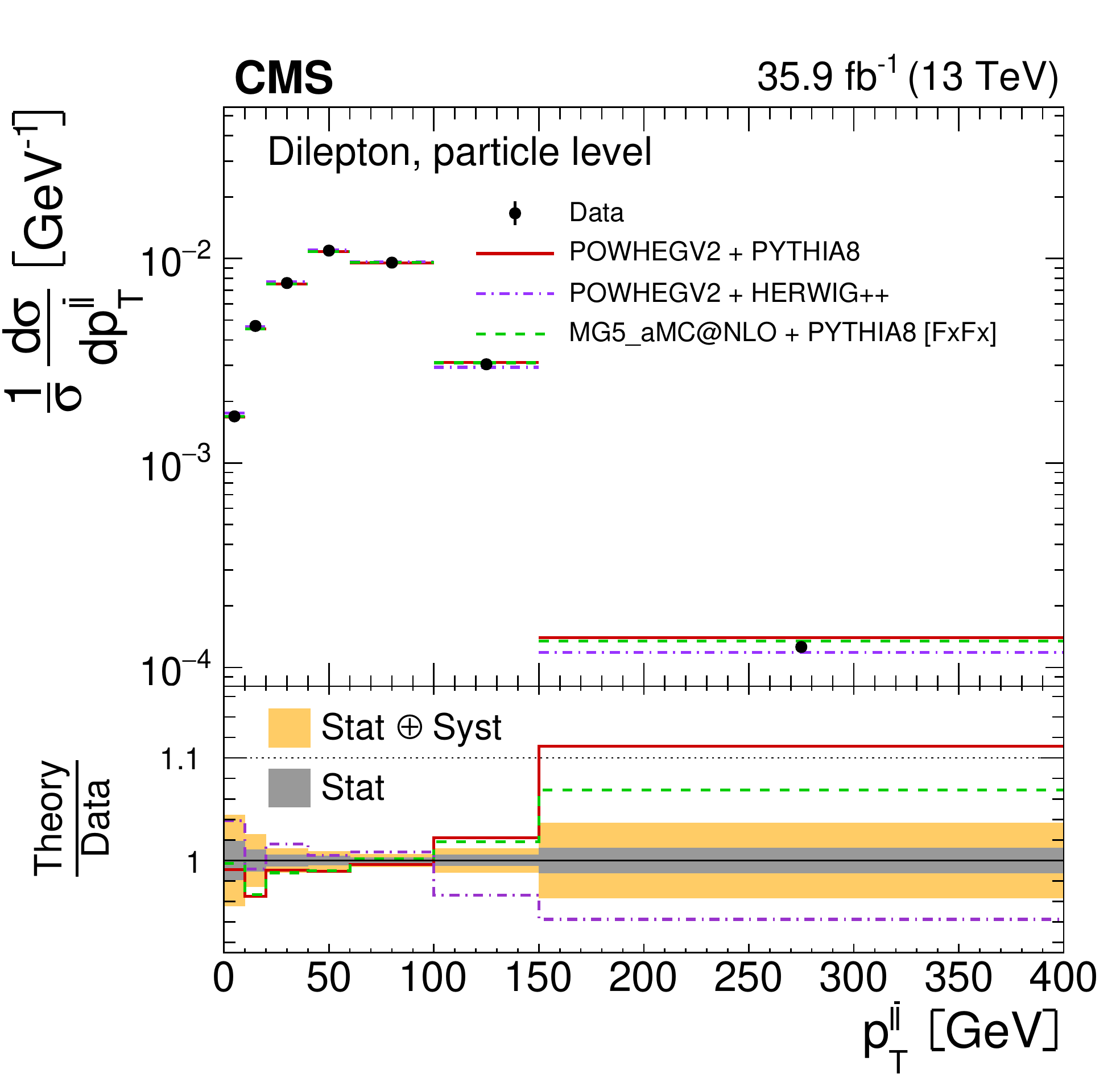}
\caption{The differential \ttbar production cross sections as a function of \ptll in a fiducial phase space at the particle level are shown for the data (points) and the MC predictions (lines). The vertical lines on the points indicate the total uncertainty in the data. The left and right plots correspond to absolute and normalised measurements, respectively. The lower panel in each plot shows the ratios of the theoretical predictions to the data. The dark and light bands show the relative statistical and total uncertainties in the data, respectively.}
\label{fig:diffxsec:res_ptll}
\end{figure*}

\begin{figure*}[!phtb]
\centering
\includegraphics[width=0.49\textwidth]{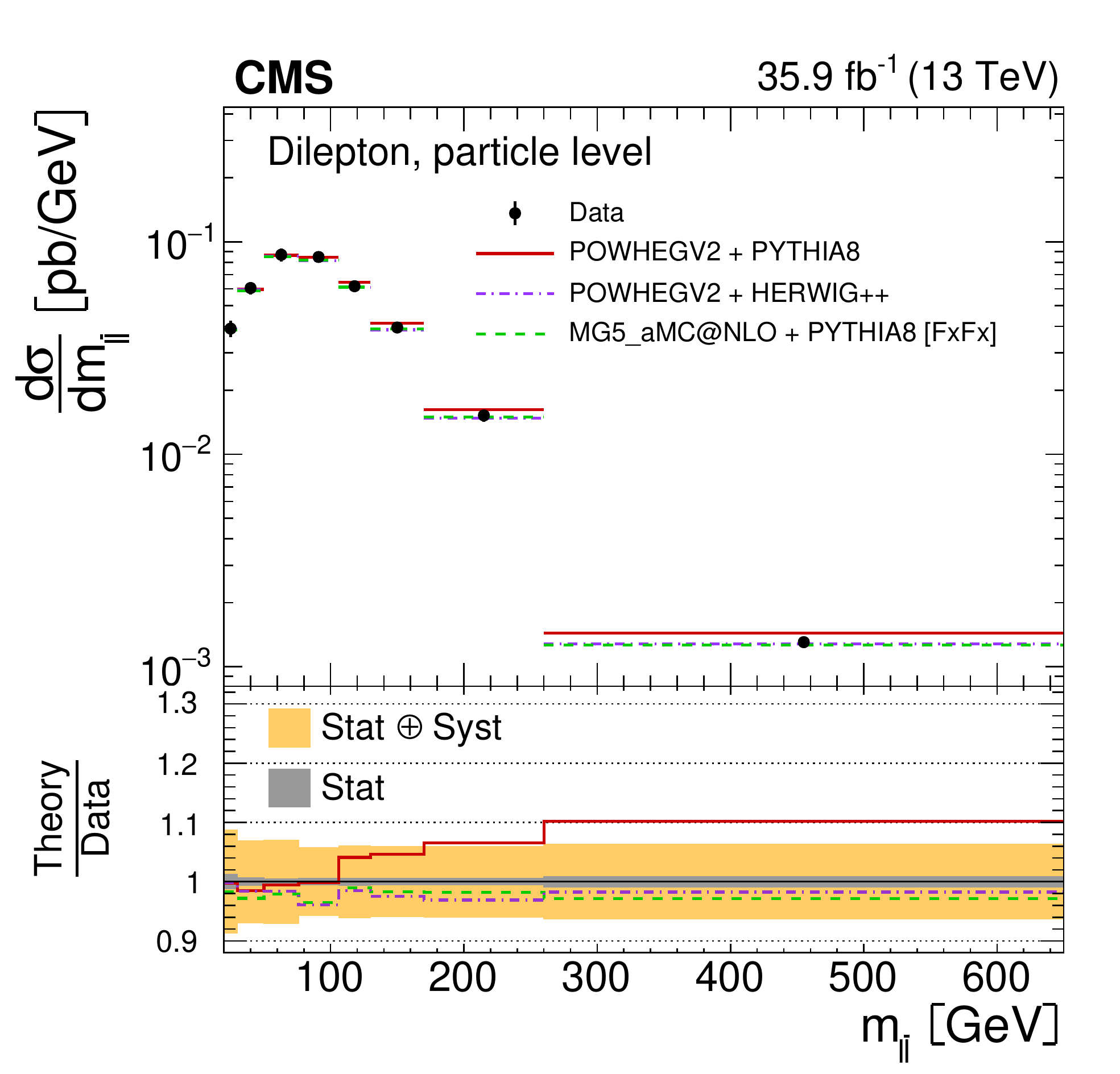}
\includegraphics[width=0.49\textwidth]{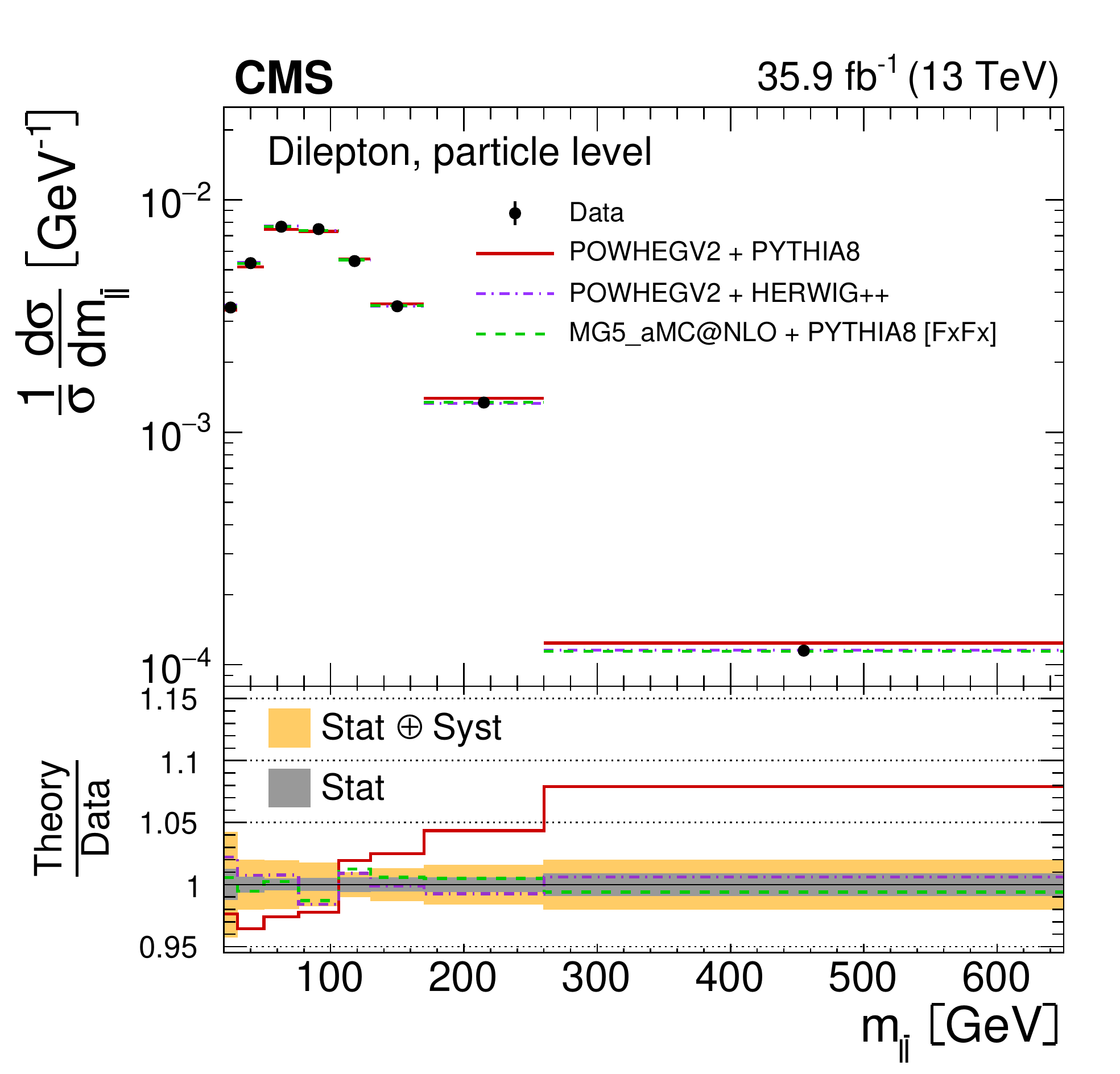}
\caption{The differential \ttbar production cross sections as a function of \mll in a fiducial phase space at the particle level are shown for the data (points) and the MC predictions (lines). The vertical lines on the points indicate the total uncertainty in the data. The left and right plots correspond to absolute and normalised measurements, respectively. The lower panel in each plot shows the ratios of the theoretical predictions to the data. The dark and light bands show the relative statistical and total uncertainties in the data, respectively.}
\label{fig:diffxsec:res_massll}
\end{figure*}

\begin{figure*}[!phtb]
\centering
\includegraphics[width=0.49\textwidth]{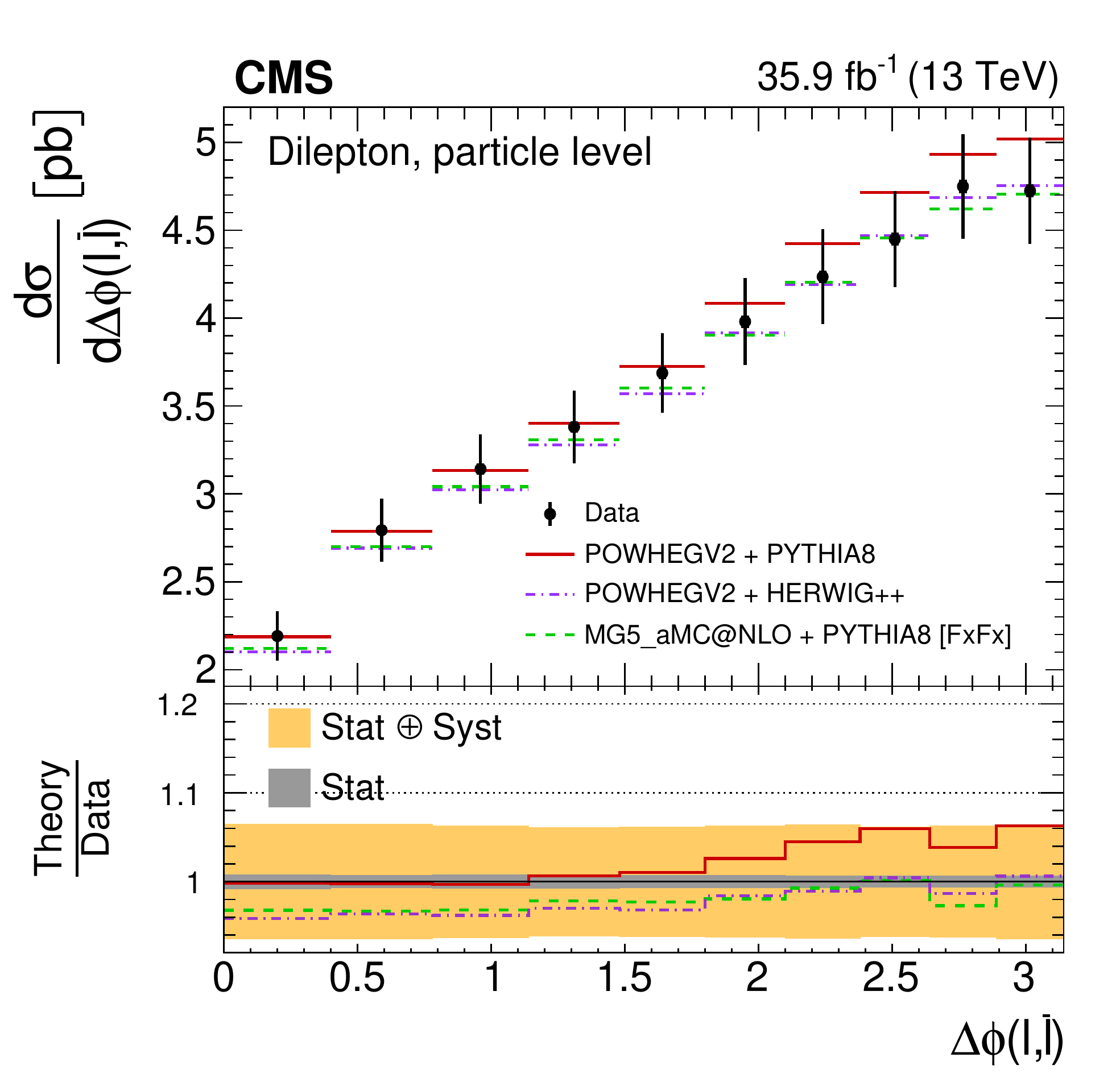}
\includegraphics[width=0.49\textwidth]{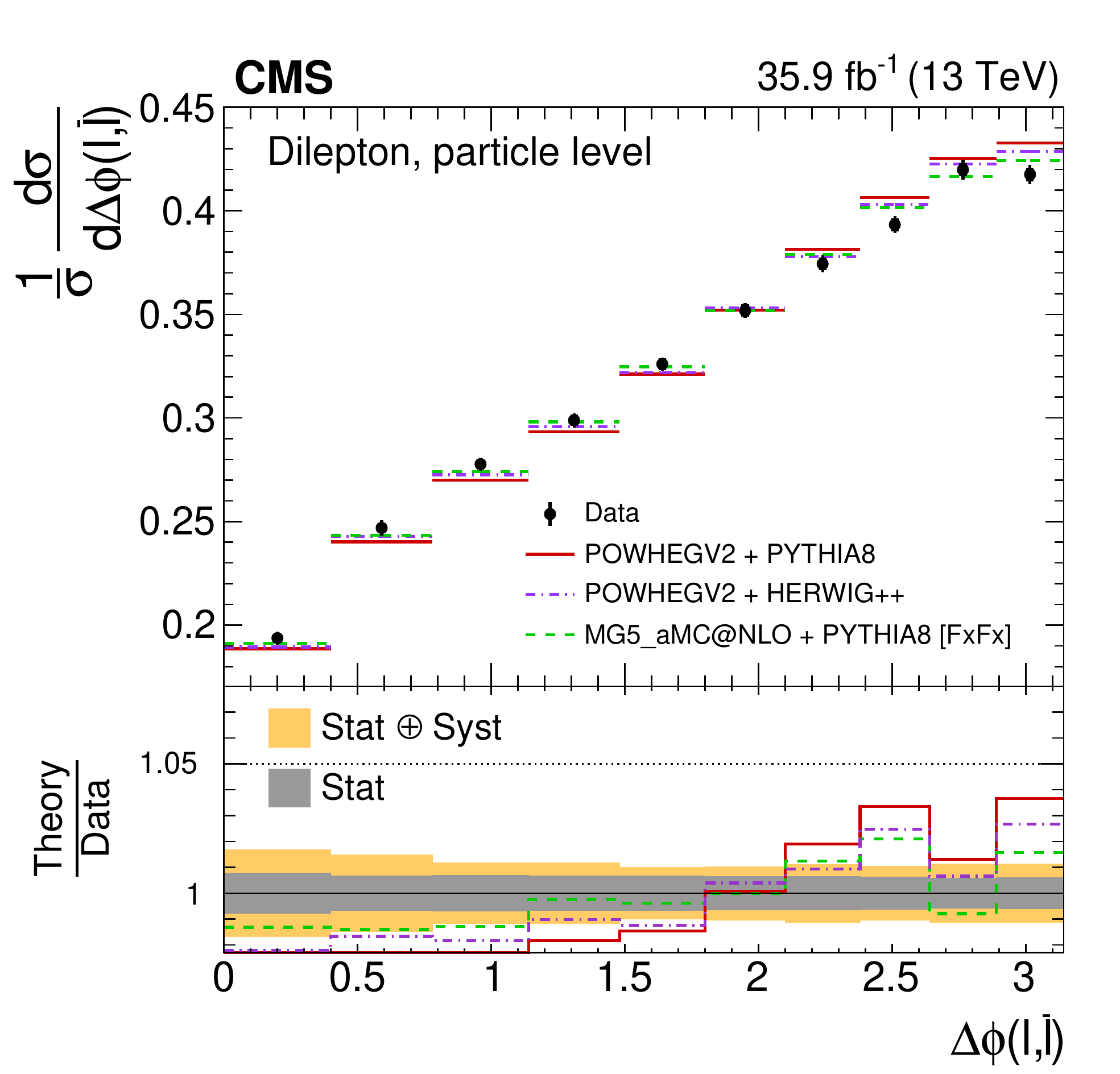}
\caption{The differential \ttbar production cross sections as a function of \delphill in a fiducial phase space at the particle level are shown for the data (points) and the MC predictions (lines). The vertical lines on the points indicate the total uncertainty in the data. The left and right plots correspond to absolute and normalised measurements, respectively. The lower panel in each plot shows the ratios of the theoretical predictions to the data. The dark and light bands show the relative statistical and total uncertainties in the data, respectively.}
\label{fig:diffxsec:res_deltaphill}
\end{figure*}

\begin{figure*}[!phtb]
\centering
\includegraphics[width=0.49\textwidth]{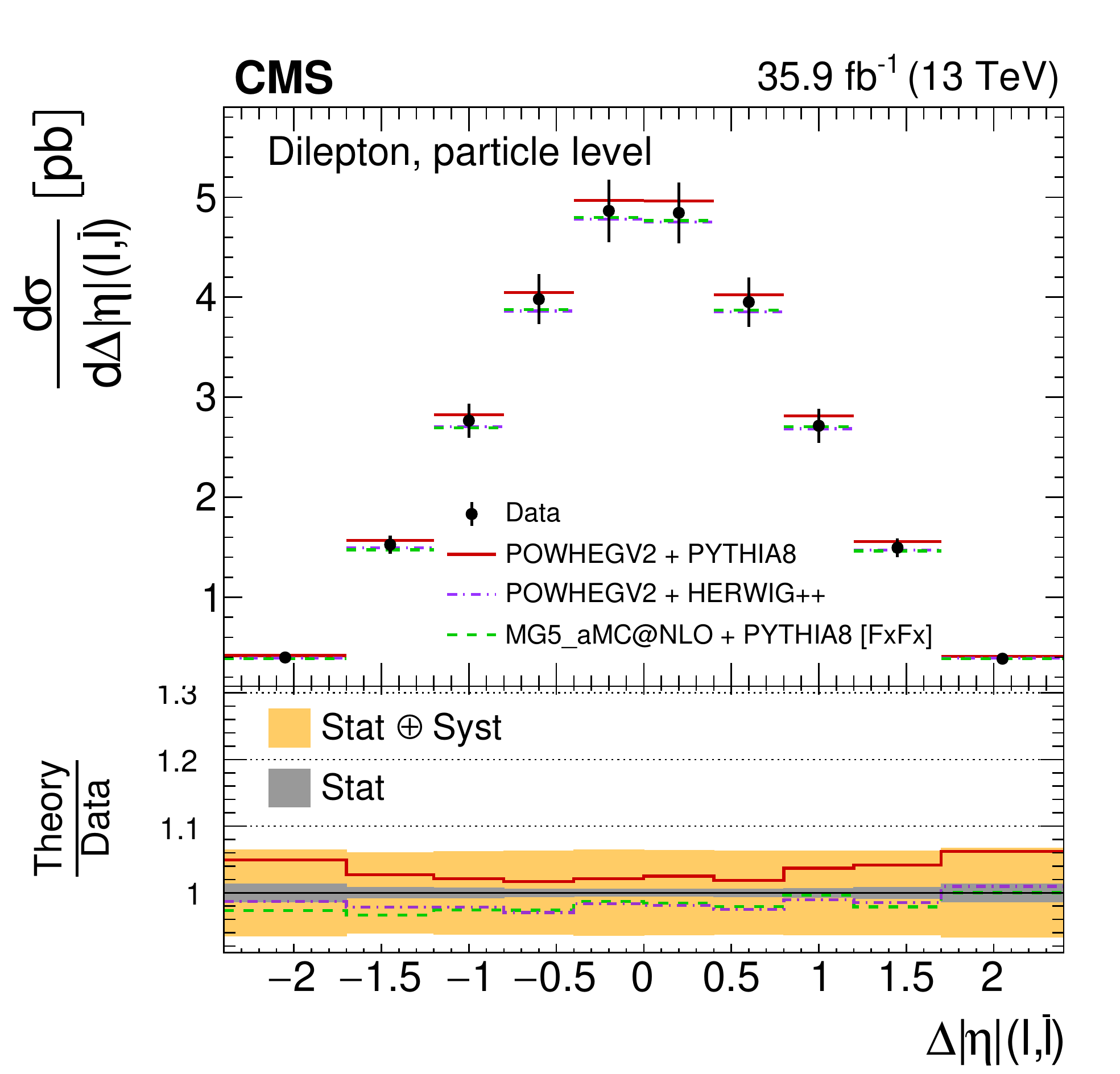}
\includegraphics[width=0.49\textwidth]{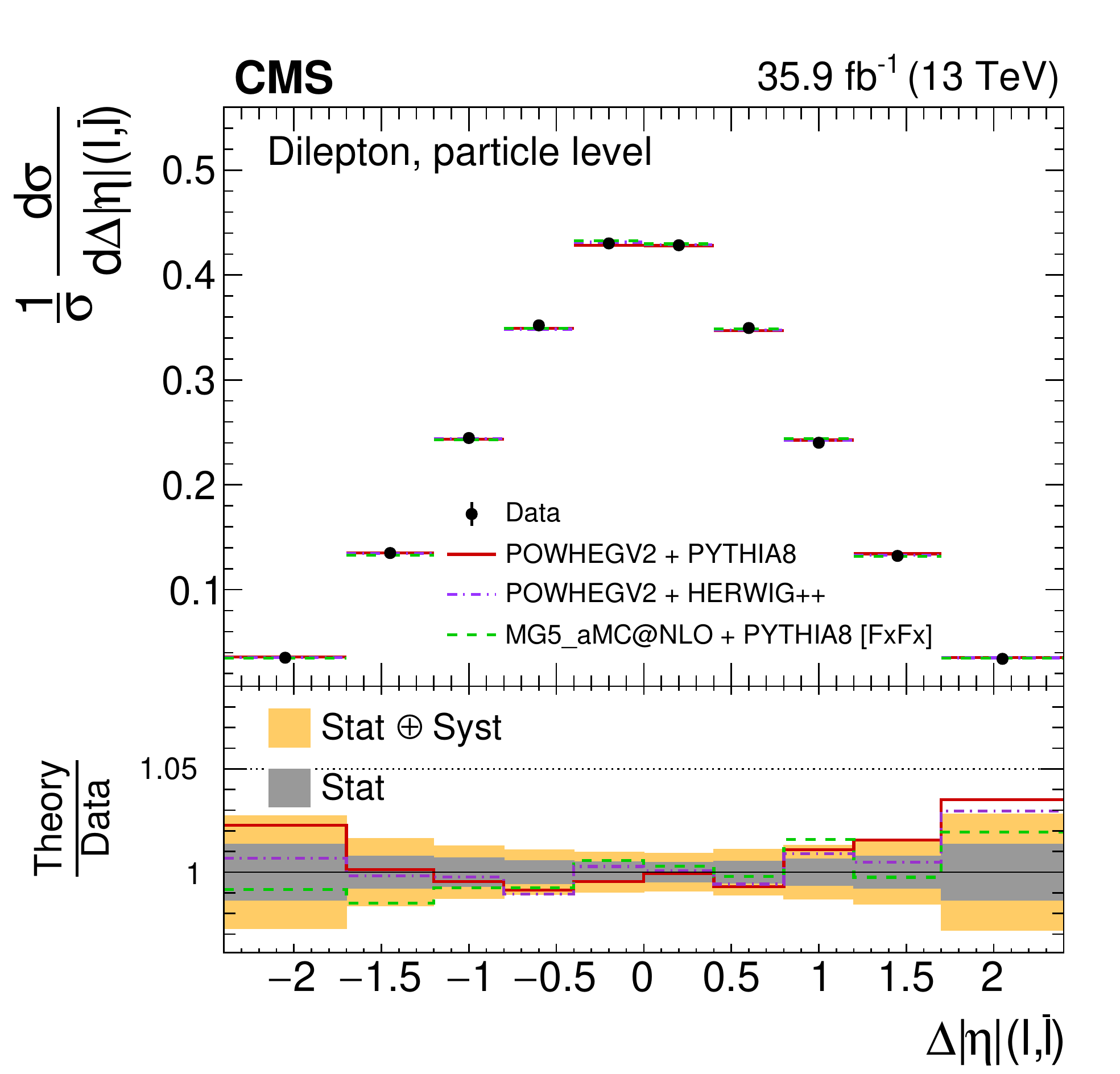}
\caption{The differential \ttbar production cross sections as a function of \deletall in a fiducial phase space at the particle level are shown for the data (points) and the MC predictions (lines). The vertical lines on the points indicate the total uncertainty in the data. The left and right plots correspond to absolute and normalised measurements, respectively. The lower panel in each plot shows the ratios of the theoretical predictions to the data. The dark and light bands show the relative statistical and total uncertainties in the data, respectively.}
\label{fig:diffxsec:res_deltaetall}
\end{figure*}

\begin{figure*}[!phtb]
\centering
\includegraphics[width=0.49\textwidth]{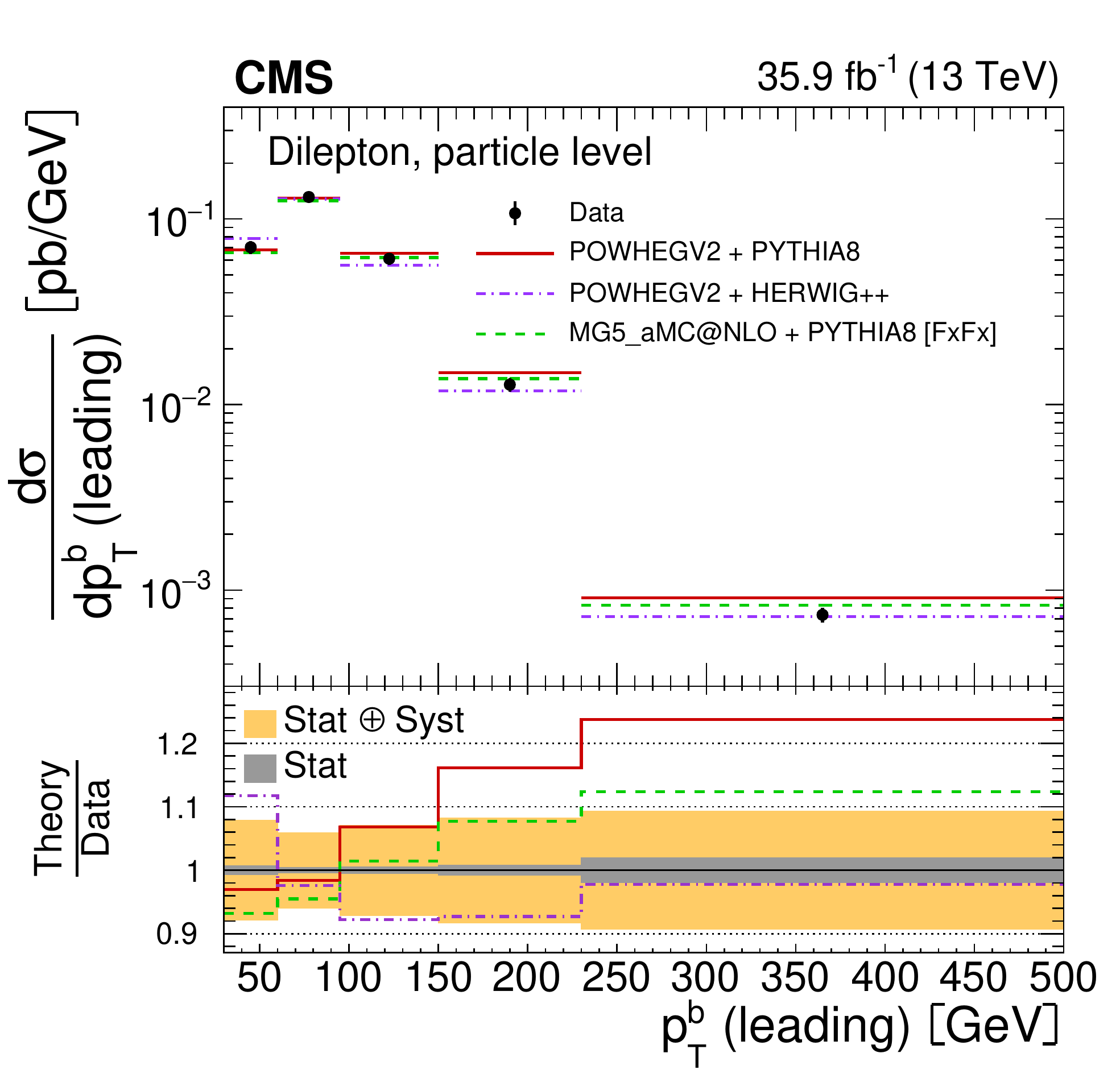}
\includegraphics[width=0.49\textwidth]{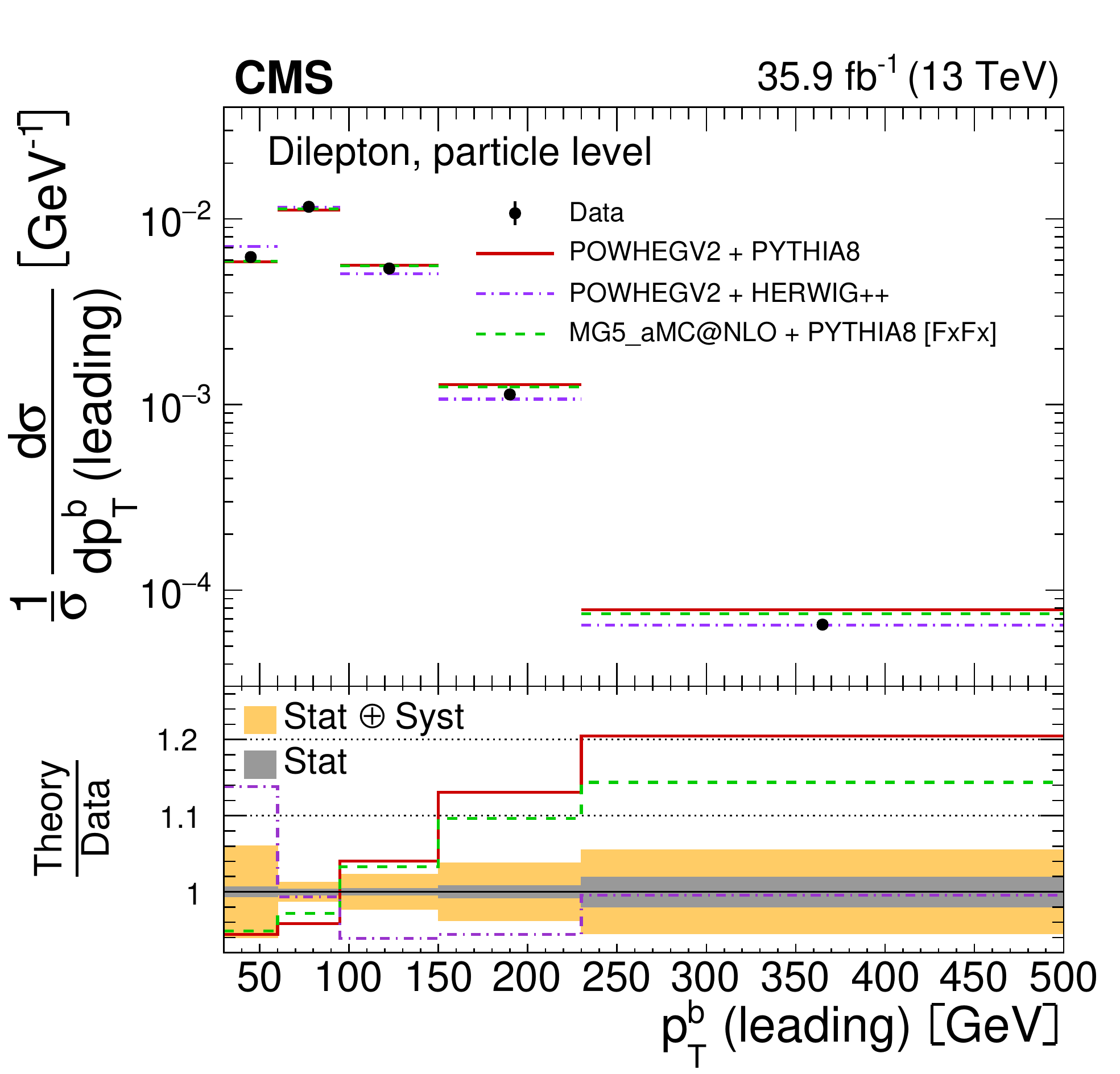}
\caption{The differential \ttbar production cross sections as a function of \ptb (leading) in a fiducial phase space at the particle level are shown for the data (points) and the MC predictions (lines). The vertical lines on the points indicate the total uncertainty in the data. The left and right plots correspond to absolute and normalised measurements, respectively. The lower panel in each plot shows the ratios of the theoretical predictions to the data. The dark and light bands show the relative statistical and total uncertainties in the data, respectively.}
\label{fig:diffxsec:res_ptbjetleading}

\end{figure*}

\begin{figure*}[!phtb]
\centering
\includegraphics[width=0.49\textwidth]{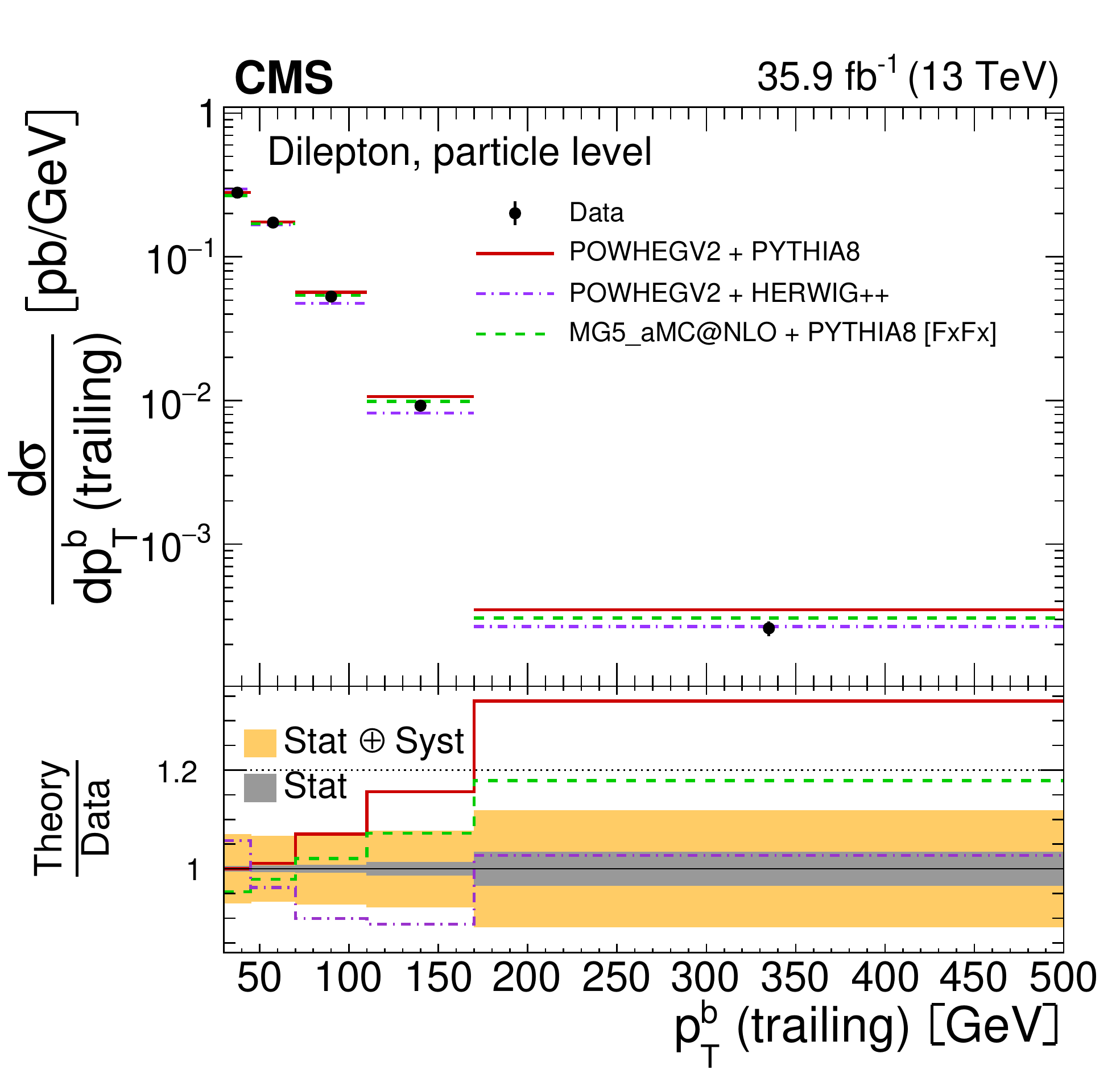}
\includegraphics[width=0.49\textwidth]{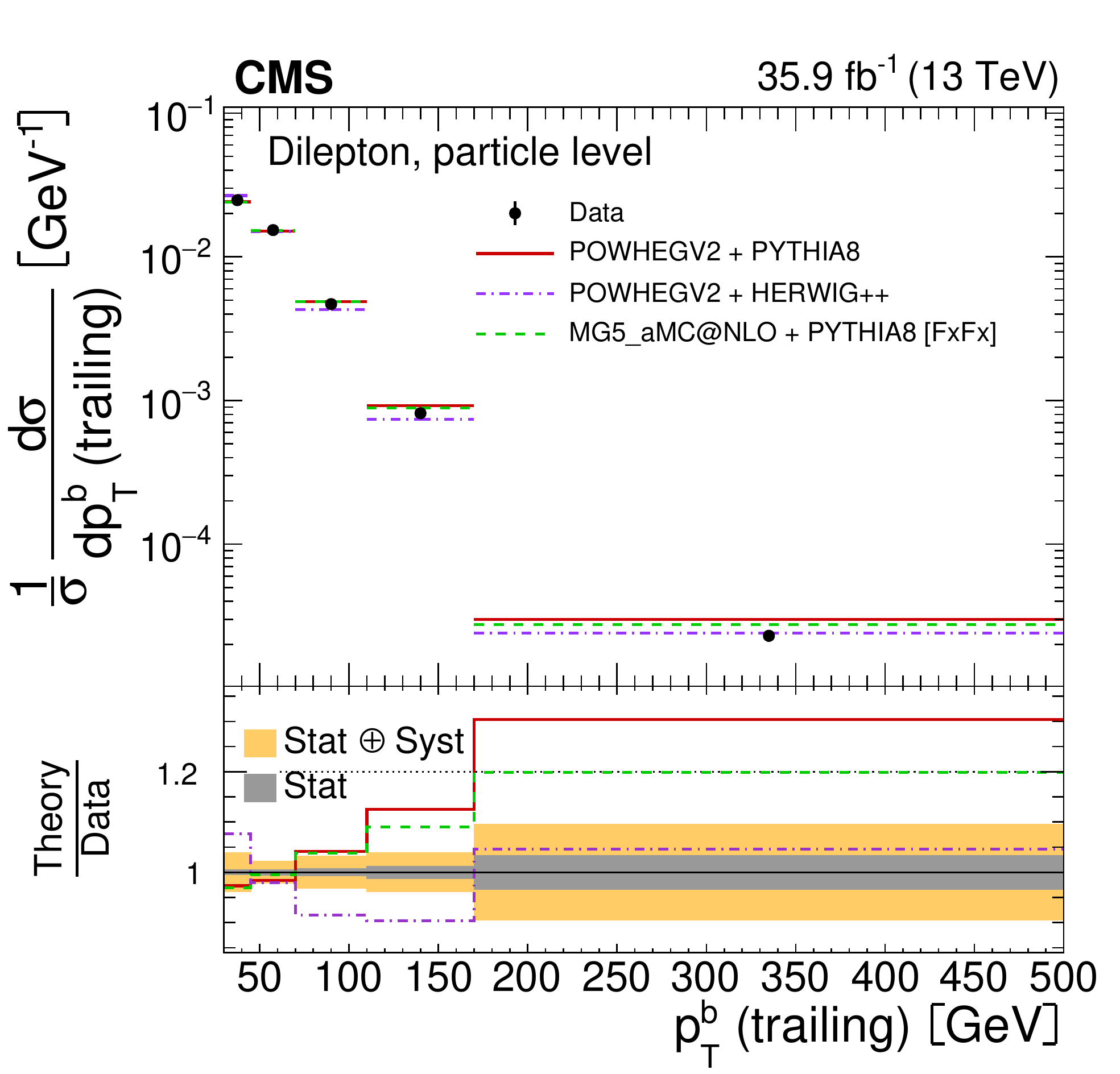}
\caption{The differential \ttbar production cross sections as a function of \ptb (trailing) in a fiducial phase space at the particle level are shown for the data (points) and the MC predictions (lines). The vertical lines on the points indicate the total uncertainty in the data. The left and right plots correspond to absolute and normalised measurements, respectively. The lower panel in each plot shows the ratios of the theoretical predictions to the data. The dark and light bands show the relative statistical and total uncertainties in the data, respectively.}
\label{fig:diffxsec:res_ptbjetsubleading}
\end{figure*}

\begin{figure*}[!phtb]
\centering
\includegraphics[width=0.49\textwidth]{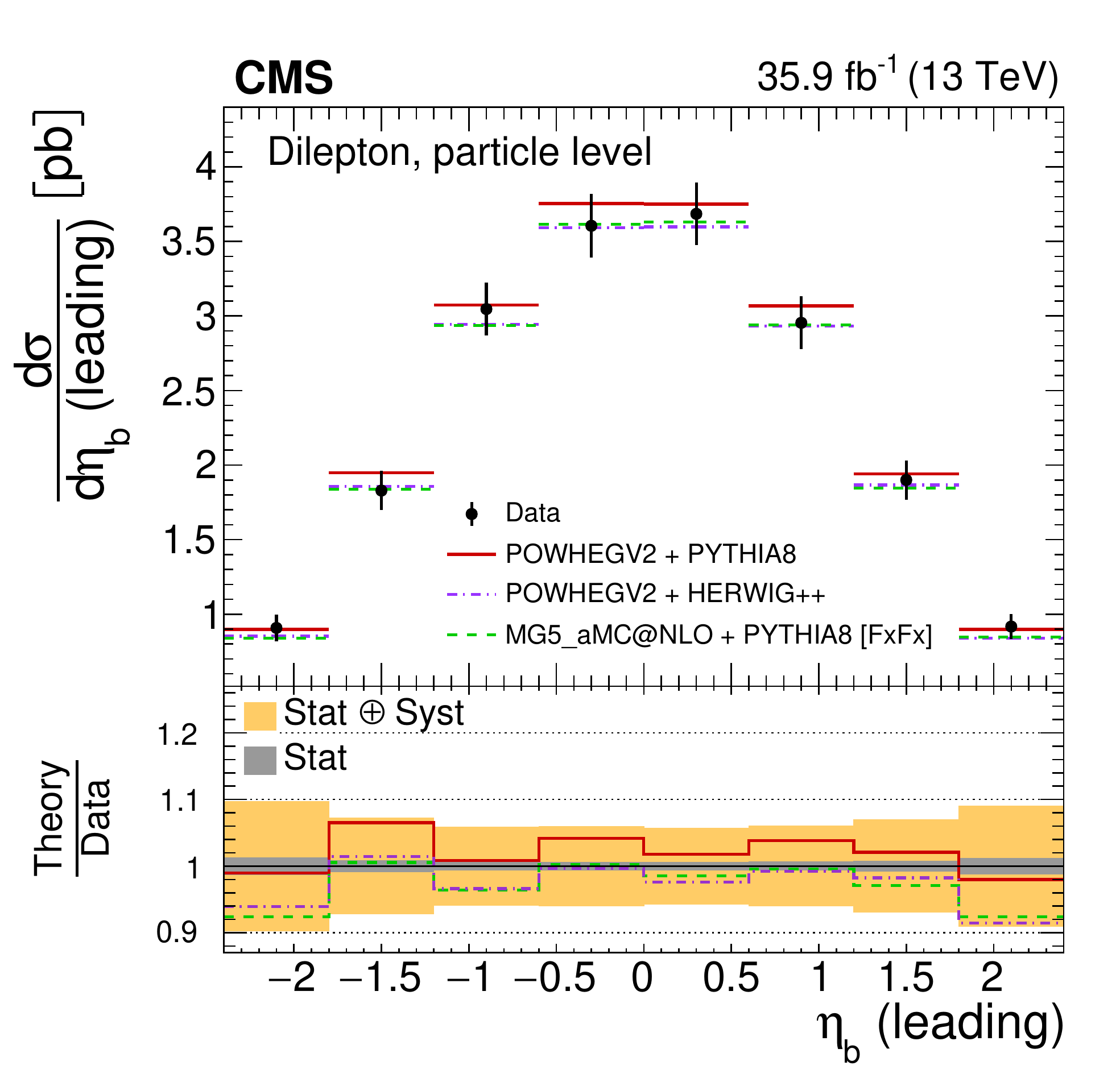}
\includegraphics[width=0.49\textwidth]{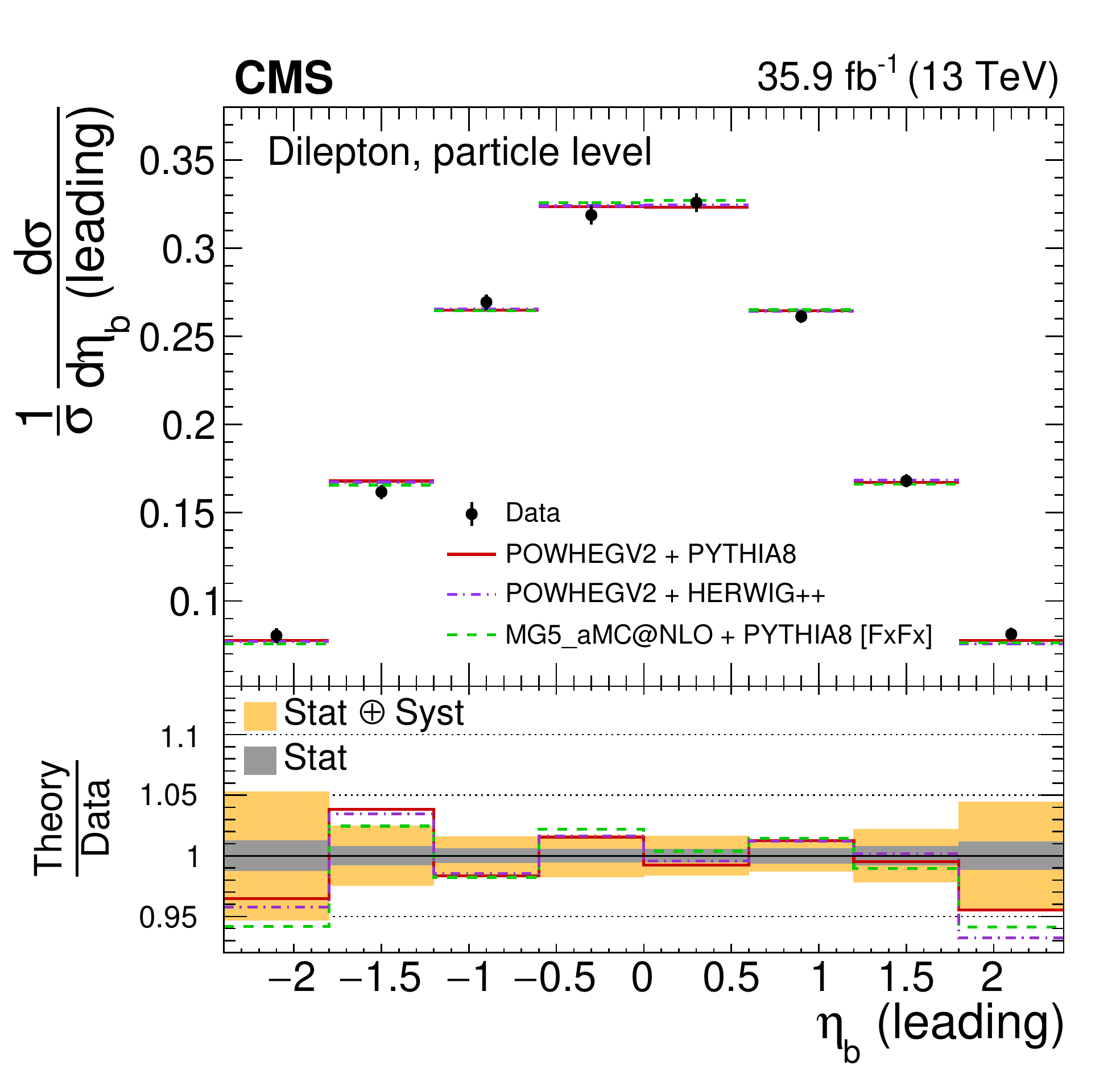}
\caption{The differential \ttbar production cross sections as a function of \etab (leading) in a fiducial phase space at the particle level are shown for the data (points) and the MC predictions (lines). The vertical lines on the points indicate the total uncertainty in the data. The left and right plots correspond to absolute and normalised measurements, respectively. The lower panel in each plot shows the ratios of the theoretical predictions to the data. The dark and light bands show the relative statistical and total uncertainties in the data, respectively.}
\label{fig:diffxsec:res_etabjetleading}
\end{figure*}

\begin{figure*}[!phtb]
\centering
\includegraphics[width=0.49\textwidth]{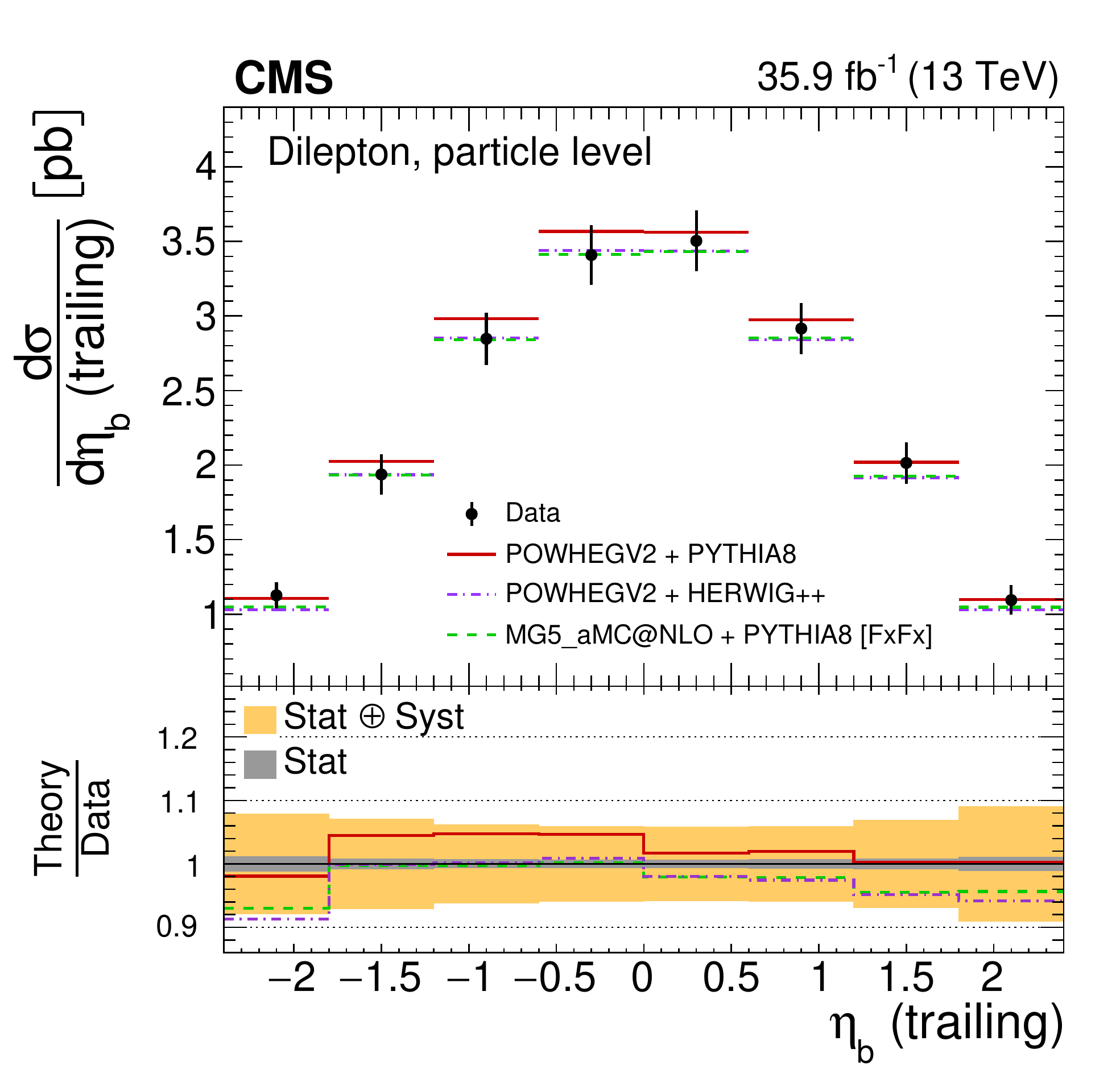}
\includegraphics[width=0.49\textwidth]{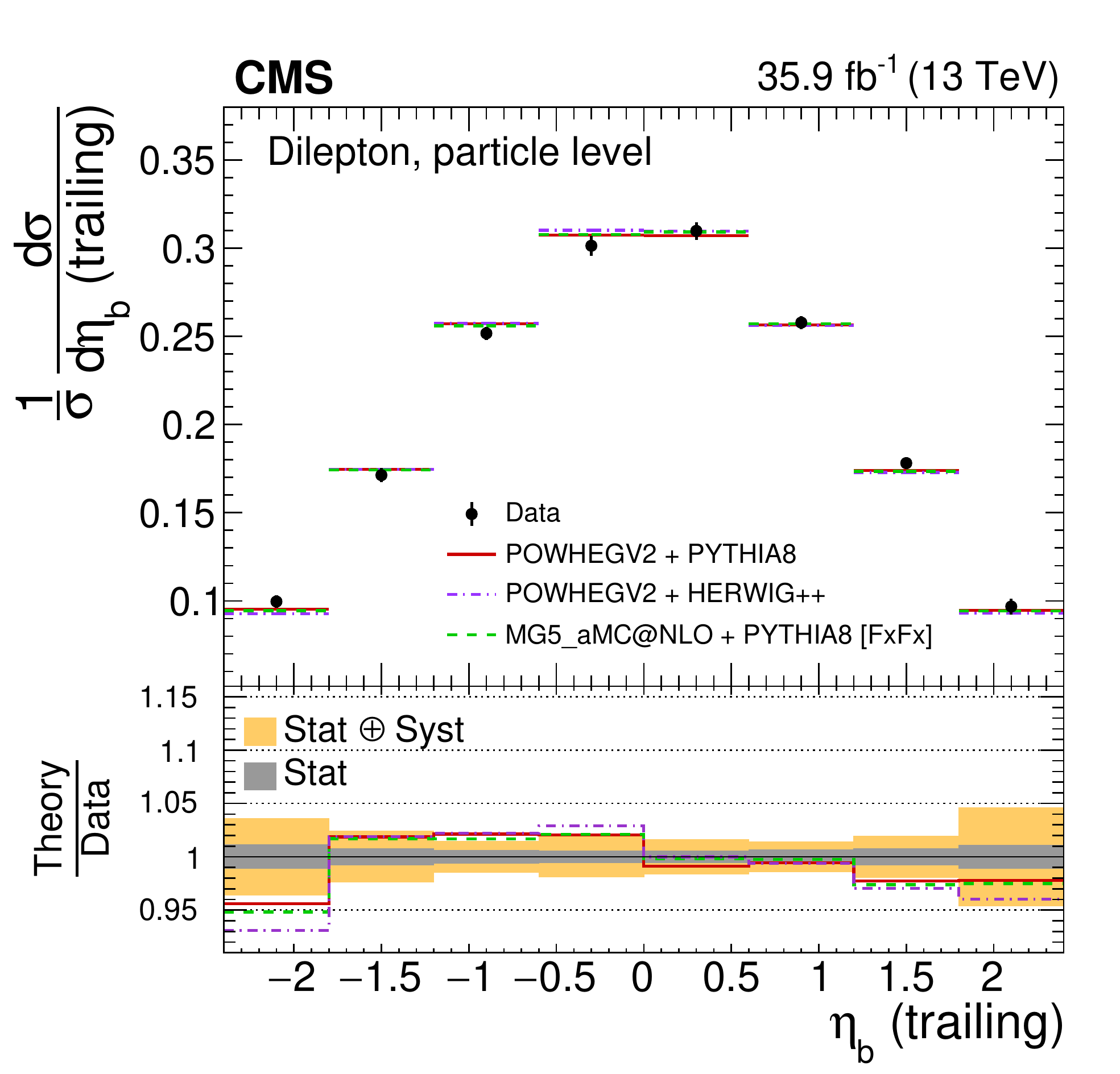}
\caption{The differential \ttbar production cross sections as a function of \etab (trailing) in a fiducial phase space at the particle level are shown for the data (points) and the MC predictions (lines). The vertical lines on the points indicate the total uncertainty in the data. The left and right plots correspond to absolute and normalised measurements, respectively. The lower panel in each plot shows the ratios of the theoretical predictions to the data. The dark and light bands show the relative statistical and total uncertainties in the data, respectively.}
\label{fig:diffxsec:res_etabjetlsubeading}
\end{figure*}

\begin{figure*}[!phtb]
\centering
\includegraphics[width=0.49\textwidth]{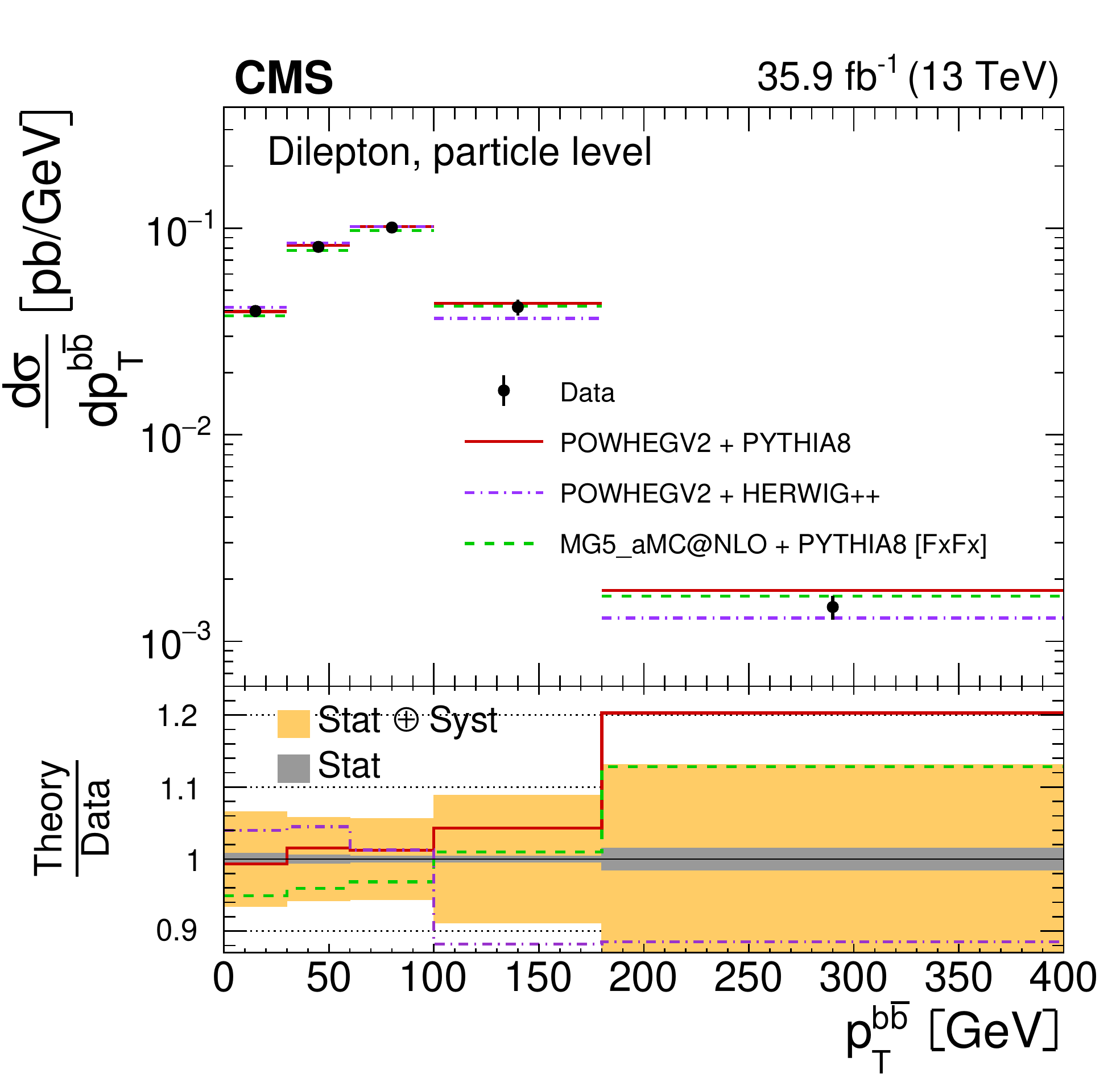}
\includegraphics[width=0.49\textwidth]{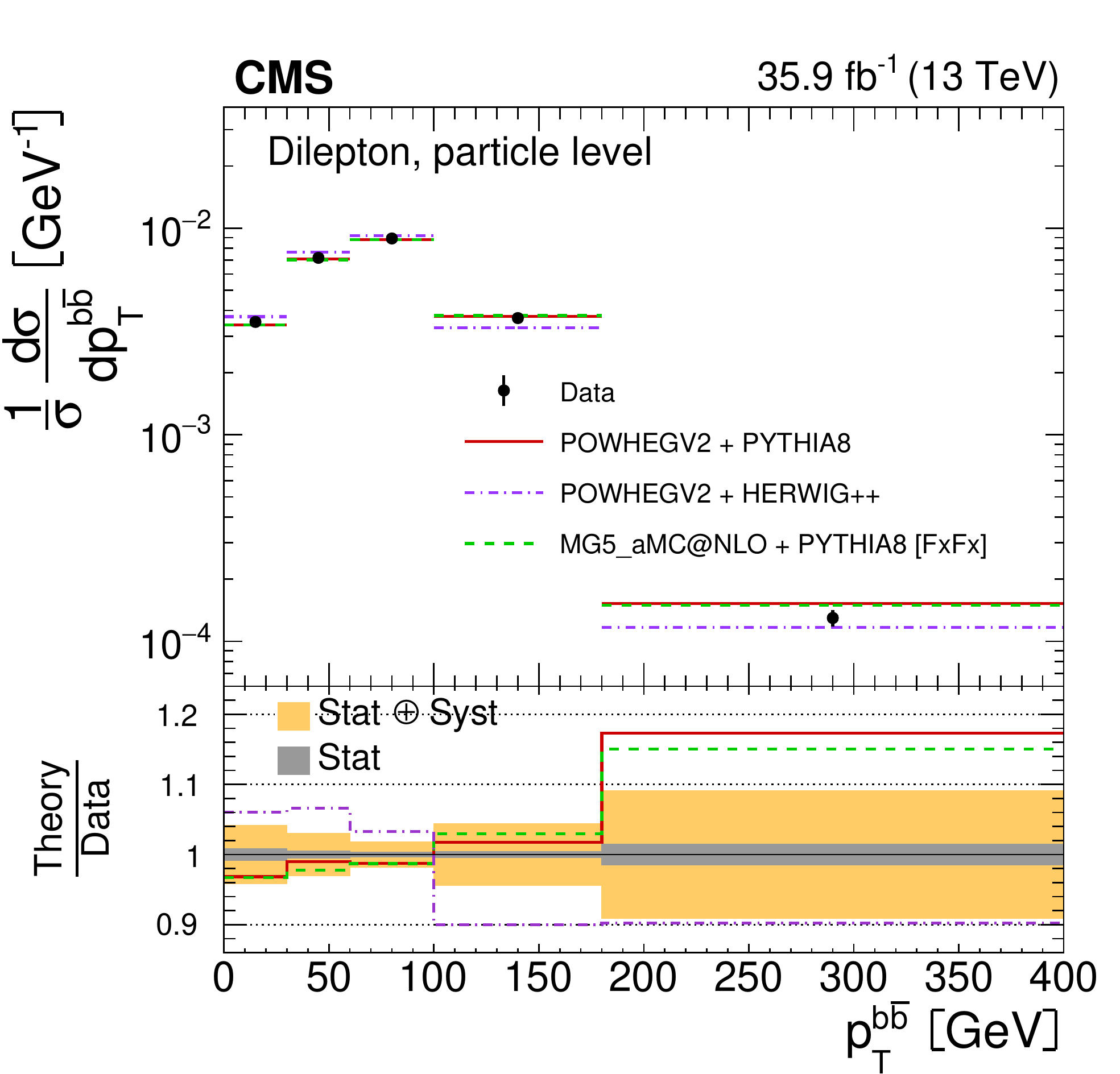}
\caption{The differential \ttbar production cross sections as a function of \ptbb in a fiducial phase space at the particle level are shown for the data (points) and the MC predictions (lines). The vertical lines on the points indicate the total uncertainty in the data. The left and right plots correspond to absolute and normalised measurements, respectively. The lower panel in each plot shows the ratios of the theoretical predictions to the data. The dark and light bands show the relative statistical and total uncertainties in the data, respectively.}
\label{fig:diffxsec:res_ptbb}
\end{figure*}

\begin{figure*}[!phtb]
\centering
\includegraphics[width=0.49\textwidth]{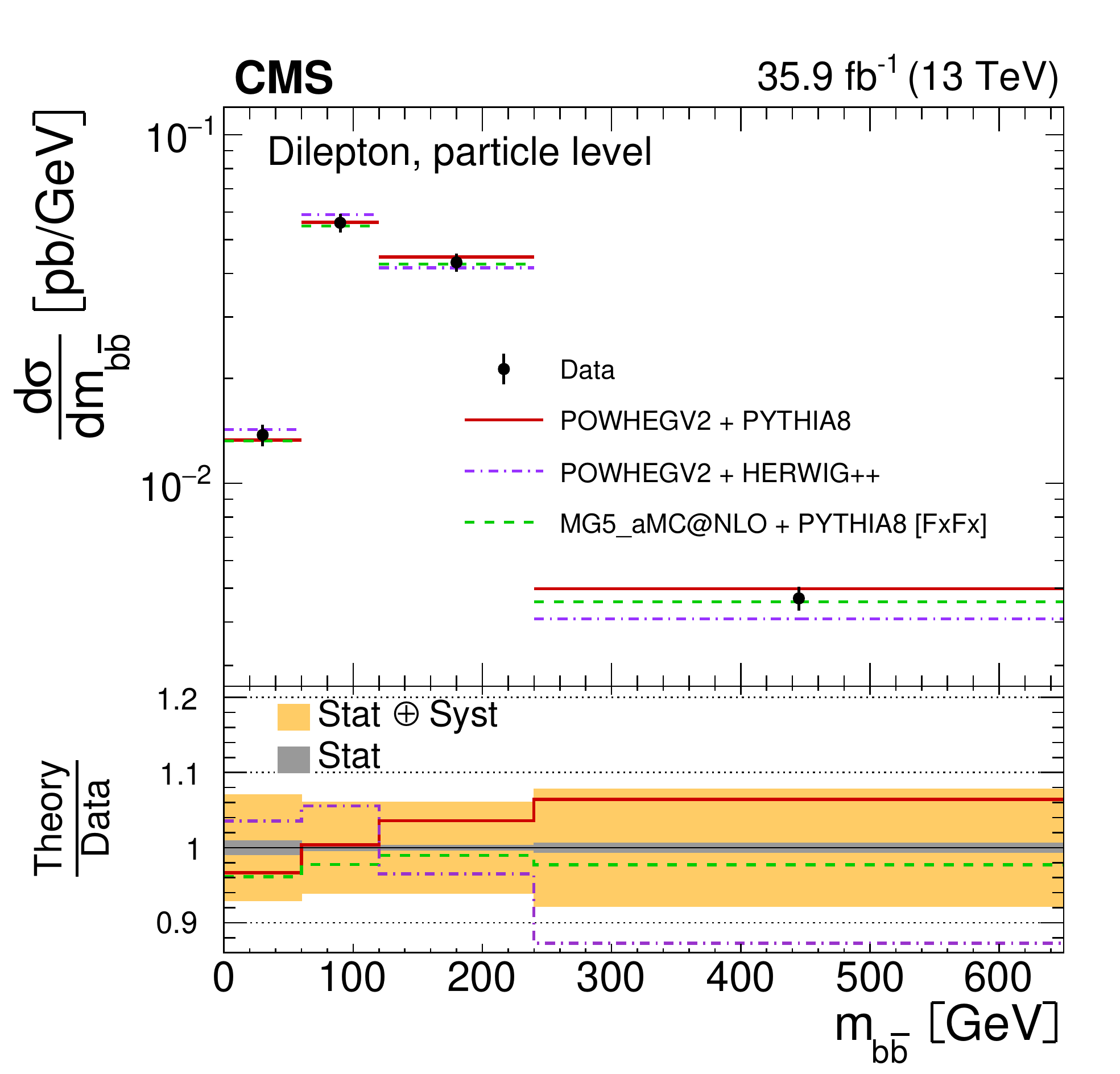}
\includegraphics[width=0.49\textwidth]{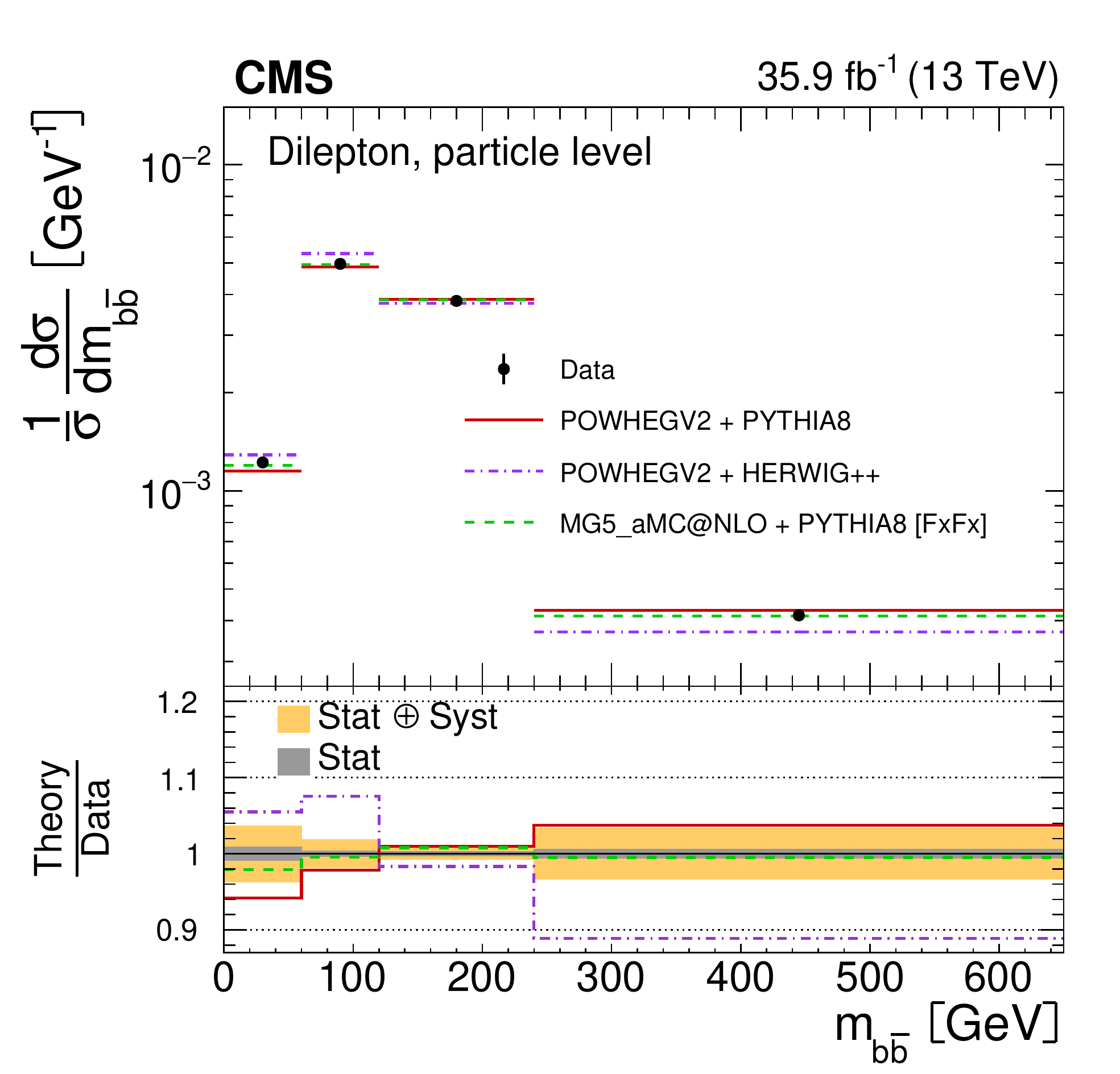}
\caption{The differential \ttbar production cross sections as a function of \mbb in a fiducial phase space at the particle level are shown for the data (points) and the MC predictions (lines). The vertical lines on the points indicate the total uncertainty in the data. The left and right plots correspond to absolute and normalised measurements, respectively. The lower panel in each plot shows the ratios of the theoretical predictions to the data. The dark and light bands show the relative statistical and total uncertainties in the data, respectively.}
\label{fig:diffxsec:res_bbbarmass}
\end{figure*}

\begin{figure*}[!phtb]
\centering
\includegraphics[width=0.49\textwidth]{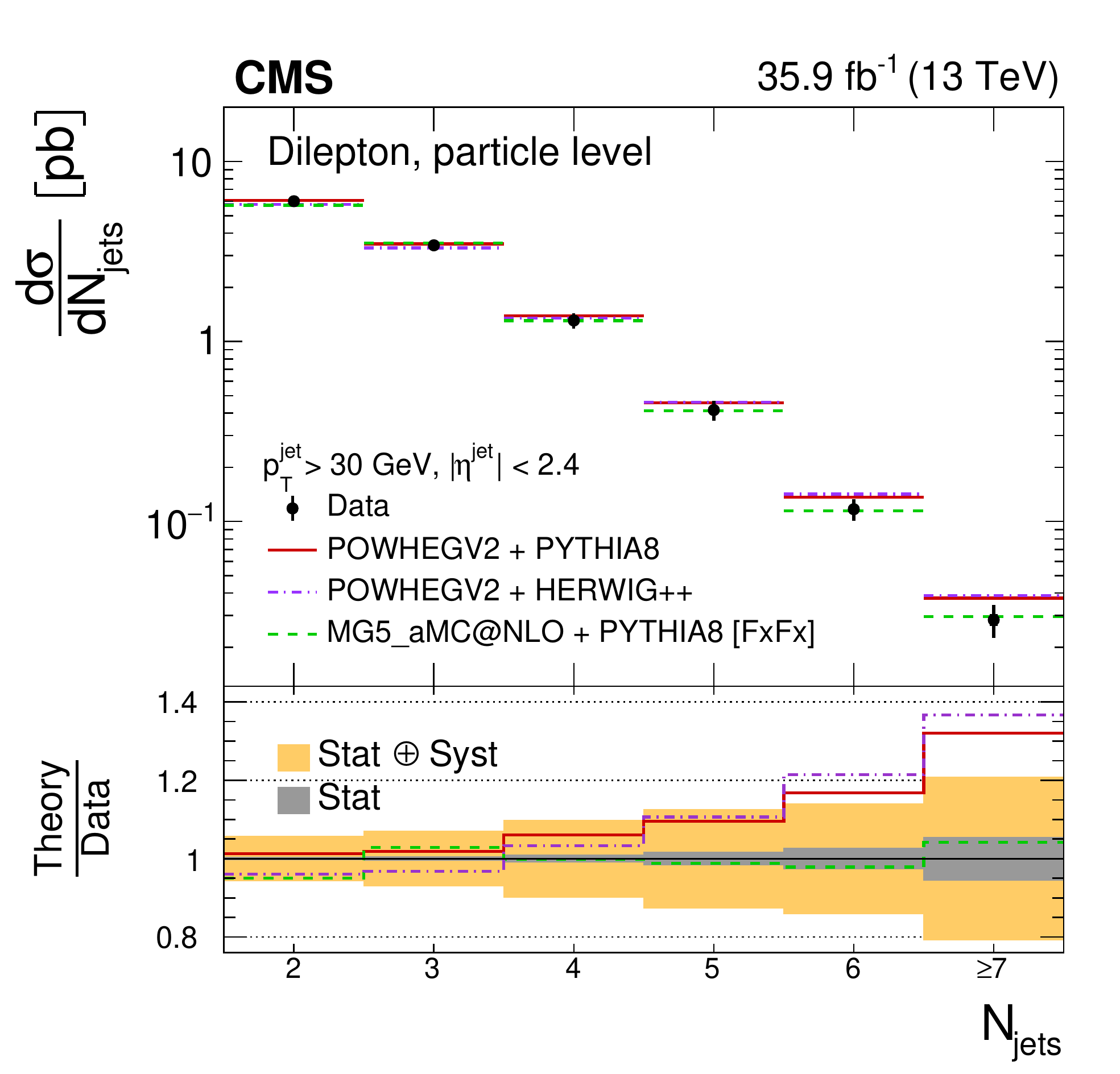}
\includegraphics[width=0.49\textwidth]{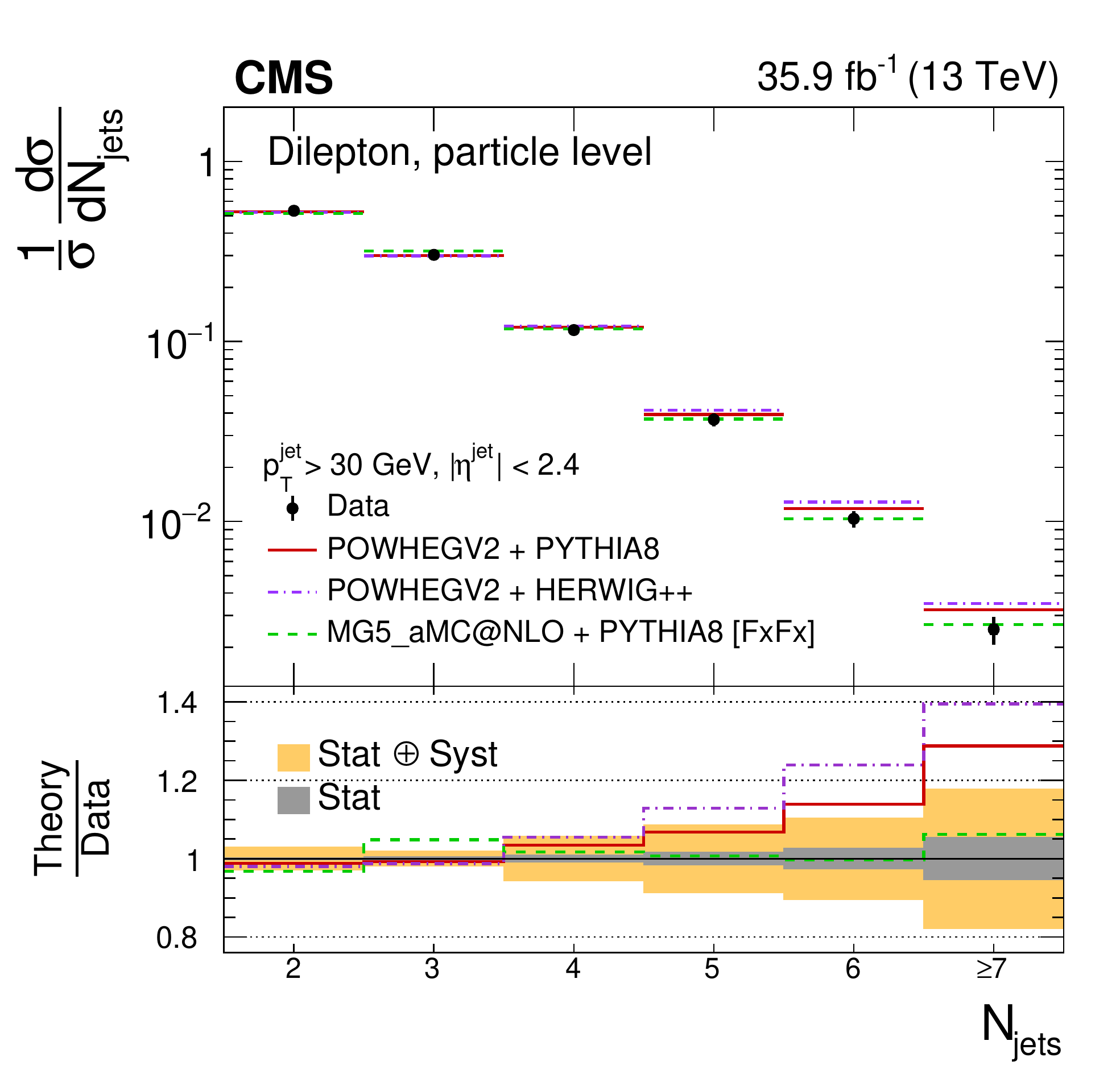}
\caption{The differential \ttbar production cross sections as a function of \Nj\ in a fiducial phase space at the particle level are shown for the data (points) and the MC predictions (lines). The vertical lines on the points indicate the total uncertainty in the data. The left and right plots correspond to absolute and normalised measurements, respectively. The lower panel in each plot shows the ratios of the theoretical predictions to the data. The dark and light bands show the relative statistical and total uncertainties in the data, respectively.}
\label{fig:diffxsec:res_njets}
\end{figure*}

The results for observables measured only at particle level are shown in Figs.~\ref{fig:diffxsec:res_ptlep}--\ref{fig:diffxsec:res_njets}. Within each figure, the left plot corresponds to the absolute measurements, and the right plot to the normalised measurements. The measurements of the kinematic properties of the leptons and {\cPqb} jets probe the modelling of the \ttbar production and top quark decay. Because of the excellent lepton energy resolution, the measurements of the lepton kinematic observables are particularly precise. The measurement of \delphill is used to constrain the CMDM of the top quark, as described in Section~\ref{sec:eft}. The measurement of \Nj\ probes higher-order corrections in the ME calculations and the modelling of radiation in the parton-shower simulations.
The \Nj\ measurement includes the integrated cross section for $\Nj > 7$ in the last bin. For all other observables, the first and last bins include the differential cross section integrated within the bin boundaries only.

\begin{figure*}[!phtb]
\centering
\includegraphics[width=0.99\textwidth]{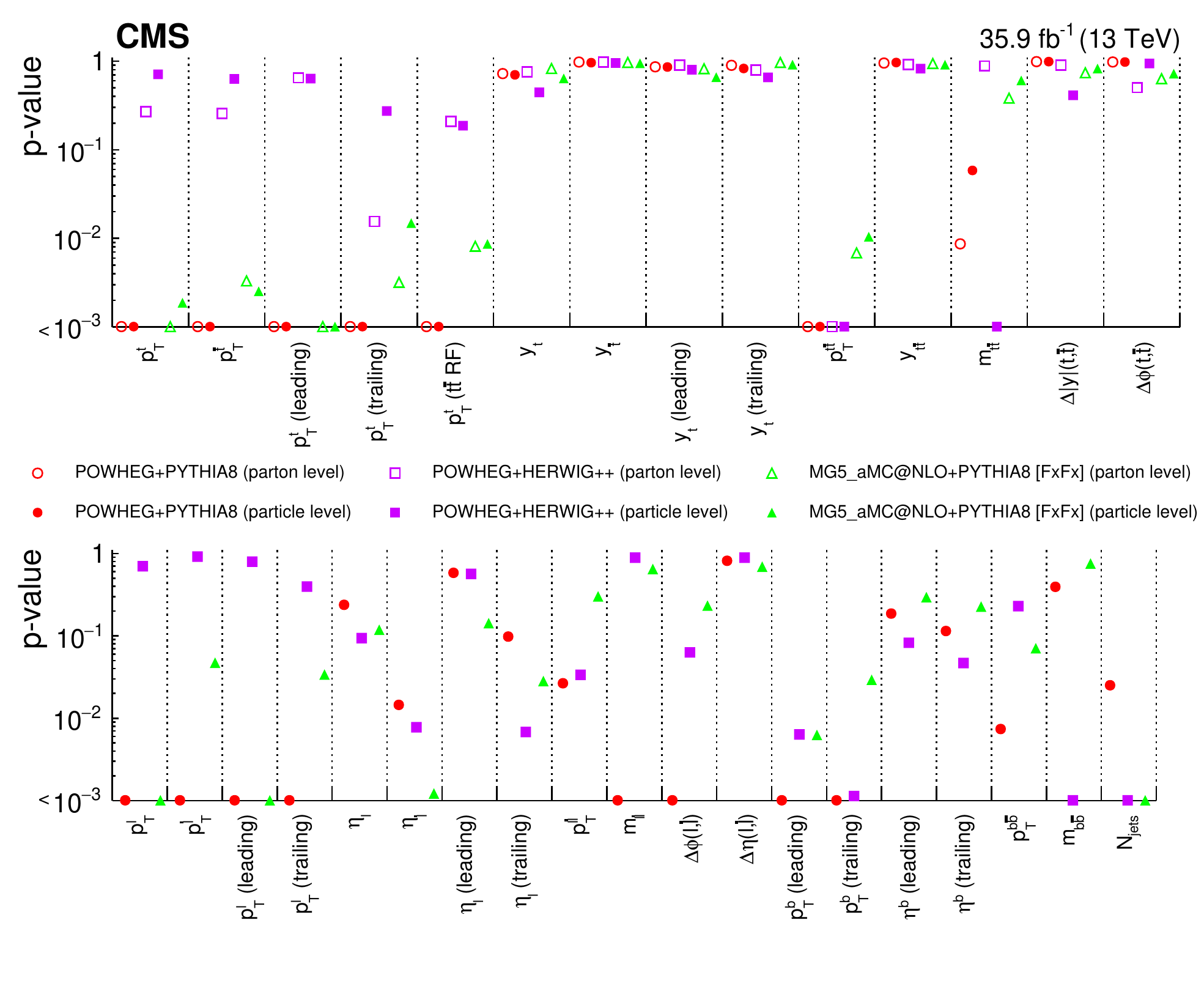}
\caption{The $p$-values quantifying the agreement between the data and MC predictions from for all normalised measurements are shown. Points situated on the horizontal axis indicate $p$-values less than 0.001. The upper panel includes distributions measured at parton and particle levels while the bottom panel includes those measured at particle level only.}
\label{fig:gof_summary}
\end{figure*}

\begin{figure*}[!phtb]
\centering
\includegraphics[width=0.99\textwidth]{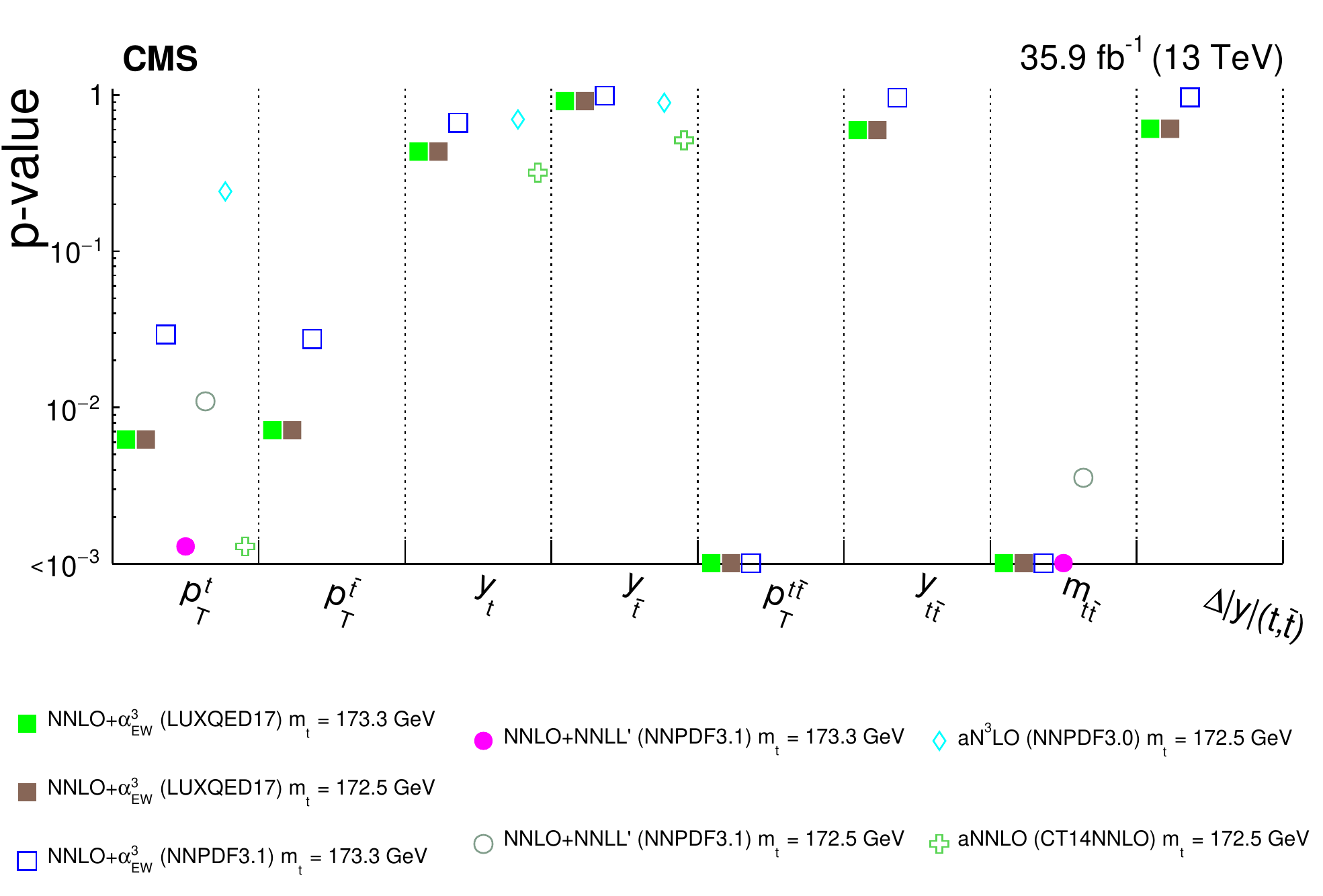}
\caption{The $p$-values quantifying the agreement between theoretical predictions with beyond-NLO precision and the data for selected normalised measurements at parton level are shown. Points situated on the horizontal axis indicate $p$-values of less than 0.001.}
\label{fig:gof_summary_bnlo}
\end{figure*}

{\tolerance=800 In Tables \ref{tab:norm_parton}--\ref{tab:abs_particle} in Appendix \ref{chi2:tables}, the \chisq per degree of freedom (dof) and corresponding $p$-value are shown, quantifying the agreement between the unfolded data and predictions for all the observables. In addition, Figs.~\ref{fig:gof_summary} and \ref{fig:gof_summary_bnlo} summarise the $p$-values for each normalised distribution. For most of the measured observables, we find generally good agreement between data and predictions, within the uncertainties in the data. The cases where significant disagreement is observed are now discussed.\par}

{\tolerance=800 Many of the different top quark \pt distributions shown in Figs.~\ref{fig:diffxsec:res_toppt}--\ref{fig:diffxsec:res_toppt_ttrestframe} exhibit significant disagreements between the data and the \pwhgpy predictions, varying smoothly from an excess of data for low \pt to a deficit for high \pt. Comparison of the data to the \mgamcpy prediction shows a similar excess of data at low \pt but a smaller deficit at high \pt . The \pwhghpp\ simulation provides a better modelling of the top quark \pt\ distributions, where a deficit of data for high \pt at the parton level is the only observed disagreement. For all MC-based predictions, the deficit at high \pt\ is most pronounced for the \pttop (trailing) distribution. Similar patterns of disagreement were observed at $\sqrt{s}=$ 7, 8, and 13\TeV by the ATLAS~\cite{bib:atlas_8TeV} and CMS~\cite{bib:TOP-12-028_paper,bib:TOP-16-008,Sirunyan:2018wem} Collaborations. The normalised and absolute \pttop and \ptantitop distributions show a similar level of disagreement with the beyond-NLO predictions.\par}

{\tolerance=1000 In Fig.~\ref{fig:diffxsec:res_ttpt}, a significant deficit of data with respect to both the \pwhgpy and \mgamcpy predictions is observed for large values of \pttt. Conversely, for the \pwhghpp\ prediction this deficit is not seen but there is an excess of data at moderate \pttt. For the \nnloew predictions shown in Fig.~\ref{fig:diffxsec:res_ttpt_bnlo}, a slight deficit of data at high \pttt is apparent. For the \mtt distributions in Figs.~\ref{fig:diffxsec:res_mtt} and \ref{fig:diffxsec:res_mtt_bnlo}, a significant excess of data with respect to all predictions in the lowest bin is observed. This excess is smaller for predictions with \mt\ = 172.5\GeV, which suggests a lower value of \mt\ could result in improved agreement for this distribution.\par}

{\tolerance=800 The distributions of kinematic properties of the leptons, {\cPqb} jets, dileptons, and {\cPqb} jet pairs (\ptlep, \ptb, \ptll, \ptbb, \mll, and \mbb) in Figs.~\ref{fig:diffxsec:res_ptlep}--\ref{fig:diffxsec:res_bbbarmass} exhibit similar disagreements with the predictions as the corresponding top-quark-based observables \pttop and \pttt with which they are correlated. In Fig.~\ref{fig:diffxsec:res_njets}, an increasing excess of data over the \pwhgpy and \pwhghpp\ predictions is observed for $\Nj\ \ge 4$. Conversely, there is good agreement between \mgamcpy and data for $\Nj\ > 3$, but disagreement for $\Nj\ = 2, 3$.\par}

\clearpage

\clearpage

\clearpage

\clearpage

\clearpage

\clearpage

\clearpage

\clearpage

\clearpage

\clearpage

\clearpage

\clearpage

\clearpage

\clearpage

\clearpage

\clearpage

\clearpage

\clearpage

\clearpage

\clearpage

\clearpage

\clearpage

\clearpage

\section{Constraining the top quark CMDM}
\label{sec:eft}
In the SM, the intrinsic spin and colour charge of the top quark give it a small magnetic dipole moment in the colour fields known as the top quark CMDM. An anomalous top quark CMDM is a feature of several BSM scenarios and can affect both the rate and kinematic properties of \ttbar production. The top quark may also have an anomalous chromoelectric dipole moment, however in this analysis it is assumed to be zero following the theoretical treatment presented in Ref. \cite{bib:zhang}. Until recently, the effect of an anomalous CMDM on \ttbar production was calculated only at LO in QCD. In Ref.~\cite{bib:zhang}, predictions for \ttbar production with anomalous CMDM at NLO in QCD in an EFT framework are provided. A comparison of the LO and NLO predictions reveals that the effect of the CMDM on \ttbar production is underestimated at LO, and that the NLO predictions have reduced  scale uncertainties with respect to those at LO. These two factors allow stronger constraints on the CMDM to be extracted using NLO predictions than with those at LO. In an EFT framework, the scale of new physics ($\Lambda$) is assumed to be large with respect to the typical scales probed at the LHC. Under this condition, BSM effects can be modelled in an EFT by adding a fixed set of dimension-6 operators to the SM Lagrangian \cite{Buchmuller:1985jz,Grzadkowski:2010es}. An operator commonly referred to as \otg\ is responsible for anomalous CMDM effects in the EFT~\cite{bib:zhang}. The contribution of \otg\ to the Lagrangian is parameterized by the dimensionless Wilson coefficient divided by the square of the BSM scale (\ctgl). The \otg\ operator results in a new \cPg\cPg\ttbar vertex, and modifies the \cPg\ttbar vertex, resulting in altered rates and kinematic properties in \ttbar production. Furthermore, changes in the chirality of the top quarks induced by \otg\ modify the spin correlation of the \ttbar pair. Thus, both the rate of \ttbar production and the difference in the azimuthal angle between the two leptons in dileptonic \ttbar events, \delphill, are sensitive to the value of \ctgl. The measurement of the absolute differential \ttbar cross section as a function of \delphill, in which the total cross section within the fiducial phase space is measured, is used to constrain \ctgl. The particle-level measurement in the fiducial phase space is the most appropriate for this purpose since it does not suffer from the model dependence introduced into the parton-level results when extrapolating to the full phase space.

To produce predictions for the \ttbar cross section as a function of \delphill and \ctgl, the model described in Ref.~\cite{bib:zhang} is implemented in the \MGaMCatNLO\ generator for the ME calculation at NLO in QCD. The parton shower and hadronisation steps are performed by interfacing this setup with \PYTHIA. The \RIVET\ framework~\cite{Buckley:2010ar} is used to apply the object definitions and requirements in order to produce particle-level predictions in the fiducial phase space identical to that of the measurements presented in this paper. The normalisations of the predictions are scaled with a $K$ factor to account for the NNLO+NNLL corrections to the inclusive \ttbar cross section calculated in Ref.~\cite{Czakon:2013goa}. However, as the acceptance of the fiducial phase space is calculated only at NLO precision, the normalisations of the predictions are not fully NNLO+NNLL precise. Since the \pttop distribution is poorly modelled by the NLO generators, the predictions are additionally corrected in order to match the \pttop prediction provided in Ref.~\cite{Czakon:2017wor} that corresponds to NNLO precision in QCD and includes electroweak corrections up to $\alphaew^{3}$.

\begin{figure*}[!phtb]
\centering
\includegraphics[width=0.47\textwidth]{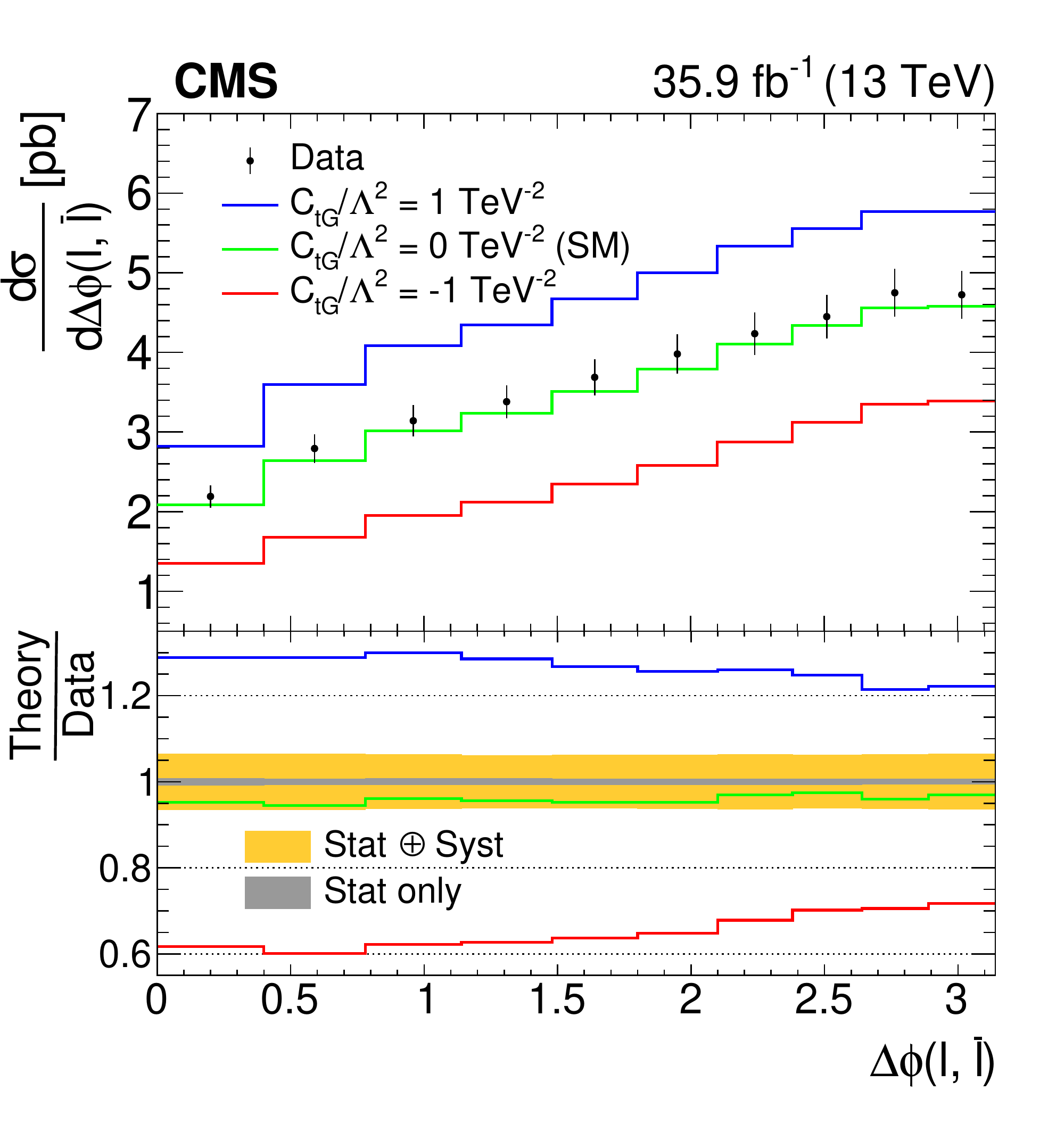}
\includegraphics[width=0.47\textwidth]{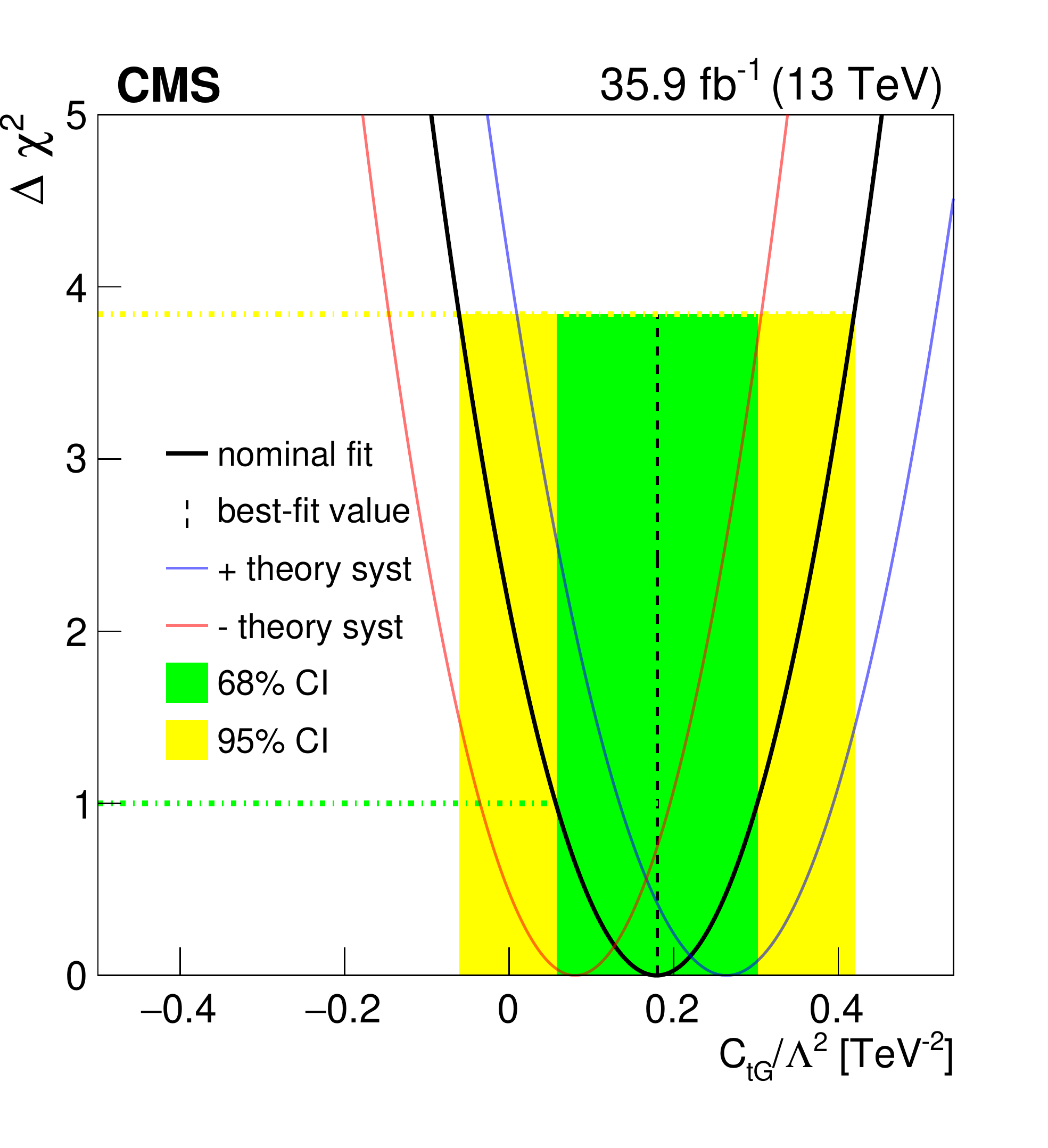}
\caption{ In the left plot, the differential \ttbar cross sections as a function of \delphill at the particle level in a fiducial phase space described in the text are shown. The points correspond to data and vertical bars on the points give the total uncertainty. The solid lines show the NLO predictions from the \MGaMCatNLO\ generator interfaced with \PYTHIA\ for \ctgl\ values of 1.0, 0.0, and $-1.0\TeV^{-2}$. The lower plot displays the ratio of the theoretical predictions to the data. In the right plot, \delchi values from the fit to the data in the left plot are shown as a function of \ctgl. The dark curve gives the result of the nominal fit, with the vertical dashed line giving the best-fit value. The two horizontal dashed lines indicate the \delchi values for the 68 and 95\% CIs. The dark and light bands correspond to those 68 and 95\% CIs, respectively. The other curves show the \delchi values for fits that give the maximally positive and negative changes in the best-fit value when the theoretical predictions are allowed to vary within their systematic uncertainties.}
\label{fig:eft_compare}
\end{figure*}

The upper left plot of Fig.~\ref{fig:eft_compare}, shows the measured differential cross section as a function of \delphill along with theoretical predictions for \ctgl\ values of 1.0, 0.0, and $-$1.0$\TeV^{-2}$. The high sensitivity of the normalisation of the measured differential cross section and the smaller sensitivity of its shape to the value of \ctgl\ are clearly seen in the lower panel of the left plot of Fig.~\ref{fig:eft_compare}, which displays the ratios of the predictions to the measurements for the three \ctgl\ values. The good agreement between the data and the \ctgl\ = 0.0 $\TeV^{-2}$ prediction corresponding to the SM is also apparent.

A \chisq minimisation technique is used to constrain \ctgl. The \chisq function is defined as:
\begin{equation*}
\chi^{2}(\ctgl)=\sum_{i=1}^{N}\sum_{j=1}^{N} (\text{data}_{i}-\text{pred}_{i} (\ctgl))(\text{data}_{j}-\text{pred}_{j}(\ctgl))\text{Cov}^{-1}_{i,j},
\end{equation*}

where $\text{data}_{i}$ and $\text{pred}_{i}$(\ctgl) are the measured and predicted differential cross section in the $i$th bin, respectively, and $\text{Cov}^{-1}_{i,j}$ is the ($i$th, $j$th) element of the inverse of the covariance matrix of the data. The covariance matrix accounts for all systematic and statistical uncertainties, as well as the inter-bin correlations introduced in the unfolding process. The minimisation results in a best fit value of 0.18$\TeV^{-2}$, corresponding to a \chidof of 0.3. Assuming Gaussian probability density functions for the uncertainties in the unfolded data, confidence intervals (CIs) can be estimated from the values of \ctgl\ for which the \delchi reaches certain values. The \delchi is defined as the difference in \chisq from the \chisq at the best fit value. This procedure yields a 95\% CI of $-0.06 < \ctgl < 0.41\TeV^{-2}$. Uncertainties arising from the theoretical predictions are separately estimated. First, the normalisations of the predictions are varied by $+5.8$\% and $-6.2$\%, corresponding to the addition in quadrature of the uncertainties from variations of the factorisation and renormalisation scales, PDFs, and \mt\ in the prediction from Ref.~\cite{Czakon:2013goa}. Second, the shapes of the predictions are varied by changing the factorisation and renormalisation scales by factors of 0.5 and 2.0 in the \MGaMCatNLO\ simulation. The \chisq minimisation is repeated for all variations, and the total theoretical uncertainty is determined from the maximally positive and negative effects on the best fit value of \ctgl. In the right plot of Fig.~\ref{fig:eft_compare}, the \delchi as a function of \ctgl\ is shown. The nominal fit to the data is represented by the solid curve with the \delchi values for the 68 and 95\% CIs indicated by the horizontal dashed lines. The dark and light regions display the corresponding 68 and 95\% CIs, respectively. Since the theoretical uncertainties do not have a clear frequentist interpretation, they are not included in the CIs. Rather, the other two curves in the figure show the results of the fits that produce the maximally positive and negative deviations from the best-fit value when the theoretical predictions are allowed to vary within their uncertainties.

In Ref.~\cite{bib:zhang}, 95\% CIs of $-0.42 < \ctgl < 0.30\TeV^{-2}$ and $-0.32 < \ctgl\ < 0.73\TeV^{-2}$ are derived using NLO predictions for the total \ttbar cross section as a function of \ctgl\ and measurements from $\sqrt{s} = 8\TeV$ CMS data \cite{CMS:2014gta} and $\sqrt{s} = 1.96\TeV$ Fermilab Tevatron data \cite{Aaltonen:2013wca}, respectively. The CMS Collaboration has previously used normalised differential \ttbar cross sections measured in the full phase space with 8\TeV data to constrain the top quark  CMDM~\cite{bib:Khachatryan:2016xws}. Using relations presented in Ref.~\cite{Bernreuther:2013aga}, these results of Ref.~\cite{bib:Khachatryan:2016xws} can be converted to a 95\% CI of $-0.89 < \ctgl < 0.43\TeV^{-2}$. Thus, the results of this work are consistent with, and improve upon, these previous constraints on \ctgl.

\section{Extraction of the top quark charge asymmetries}
\label{sec:ca}
\begin{figure}[phtb!]
\centering
\includegraphics[width=0.65\textwidth]{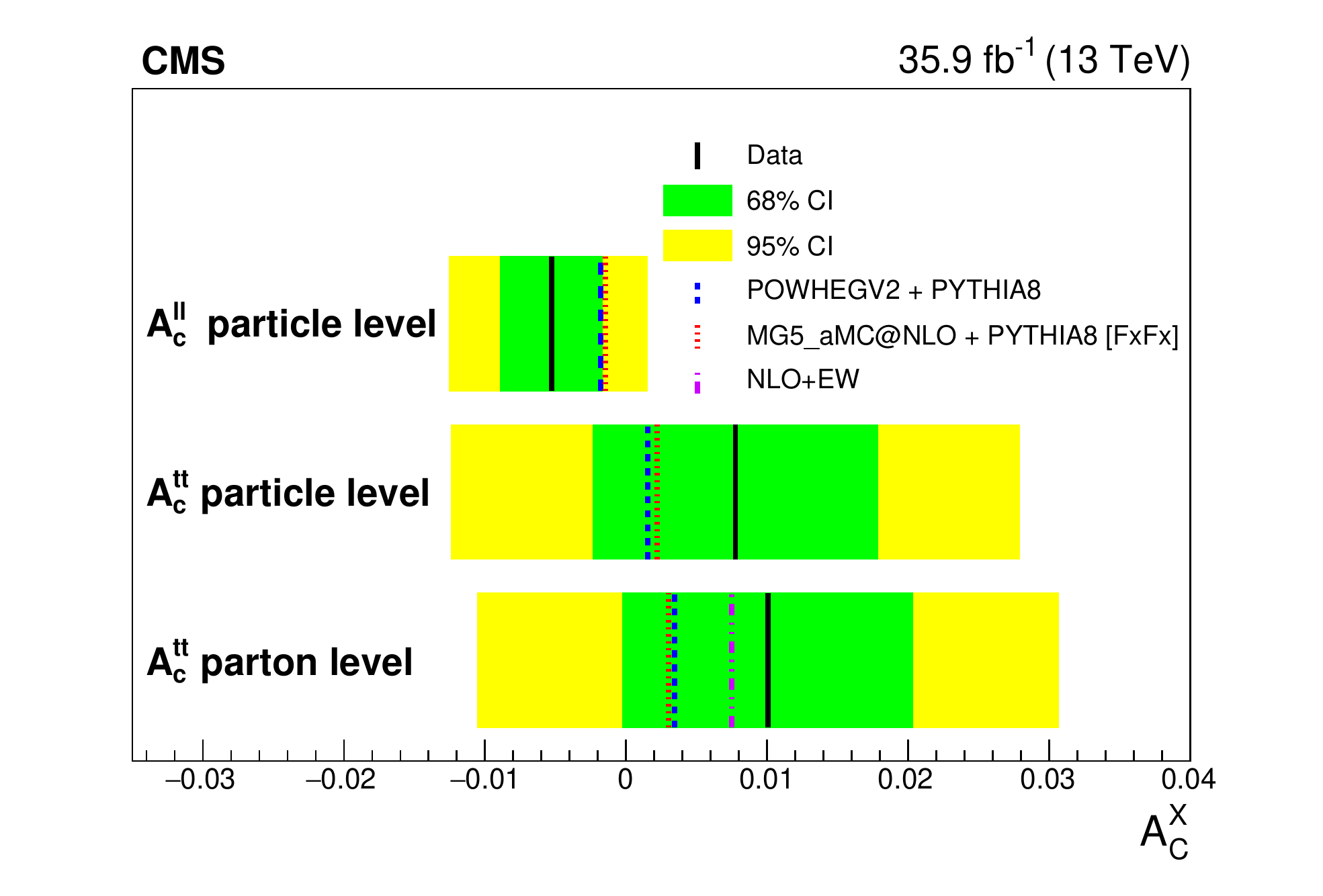}
\caption{The results of the \acx\ extraction (x = \ttbar or \llbar) from integrating the normalised parton- and particle-level differential cross section measurements as a function of \delytt and \deletall are shown. The central values for the data are indicated by the solid lines with the 68 and 95\% CIs represented by the dark and light shaded bands, respectively. The three types of dashed lines indicate the SM predictions produced with the \MGaMCatNLO\ and \POWHEG\ generators, both interfaced with \PYTHIA, and a calculation at NLO precision in QCD and including corrections arising from mixing between QCD and electroweak diagrams, and between QCD and QED diagrams~\cite{Bernreuther:2015yna}.}
\label{fig:ca}
\end{figure}

The measurements of normalised differential cross sections as a function of \delytt at parton and particle levels, and as a function of \deletall at particle level shown in Figs.~\ref{fig:diffxsec:res_deltay} and \ref{fig:diffxsec:res_deltaetall}, respectively, allow the extraction of the \ttbar and leptonic charge asymmetries, \act\ and \acl. These observables are sensitive to a number of BSM scenarios such as axigluon, \cPZpr, and \PWpr\ states coupling to top quarks~\cite{Aguilar-Saavedra2011}.
The \act\ and \acl\ asymmetries are defined as:
\begin{equation*}
\act = \frac{\sigma_{\ttbar}( \delytt  > 0) - \sigma_{\ttbar}( \delytt  < 0)}{\sigma_{\ttbar}( \delytt > 0) + \sigma_{\ttbar}( \delytt  < 0)}, \; \; \acl = \frac{\sigma_{\ttbar}(\deletall > 0)- \sigma_{\ttbar}(\deletall < 0)}{\sigma_{\ttbar}(\deletall > 0) + \sigma_{\ttbar}(\deletall < 0)},
\end{equation*}
where $\sigma_{\ttbar}$ represents the measured integrated \ttbar cross section in the specified range~\cite{bib:Kuhn:2011ri}. After the extraction of \act\ and \acl\ from the data, the uncertainties in \act\ and \acl\ are derived by combining the statistical and systematic uncertainties in the data in each bin, while accounting for the inter-bin correlations introduced during the unfolding procedure. The measured charge asymmetries and corresponding uncertainties are: $\act \text{(parton level)} = 0.01 \pm 0.009$, $\act \text{(particle level)} = 0.008 \pm 0.009$, and $\acl \text{(particle level)} = -0.005 \pm 0.004$. In Fig.~\ref{fig:ca}, the central values and the 68 and 95\% CI bands are compared with the SM predictions produced with the \POWHEG\ and \MGaMCatNLO\ generators interfaced with \PYTHIA, and a calculation at NLO precision in QCD and including corrections arising from mixing between QCD and electroweak diagrams, and between QCD and quantum electrodynamics (QED) diagrams taken from Ref.~\cite{Bernreuther:2015yna}. The results are in good agreement with the SM predictions and represent the first measurement of \act\ and \acl\ with 13\TeV data.

\section{Summary}
\label{sec:summary}
Measurements of differential \ttbar cross sections using events containing two oppositely charged leptons produced in pp~collisions at a centre-of-mass energy of 13\TeV are presented. The data were recorded with the CMS detector in 2016 and correspond to a integrated luminosity of \lumivalue. The differential cross sections are presented as functions of numerous observables related to \ttbar production and decay and are based on both particle-level objects in a phase space close to that of the detector acceptance and parton-level top quarks in the full phase space. For each observable, absolute and normalised differential cross sections are presented.  Most measured differential cross sections are well modelled by theoretical predictions. However, significant disagreement between the data and Monte Carlo simulation with next-to-leading-order (NLO) precision in quantum chromodynamics is observed for the transverse momentum of top quarks, leptons, {\cPqb} jets, \ttbar, \llbar, and \bbbar\ systems, and the invariant mass of the \ttbar, \llbar, and \bbbar\ systems. Predictions with beyond-NLO precision are generally in closer agreement with the data although some significant discrepancies remain. The jet multiplicity distribution is not well described by any of the Monte Carlo predictions. The absolute particle-level differential cross section as a function of \delphill is used to constrain the top quark chromomagnetic dipole moment at NLO precision in quantum chromodynamics using an effective field theory framework. The \ttbar and leptonic charge asymmetries are measured using 13\TeV data for the first time and found to be in agreement with standard model predictions.

\clearpage

\begin{acknowledgments}
\hyphenation{Bundes-ministerium Forschungs-gemeinschaft Forschungs-zentren Rachada-pisek} We congratulate our colleagues in the CERN accelerator departments for the excellent performance of the LHC and thank the technical and administrative staffs at CERN and at other CMS institutes for their contributions to the success of the CMS effort. In addition, we gratefully acknowledge the computing centres and personnel of the Worldwide LHC Computing Grid for delivering so effectively the computing infrastructure essential to our analyses. Finally, we acknowledge the enduring support for the construction and operation of the LHC and the CMS detector provided by the following funding agencies: the Austrian Federal Ministry of Education, Science and Research and the Austrian Science Fund; the Belgian Fonds de la Recherche Scientifique, and Fonds voor Wetenschappelijk Onderzoek; the Brazilian Funding Agencies (CNPq, CAPES, FAPERJ, FAPERGS, and FAPESP); the Bulgarian Ministry of Education and Science; CERN; the Chinese Academy of Sciences, Ministry of Science and Technology, and National Natural Science Foundation of China; the Colombian Funding Agency (COLCIENCIAS); the Croatian Ministry of Science, Education and Sport, and the Croatian Science Foundation; the Research Promotion Foundation, Cyprus; the Secretariat for Higher Education, Science, Technology and Innovation, Ecuador; the Ministry of Education and Research, Estonian Research Council via IUT23-4 and IUT23-6 and European Regional Development Fund, Estonia; the Academy of Finland, Finnish Ministry of Education and Culture, and Helsinki Institute of Physics; the Institut National de Physique Nucl\'eaire et de Physique des Particules~/~CNRS, and Commissariat \`a l'\'Energie Atomique et aux \'Energies Alternatives~/~CEA, France; the Bundesministerium f\"ur Bildung und Forschung, Deutsche Forschungsgemeinschaft, and Helmholtz-Gemeinschaft Deutscher Forschungszentren, Germany; the General Secretariat for Research and Technology, Greece; the National Research, Development and Innovation Fund, Hungary; the Department of Atomic Energy and the Department of Science and Technology, India; the Institute for Studies in Theoretical Physics and Mathematics, Iran; the Science Foundation, Ireland; the Istituto Nazionale di Fisica Nucleare, Italy; the Ministry of Science, ICT and Future Planning, and National Research Foundation (NRF), Republic of Korea; the Ministry of Education and Science of the Republic of Latvia; the Lithuanian Academy of Sciences; the Ministry of Education, and University of Malaya (Malaysia); the Ministry of Science of Montenegro; the Mexican Funding Agencies (BUAP, CINVESTAV, CONACYT, LNS, SEP, and UASLP-FAI); the Ministry of Business, Innovation and Employment, New Zealand; the Pakistan Atomic Energy Commission; the Ministry of Science and Higher Education and the National Science Centre, Poland; the Funda\c{c}\~ao para a Ci\^encia e a Tecnologia, Portugal; JINR, Dubna; the Ministry of Education and Science of the Russian Federation, the Federal Agency of Atomic Energy of the Russian Federation, Russian Academy of Sciences, the Russian Foundation for Basic Research, and the National Research Center ``Kurchatov Institute"; the Ministry of Education, Science and Technological Development of Serbia; the Secretar\'{\i}a de Estado de Investigaci\'on, Desarrollo e Innovaci\'on, Programa Consolider-Ingenio 2010, Plan Estatal de Investigaci\'on Cient\'{\i}fica y T\'ecnica y de Innovaci\'on 2013-2016, Plan de Ciencia, Tecnolog\'{i}a e Innovaci\'on 2013-2017 del Principado de Asturias, and Fondo Europeo de Desarrollo Regional, Spain; the Ministry of Science, Technology and Research, Sri Lanka; the Swiss Funding Agencies (ETH Board, ETH Zurich, PSI, SNF, UniZH, Canton Zurich, and SER); the Ministry of Science and Technology, Taipei; the Thailand Center of Excellence in Physics, the Institute for the Promotion of Teaching Science and Technology of Thailand, Special Task Force for Activating Research and the National Science and Technology Development Agency of Thailand; the Scientific and Technical Research Council of Turkey, and Turkish Atomic Energy Authority; the National Academy of Sciences of Ukraine, and State Fund for Fundamental Researches, Ukraine; the Science and Technology Facilities Council, UK; the US Department of Energy, and the US National Science Foundation.

Individuals have received support from the Marie-Curie programme and the European Research Council and Horizon 2020 Grant, contract No. 675440 (European Union); the Leventis Foundation; the A. P. Sloan Foundation; the Alexander von Humboldt Foundation; the Belgian Federal Science Policy Office; the Fonds pour la Formation \`a la Recherche dans l'Industrie et dans l'Agriculture (FRIA-Belgium); the Agentschap voor Innovatie door Wetenschap en Technologie (IWT-Belgium); the F.R.S.-FNRS and FWO (Belgium) under the ``Excellence of Science - EOS" - be.h project n. 30820817; the Ministry of Education, Youth and Sports (MEYS) of the Czech Republic; the Lend\"ulet (``Momentum") Programme and the J\'anos Bolyai Research Scholarship of the Hungarian Academy of Sciences, the New National Excellence Program \'UNKP, the NKFIA research grants 123842, 123959, 124845, 124850 and 125105 (Hungary); the Council of Scientific and Industrial Research, India; the HOMING PLUS programme of the Foundation for Polish Science, cofinanced from European Union, Regional Development Fund, the Mobility Plus programme of the Ministry of Science and Higher Education, the National Science Center (Poland), contracts Harmonia 2014/14/M/ST2/00428, Opus 2014/13/B/ST2/02543, 2014/15/B/ST2/03998, and 2015/19/B/ST2/02861, Sonata-bis 2012/07/E/ST2/01406; the National Priorities Research Program by Qatar National Research Fund; the Programa de Excelencia Mar\'{i}a de Maeztu, and the Programa Severo Ochoa del Principado de Asturias; the Thalis and Aristeia programmes cofinanced by EU-ESF, and the Greek NSRF; the Rachadapisek Sompot Fund for Postdoctoral Fellowship, Chulalongkorn University, and the Chulalongkorn Academic into Its 2nd Century Project Advancement Project (Thailand); the Welch Foundation, contract C-1845; and the Weston Havens Foundation (USA).

\end{acknowledgments}

\clearpage
\newpage
\bibliography{auto_generated}

\providecommand{\href}[2]{#2}\begingroup\raggedright\begin{thebibliography}{10}%
\makeatletter
\providecommand{\hrefCMSnoop }[0]{\@secondoftwo}%
\makeatother
\providecommand{\doi}{\texttt{doi:}\begingroup \urlstyle{tt}\Url}

\bibitem{bib:ATLASnew}
\hrefCMSnoop {}{{ATLAS Collaboration}, ``Measurements of top quark pair
  relative differential cross-sections with {ATLAS} in pp collisions at
  $\sqrt{s}$ = 7 {TeV}'',} \textit{ Eur. Phys. J. C} \textbf{ 73} (2013) 2261,
  \href{http://dx.doi.org/10.1140/epjc/s10052-012-2261-1}{\doi{10.1140/epjc/s10052-012-2261-1}},
  \href{http://www.arXiv.org/abs/1207.5644}{\texttt{arXiv:1207.5644}}.

\bibitem{Aad:2014iaa}
\hrefCMSnoop {}{{ATLAS Collaboration}, ``Measurement of the \ttbar\ production
  cross-section as a function of jet multiplicity and jet transverse momentum
  in 7 {TeV} proton-proton collisions with the {ATLAS} detector'',} \textit{
  JHEP} \textbf{ 01} (2015) 020,
  \href{http://dx.doi.org/10.1007/JHEP01(2015)020}{\doi{10.1007/JHEP01(2015)020}},
  \href{http://www.arXiv.org/abs/1407.0891}{\texttt{arXiv:1407.0891}}.

\bibitem{bib:TOP-11-013_paper}
\hrefCMSnoop {}{{CMS Collaboration}, ``Measurement of differential top-quark
  pair production cross sections in pp collisions at $\sqrt{s}$ = 7 {TeV}'',}
  \textit{ Eur. Phys. J. C} \textbf{ 73} (2013) 2339,
  \href{http://dx.doi.org/10.1140/epjc/s10052-013-2339-4}{\doi{10.1140/epjc/s10052-013-2339-4}},
  \href{http://www.arXiv.org/abs/1211.2220}{\texttt{arXiv:1211.2220}}.

\bibitem{Khachatryan:2016oou}
\hrefCMSnoop {}{{CMS Collaboration}, ``Measurement of the differential cross
  sections for top quark pair production as a function of kinematic event
  variables in pp collisions at $\sqrt{s}$ = 7 and 8 {TeV}'',} \textit{ Phys.
  Rev. D} \textbf{ 94} (2016) 052006,
  \href{http://dx.doi.org/10.1103/PhysRevD.94.052006}{\doi{10.1103/PhysRevD.94.052006}},
  \href{http://www.arXiv.org/abs/1607.00837}{\texttt{arXiv:1607.00837}}.

\bibitem{bib:atlas_8TeV}
\hrefCMSnoop {}{{ATLAS Collaboration}, ``Measurements of top-quark pair
  differential cross-sections in the lepton+jets channel in pp collisions at
  $\sqrt{s}=8$ {TeV} using the {ATLAS} detector'',} \textit{ Eur. Phys. J. C}
  \textbf{ 76} (2016) 538,
  \href{http://dx.doi.org/10.1140/epjc/s10052-016-4366-4}{\doi{10.1140/epjc/s10052-016-4366-4}},
  \href{http://www.arXiv.org/abs/1511.04716}{\texttt{arXiv:1511.04716}}.

\bibitem{Aaboud:2017ujq}
\hrefCMSnoop {}{{ATLAS Collaboration}, ``Measurement of lepton differential
  distributions and the top quark mass in \ttbar\ production in pp collisions
  at $\sqrt{s}=8$ {TeV} with the {ATLAS} detector'',} \textit{ Eur. Phys. J. C}
  \textbf{ 77} (2017) 804,
  \href{http://dx.doi.org/10.1140/epjc/s10052-017-5349-9}{\doi{10.1140/epjc/s10052-017-5349-9}},
  \href{http://www.arXiv.org/abs/1709.09407}{\texttt{arXiv:1709.09407}}.

\bibitem{Aad:2015hna}
\hrefCMSnoop {}{{ATLAS Collaboration}, ``{Measurement of the differential
  cross-section of highly boosted top quarks as a function of their transverse
  momentum in $\sqrt{s}$ = 8 {TeV} proton-proton collisions using the {ATLAS}
  detector}'',} \textit{ Phys. Rev. D} \textbf{ 93} (2016) 032009,
  \href{http://dx.doi.org/10.1103/PhysRevD.93.032009}{\doi{10.1103/PhysRevD.93.032009}},
  \href{http://www.arXiv.org/abs/1510.03818}{\texttt{arXiv:1510.03818}}.

\bibitem{bib:TOP-12-028_paper}
\hrefCMSnoop {}{{CMS Collaboration}, ``{Measurement of the differential cross
  section for top quark pair production in pp collisions at $\sqrt{s} = 8$
  {TeV}}'',} \textit{ Eur. Phys. J. C} \textbf{ 75} (2015) 542,
  \href{http://dx.doi.org/10.1140/epjc/s10052-015-3709-x}{\doi{10.1140/epjc/s10052-015-3709-x}},
  \href{http://www.arXiv.org/abs/1505.04480}{\texttt{arXiv:1505.04480}}.

\bibitem{bib:TOP-12-041_paper}
\hrefCMSnoop {}{{CMS Collaboration}, ``Measurement of $\mathrm
  {t}\overline{\mathrm {t}}$ production with additional jet activity, including
  $\mathrm{b}$ quark jets, in the dilepton decay channel using pp collisions at
  $\sqrt{s} = 8$ {TeV}'',} \textit{ Eur. Phys. J. C} \textbf{ 76} (2016) 379,
  \href{http://dx.doi.org/10.1140/epjc/s10052-016-4105-x}{\doi{10.1140/epjc/s10052-016-4105-x}},
  \href{http://www.arXiv.org/abs/1510.03072}{\texttt{arXiv:1510.03072}}.

\bibitem{Khachatryan:2015fwh}
\hrefCMSnoop {}{{CMS Collaboration}, ``Measurement of the
  $\mathrm{t}\overline{{\mathrm{t}}}$ production cross section in the all-jets
  final state in pp collisions at $\sqrt{s}=8$ {TeV}'',} \textit{ Eur. Phys. J.
  C} \textbf{ 76} (2016) 128,
  \href{http://dx.doi.org/10.1140/epjc/s10052-016-3956-5}{\doi{10.1140/epjc/s10052-016-3956-5}},
  \href{http://www.arXiv.org/abs/Khachatryan:2016gxp}{\texttt{arXiv:Khachatryan:2016gxp}}.

\bibitem{Khachatryan:2016gxp}
\hrefCMSnoop {}{{CMS Collaboration}, ``Measurement of the integrated and
  differential \ttbar\ production cross sections for high-$p_t$ top quarks in
  pp collisions at $\sqrt s =$ 8 {TeV}'',} \textit{ Phys. Rev. D} \textbf{ 94}
  (2016) 072002,
  \href{http://dx.doi.org/10.1103/PhysRevD.94.072002}{\doi{10.1103/PhysRevD.94.072002}},
  \href{http://www.arXiv.org/abs/1605.00116}{\texttt{arXiv:1605.00116}}.

\bibitem{Sirunyan:2017azo}
\hrefCMSnoop {}{{CMS Collaboration}, ``Measurement of double-differential cross
  sections for top quark pair production in pp collisions at $\sqrt{s} = 8$
  $\,\text {TeV}$ and impact on parton distribution functions'',} \textit{ Eur.
  Phys. J. C} \textbf{ 77} (2017) 459,
  \href{http://dx.doi.org/10.1140/epjc/s10052-017-4984-5}{\doi{10.1140/epjc/s10052-017-4984-5}},
  \href{http://www.arXiv.org/abs/1703.01630}{\texttt{arXiv:1703.01630}}.

\bibitem{Aaboud:2017fha}
\hrefCMSnoop {}{{ATLAS Collaboration}, ``{Measurements of top-quark pair
  differential cross-sections in the lepton+jets channel in pp collisions at
  $\sqrt{s}$ = 13 {TeV} using the {ATLAS} detector}'',} \textit{ JHEP} \textbf{
  11} (2017) 191,
  \href{http://dx.doi.org/10.1007/JHEP11(2017)191}{\doi{10.1007/JHEP11(2017)191}},
  \href{http://www.arXiv.org/abs/1708.00727}{\texttt{arXiv:1708.00727}}.

\bibitem{Aaboud:2016xii}
\hrefCMSnoop {}{{ATLAS Collaboration}, ``Measurement of jet activity produced
  in top-quark events with an electron, a muon and two b-tagged jets in the
  final state in pp collisions at $\sqrt{s}=13$ {TeV} with the {ATLAS}
  detector'',} \textit{ Eur. Phys. J. C} \textbf{ 77} (2017) 220,
  \href{http://dx.doi.org/10.1140/epjc/s10052-017-4766-0}{\doi{10.1140/epjc/s10052-017-4766-0}},
  \href{http://www.arXiv.org/abs/1610.09978}{\texttt{arXiv:1610.09978}}.

\bibitem{bib:Aaboud:2016syx}
\hrefCMSnoop {}{{ATLAS Collaboration}, ``Measurements of top-quark pair
  differential cross-sections in the e$\mu$ channel in pp collisions at
  $\sqrt{s} = 13$ {TeV} using the {ATLAS} detector'',} \textit{ Eur. Phys. J.
  C} \textbf{ 77} (2017) 292,
  \href{http://dx.doi.org/10.1140/epjc/s10052-017-4821-x}{\doi{10.1140/epjc/s10052-017-4821-x}},
  \href{http://www.arXiv.org/abs/1612.05220}{\texttt{arXiv:1612.05220}}.

\bibitem{bib:TOP-16-008}
\hrefCMSnoop {}{{CMS Collaboration}, ``Measurement of differential cross
  sections for top quark pair production using the lepton+jets final state in
  proton-proton collisions at 13 {TeV}'',} \textit{ Phys. Rev. D} \textbf{ 95}
  (2017) 092001,
  \href{http://dx.doi.org/10.1103/PhysRevD.95.092001}{\doi{10.1103/PhysRevD.95.092001}},
  \href{http://www.arXiv.org/abs/1610.04191}{\texttt{arXiv:1610.04191}}.

\bibitem{Sirunyan:2018wem}
\hrefCMSnoop {}{{CMS Collaboration}, ``Measurement of differential cross
  sections for the production of top quark pairs and of additional jets in
  lepton+jets events from pp collisions at $\sqrt{s} =$ 13 {TeV}'',} \textit{
  Phys. Rev. D} \textbf{ 97} (2018) 112003,
  \href{http://dx.doi.org/10.1103/PhysRevD.97.112003}{\doi{10.1103/PhysRevD.97.112003}},
  \href{http://www.arXiv.org/abs/1803.08856}{\texttt{arXiv:1803.08856}}.

\bibitem{Sirunyan:2017mzl}
\hrefCMSnoop {}{{CMS Collaboration}, ``Measurement of normalized differential
  $\mathrm{t}\overline{\mathrm{t}}$ cross sections in the dilepton channel from
  pp collisions at $ \sqrt{s}=13 $ {TeV}'',} \textit{ JHEP} \textbf{ 04} (2018)
  060,
  \href{http://dx.doi.org/10.1007/JHEP04(2018)060}{\doi{10.1007/JHEP04(2018)060}},
  \href{http://www.arXiv.org/abs/1708.07638}{\texttt{arXiv:1708.07638}}.

\bibitem{Sirunyan:2018ptc}
\hrefCMSnoop {}{{CMS Collaboration}, ``Measurements of differential cross
  sections of top quark pair production as a function of kinematic event
  variables in proton-proton collisions at $ \sqrt{s}=13 $ {TeV}'',} \textit{
  JHEP} \textbf{ 06} (2018) 002,
  \href{http://dx.doi.org/10.1007/JHEP06(2018)002}{\doi{10.1007/JHEP06(2018)002}},
  \href{http://www.arXiv.org/abs/1803.03991}{\texttt{arXiv:1803.03991}}.

\bibitem{bib:CMS-PAS-LUM-17-001}
\href {https://cds.cern.ch/record/2257069}{{CMS Collaboration}, ``{CMS}
  luminosity measurements for the 2016 data taking period'',} Technical Report
  CMS-PAS-LUM-17-001, CERN, Geneva, 2017.

\bibitem{bib:difftop}
\hrefCMSnoop {}{M.~Guzzi, K.~Lipka, and S.-O. Moch, ``Top-quark pair production
  at hadron colliders: differential cross section and phenomenological
  applications with difftop'',} \textit{ JHEP} \textbf{ 01} (2015) 082,
  \href{http://dx.doi.org/10.1007/JHEP01(2015)082}{\doi{10.1007/JHEP01(2015)082}},
  \href{http://www.arXiv.org/abs/1406.0386}{\texttt{arXiv:1406.0386}}.

\bibitem{bib:kidonakis_13TeV}
\hrefCMSnoop {}{N.~Kidonakis, ``{NNNLO} soft-gluon corrections for the
  top-quark \pt and rapidity distributions'',} \textit{ Phys. Rev. D} \textbf{
  91} (2015) 031501,
  \href{http://dx.doi.org/10.1103/PhysRevD.91.031501}{\doi{10.1103/PhysRevD.91.031501}},
  \href{http://www.arXiv.org/abs/1411.2633}{\texttt{arXiv:1411.2633}}.

\bibitem{bib:mitov}
\hrefCMSnoop {}{M.~Czakon, D.~Heymes, and A.~Mitov, ``High-precision
  differential predictions for top-quark pairs at the {LHC}'',} \textit{ Phys.
  Rev. Lett.} \textbf{ 116} (2016) 082003,
  \href{http://dx.doi.org/10.1103/PhysRevLett.116.082003}{\doi{10.1103/PhysRevLett.116.082003}},
  \href{http://www.arXiv.org/abs/1511.00549}{\texttt{arXiv:1511.00549}}.

\bibitem{bib:Pecjak:2016nee}
\hrefCMSnoop {}{B.~D. Pecjak, D.~J. Scott, X.~Wang, and L.~L. Yang, ``Resummed
  differential cross sections for top-quark pairs at the {LHC}'',} \textit{
  Phys. Rev. Lett.} \textbf{ 116} (2016) 202001,
  \href{http://dx.doi.org/10.1103/PhysRevLett.116.202001}{\doi{10.1103/PhysRevLett.116.202001}},
  \href{http://www.arXiv.org/abs/1601.07020}{\texttt{arXiv:1601.07020}}.

\bibitem{Li:2013uma}
X.-Q. Li\hrefCMSnoop {}{ {et~al.}, ``Light top squark in precision top quark
  sample'',} \textit{ Phys. Rev. D} \textbf{ 89} (2014) 077703,
  \href{http://dx.doi.org/10.1103/PhysRevD.89.077703}{\doi{10.1103/PhysRevD.89.077703}},
  \href{http://www.arXiv.org/abs/1311.6874}{\texttt{arXiv:1311.6874}}.

\bibitem{Frederix:2007gi}
\hrefCMSnoop {}{R.~Frederix and F.~Maltoni, ``Top pair invariant mass
  distribution: a window on new physics'',} \textit{ JHEP} \textbf{ 01} (2009)
  047,
  \href{http://dx.doi.org/10.1088/1126-6708/2009/01/047}{\doi{10.1088/1126-6708/2009/01/047}},
  \href{http://www.arXiv.org/abs/0712.2355}{\texttt{arXiv:0712.2355}}.

\bibitem{Harris:2011ez}
\hrefCMSnoop {}{R.~M. Harris and S.~Jain, ``Cross sections for leptophobic
  topcolor {Z'} decaying to top-antitop'',} \textit{ Eur. Phys. J. C} \textbf{
  72} (2012) 2072,
  \href{http://dx.doi.org/10.1140/epjc/s10052-012-2072-4}{\doi{10.1140/epjc/s10052-012-2072-4}},
  \href{http://www.arXiv.org/abs/1112.4928}{\texttt{arXiv:1112.4928}}.

\bibitem{bib:zhang}
\hrefCMSnoop {}{D.~Buarque~Franzosi and C.~Zhang, ``Probing the top-quark
  chromomagnetic dipole moment at next-to-leading order in {QCD}'',} \textit{
  Phys. Rev. D} \textbf{ 91} (2015) 114010,
  \href{http://dx.doi.org/10.1103/PhysRevD.91.114010}{\doi{10.1103/PhysRevD.91.114010}},
  \href{http://www.arXiv.org/abs/1503.08841}{\texttt{arXiv:1503.08841}}.

\bibitem{Martinez:2007qf}
\hrefCMSnoop {}{R.~Martinez, M.~A. Perez, and N.~Poveda, ``Chromomagnetic
  dipole moment of the top quark revisited'',} \textit{ Eur. Phys. J. C}
  \textbf{ 53} (2008) 221,
  \href{http://dx.doi.org/10.1140/epjc/s10052-007-0457-6}{\doi{10.1140/epjc/s10052-007-0457-6}},
  \href{http://www.arXiv.org/abs/hep-ph/0701098}{\texttt{arXiv:hep-ph/0701098}}.

\bibitem{Khachatryan:2016bia}
\hrefCMSnoop {}{{CMS Collaboration}, ``The {CMS} trigger system'',} \textit{
  JINST} \textbf{ 12} (2017) P01020,
  \href{http://dx.doi.org/10.1088/1748-0221/12/01/P01020}{\doi{10.1088/1748-0221/12/01/P01020}},
  \href{http://www.arXiv.org/abs/1609.02366}{\texttt{arXiv:1609.02366}}.

\bibitem{bib:Chatrchyan:2008zzk}
\hrefCMSnoop {}{{CMS Collaboration}, ``The {CMS} experiment at the {CERN}
  {LHC}'',} \textit{ JINST} \textbf{ 3} (2008) S08004,
  \href{http://dx.doi.org/10.1088/1748-0221/3/08/S08004}{\doi{10.1088/1748-0221/3/08/S08004}}.

\bibitem{Frixione:2007nw}
\hrefCMSnoop {}{S.~Frixione, P.~Nason, and G.~Ridolfi, ``A positive-weight
  next-to-leading-order {Monte Carlo} for heavy flavour hadroproduction'',}
  \textit{ JHEP} \textbf{ 09} (2007) 126,
  \href{http://dx.doi.org/10.1088/1126-6708/2007/09/126}{\doi{10.1088/1126-6708/2007/09/126}},
  \href{http://www.arXiv.org/abs/0707.3088}{\texttt{arXiv:0707.3088}}.

\bibitem{bib:powheg0}
\hrefCMSnoop {}{P.~Nason, ``A new method for combining {NLO} {QCD} with shower
  {Monte} {Carlo} algorithms'',} \textit{ JHEP} \textbf{ 11} (2004) 040,
  \href{http://dx.doi.org/10.1088/1126-6708/2004/11/040}{\doi{10.1088/1126-6708/2004/11/040}},
  \href{http://www.arXiv.org/abs/hep-ph/0409146}{\texttt{arXiv:hep-ph/0409146}}.

\bibitem{bib:powheg}
\hrefCMSnoop {}{S.~Frixione, P.~Nason, and C.~Oleari, ``Matching {NLO} {QCD}
  computations with parton shower simulations: the {POWHEG} method'',} \textit{
  JHEP} \textbf{ 11} (2007) 070,
  \href{http://dx.doi.org/10.1088/1126-6708/2007/11/070}{\doi{10.1088/1126-6708/2007/11/070}},
  \href{http://www.arXiv.org/abs/0709.2092}{\texttt{arXiv:0709.2092}}.

\bibitem{bib:powheg2}
\hrefCMSnoop {}{S.~Alioli, P.~Nason, C.~Oleari, and E.~Re, ``A general
  framework for implementing {NLO} calculations in shower {Monte} {Carlo}
  programs: the {POWHEG} {BOX}'',} \textit{ JHEP} \textbf{ 06} (2010) 043,
  \href{http://dx.doi.org/10.1007/JHEP06(2010)043}{\doi{10.1007/JHEP06(2010)043}},
  \href{http://www.arXiv.org/abs/1002.2581}{\texttt{arXiv:1002.2581}}.

\bibitem{bib:CMS:2016kle}
\href {https://cds.cern.ch/record/2235192}{{CMS Collaboration},
  ``Investigations of the impact of the parton shower tuning in {PYTHIA} in the
  modelling of \ttbar\ at $\sqrt{s}=8$ and 13 {TeV}'',} CMS Physics Analysis
  Summary CMS-PAS-TOP-16-021, CERN, 2016.

\bibitem{Sjostrand:2007gs}
T.~Sj{\"o}strand\hrefCMSnoop {}{ {et~al.}, ``An introduction to {PYTHIA}
  8.2'',} \textit{ Comput. Phys. Commun.} \textbf{ 191} (2015) 159,
  \href{http://dx.doi.org/10.1016/j.cpc.2015.01.024}{\doi{10.1016/j.cpc.2015.01.024}},
  \href{http://www.arXiv.org/abs/1410.3012}{\texttt{arXiv:1410.3012}}.

\bibitem{bib:CUETP8tune}
\hrefCMSnoop {}{{CMS Collaboration}, ``Event generator tunes obtained from
  underlying event and multiparton scattering measurements'',} \textit{ Eur.
  Phys. J. C} \textbf{ 76} (2016) 155,
  \href{http://dx.doi.org/10.1140/epjc/s10052-016-3988-x}{\doi{10.1140/epjc/s10052-016-3988-x}},
  \href{http://www.arXiv.org/abs/1512.00815}{\texttt{arXiv:1512.00815}}.

\bibitem{Skands:2014pea}
\hrefCMSnoop {}{P.~Skands, S.~Carrazza, and J.~Rojo, ``Tuning {PYTHIA} 8.1: the
  {Monash} 2013 tune'',} \textit{ Eur. Phys. J. C} \textbf{ 74} (2014) 3024,
  \href{http://dx.doi.org/10.1140/epjc/s10052-014-3024-y}{\doi{10.1140/epjc/s10052-014-3024-y}},
  \href{http://www.arXiv.org/abs/1404.5630}{\texttt{arXiv:1404.5630}}.

\bibitem{Alwall:2014hca}
J.~Alwall\hrefCMSnoop {}{ {et~al.}, ``The automated computation of tree-level
  and next-to-leading order differential cross sections, and their matching to
  parton shower simulations'',} \textit{ JHEP} \textbf{ 07} (2014) 079,
  \href{http://dx.doi.org/10.1007/JHEP07(2014)079}{\doi{10.1007/JHEP07(2014)079}},
  \href{http://www.arXiv.org/abs/1405.0301}{\texttt{arXiv:1405.0301}}.

\bibitem{bib:madspin}
\hrefCMSnoop {}{P.~Artoisenet, R.~Frederix, O.~Mattelaer, and R.~Rietkerk,
  ``Automatic spin-entangled decays of heavy resonances in {M}onte {C}arlo
  simulations'',} \textit{ JHEP} \textbf{ 03} (2013) 015,
  \href{http://dx.doi.org/10.1007/JHEP03(2013)015}{\doi{10.1007/JHEP03(2013)015}},
  \href{http://www.arXiv.org/abs/1212.3460}{\texttt{arXiv:1212.3460}}.

\bibitem{Frederix:2012ps}
\hrefCMSnoop {}{R.~Frederix and S.~Frixione, ``Merging meets matching in
  {MC@NLO}'',} \textit{ JHEP} \textbf{ 12} (2012) 061,
  \href{http://dx.doi.org/10.1007/JHEP12(2012)061}{\doi{10.1007/JHEP12(2012)061}},
  \href{http://www.arXiv.org/abs/1209.6215}{\texttt{arXiv:1209.6215}}.

\bibitem{bib:herwigpp}
\hrefCMSnoop {}{M.~Bahr {et~al.}, ``Herwig++ physics and manual'',} \textit{
  Eur. Phys. J. C} \textbf{ 58} (2008) 639,
  \href{http://dx.doi.org/10.1140/epjc/s10052-008-0798-9}{\doi{10.1140/epjc/s10052-008-0798-9}},
  \href{http://www.arXiv.org/abs/0803.0883}{\texttt{arXiv:0803.0883}}.

\bibitem{bib:EE5Ctune}
\hrefCMSnoop {}{M.~H. Seymour and A.~Siodmok, ``Constraining {MPI} models using
  $\sigma_{\mathrm{eff}}$ and recent {T}evatron and {LHC} underlying event
  data'',} \textit{ JHEP} \textbf{ 10} (2013) 113,
  \href{http://dx.doi.org/10.1007/JHEP10(2013)113}{\doi{10.1007/JHEP10(2013)113}},
  \href{http://www.arXiv.org/abs/1307.5015}{\texttt{arXiv:1307.5015}}.

\bibitem{Alwall:2007fs}
\hrefCMSnoop {}{J.~Alwall {et~al.}, ``Comparative study of various algorithms
  for the merging of parton showers and matrix elements in hadronic
  collisions'',} \textit{ Eur. Phys. J. C} \textbf{ 53} (2008) 473,
  \href{http://dx.doi.org/10.1140/epjc/s10052-007-0490-5}{\doi{10.1140/epjc/s10052-007-0490-5}},
  \href{http://www.arXiv.org/abs/0706.2569}{\texttt{arXiv:0706.2569}}.

\bibitem{bib:powheg1}
\hrefCMSnoop {}{S.~Alioli, P.~Nason, C.~Oleari, and E.~Re, ``{NLO} single-top
  production matched with shower in {POWHEG}: $s$- and $t$-channel
  contributions'',} \textit{ JHEP} \textbf{ 09} (2009) 111,
  \href{http://dx.doi.org/10.1088/1126-6708/2009/09/111}{\doi{10.1088/1126-6708/2009/09/111}},
  \href{http://www.arXiv.org/abs/0907.4076}{\texttt{arXiv:0907.4076}}.

\bibitem{bib:powheg3}
\hrefCMSnoop {}{E.~Re, ``Single-top {W}t-channel production matched with parton
  showers using the {POWHEG} method'',} \textit{ Eur. Phys. J. C} \textbf{ 71}
  (2011) 1547,
  \href{http://dx.doi.org/10.1140/epjc/s10052-011-1547-z}{\doi{10.1140/epjc/s10052-011-1547-z}},
  \href{http://www.arXiv.org/abs/1009.2450}{\texttt{arXiv:1009.2450}}.

\bibitem{bib:NNPDF}
\hrefCMSnoop {}{{NNPDF} Collaboration, ``Unbiased global determination of
  parton distributions and their uncertainties at {NNLO} and {LO}'',} \textit{
  Nucl. Phys. B} \textbf{ 855} (2012) 153,
  \href{http://dx.doi.org/10.1016/j.nuclphysb.2011.09.024}{\doi{10.1016/j.nuclphysb.2011.09.024}},
  \href{http://www.arXiv.org/abs/1107.2652}{\texttt{arXiv:1107.2652}}.

\bibitem{Li:2012wna}
\hrefCMSnoop {}{Y.~Li and F.~Petriello, ``Combining {QCD} and electroweak
  corrections to dilepton production in {FEWZ}'',} \textit{ Phys. Rev. D}
  \textbf{ 86} (2012) 094034,
  \href{http://dx.doi.org/10.1103/PhysRevD.86.094034}{\doi{10.1103/PhysRevD.86.094034}},
  \href{http://www.arXiv.org/abs/1208.5967}{\texttt{arXiv:1208.5967}}.

\bibitem{bib:twchan}
\hrefCMSnoop {}{N.~Kidonakis, ``Two-loop soft anomalous dimensions for single
  top quark associated production with $\mathrm{W^-}$ or $\mathrm{H^-}$'',}
  \textit{ Phys. Rev. D} \textbf{ 82} (2010) 054018,
  \href{http://dx.doi.org/10.1103/PhysRevD.82.054018}{\doi{10.1103/PhysRevD.82.054018}},
  \href{http://www.arXiv.org/abs/hep-ph/1005.4451}{\texttt{arXiv:hep-ph/1005.4451}}.

\bibitem{bib:mcfm:diboson}
\hrefCMSnoop {}{J.~M. Campbell, R.~K. Ellis, and C.~Williams, ``Vector boson
  pair production at the {LHC}'',} \textit{ JHEP} \textbf{ 07} (2011) 018,
  \href{http://dx.doi.org/10.1007/JHEP07(2011)018}{\doi{10.1007/JHEP07(2011)018}},
  \href{http://www.arXiv.org/abs/1105.0020}{\texttt{arXiv:1105.0020}}.

\bibitem{bib:Maltoni:2015ena}
\hrefCMSnoop {}{F.~Maltoni, D.~Pagani, and I.~Tsinikos, ``Associated production
  of a top-quark pair with vector bosons at {NLO} in {QCD}: impact on
  $\mathrm{t}\overline{\mathrm{t}}\mathrm{H} $ searches at the {LHC}'',}
  \textit{ JHEP} \textbf{ 02} (2016) 113,
  \href{http://dx.doi.org/10.1007/JHEP02(2016)113}{\doi{10.1007/JHEP02(2016)113}},
  \href{http://www.arXiv.org/abs/1507.05640}{\texttt{arXiv:1507.05640}}.

\bibitem{Czakon:2011xx}
\hrefCMSnoop {}{M.~Czakon and A.~Mitov, ``Top++: a program for the calculation
  of the top-pair cross-section at hadron colliders'',} \textit{ Comput. Phys.
  Commun.} \textbf{ 185} (2014) 2930,
  \href{http://dx.doi.org/10.1016/j.cpc.2014.06.021}{\doi{10.1016/j.cpc.2014.06.021}},
  \href{http://www.arXiv.org/abs/1112.5675}{\texttt{arXiv:1112.5675}}.

\bibitem{bib:geant}
\hrefCMSnoop {}{{GEANT4} Collaboration, ``{GEANT4}---a simulation toolkit'',}
  \textit{ Nucl. Instrum. Meth. A} \textbf{ 506} (2003) 250,
  \href{http://dx.doi.org/10.1016/S0168-9002(03)01368-8}{\doi{10.1016/S0168-9002(03)01368-8}}.

\bibitem{bib:Sirunyan:2017ulk}
\hrefCMSnoop {}{{CMS Collaboration}, ``Particle-flow reconstruction and global
  event description with the {CMS} detector'',} \textit{ JINST} \textbf{ 12}
  (2017) P10003,
  \href{http://dx.doi.org/10.1088/1748-0221/12/10/P10003}{\doi{10.1088/1748-0221/12/10/P10003}},
  \href{http://www.arXiv.org/abs/1706.04965}{\texttt{arXiv:1706.04965}}.

\bibitem{1748-0221-10-06-P06005}
\hrefCMSnoop {}{{CMS Collaboration}, ``Performance of electron reconstruction
  and selection with the {CMS} detector in proton-proton collisions at
  $\sqrt{s}$ = 8 {TeV}'',} \textit{ JINST} \textbf{ 10} (2015) P06005,
  \href{http://dx.doi.org/10.1088/1748-0221/10/06/P06005}{\doi{10.1088/1748-0221/10/06/P06005}},
  \href{http://www.arXiv.org/abs/1502.02701}{\texttt{arXiv:1502.02701}}.

\bibitem{Sirunyan:2018fpa}
\hrefCMSnoop {}{{CMS Collaboration}, ``Performance of the {CMS} muon detector
  and muon reconstruction with proton-proton collisions at $\sqrt{s}=$ 13
  {TeV}'',} \textit{ JINST} \textbf{ 13} (2018) P06015,
  \href{http://dx.doi.org/10.1088/1748-0221/13/06/P06015}{\doi{10.1088/1748-0221/13/06/P06015}},
  \href{http://www.arXiv.org/abs/1804.04528}{\texttt{arXiv:1804.04528}}.

\bibitem{bib:antikt}
\hrefCMSnoop {}{M.~Cacciari, G.~P. Salam, and G.~Soyez, ``The anti-\kt jet
  clustering algorithm'',} \textit{ JHEP} \textbf{ 04} (2008) 063,
  \href{http://dx.doi.org/10.1088/1126-6708/2008/04/063}{\doi{10.1088/1126-6708/2008/04/063}},
  \href{http://www.arXiv.org/abs/0802.1189}{\texttt{arXiv:0802.1189}}.

\bibitem{bib:Cacciari:2011ma}
\hrefCMSnoop {}{M.~Cacciari, G.~P. Salam, and G.~Soyez, ``Fast{J}et user
  manual'',} \textit{ Eur. Phys. J. C} \textbf{ 72} (2012) 1896,
  \href{http://dx.doi.org/10.1140/epjc/s10052-012-1896-2}{\doi{10.1140/epjc/s10052-012-1896-2}},
  \href{http://www.arXiv.org/abs/1111.6097}{\texttt{arXiv:1111.6097}}.

\bibitem{Sirunyan:2017ezt}
\hrefCMSnoop {}{{CMS Collaboration}, ``Identification of heavy-flavour jets
  with the cms detector in pp collisions at 13 {TeV}'',} \textit{ JINST}
  \textbf{ 13} (2018) P05011,
  \href{http://dx.doi.org/10.1088/1748-0221/13/05/P05011}{\doi{10.1088/1748-0221/13/05/P05011}},
  \href{http://www.arXiv.org/abs/1712.07158}{\texttt{arXiv:1712.07158}}.

\bibitem{bib:TOP-15-003_paper}
\hrefCMSnoop {}{{CMS Collaboration}, ``Measurement of the top quark pair
  production cross section in proton-proton collisions at $\sqrt{s}$ = 13 {TeV}
  with the {CMS} detector'',} \textit{ Phys. Rev. Lett.} \textbf{ 116} (2016)
  052002,
  \href{http://dx.doi.org/10.1103/PhysRevLett.116.052002}{\doi{10.1103/PhysRevLett.116.052002}},
  \href{http://www.arXiv.org/abs/1510.05302}{\texttt{arXiv:1510.05302}}.

\bibitem{Patrignani:2016xqp}
\hrefCMSnoop {}{{Particle Data Group}, C.~Patrignani {et~al.}, ``Review of
  particle physics'',} \textit{ Chin. Phys. C} \textbf{ 40} (2016) 100001,
  \href{http://dx.doi.org/10.1088/1674-1137/40/10/100001}{\doi{10.1088/1674-1137/40/10/100001}}.

\bibitem{bib:svd}
\hrefCMSnoop {}{A.~Hoecker and V.~Kartvelishvili, ``{SVD} approach to data
  unfolding'',} \textit{ Nucl. Instrum. Meth. A} \textbf{ 372} (1996) 469,
  \href{http://dx.doi.org/10.1016/0168-9002(95)01478-0}{\doi{10.1016/0168-9002(95)01478-0}},
  \href{http://www.arXiv.org/abs/hep-ph/9509307}{\texttt{arXiv:hep-ph/9509307}}.

\bibitem{bib:blobel}
\href
  {http://www.ippp.dur.ac.uk/Workshops/02/statistics/proceedings//blobel2.pdf}{V.~Blobel,
  ``{An unfolding method for high-energy physics experiments}'',} in \textit{
  {Advanced Statistical Techniques in Particle Physics. Proceedings,
  Conference, Durham, UK, March 18-22, 2002}}, p.~258.
\newblock 2002.
\newblock
  \href{http://www.arXiv.org/abs/hep-ex/0208022}{\texttt{arXiv:hep-ex/0208022}}.

\bibitem{bib:james}
F.~James, ``Statistical methods in experimental physics''.
\newblock World Scientific, second edition, 2006.

\bibitem{bib:CMS-NOTE-2017-004}
\href {https://cds.cern.ch/record/2267573/}{{CMS Collaboration}, ``{Object
  definitions for top quark analyses at the particle level}'',} Technical
  Report CMS-NOTE-2017-004, CERN, Geneva, 2017.

\bibitem{bib:tp}
\hrefCMSnoop {}{{CMS Collaboration}, ``Measurement of the {Drell--Yan} cross
  sections in pp collisions at $\sqrt{s} = 7$ {TeV} with the {CMS}
  experiment'',} \textit{ JHEP} \textbf{ 10} (2011) 007,
  \href{http://dx.doi.org/10.1007/JHEP10(2011)007}{\doi{10.1007/JHEP10(2011)007}},
  \href{http://www.arXiv.org/abs/1108.0566}{\texttt{arXiv:1108.0566}}.

\bibitem{Khachatryan:2016kdb}
\hrefCMSnoop {}{{CMS Collaboration}, ``Jet energy scale and resolution in the
  {CMS} experiment in pp collisions at 8 {TeV}'',} \textit{ JINST} \textbf{ 12}
  (2017) P02014,
  \href{http://dx.doi.org/10.1088/1748-0221/12/02/P02014}{\doi{10.1088/1748-0221/12/02/P02014}},
  \href{http://www.arXiv.org/abs/1607.03663}{\texttt{arXiv:1607.03663}}.

\bibitem{Aaboud:2016mmw}
\hrefCMSnoop {}{{ATLAS Collaboration}, ``Measurement of the inelastic
  proton-proton cross section at $\sqrt{s} = 13$ {TeV} with the {ATLAS}
  detector at the {LHC}'',} \textit{ Phys. Rev. Lett.} \textbf{ 117} (2016)
  182002,
  \href{http://dx.doi.org/10.1103/PhysRevLett.117.182002}{\doi{10.1103/PhysRevLett.117.182002}},
  \href{http://www.arXiv.org/abs/1606.02625}{\texttt{arXiv:1606.02625}}.

\bibitem{Argyropoulos:2014zoa}
\hrefCMSnoop {}{S.~Argyropoulos and T.~Sj{\"o}strand, ``Effects of color
  reconnection on \ttbar\ final states at the {LHC}'',} \textit{ JHEP} \textbf{
  11} (2014) 043,
  \href{http://dx.doi.org/10.1007/JHEP11(2014)043}{\doi{10.1007/JHEP11(2014)043}},
  \href{http://www.arXiv.org/abs/1407.6653}{\texttt{arXiv:1407.6653}}.

\bibitem{Christiansen:2015yqa}
\hrefCMSnoop {}{J.~R. Christiansen and P.~Z. Skands, ``String formation beyond
  leading colour'',} \textit{ JHEP} \textbf{ 08} (2015) 003,
  \href{http://dx.doi.org/10.1007/JHEP08(2015)003}{\doi{10.1007/JHEP08(2015)003}},
  \href{http://www.arXiv.org/abs/1505.01681}{\texttt{arXiv:1505.01681}}.

\bibitem{Bowler:1981sb}
\hrefCMSnoop {}{M.~G. Bowler, ``$\text{e}^{+}\text{e}^{-}$ production of heavy
  quarks in the string model'',} \textit{ Z. Phys. C} \textbf{ 11} (1981) 169,
  \href{http://dx.doi.org/10.1007/BF01574001}{\doi{10.1007/BF01574001}}.

\bibitem{PhysRevD.27.105}
\hrefCMSnoop {}{C.~Peterson, D.~Schlatter, I.~Schmitt, and P.~M. Zerwas,
  ``Scaling violations in inclusive $\text{e}^{+}\text{e}^{\ensuremath{-}}$
  annihilation spectra'',} \textit{ Phys. Rev. D} \textbf{ 27} (1983) 105,
  \href{http://dx.doi.org/10.1103/PhysRevD.27.105}{\doi{10.1103/PhysRevD.27.105}}.

\bibitem{Czakon:2017wor}
M.~Czakon\hrefCMSnoop {}{ {et~al.}, ``Top-pair production at the {LHC} through
  {NNLO} {QCD} and {NLO} {EW}'',} \textit{ JHEP} \textbf{ 10} (2017) 186,
  \href{http://dx.doi.org/10.1007/JHEP10(2017)186}{\doi{10.1007/JHEP10(2017)186}},
  \href{http://www.arXiv.org/abs/1705.04105}{\texttt{arXiv:1705.04105}}.

\bibitem{Manohar:2016nzj}
\hrefCMSnoop {}{A.~Manohar, P.~Nason, G.~P. Salam, and G.~Zanderighi, ``How
  bright is the proton? a precise determination of the photon parton
  distribution function'',} \textit{ Phys. Rev. Lett.} \textbf{ 117} (2016)
  242002,
  \href{http://dx.doi.org/10.1103/PhysRevLett.117.242002}{\doi{10.1103/PhysRevLett.117.242002}},
  \href{http://www.arXiv.org/abs/1607.04266}{\texttt{arXiv:1607.04266}}.

\bibitem{Bertone:2017bme}
\hrefCMSnoop {}{{NNPDF} Collaboration, ``Illuminating the photon content of the
  proton within a global {PDF} analysis'',} \textit{ SciPost Phys.} \textbf{ 5}
  (2018) 008,
  \href{http://dx.doi.org/10.21468/SciPostPhys.5.1.008}{\doi{10.21468/SciPostPhys.5.1.008}},
  \href{http://www.arXiv.org/abs/1712.07053}{\texttt{arXiv:1712.07053}}.

\bibitem{Czakon:2018nun}
M.~Czakon\hrefCMSnoop {}{ {et~al.}, ``Resummation for (boosted) top-quark pair
  production at {NNLO+NNLL'} in {QCD}'',} \textit{ JHEP} \textbf{ 05} (2018)
  149,
  \href{http://dx.doi.org/10.1007/JHEP05(2018)149}{\doi{10.1007/JHEP05(2018)149}},
  \href{http://www.arXiv.org/abs/1803.07623}{\texttt{arXiv:1803.07623}}.

\bibitem{Ball:2017nwa}
\hrefCMSnoop {}{{NNPDF} Collaboration, ``Parton distributions from
  high-precision collider data'',} \textit{ Eur. Phys. J. C} \textbf{ 77}
  (2017) 663,
  \href{http://dx.doi.org/10.1140/epjc/s10052-017-5199-5}{\doi{10.1140/epjc/s10052-017-5199-5}},
  \href{http://www.arXiv.org/abs/1706.00428}{\texttt{arXiv:1706.00428}}.

\bibitem{bib:Dulat:2015mca}
S.~Dulat\hrefCMSnoop {}{ {et~al.}, ``New parton distribution functions from a
  global analysis of quantum chromodynamics'',} \textit{ Phys. Rev. D} \textbf{
  93} (2016) 033006,
  \href{http://dx.doi.org/10.1103/PhysRevD.93.033006}{\doi{10.1103/PhysRevD.93.033006}},
  \href{http://www.arXiv.org/abs/1506.07443}{\texttt{arXiv:1506.07443}}.

\bibitem{Buchmuller:1985jz}
\hrefCMSnoop {}{W.~Buchmuller and D.~Wyler, ``Effective lagrangian analysis of
  new interactions and flavor conservation'',} \textit{ Nucl. Phys. B} \textbf{
  268} (1986) 621,
  \href{http://dx.doi.org/10.1016/0550-3213(86)90262-2}{\doi{10.1016/0550-3213(86)90262-2}}.

\bibitem{Grzadkowski:2010es}
\hrefCMSnoop {}{B.~Grzadkowski, M.~Iskrzynski, M.~Misiak, and J.~Rosiek,
  ``Dimension-six terms in the standard model lagrangian'',} \textit{ JHEP}
  \textbf{ 10} (2010) 085,
  \href{http://dx.doi.org/10.1007/JHEP10(2010)085}{\doi{10.1007/JHEP10(2010)085}},
  \href{http://www.arXiv.org/abs/1008.4884}{\texttt{arXiv:1008.4884}}.

\bibitem{Buckley:2010ar}
A.~Buckley\hrefCMSnoop {}{ {et~al.}, ``Rivet user manual'',} \textit{ Comput.
  Phys. Commun.} \textbf{ 184} (2013) 2803,
  \href{http://dx.doi.org/10.1016/j.cpc.2013.05.021}{\doi{10.1016/j.cpc.2013.05.021}},
  \href{http://www.arXiv.org/abs/1003.0694}{\texttt{arXiv:1003.0694}}.

\bibitem{Czakon:2013goa}
\hrefCMSnoop {}{M.~Czakon, P.~Fiedler, and A.~Mitov, ``Total top-quark
  pair-production cross section at hadron colliders through
  $o({\alpha_s}^4)$'',} \textit{ Phys. Rev. Lett.} \textbf{ 110} (2013) 252004,
  \href{http://dx.doi.org/10.1103/PhysRevLett.110.252004}{\doi{10.1103/PhysRevLett.110.252004}},
  \href{http://www.arXiv.org/abs/1303.6254}{\texttt{arXiv:1303.6254}}.

\bibitem{CMS:2014gta}
\href {http://cds.cern.ch/record/1950834}{{CMS Collaboration}, ``Combination of
  {ATLAS} and {CMS} top quark pair cross section measurements in the e$\mu$
  final state using proton-proton collisions at 8 {TeV}'',} Technical Report
  CMS-PAS-TOP-14-016, CERN, Geneva, 2014.

\bibitem{Aaltonen:2013wca}
\hrefCMSnoop {}{{CDF, D0} Collaboration, ``Combination of measurements of the
  top-quark pair production cross section from the tevatron collider'',}
  \textit{ Phys. Rev. D} \textbf{ 89} (2014) 072001,
  \href{http://dx.doi.org/10.1103/PhysRevD.89.072001}{\doi{10.1103/PhysRevD.89.072001}},
  \href{http://www.arXiv.org/abs/1309.7570}{\texttt{arXiv:1309.7570}}.

\bibitem{bib:Khachatryan:2016xws}
\hrefCMSnoop {}{{CMS Collaboration}, ``Measurements of \ttbar\ spin
  correlations and top quark polarization using dilepton final states in pp
  collisions at $\sqrt{s}$ = 8 {TeV}'',} \textit{ Phys. Rev. D} \textbf{ 93}
  (2016) 052007,
  \href{http://dx.doi.org/10.1103/PhysRevD.93.052007}{\doi{10.1103/PhysRevD.93.052007}},
  \href{http://www.arXiv.org/abs/1601.01107}{\texttt{arXiv:1601.01107}}.

\bibitem{Bernreuther:2013aga}
\hrefCMSnoop {}{W.~Bernreuther and Z.-G. Si, ``Top quark spin correlations and
  polarization at the {LHC}: standard model predictions and effects of
  anomalous top chromo moments'',} \textit{ Phys. Lett. B} \textbf{ 725} (2013)
  115,
  \href{http://dx.doi.org/10.1016/j.physletb.2015.03.035}{\doi{10.1016/j.physletb.2015.03.035}},
  \href{http://www.arXiv.org/abs/1305.2066}{\texttt{arXiv:1305.2066}}.
  [Erratum: \emph{Phys. Lett. B} {\bf{744}} (2015) 413].

\bibitem{Aguilar-Saavedra2011}
\hrefCMSnoop {}{J.~A. Aguilar-Saavedra and M.~P{\'e}rez-Victoria, ``Simple
  models for the top asymmetry: constraints and predictions'',} \textit{ JHEP}
  \textbf{ 09} (2011) 097,
  \href{http://dx.doi.org/10.1007/JHEP09(2011)097}{\doi{10.1007/JHEP09(2011)097}},
  \href{http://www.arXiv.org/abs/1107.0841}{\texttt{arXiv:1107.0841}}.

\bibitem{bib:Kuhn:2011ri}
\hrefCMSnoop {}{J.~H. Kuhn and G.~Rodrigo, ``Charge asymmetries of top quarks
  at hadron colliders revisited'',} \textit{ JHEP} \textbf{ 01} (2012) 063,
  \href{http://dx.doi.org/10.1007/JHEP01(2012)063}{\doi{10.1007/JHEP01(2012)063}},
  \href{http://www.arXiv.org/abs/1109.6830}{\texttt{arXiv:1109.6830}}.

\bibitem{Bernreuther:2015yna}
\hrefCMSnoop {}{W.~Bernreuther, D.~Heisler, and Z.-G. Si, ``A set of top quark
  spin correlation and polarization observables for the {LHC}: Standard model
  predictions and new physics contributions'',} \textit{ JHEP} \textbf{ 12}
  (2015) 026,
  \href{http://dx.doi.org/10.1007/JHEP12(2015)026}{\doi{10.1007/JHEP12(2015)026}},
  \href{http://www.arXiv.org/abs/1508.05271}{\texttt{arXiv:1508.05271}}.

\end{thebibliography}\endgroup

\clearpage

\appendix
\section{Tables of parton-level differential cross sections}
\label{app:tables}
All the measured differential cross sections at the parton level are tabulated in Tables \ref{tab:norm_parton0}--\ref{tab:norm_parton13}. The statistical and systematic uncertainties are quoted separately for each bin.

\begin{table}[!htpb]
\centering
\topcaption{The measured differential cross section and bin boundaries for each bin of the normalized and absolute measurements of the \ttbar differential cross section at parton level in the full phase space as a function of \pttop are tabulated.}
\begin{tabular}{ c  c  c }
\pttop [\GeVns{}]  & $\frac{1}{\sigma}$ $\frac{\rd\sigma}{\rd\pttop} $[\GeVns$^{-1}$] & $\frac{\rd\sigma}{\rd\pttop}$ [pb/\GeVns{}] \\
\hline
$[0, 65]$ & (4.118 $\pm$ 0.044 $\pm$ 0.194) $ \times 10^{-3}$  & 3.487 $\pm$ 0.039 $\pm$ 0.278 \\
$[65, 125]$ & (6.016 $\pm$ 0.059 $\pm$ 0.282) $ \times 10^{-3}$  & 5.094 $\pm$ 0.05 $\pm$ 0.467 \\
$[125, 200]$ & (3.352 $\pm$ 0.03 $\pm$ 0.116) $ \times 10^{-3}$  & 2.838 $\pm$ 0.026 $\pm$ 0.211 \\
$[200, 290]$ & (9.948 $\pm$ 0.125 $\pm$ 0.386) $ \times 10^{-4}$  & 0.842 $\pm$ 0.011 $\pm$ 0.051 \\
$[290, 400]$ & (2.213 $\pm$ 0.035 $\pm$ 0.106) $ \times 10^{-4}$  & 0.187 $\pm$ 0.003 $\pm$ 0.013 \\
$[400, 550]$ & (4.074 $\pm$ 0.155 $\pm$ 0.296) $ \times 10^{-5}$  & (3.45 $\pm$ 0.131 $\pm$ 0.302)  $ \times 10^{-2}$  \\
\end{tabular}
\label{tab:norm_parton0}
\end{table}

\begin{table}[!htpb]
\centering
\topcaption{The measured differential cross section and bin boundaries for each bin of the normalized and absolute measurements of the \ttbar differential cross section at parton level in the full phase space as a function of \ptantitop are tabulated.}
\begin{tabular}{ c  c  c }
\ptantitop [\GeVns{}]  & $\frac{1}{\sigma}$ $\frac{\rd\sigma}{\rd\ptantitop}$ [\GeVns$^{-1}$] & $\frac{\rd\sigma}{\rd\ptantitop}$ [pb/\GeVns{}] \\
\hline
$[0, 65]$ & (4.172 $\pm$ 0.044 $\pm$ 0.244) $ \times 10^{-3}$  & 3.532 $\pm$ 0.039 $\pm$ 0.313 \\
$[65, 125]$ & (6.031 $\pm$ 0.059 $\pm$ 0.224) $ \times 10^{-3}$  & 5.105 $\pm$ 0.05 $\pm$ 0.436 \\
$[125, 200]$ & (3.254 $\pm$ 0.03 $\pm$ 0.12) $ \times 10^{-3}$  & 2.755 $\pm$ 0.026 $\pm$ 0.212 \\
$[200, 290]$ & (1.027 $\pm$ 0.013 $\pm$ 0.051) $ \times 10^{-3}$  & 0.869 $\pm$ 0.011 $\pm$ 0.059 \\
$[290, 400]$ & (2.239 $\pm$ 0.035 $\pm$ 0.11) $ \times 10^{-4}$  & 0.19 $\pm$ 0.003 $\pm$ 0.012 \\
$[400, 550]$ & (3.941 $\pm$ 0.152 $\pm$ 0.506) $ \times 10^{-5}$  & (3.336 $\pm$ 0.129 $\pm$ 0.466)  $ \times 10^{-2}$  \\
\end{tabular}
\label{tab:norm_parton1}
\end{table}

\begin{table}[!htpb]
\centering
\topcaption{The measured differential cross section and bin boundaries for each bin of the normalized and absolute measurements of the \ttbar differential cross section at parton level in the full phase space as a function of \pttop (leading) are tabulated.}
\begin{tabular}{ c  c  c }
\pttop (leading) [\GeVns{}]  & $\frac{1}{\sigma}$ $\frac{\rd\sigma}{\rd\pttop (\text{leading})}$ [\GeVns$^{-1}$] & $\frac{\rd\sigma}{\rd\pttop (\text{leading}) }$ [pb/\GeVns{}] \\
\hline
$[0, 65]$ & (2.618 $\pm$ 0.033 $\pm$ 0.162) $ \times 10^{-3}$  & 2.221 $\pm$ 0.029 $\pm$ 0.195 \\
$[65, 125]$ & (6.046 $\pm$ 0.042 $\pm$ 0.155) $ \times 10^{-3}$  & 5.129 $\pm$ 0.037 $\pm$ 0.403 \\
$[125, 200]$ & (3.981 $\pm$ 0.027 $\pm$ 0.111) $ \times 10^{-3}$  & 3.377 $\pm$ 0.024 $\pm$ 0.265 \\
$[200, 290]$ & (1.377 $\pm$ 0.013 $\pm$ 0.044) $ \times 10^{-3}$  & 1.168 $\pm$ 0.011 $\pm$ 0.071 \\
$[290, 400]$ & (3.227 $\pm$ 0.038 $\pm$ 0.124) $ \times 10^{-4}$  & 0.274 $\pm$ 0.003 $\pm$ 0.017 \\
$[400, 550]$ & (6.076 $\pm$ 0.171 $\pm$ 0.439) $ \times 10^{-5}$  & (5.154 $\pm$ 0.145 $\pm$ 0.466)  $ \times 10^{-2}$  \\
\end{tabular}
\label{tab:norm_parton2}
\end{table}

\begin{table}[!htpb]
\centering
\topcaption{The measured differential cross section and bin boundaries for each bin of the normalized and absolute measurements of the \ttbar differential cross section at parton level in the full phase space as a function of \pttop (trailing) are tabulated.}
\begin{tabular}{ c  c  c }
\pttop (trailing) [\GeVns{}]  & $\frac{1}{\sigma}$ $\frac{\rd\sigma}{\rd\pttop (\text{trailing}) }$ [\GeVns$^{-1}$] & $\frac{\rd\sigma}{\rd\pttop (\text{trailing}) }$ [pb/\GeVns{}] \\
\hline
$[0, 65]$ & (5.659 $\pm$ 0.048 $\pm$ 0.232) $ \times 10^{-3}$  & 4.79 $\pm$ 0.043 $\pm$ 0.371 \\
$[65, 125]$ & (5.992 $\pm$ 0.068 $\pm$ 0.289) $ \times 10^{-3}$  & 5.071 $\pm$ 0.058 $\pm$ 0.486 \\
$[125, 200]$ & (2.641 $\pm$ 0.032 $\pm$ 0.125) $ \times 10^{-3}$  & 2.235 $\pm$ 0.027 $\pm$ 0.162 \\
$[200, 290]$ & (6.452 $\pm$ 0.121 $\pm$ 0.441) $ \times 10^{-4}$  & 0.546 $\pm$ 0.01 $\pm$ 0.043 \\
$[290, 400]$ & (1.235 $\pm$ 0.03 $\pm$ 0.083) $ \times 10^{-4}$  & 0.105 $\pm$ 0.003 $\pm$ 0.009 \\
$[400, 550]$ & (1.949 $\pm$ 0.126 $\pm$ 0.306) $ \times 10^{-5}$  & (1.65 $\pm$ 0.107 $\pm$ 0.267)  $ \times 10^{-2}$  \\
\end{tabular}
\label{tab:norm_parton3}
\end{table}

\begin{table}[!htpb]
\centering
\topcaption{The measured differential cross section and bin boundaries for each bin of the normalized and absolute measurements of the \ttbar differential cross section at parton level in the full phase space as a function of \pttop (\ttbar RF) are tabulated.}
\begin{tabular}{ c  c  c }
\pttop (\ttbar RF) [\GeVns{}]  & $\frac{1}{\sigma}$ $\frac{\rd\sigma}{\rd\pttop (\ttbar~\mathrm{RF})}$ [\GeVns$^{-1}$] & $\frac{\rd\sigma}{\rd\pttop (\ttbar~\mathrm{RF})}$ [pb/\GeVns{}] \\
\hline
$[0, 65]$ & (4.505 $\pm$ 0.045 $\pm$ 0.218) $ \times 10^{-3}$  & 3.814 $\pm$ 0.041 $\pm$ 0.296 \\
$[65, 125]$ & (6.24 $\pm$ 0.065 $\pm$ 0.263) $ \times 10^{-3}$  & 5.282 $\pm$ 0.056 $\pm$ 0.489 \\
$[125, 200]$ & (3.149 $\pm$ 0.032 $\pm$ 0.102) $ \times 10^{-3}$  & 2.666 $\pm$ 0.028 $\pm$ 0.191 \\
$[200, 290]$ & (8.229 $\pm$ 0.126 $\pm$ 0.405) $ \times 10^{-4}$  & 0.697 $\pm$ 0.011 $\pm$ 0.045 \\
$[290, 400]$ & (1.682 $\pm$ 0.032 $\pm$ 0.091) $ \times 10^{-4}$  & 0.142 $\pm$ 0.003 $\pm$ 0.009 \\
$[400, 550]$ & (2.64 $\pm$ 0.136 $\pm$ 0.284) $ \times 10^{-5}$  & (2.235 $\pm$ 0.115 $\pm$ 0.266)  $ \times 10^{-2}$  \\
\end{tabular}
\label{tab:norm_parton4}
\end{table}

\begin{table}[!htpb]
\centering
\topcaption{The measured differential cross section and bin boundaries for each bin of the normalized and absolute measurements of the \ttbar differential cross section at parton level in the full phase space as a function of \ytop are tabulated.}
\begin{tabular}{ c  c  c }
\ytop  & $\frac{1}{\sigma}$ $\frac{\rd\sigma}{\rd \ytop}$ & $\frac{\rd\sigma}{\rd \ytop}$ [pb] \\
\hline
$[-2.6, -1.8]$ & (7.371 $\pm$ 0.137 $\pm$ 0.395) $ \times 10^{-2}$  & (6 $\pm$ 0.117 $\pm$ 0.614)  $ \times 10$  \\
$[-1.8, -1.35]$ & 0.162 $\pm$ 0.002 $\pm$ 0.004 & (1.32 $\pm$ 0.014 $\pm$ 0.098)  $ \times 10^{2}$  \\
$[-1.35, -0.9]$ & 0.231 $\pm$ 0.002 $\pm$ 0.006 & (1.884 $\pm$ 0.019 $\pm$ 0.131)  $ \times 10^{2}$  \\
$[-0.9, -0.45]$ & 0.279 $\pm$ 0.003 $\pm$ 0.006 & (2.274 $\pm$ 0.022 $\pm$ 0.157)  $ \times 10^{2}$  \\
$[-0.45, 0]$ & 0.301 $\pm$ 0.003 $\pm$ 0.009 & (2.452 $\pm$ 0.023 $\pm$ 0.172)  $ \times 10^{2}$  \\
$[0, 0.45]$ & 0.304 $\pm$ 0.003 $\pm$ 0.011 & (2.474 $\pm$ 0.023 $\pm$ 0.191)  $ \times 10^{2}$  \\
$[0.45, 0.9]$ & 0.286 $\pm$ 0.003 $\pm$ 0.009 & (2.326 $\pm$ 0.021 $\pm$ 0.163)  $ \times 10^{2}$  \\
$[0.9, 1.35]$ & 0.227 $\pm$ 0.002 $\pm$ 0.005 & (1.844 $\pm$ 0.018 $\pm$ 0.131)  $ \times 10^{2}$  \\
$[1.35, 1.8]$ & 0.164 $\pm$ 0.002 $\pm$ 0.005 & (1.331 $\pm$ 0.014 $\pm$ 0.103)  $ \times 10^{2}$  \\
$[1.8, 2.6]$ & (7.737 $\pm$ 0.135 $\pm$ 0.328) $ \times 10^{-2}$  & (6.298 $\pm$ 0.116 $\pm$ 0.561)  $ \times 10$  \\
\end{tabular}
\label{tab:norm_parton5}
\end{table}

\begin{table}[!htpb]
\centering
\topcaption{The measured differential cross section and bin boundaries for each bin of the normalized and absolute measurements of the \ttbar differential cross section at parton level in the full phase space as a function of \yantitop are tabulated.}
\begin{tabular}{ c  c  c }
\yantitop  & $\frac{1}{\sigma}$ $\frac{\rd\sigma}{\rd \yantitop}$ & $\frac{\rd\sigma}{\rd \yantitop}$ [pb] \\
\hline
$[-2.6, -1.8]$ & (7.496 $\pm$ 0.137 $\pm$ 0.35) $ \times 10^{-2}$  & (6.106 $\pm$ 0.117 $\pm$ 0.587)  $ \times 10$  \\
$[-1.8, -1.35]$ & 0.159 $\pm$ 0.002 $\pm$ 0.003 & (1.299 $\pm$ 0.014 $\pm$ 0.099)  $ \times 10^{2}$  \\
$[-1.35, -0.9]$ & 0.231 $\pm$ 0.002 $\pm$ 0.006 & (1.881 $\pm$ 0.019 $\pm$ 0.129)  $ \times 10^{2}$  \\
$[-0.9, -0.45]$ & 0.279 $\pm$ 0.003 $\pm$ 0.007 & (2.273 $\pm$ 0.022 $\pm$ 0.157)  $ \times 10^{2}$  \\
$[-0.45, 0]$ & 0.307 $\pm$ 0.003 $\pm$ 0.006 & (2.499 $\pm$ 0.023 $\pm$ 0.18)  $ \times 10^{2}$  \\
$[0, 0.45]$ & 0.307 $\pm$ 0.003 $\pm$ 0.009 & (2.501 $\pm$ 0.023 $\pm$ 0.171)  $ \times 10^{2}$  \\
$[0.45, 0.9]$ & 0.277 $\pm$ 0.003 $\pm$ 0.006 & (2.255 $\pm$ 0.021 $\pm$ 0.155)  $ \times 10^{2}$  \\
$[0.9, 1.35]$ & 0.236 $\pm$ 0.002 $\pm$ 0.009 & (1.921 $\pm$ 0.018 $\pm$ 0.141)  $ \times 10^{2}$  \\
$[1.35, 1.8]$ & 0.161 $\pm$ 0.002 $\pm$ 0.009 & (1.307 $\pm$ 0.014 $\pm$ 0.12)  $ \times 10^{2}$  \\
$[1.8, 2.6]$ & (7.454 $\pm$ 0.135 $\pm$ 0.428) $ \times 10^{-2}$  & (6.072 $\pm$ 0.116 $\pm$ 0.602)  $ \times 10$  \\
\end{tabular}
\label{tab:norm_parton6}
\end{table}

\begin{table}[!htpb]
\centering
\topcaption{The measured differential cross section and bin boundaries for each bin of the normalized and absolute measurements of the \ttbar differential cross section at parton level in the full phase space as a function of \ytop (leading) are tabulated.}
\begin{tabular}{ c  c  c }
\ytop (leading)  & $\frac{1}{\sigma}$ $\frac{\rd\sigma}{\rd \ytop (\text{leading})}$ & $\frac{\rd\sigma}{\rd \ytop (\text{leading})}$ [pb] \\
\hline
$[-2.6, -1.65]$ & (8.214 $\pm$ 0.137 $\pm$ 0.385) $ \times 10^{-2}$  & (6.707 $\pm$ 0.119 $\pm$ 0.65)  $ \times 10$  \\
$[-1.65, -1.1]$ & 0.191 $\pm$ 0.002 $\pm$ 0.005 & (1.56 $\pm$ 0.015 $\pm$ 0.119)  $ \times 10^{2}$  \\
$[-1.1, -0.55]$ & 0.27 $\pm$ 0.002 $\pm$ 0.007 & (2.208 $\pm$ 0.018 $\pm$ 0.147)  $ \times 10^{2}$  \\
$[-0.55, 0]$ & 0.303 $\pm$ 0.003 $\pm$ 0.007 & (2.473 $\pm$ 0.021 $\pm$ 0.168)  $ \times 10^{2}$  \\
$[0, 0.55]$ & 0.304 $\pm$ 0.003 $\pm$ 0.007 & (2.481 $\pm$ 0.021 $\pm$ 0.167)  $ \times 10^{2}$  \\
$[0.55, 1.1]$ & 0.27 $\pm$ 0.002 $\pm$ 0.006 & (2.206 $\pm$ 0.019 $\pm$ 0.147)  $ \times 10^{2}$  \\
$[1.1, 1.65]$ & 0.189 $\pm$ 0.002 $\pm$ 0.008 & (1.546 $\pm$ 0.015 $\pm$ 0.131)  $ \times 10^{2}$  \\
$[1.65, 2.6]$ & (8.621 $\pm$ 0.131 $\pm$ 0.484) $ \times 10^{-2}$  & (7.04 $\pm$ 0.114 $\pm$ 0.674)  $ \times 10$  \\
\end{tabular}
\label{tab:norm_parton7}
\end{table}

\begin{table}[!htpb]
\centering
\topcaption{The measured differential cross section and bin boundaries for each bin of the normalized and absolute measurements of the \ttbar differential cross section at parton level in the full phase space as a function of \ytop (trailing) are tabulated.}
\begin{tabular}{ c  c  c }
\ytop (trailing) & $\frac{1}{\sigma}$ $\frac{\rd\sigma}{\rd \ytop (\text{trailing})}$ & $\frac{\rd\sigma}{\rd\ytop (\text{trailing}) }$ [pb] \\
\hline
$[-2.6, -1.65]$ & (8.68 $\pm$ 0.152 $\pm$ 0.446) $ \times 10^{-2}$  & (7.056 $\pm$ 0.131 $\pm$ 0.73)  $ \times 10$  \\
$[-1.65, -1.1]$ & 0.191 $\pm$ 0.002 $\pm$ 0.006 & (1.552 $\pm$ 0.016 $\pm$ 0.119)  $ \times 10^{2}$  \\
$[-1.1, -0.55]$ & 0.265 $\pm$ 0.003 $\pm$ 0.009 & (2.158 $\pm$ 0.02 $\pm$ 0.146)  $ \times 10^{2}$  \\
$[-0.55, 0]$ & 0.299 $\pm$ 0.003 $\pm$ 0.007 & (2.427 $\pm$ 0.022 $\pm$ 0.174)  $ \times 10^{2}$  \\
$[0, 0.55]$ & 0.297 $\pm$ 0.003 $\pm$ 0.012 & (2.418 $\pm$ 0.023 $\pm$ 0.19)  $ \times 10^{2}$  \\
$[0.55, 1.1]$ & 0.27 $\pm$ 0.003 $\pm$ 0.008 & (2.194 $\pm$ 0.021 $\pm$ 0.151)  $ \times 10^{2}$  \\
$[1.1, 1.65]$ & 0.197 $\pm$ 0.002 $\pm$ 0.006 & (1.598 $\pm$ 0.016 $\pm$ 0.119)  $ \times 10^{2}$  \\
$[1.65, 2.6]$ & (8.644 $\pm$ 0.144 $\pm$ 0.387) $ \times 10^{-2}$  & (7.027 $\pm$ 0.125 $\pm$ 0.62)  $ \times 10$  \\
\end{tabular}
\label{tab:norm_parton8}
\end{table}

\begin{table}[!htpb]
\centering
\topcaption{The measured differential cross section and bin boundaries for each bin of the normalized and absolute measurements of the \ttbar differential cross section at parton level in the full phase space as a function of \pttt are tabulated.}
\begin{tabular}{ c  c  c }
\pttt & $\frac{1}{\sigma}$ $\frac{\rd\sigma}{\rd\pttt}$ & $\frac{\rd\sigma}{\rd\pttt}$ [pb] \\
\hline
$[0, 40]$ & (1.275 $\pm$ 0.006 $\pm$ 0.085) $ \times 10^{-2}$  & (1.074 $\pm$ 0.006 $\pm$ 0.101)  $ \times 10$  \\
$[40, 100]$ & (4.815 $\pm$ 0.047 $\pm$ 0.607) $ \times 10^{-3}$  & 4.053 $\pm$ 0.039 $\pm$ 0.575 \\
$[100, 200]$ & (1.507 $\pm$ 0.013 $\pm$ 0.052) $ \times 10^{-3}$  & 1.269 $\pm$ 0.012 $\pm$ 0.102 \\
$[200, 310]$ & (3.368 $\pm$ 0.049 $\pm$ 0.126) $ \times 10^{-4}$  & 0.284 $\pm$ 0.004 $\pm$ 0.023 \\
$[310, 420]$ & (8.932 $\pm$ 0.197 $\pm$ 0.457) $ \times 10^{-5}$  & (7.519 $\pm$ 0.166 $\pm$ 0.665)  $ \times 10^{-2}$  \\
$[420, 570]$ & (2.296 $\pm$ 0.085 $\pm$ 0.146) $ \times 10^{-5}$  & (1.933 $\pm$ 0.072 $\pm$ 0.182)  $ \times 10^{-2}$  \\
\end{tabular}
\label{tab:norm_parton9}
\end{table}

\begin{table}[!htpb]
\centering
\topcaption{The measured differential cross section and bin boundaries for each bin of the normalized and absolute measurements of the \ttbar differential cross section at parton level in the full phase space as a function of \ytt are tabulated.}
\begin{tabular}{ c  c  c }
\ytt & $\frac{1}{\sigma}$ $\frac{\rd\sigma}{\rd \ytt}$ & $\frac{\rd\sigma}{\rd \ytt}$ [pb] \\
\hline
$[-2.6, -1.6]$ & (5.394 $\pm$ 0.127 $\pm$ 0.586) $ \times 10^{-2}$  & (4.443 $\pm$ 0.109 $\pm$ 0.647)  $ \times 10$  \\
$[-1.6, -1.2]$ & 0.173 $\pm$ 0.002 $\pm$ 0.004 & (1.424 $\pm$ 0.016 $\pm$ 0.103)  $ \times 10^{2}$  \\
$[-1.2, -0.8]$ & 0.256 $\pm$ 0.003 $\pm$ 0.009 & (2.108 $\pm$ 0.022 $\pm$ 0.157)  $ \times 10^{2}$  \\
$[-0.8, -0.4]$ & 0.316 $\pm$ 0.003 $\pm$ 0.008 & (2.602 $\pm$ 0.024 $\pm$ 0.189)  $ \times 10^{2}$  \\
$[-0.4, 0]$ & 0.369 $\pm$ 0.003 $\pm$ 0.01 & (3.04 $\pm$ 0.026 $\pm$ 0.191)  $ \times 10^{2}$  \\
$[0, 0.4]$ & 0.353 $\pm$ 0.003 $\pm$ 0.008 & (2.91 $\pm$ 0.025 $\pm$ 0.209)  $ \times 10^{2}$  \\
$[0.4, 0.8]$ & 0.321 $\pm$ 0.003 $\pm$ 0.007 & (2.642 $\pm$ 0.023 $\pm$ 0.189)  $ \times 10^{2}$  \\
$[0.8, 1.2]$ & 0.261 $\pm$ 0.002 $\pm$ 0.008 & (2.149 $\pm$ 0.02 $\pm$ 0.148)  $ \times 10^{2}$  \\
$[1.2, 1.6]$ & 0.169 $\pm$ 0.002 $\pm$ 0.005 & (1.396 $\pm$ 0.016 $\pm$ 0.112)  $ \times 10^{2}$  \\
$[1.6, 2.6]$ & (5.885 $\pm$ 0.125 $\pm$ 0.336) $ \times 10^{-2}$  & (4.848 $\pm$ 0.108 $\pm$ 0.514)  $ \times 10$  \\
\end{tabular}
\label{tab:norm_parton10}
\end{table}

\begin{table}[!htpb]
\centering
\topcaption{The measured differential cross section and bin boundaries for each bin of the normalized and absolute measurements of the \ttbar differential cross section at parton level in the full phase space as a function of \mtt are tabulated.}
\begin{tabular}{ c  c  c }
\mtt [\GeVns{}] & $\frac{1}{\sigma}$ $\frac{\rd\sigma}{\rd\mtt}$ [\GeVns$^{-1}$] & $\frac{\rd\sigma}{\rd\mtt}$ [pb/\GeVns{}] \\
\hline
$[300, 380]$ & (1.981 $\pm$ 0.036 $\pm$ 0.18) $ \times 10^{-3}$  & 1.664 $\pm$ 0.031 $\pm$ 0.163 \\
$[380, 470]$ & (3.992 $\pm$ 0.049 $\pm$ 0.183) $ \times 10^{-3}$  & 3.354 $\pm$ 0.041 $\pm$ 0.324 \\
$[470, 620]$ & (2.009 $\pm$ 0.023 $\pm$ 0.057) $ \times 10^{-3}$  & 1.688 $\pm$ 0.019 $\pm$ 0.122 \\
$[620, 820]$ & (6.363 $\pm$ 0.108 $\pm$ 0.355) $ \times 10^{-4}$  & 0.535 $\pm$ 0.009 $\pm$ 0.038 \\
$[820, 1100]$ & (1.438 $\pm$ 0.041 $\pm$ 0.105) $ \times 10^{-4}$  & 0.121 $\pm$ 0.003 $\pm$ 0.012 \\
$[1100, 1500]$ & (2.72 $\pm$ 0.106 $\pm$ 0.206) $ \times 10^{-5}$  & (2.285 $\pm$ 0.089 $\pm$ 0.21)  $ \times 10^{-2}$  \\
$[1500, 2500]$ & (2.45 $\pm$ 0.24 $\pm$ 0.464) $ \times 10^{-6}$  & (2.059 $\pm$ 0.201 $\pm$ 0.383)  $ \times 10^{-3}$  \\
\end{tabular}
\label{tab:norm_parton11}
\end{table}

\begin{table}[!htpb]
\centering
\topcaption{The measured differential cross section and bin boundaries for each bin of the normalized and absolute measurements of the \ttbar differential cross section at parton level in the full phase space as a function of \delytt are tabulated.}
\begin{tabular}{ c  c  c }
\delytt & $\frac{1}{\sigma}$ $\frac{\rd\sigma}{\rd\delytt}$ & $\frac{\rd\sigma}{\rd\delytt}$ [pb] \\
\hline
$[-2.6, -1.4]$ & (5.24 $\pm$ 0.105 $\pm$ 0.272) $ \times 10^{-2}$  & (4.365 $\pm$ 0.09 $\pm$ 0.408)  $ \times 10$  \\
$[-1.4, -0.9]$ & 0.193 $\pm$ 0.002 $\pm$ 0.009 & (1.61 $\pm$ 0.019 $\pm$ 0.137)  $ \times 10^{2}$  \\
$[-0.9, -0.4]$ & 0.321 $\pm$ 0.003 $\pm$ 0.007 & (2.675 $\pm$ 0.028 $\pm$ 0.202)  $ \times 10^{2}$  \\
$[-0.4, 0]$ & 0.436 $\pm$ 0.004 $\pm$ 0.014 & (3.63 $\pm$ 0.037 $\pm$ 0.246)  $ \times 10^{2}$  \\
$[0, 0.4]$ & 0.443 $\pm$ 0.004 $\pm$ 0.012 & (3.693 $\pm$ 0.038 $\pm$ 0.241)  $ \times 10^{2}$  \\
$[0.4, 0.9]$ & 0.325 $\pm$ 0.003 $\pm$ 0.013 & (2.71 $\pm$ 0.029 $\pm$ 0.217)  $ \times 10^{2}$  \\
$[0.9, 1.4]$ & 0.199 $\pm$ 0.002 $\pm$ 0.006 & (1.661 $\pm$ 0.019 $\pm$ 0.134)  $ \times 10^{2}$  \\
$[1.4, 2.6]$ & (5.489 $\pm$ 0.103 $\pm$ 0.422) $ \times 10^{-2}$  & (4.572 $\pm$ 0.088 $\pm$ 0.47)  $ \times 10$  \\
\end{tabular}
\label{tab:norm_parton12}
\end{table}

\begin{table}[!htpb]
\centering
\topcaption{The measured differential cross section and bin boundaries for each bin of the normalized and absolute measurements of the \ttbar differential cross section at parton level in the full phase space as a function of \delphitt are tabulated.}
\begin{tabular}{ c  c  c }
\delphitt [\GeVns{}]  & $\frac{1}{\sigma}$ $\frac{\rd\sigma}{\rd\delphitt}$ [\GeVns$^{-1}$] & $\frac{\rd\sigma}{\rd\delphitt}$ [pb/\GeVns{}] \\
\hline
$[0, 1.57]$ & (6.336 $\pm$ 0.072 $\pm$ 0.336) $ \times 10^{-2}$  & (5.293 $\pm$ 0.061 $\pm$ 0.49)  $ \times 10$  \\
$[1.57, 2.67]$ & 0.218 $\pm$ 0.001 $\pm$ 0.009 & (1.818 $\pm$ 0.011 $\pm$ 0.159)  $ \times 10^{2}$  \\
$[2.67, 3.02]$ & 1.016 $\pm$ 0.006 $\pm$ 0.027 & (8.49 $\pm$ 0.051 $\pm$ 0.63)  $ \times 10^{2}$  \\
$[3.02, 3.142]$ & 2.504 $\pm$ 0.019 $\pm$ 0.151 & (2.092 $\pm$ 0.016 $\pm$ 0.168)  $ \times 10^{3}$  \\
\end{tabular}
\label{tab:norm_parton13}
\end{table}

\clearpage

\section{Tables of particle-level differential cross sections}
\label{app:tables_particle}
All the measured differential cross sections at the particle level are tabulated in Tables \ref{tab:norm_particle0}--\ref{tab:norm_particle32}. The statistical and systematic uncertainties are quoted separately for each bin.

\begin{table}[!htpb]
\centering
\topcaption{The measured differential cross section and bin boundaries for each bin of the normalized and absolute measurements of the \ttbar differential cross section at particle level in the fiducial phase space as a function of \pttop are tabulated.}
\begin{tabular}{ c  c  c }
\pttop [\GeVns{}]  & $\frac{1}{\sigma}$ $\frac{\rd\sigma}{\rd\pttop}$ [\GeVns$^{-1}$] & $\frac{\rd\sigma}{\rd\pttop}$ [pb/\GeVns{}] \\
\hline
$[0, 65]$ & (3.991 $\pm$ 0.037 $\pm$ 0.162) $ \times 10^{-3}$  & (4.529 $\pm$ 0.043 $\pm$ 0.328)  $ \times 10^{-2}$  \\
$[65, 125]$ & (5.734 $\pm$ 0.053 $\pm$ 0.251) $ \times 10^{-3}$  & (6.507 $\pm$ 0.061 $\pm$ 0.538)  $ \times 10^{-2}$  \\
$[125, 200]$ & (3.369 $\pm$ 0.03 $\pm$ 0.114) $ \times 10^{-3}$  & (3.823 $\pm$ 0.035 $\pm$ 0.255)  $ \times 10^{-2}$  \\
$[200, 290]$ & (1.152 $\pm$ 0.014 $\pm$ 0.041) $ \times 10^{-3}$  & (1.307 $\pm$ 0.016 $\pm$ 0.077)  $ \times 10^{-2}$  \\
$[290, 400]$ & (2.907 $\pm$ 0.044 $\pm$ 0.134) $ \times 10^{-4}$  & (3.299 $\pm$ 0.051 $\pm$ 0.236)  $ \times 10^{-3}$  \\
$[400, 550]$ & (5.515 $\pm$ 0.206 $\pm$ 0.36) $ \times 10^{-5}$  & (6.259 $\pm$ 0.234 $\pm$ 0.555)  $ \times 10^{-4}$  \\
\end{tabular}
\label{tab:norm_particle0}
\end{table}

\begin{table}[!htpb]
\centering
\topcaption{The measured differential cross section and bin boundaries for each bin of the normalized and absolute measurements of the \ttbar differential cross section at particle level in the fiducial phase space as a function of \ptantitop are tabulated.}
\begin{tabular}{ c  c  c }
\ptantitop [\GeVns{}]  & $\frac{1}{\sigma}$ $\frac{\rd\sigma}{\rd\ptantitop}$ [\GeVns$^{-1}$] & $\frac{\rd\sigma}{\rd\ptantitop}$ [pb/\GeVns{}] \\
\hline
$[0, 65]$ & (4.038 $\pm$ 0.037 $\pm$ 0.205) $ \times 10^{-3}$  & (4.581 $\pm$ 0.044 $\pm$ 0.368)  $ \times 10^{-2}$  \\
$[65, 125]$ & (5.746 $\pm$ 0.053 $\pm$ 0.192) $ \times 10^{-3}$  & (6.519 $\pm$ 0.061 $\pm$ 0.487)  $ \times 10^{-2}$  \\
$[125, 200]$ & (3.277 $\pm$ 0.03 $\pm$ 0.107) $ \times 10^{-3}$  & (3.717 $\pm$ 0.035 $\pm$ 0.258)  $ \times 10^{-2}$  \\
$[200, 290]$ & (1.186 $\pm$ 0.014 $\pm$ 0.053) $ \times 10^{-3}$  & (1.345 $\pm$ 0.016 $\pm$ 0.087)  $ \times 10^{-2}$  \\
$[290, 400]$ & (2.937 $\pm$ 0.045 $\pm$ 0.122) $ \times 10^{-4}$  & (3.332 $\pm$ 0.051 $\pm$ 0.221)  $ \times 10^{-3}$  \\
$[400, 550]$ & (5.37 $\pm$ 0.204 $\pm$ 0.672) $ \times 10^{-5}$  & (6.093 $\pm$ 0.231 $\pm$ 0.865)  $ \times 10^{-4}$  \\
\end{tabular}
\label{tab:norm_particle1}
\end{table}

\begin{table}[!htpb]
\centering
\topcaption{The measured differential cross section and bin boundaries for each bin of the normalized and absolute measurements of the \ttbar differential cross section at particle level in the fiducial phase space as a function of \pttop (leading) are tabulated.}
\begin{tabular}{ c  c  c }
\pttop (leading) [\GeVns{}]  & $\frac{1}{\sigma}$ $\frac{\rd\sigma}{\rd\pttop (\text{leading}) }$ [\GeVns$^{-1}$] & $\frac{\rd\sigma}{\rd\pttop (\text{leading}) }$ [pb/\GeVns{}] \\
\hline
$[0, 65]$ & (2.474 $\pm$ 0.028 $\pm$ 0.136) $ \times 10^{-3}$  & (2.805 $\pm$ 0.033 $\pm$ 0.227)  $ \times 10^{-2}$  \\
$[65, 125]$ & (5.735 $\pm$ 0.041 $\pm$ 0.145) $ \times 10^{-3}$  & (6.504 $\pm$ 0.049 $\pm$ 0.462)  $ \times 10^{-2}$  \\
$[125, 200]$ & (3.937 $\pm$ 0.027 $\pm$ 0.102) $ \times 10^{-3}$  & (4.465 $\pm$ 0.033 $\pm$ 0.315)  $ \times 10^{-2}$  \\
$[200, 290]$ & (1.568 $\pm$ 0.015 $\pm$ 0.046) $ \times 10^{-3}$  & (1.778 $\pm$ 0.017 $\pm$ 0.102)  $ \times 10^{-2}$  \\
$[290, 400]$ & (4.199 $\pm$ 0.049 $\pm$ 0.141) $ \times 10^{-4}$  & (4.761 $\pm$ 0.057 $\pm$ 0.297)  $ \times 10^{-3}$  \\
$[400, 550]$ & (8.319 $\pm$ 0.232 $\pm$ 0.532) $ \times 10^{-5}$  & (9.434 $\pm$ 0.264 $\pm$ 0.839)  $ \times 10^{-4}$  \\
\end{tabular}
\label{tab:norm_particle2}
\end{table}

\begin{table}[!htpb]
\centering
\topcaption{The measured differential cross section and bin boundaries for each bin of the normalized and absolute measurements of the \ttbar differential cross section at particle level in the fiducial phase space as a function of \pttop (trailing) are tabulated.}
\begin{tabular}{ c  c  c }
\pttop (trailing) [\GeVns{}]  & $\frac{1}{\sigma}$ $\frac{\rd\sigma}{\rd\pttop (\text{trailing}) }$ [\GeVns$^{-1}$] & $\frac{\rd\sigma}{\rd\pttop (\text{trailing}) }$ [pb/\GeVns{}] \\
\hline
$[0, 65]$ & (5.535 $\pm$ 0.039 $\pm$ 0.193) $ \times 10^{-3}$  & (6.287 $\pm$ 0.047 $\pm$ 0.445)  $ \times 10^{-2}$  \\
$[65, 125]$ & (5.737 $\pm$ 0.058 $\pm$ 0.244) $ \times 10^{-3}$  & (6.516 $\pm$ 0.067 $\pm$ 0.545)  $ \times 10^{-2}$  \\
$[125, 200]$ & (2.726 $\pm$ 0.032 $\pm$ 0.114) $ \times 10^{-3}$  & (3.096 $\pm$ 0.037 $\pm$ 0.202)  $ \times 10^{-2}$  \\
$[200, 290]$ & (7.709 $\pm$ 0.142 $\pm$ 0.511) $ \times 10^{-4}$  & (8.756 $\pm$ 0.161 $\pm$ 0.689)  $ \times 10^{-3}$  \\
$[290, 400]$ & (1.665 $\pm$ 0.038 $\pm$ 0.119) $ \times 10^{-4}$  & (1.891 $\pm$ 0.044 $\pm$ 0.173)  $ \times 10^{-3}$  \\
$[400, 550]$ & (2.593 $\pm$ 0.162 $\pm$ 0.43) $ \times 10^{-5}$  & (2.945 $\pm$ 0.184 $\pm$ 0.523)  $ \times 10^{-4}$  \\
\end{tabular}
\label{tab:norm_particle3}
\end{table}

\begin{table}[!htpb]
\centering
\topcaption{The measured differential cross section and bin boundaries for each bin of the normalized and absolute measurements of the \ttbar differential cross section at particle level in the fiducial phase space as a function of \pttop (\ttbar RF) are tabulated.}
\begin{tabular}{ c  c  c }
\pttop (\ttbar RF) [\GeVns{}]  & $\frac{1}{\sigma}$ $\frac{\rd\sigma}{\rd\pttop (\ttbar \mathrm{RF})}$ [\GeVns$^{-1}$] & $\frac{\rd\sigma}{\rd\pttop (\ttbar \mathrm{RF})}$ [pb/\GeVns{}] \\
\hline
$[0, 65]$ & (4.38 $\pm$ 0.037 $\pm$ 0.174) $ \times 10^{-3}$  & (4.973 $\pm$ 0.044 $\pm$ 0.355)  $ \times 10^{-2}$  \\
$[65, 125]$ & (5.94 $\pm$ 0.056 $\pm$ 0.217) $ \times 10^{-3}$  & (6.744 $\pm$ 0.065 $\pm$ 0.538)  $ \times 10^{-2}$  \\
$[125, 200]$ & (3.207 $\pm$ 0.031 $\pm$ 0.097) $ \times 10^{-3}$  & (3.641 $\pm$ 0.037 $\pm$ 0.236)  $ \times 10^{-2}$  \\
$[200, 290]$ & (9.766 $\pm$ 0.148 $\pm$ 0.417) $ \times 10^{-4}$  & (1.109 $\pm$ 0.017 $\pm$ 0.069)  $ \times 10^{-2}$  \\
$[290, 400]$ & (2.273 $\pm$ 0.042 $\pm$ 0.108) $ \times 10^{-4}$  & (2.58 $\pm$ 0.048 $\pm$ 0.178)  $ \times 10^{-3}$  \\
$[400, 550]$ & (3.628 $\pm$ 0.181 $\pm$ 0.385) $ \times 10^{-5}$  & (4.119 $\pm$ 0.205 $\pm$ 0.515)  $ \times 10^{-4}$  \\
\end{tabular}
\label{tab:norm_particle4}
\end{table}

\begin{table}[!htpb]
\centering
\topcaption{The measured differential cross section and bin boundaries for each bin of the normalized and absolute measurements of the \ttbar differential cross section at particle level in the fiducial phase space as a function of \ytop are tabulated.}
\begin{tabular}{ c  c  c }
\ytop  & $\frac{1}{\sigma}$ $\frac{\rd\sigma}{\rd \ytop}$ & $\frac{\rd\sigma}{\rd \ytop}$ [pb] \\
\hline
$[-2.6, -1.8]$ & (3.154 $\pm$ 0.061 $\pm$ 0.171) $ \times 10^{-2}$  & 0.356 $\pm$ 0.007 $\pm$ 0.034 \\
$[-1.8, -1.35]$ & 0.145 $\pm$ 0.001 $\pm$ 0.004 & 1.643 $\pm$ 0.016 $\pm$ 0.11 \\
$[-1.35, -0.9]$ & 0.248 $\pm$ 0.002 $\pm$ 0.005 & 2.806 $\pm$ 0.027 $\pm$ 0.176 \\
$[-0.9, -0.45]$ & 0.314 $\pm$ 0.003 $\pm$ 0.006 & 3.548 $\pm$ 0.032 $\pm$ 0.228 \\
$[-0.45, 0]$ & 0.343 $\pm$ 0.003 $\pm$ 0.01 & 3.878 $\pm$ 0.034 $\pm$ 0.248 \\
$[0, 0.45]$ & 0.346 $\pm$ 0.003 $\pm$ 0.011 & 3.913 $\pm$ 0.034 $\pm$ 0.273 \\
$[0.45, 0.9]$ & 0.32 $\pm$ 0.003 $\pm$ 0.01 & 3.622 $\pm$ 0.031 $\pm$ 0.235 \\
$[0.9, 1.35]$ & 0.244 $\pm$ 0.002 $\pm$ 0.005 & 2.753 $\pm$ 0.025 $\pm$ 0.178 \\
$[1.35, 1.8]$ & 0.146 $\pm$ 0.001 $\pm$ 0.005 & 1.655 $\pm$ 0.016 $\pm$ 0.12 \\
$[1.8, 2.6]$ & (3.31 $\pm$ 0.06 $\pm$ 0.144) $ \times 10^{-2}$  & 0.374 $\pm$ 0.007 $\pm$ 0.031 \\
\end{tabular}
\label{tab:norm_particle5}
\end{table}

\begin{table}[!htpb]
\centering
\topcaption{The measured differential cross section and bin boundaries for each bin of the normalized and absolute measurements of the \ttbar differential cross section at particle level in the fiducial phase space as a function of \yantitop are tabulated.}
\begin{tabular}{ c  c  c }
\yantitop  & $\frac{1}{\sigma}$ $\frac{\rd\sigma}{\rd \yantitop}$ & $\frac{\rd\sigma}{\rd \yantitop}$ [pb] \\
\hline
$[-2.6, -1.8]$ & (3.21 $\pm$ 0.061 $\pm$ 0.163) $ \times 10^{-2}$  & 0.363 $\pm$ 0.007 $\pm$ 0.033 \\
$[-1.8, -1.35]$ & 0.143 $\pm$ 0.001 $\pm$ 0.003 & 1.619 $\pm$ 0.016 $\pm$ 0.115 \\
$[-1.35, -0.9]$ & 0.246 $\pm$ 0.002 $\pm$ 0.007 & 2.785 $\pm$ 0.027 $\pm$ 0.182 \\
$[-0.9, -0.45]$ & 0.313 $\pm$ 0.003 $\pm$ 0.007 & 3.537 $\pm$ 0.032 $\pm$ 0.222 \\
$[-0.45, 0]$ & 0.35 $\pm$ 0.003 $\pm$ 0.007 & 3.961 $\pm$ 0.035 $\pm$ 0.258 \\
$[0, 0.45]$ & 0.349 $\pm$ 0.003 $\pm$ 0.009 & 3.943 $\pm$ 0.034 $\pm$ 0.257 \\
$[0.45, 0.9]$ & 0.312 $\pm$ 0.003 $\pm$ 0.006 & 3.522 $\pm$ 0.031 $\pm$ 0.218 \\
$[0.9, 1.35]$ & 0.252 $\pm$ 0.002 $\pm$ 0.009 & 2.846 $\pm$ 0.025 $\pm$ 0.189 \\
$[1.35, 1.8]$ & 0.143 $\pm$ 0.001 $\pm$ 0.008 & 1.62 $\pm$ 0.016 $\pm$ 0.138 \\
$[1.8, 2.6]$ & (3.187 $\pm$ 0.06 $\pm$ 0.206) $ \times 10^{-2}$  & 0.36 $\pm$ 0.007 $\pm$ 0.034 \\
\end{tabular}
\label{tab:norm_particle6}
\end{table}

\begin{table}[!htpb]
\centering
\topcaption{The measured differential cross section and bin boundaries for each bin of the normalized and absolute measurements of the \ttbar differential cross section at particle level in the fiducial phase space as a function of \ytop (leading) are tabulated.}
\begin{tabular}{ c  c  c }
\ytop (leading)  & $\frac{1}{\sigma}$ $\frac{\rd\sigma}{\rd \ytop (\text{leading})}$ & $\frac{\rd \sigma}{\rd \ytop (\text{leading})}$ [pb] \\
\hline
$[-2.6, -1.65]$ & (4.328 $\pm$ 0.075 $\pm$ 0.229) $ \times 10^{-2}$  & 0.489 $\pm$ 0.009 $\pm$ 0.046 \\
$[-1.65, -1.1]$ & 0.19 $\pm$ 0.002 $\pm$ 0.005 & 2.148 $\pm$ 0.02 $\pm$ 0.153 \\
$[-1.1, -0.55]$ & 0.299 $\pm$ 0.002 $\pm$ 0.006 & 3.378 $\pm$ 0.027 $\pm$ 0.206 \\
$[-0.55, 0]$ & 0.344 $\pm$ 0.003 $\pm$ 0.007 & 3.891 $\pm$ 0.031 $\pm$ 0.237 \\
$[0, 0.55]$ & 0.344 $\pm$ 0.003 $\pm$ 0.007 & 3.89 $\pm$ 0.032 $\pm$ 0.237 \\
$[0.55, 1.1]$ & 0.3 $\pm$ 0.002 $\pm$ 0.004 & 3.389 $\pm$ 0.028 $\pm$ 0.206 \\
$[1.1, 1.65]$ & 0.188 $\pm$ 0.002 $\pm$ 0.007 & 2.126 $\pm$ 0.02 $\pm$ 0.167 \\
$[1.65, 2.6]$ & (4.542 $\pm$ 0.071 $\pm$ 0.274) $ \times 10^{-2}$  & 0.513 $\pm$ 0.008 $\pm$ 0.045 \\
\end{tabular}
\label{tab:norm_particle7}
\end{table}

\begin{table}[!htpb]
\centering
\topcaption{The measured differential cross section and bin boundaries for each bin of the normalized and absolute measurements of the \ttbar differential cross section at particle level in the fiducial phase space as a function of \ytop (trailing) are tabulated.}
\begin{tabular}{ c  c  c }
\ytop (trailing)  & $\frac{1}{\sigma}$ $\frac{\rd\sigma}{\rd \ytop (\text{trailing})}$ & $\frac{\rd\sigma}{\rd \ytop (\text{trailing})}$ [pb] \\
\hline
$[-2.6, -1.65]$ & (4.461 $\pm$ 0.08 $\pm$ 0.233) $ \times 10^{-2}$  & 0.504 $\pm$ 0.009 $\pm$ 0.048 \\
$[-1.65, -1.1]$ & 0.19 $\pm$ 0.002 $\pm$ 0.006 & 2.151 $\pm$ 0.021 $\pm$ 0.15 \\
$[-1.1, -0.55]$ & 0.296 $\pm$ 0.002 $\pm$ 0.009 & 3.351 $\pm$ 0.029 $\pm$ 0.209 \\
$[-0.55, 0]$ & 0.342 $\pm$ 0.003 $\pm$ 0.007 & 3.86 $\pm$ 0.033 $\pm$ 0.253 \\
$[0, 0.55]$ & 0.34 $\pm$ 0.003 $\pm$ 0.013 & 3.845 $\pm$ 0.034 $\pm$ 0.278 \\
$[0.55, 1.1]$ & 0.301 $\pm$ 0.003 $\pm$ 0.008 & 3.403 $\pm$ 0.03 $\pm$ 0.213 \\
$[1.1, 1.65]$ & 0.195 $\pm$ 0.002 $\pm$ 0.005 & 2.203 $\pm$ 0.021 $\pm$ 0.148 \\
$[1.65, 2.6]$ & (4.423 $\pm$ 0.076 $\pm$ 0.201) $ \times 10^{-2}$  & 0.5 $\pm$ 0.009 $\pm$ 0.041 \\
\end{tabular}
\label{tab:norm_particle8}
\end{table}

\begin{table}[!htpb]
\centering
\topcaption{The measured differential cross section and bin boundaries for each bin of the normalized and absolute measurements of the \ttbar differential cross section at particle level in the fiducial phase space as a function of \pttt are tabulated.}
\begin{tabular}{ c  c  c }
\pttt & $\frac{1}{\sigma}$ $\frac{\rd\sigma}{\rd\pttt}$ & $\frac{\rd\sigma}{\rd\pttt}$ [pb] \\
\hline
$[0, 40]$ & (1.171 $\pm$ 0.006 $\pm$ 0.078) $ \times 10^{-2}$  & 0.132 $\pm$ 0.001 $\pm$ 0.011 \\
$[40, 100]$ & (5.35 $\pm$ 0.045 $\pm$ 0.582) $ \times 10^{-3}$  & (6.05 $\pm$ 0.052 $\pm$ 0.781)  $ \times 10^{-2}$  \\
$[100, 200]$ & (1.579 $\pm$ 0.014 $\pm$ 0.06) $ \times 10^{-3}$  & (1.785 $\pm$ 0.016 $\pm$ 0.128)  $ \times 10^{-2}$  \\
$[200, 310]$ & (3.509 $\pm$ 0.053 $\pm$ 0.128) $ \times 10^{-4}$  & (3.968 $\pm$ 0.06 $\pm$ 0.305)  $ \times 10^{-3}$  \\
$[310, 420]$ & (9.53 $\pm$ 0.218 $\pm$ 0.562) $ \times 10^{-5}$  & (1.078 $\pm$ 0.025 $\pm$ 0.096)  $ \times 10^{-3}$  \\
$[420, 570]$ & (2.45 $\pm$ 0.099 $\pm$ 0.158) $ \times 10^{-5}$  & (2.771 $\pm$ 0.112 $\pm$ 0.265)  $ \times 10^{-4}$  \\
\end{tabular}
\label{tab:norm_particle9}
\end{table}

\begin{table}[!htpb]
\centering
\topcaption{The measured differential cross section and bin boundaries for each bin of the normalized and absolute measurements of the \ttbar differential cross section at particle level in the fiducial phase space as a function of \ytt are tabulated.}
\begin{tabular}{ c  c  c }
\ytt  & $\frac{1}{\sigma}$ $\frac{\rd\sigma}{\rd \ytt}$ & $\frac{\rd\sigma}{\rd \ytt}$ [pb] \\
\hline
$[-2.6, -1.6]$ & (1.413 $\pm$ 0.035 $\pm$ 0.165) $ \times 10^{-2}$  & 0.16 $\pm$ 0.004 $\pm$ 0.023 \\
$[-1.6, -1.2]$ & 0.128 $\pm$ 0.001 $\pm$ 0.003 & 1.446 $\pm$ 0.016 $\pm$ 0.098 \\
$[-1.2, -0.8]$ & 0.26 $\pm$ 0.002 $\pm$ 0.008 & 2.941 $\pm$ 0.029 $\pm$ 0.192 \\
$[-0.8, -0.4]$ & 0.373 $\pm$ 0.003 $\pm$ 0.009 & 4.213 $\pm$ 0.036 $\pm$ 0.279 \\
$[-0.4, 0]$ & 0.458 $\pm$ 0.004 $\pm$ 0.009 & 5.175 $\pm$ 0.041 $\pm$ 0.299 \\
$[0, 0.4]$ & 0.44 $\pm$ 0.003 $\pm$ 0.009 & 4.971 $\pm$ 0.04 $\pm$ 0.331 \\
$[0.4, 0.8]$ & 0.378 $\pm$ 0.003 $\pm$ 0.007 & 4.271 $\pm$ 0.035 $\pm$ 0.277 \\
$[0.8, 1.2]$ & 0.265 $\pm$ 0.002 $\pm$ 0.007 & 2.995 $\pm$ 0.027 $\pm$ 0.186 \\
$[1.2, 1.6]$ & 0.125 $\pm$ 0.001 $\pm$ 0.004 & 1.416 $\pm$ 0.015 $\pm$ 0.105 \\
$[1.6, 2.6]$ & (1.545 $\pm$ 0.035 $\pm$ 0.087) $ \times 10^{-2}$  & 0.175 $\pm$ 0.004 $\pm$ 0.017 \\
\end{tabular}
\label{tab:norm_particle10}
\end{table}

\begin{table}[!htpb]
\centering
\topcaption{The measured differential cross section and bin boundaries for each bin of the normalized and absolute measurements of the \ttbar differential cross section at particle level in the fiducial phase space as a function of \mtt are tabulated.}
\begin{tabular}{ c  c  c }
\mtt [\GeVns{}]  & $\frac{1}{\sigma}$ $\frac{\rd\sigma}{\rd\mtt}$ [\GeVns$^{-1}$] & $\frac{\rd\sigma}{\rd \mtt}$ [pb/\GeVns{}] \\
\hline
$[300, 380]$ & (2.719 $\pm$ 0.032 $\pm$ 0.161) $ \times 10^{-3}$  & (3.077 $\pm$ 0.038 $\pm$ 0.228)  $ \times 10^{-2}$  \\
$[380, 470]$ & (3.639 $\pm$ 0.047 $\pm$ 0.193) $ \times 10^{-3}$  & (4.118 $\pm$ 0.054 $\pm$ 0.392)  $ \times 10^{-2}$  \\
$[470, 620]$ & (1.924 $\pm$ 0.023 $\pm$ 0.056) $ \times 10^{-3}$  & (2.178 $\pm$ 0.026 $\pm$ 0.137)  $ \times 10^{-2}$  \\
$[620, 820]$ & (5.977 $\pm$ 0.103 $\pm$ 0.3) $ \times 10^{-4}$  & (6.764 $\pm$ 0.117 $\pm$ 0.47)  $ \times 10^{-3}$  \\
$[820, 1100]$ & (1.305 $\pm$ 0.036 $\pm$ 0.097) $ \times 10^{-4}$  & (1.477 $\pm$ 0.041 $\pm$ 0.139)  $ \times 10^{-3}$  \\
$[1100, 1500]$ & (2.19 $\pm$ 0.081 $\pm$ 0.176) $ \times 10^{-5}$  & (2.478 $\pm$ 0.091 $\pm$ 0.245)  $ \times 10^{-4}$  \\
$[1500, 2500]$ & (1.526 $\pm$ 0.142 $\pm$ 0.33) $ \times 10^{-6}$  & (1.727 $\pm$ 0.16 $\pm$ 0.376)  $ \times 10^{-5}$  \\
\end{tabular}
\label{tab:norm_particle11}
\end{table}

\begin{table}[!htpb]
\centering
\topcaption{The measured differential cross section and bin boundaries for each bin of the normalized and absolute measurements of the \ttbar differential cross section at particle level in the fiducial phase space as a function of \delytt are tabulated.}
\begin{tabular}{ c  c  c }
\delytt & $\frac{1}{\sigma}$ $\frac{\rd\sigma}{\rd\delytt}$ & $\frac{\rd\sigma}{\rd\delytt}$ [pb] \\
\hline
$[-2.6, -1.4]$ & (2.821 $\pm$ 0.054 $\pm$ 0.148) $ \times 10^{-2}$  & 0.319 $\pm$ 0.006 $\pm$ 0.027 \\
$[-1.4, -0.9]$ & 0.179 $\pm$ 0.002 $\pm$ 0.007 & 2.021 $\pm$ 0.022 $\pm$ 0.153 \\
$[-0.9, -0.4]$ & 0.345 $\pm$ 0.003 $\pm$ 0.008 & 3.903 $\pm$ 0.036 $\pm$ 0.271 \\
$[-0.4, 0]$ & 0.5 $\pm$ 0.004 $\pm$ 0.014 & 5.655 $\pm$ 0.052 $\pm$ 0.355 \\
$[0, 0.4]$ & 0.507 $\pm$ 0.005 $\pm$ 0.011 & 5.731 $\pm$ 0.054 $\pm$ 0.347 \\
$[0.4, 0.9]$ & 0.349 $\pm$ 0.003 $\pm$ 0.011 & 3.949 $\pm$ 0.037 $\pm$ 0.282 \\
$[0.9, 1.4]$ & 0.183 $\pm$ 0.002 $\pm$ 0.005 & 2.072 $\pm$ 0.022 $\pm$ 0.148 \\
$[1.4, 2.6]$ & (2.931 $\pm$ 0.052 $\pm$ 0.209) $ \times 10^{-2}$  & 0.331 $\pm$ 0.006 $\pm$ 0.03 \\
\end{tabular}
\label{tab:norm_particle12}
\end{table}

\begin{table}[!htpb]
\centering
\topcaption{The measured differential cross section and bin boundaries for each bin of the normalized and absolute measurements of the \ttbar differential cross section at particle level in the fiducial phase space as a function of \delphitt are tabulated.}
\begin{tabular}{ c  c  c }
\delphitt [\GeVns{}]  & $\frac{1}{\sigma}$ $\frac{\rd\sigma}{\rd\delphitt}$ [\GeVns$^{-1}$] & $\frac{\rd\sigma}{\rd\delphitt}$ [pb/\GeVns{}] \\
\hline
$[0, 1.57]$ & (6.284 $\pm$ 0.067 $\pm$ 0.309) $ \times 10^{-2}$  & 0.711 $\pm$ 0.008 $\pm$ 0.059 \\
$[1.57, 2.67]$ & 0.223 $\pm$ 0.001 $\pm$ 0.008 & 2.525 $\pm$ 0.014 $\pm$ 0.2 \\
$[2.67, 3.02]$ & 1.051 $\pm$ 0.005 $\pm$ 0.025 & (1.189 $\pm$ 0.007 $\pm$ 0.081)  $ \times 10$  \\
$[3.02, 3.142]$ & 2.362 $\pm$ 0.017 $\pm$ 0.134 & (2.672 $\pm$ 0.02 $\pm$ 0.194)  $ \times 10$  \\
\end{tabular}
\label{tab:norm_particle13}
\end{table}

\begin{table}[!htpb]
\centering
\topcaption{The measured differential cross section and bin boundaries for each bin of the normalized and absolute measurements of the \ttbar differential cross section at particle level in the fiducial phase space as a function of \ptlep are tabulated.}
\begin{tabular}{ c  c  c }
 \ptlep [\GeVns{}]  & $\frac{1}{\sigma}$ $\frac{\rd\sigma}{\rd\ptlep}$ [\GeVns$^{-1}$] & $\frac{\rd\sigma}{\rd\ptlep}$ [pb/\GeVns{}] \\
\hline
$[20, 40]$ & (1.848 $\pm$ 0.006 $\pm$ 0.02) $ \times 10^{-2}$  & 0.209 $\pm$ 0.001 $\pm$ 0.014 \\
$[40, 70]$ & (1.251 $\pm$ 0.004 $\pm$ 0.011) $ \times 10^{-2}$  & 0.142 $\pm$ 0.001 $\pm$ 0.008 \\
$[70, 120]$ & (4.011 $\pm$ 0.017 $\pm$ 0.044) $ \times 10^{-3}$  & (4.548 $\pm$ 0.022 $\pm$ 0.275)  $ \times 10^{-2}$  \\
$[120, 180]$ & (7.254 $\pm$ 0.068 $\pm$ 0.126) $ \times 10^{-4}$  & (8.224 $\pm$ 0.079 $\pm$ 0.527)  $ \times 10^{-3}$  \\
$[180, 400]$ & (5.06 $\pm$ 0.102 $\pm$ 0.178) $ \times 10^{-5}$  & (5.737 $\pm$ 0.117 $\pm$ 0.417)  $ \times 10^{-4}$  \\
\end{tabular}
\label{tab:norm_particle14}
\end{table}

\begin{table}[!htpb]
\centering
\topcaption{The measured differential cross section and bin boundaries for each bin of the normalized and absolute measurements of the \ttbar differential cross section at particle level in the fiducial phase space as a function of \ptalep are tabulated.}
\begin{tabular}{ c  c  c }
 \ptalep [\GeVns{}]  & $\frac{1}{\sigma}$ $\frac{\rd\sigma}{\rd\ptalep}$ [\GeVns$^{-1}$] & $\frac{\rd\sigma}{\rd\ptalep}$ [pb/\GeVns{}] \\
\hline
$[20, 40]$ & (1.838 $\pm$ 0.006 $\pm$ 0.019) $ \times 10^{-2}$  & 0.208 $\pm$ 0.001 $\pm$ 0.014 \\
$[40, 70]$ & (1.26 $\pm$ 0.004 $\pm$ 0.006) $ \times 10^{-2}$  & 0.143 $\pm$ 0.001 $\pm$ 0.008 \\
$[70, 120]$ & (3.999 $\pm$ 0.017 $\pm$ 0.044) $ \times 10^{-3}$  & (4.534 $\pm$ 0.022 $\pm$ 0.271)  $ \times 10^{-2}$  \\
$[120, 180]$ & (7.292 $\pm$ 0.068 $\pm$ 0.14) $ \times 10^{-4}$  & (8.266 $\pm$ 0.079 $\pm$ 0.55)  $ \times 10^{-3}$  \\
$[180, 400]$ & (4.895 $\pm$ 0.101 $\pm$ 0.197) $ \times 10^{-5}$  & (5.549 $\pm$ 0.116 $\pm$ 0.441)  $ \times 10^{-4}$  \\
\end{tabular}
\label{tab:norm_particle15}
\end{table}

\begin{table}[!htpb]
\centering
\topcaption{The measured differential cross section and bin boundaries for each bin of the normalized and absolute measurements of the \ttbar differential cross section at particle level in the fiducial phase space as a function of \ptlep (leading) are tabulated.}
\begin{tabular}{ c  c  c }
\ptlep (leading) [\GeVns{}]  & $\frac{1}{\sigma}$ $\frac{\rd\sigma}{\rd\ptlep\text{(leading)}}$ [\GeVns$^{-1}$] & $\frac{\rd\sigma}{\rd\ptlep\text{(leading)}}$ [pb/\GeVns{}] \\
\hline
$[20, 40]$ & (7.01 $\pm$ 0.042 $\pm$ 0.177) $ \times 10^{-3}$  & (7.959 $\pm$ 0.051 $\pm$ 0.601)  $ \times 10^{-2}$  \\
$[40, 70]$ & (1.441 $\pm$ 0.004 $\pm$ 0.009) $ \times 10^{-2}$  & 0.164 $\pm$ 0.001 $\pm$ 0.01 \\
$[70, 120]$ & (6.562 $\pm$ 0.021 $\pm$ 0.061) $ \times 10^{-3}$  & (7.45 $\pm$ 0.028 $\pm$ 0.445)  $ \times 10^{-2}$  \\
$[120, 180]$ & (1.312 $\pm$ 0.009 $\pm$ 0.02) $ \times 10^{-3}$  & (1.489 $\pm$ 0.011 $\pm$ 0.096)  $ \times 10^{-2}$  \\
$[180, 400]$ & (9.39 $\pm$ 0.138 $\pm$ 0.32) $ \times 10^{-5}$  & (1.066 $\pm$ 0.016 $\pm$ 0.079)  $ \times 10^{-3}$  \\
\end{tabular}
\label{tab:norm_particle16}
\end{table}

\begin{table}[!htpb]
\centering
\topcaption{The measured differential cross section and bin boundaries for each bin of the normalized and absolute measurements of the \ttbar differential cross section at particle level in the fiducial phase space as a function of \ptlep (trailing) are tabulated.}
\begin{tabular}{ c  c  c }
 \ptlep (trailing)  [\GeVns{}]  & $\frac{1}{\sigma}$ $\frac{\rd\sigma}{\rd\ptlep\text{(trailing)}}$ [\GeVns$^{-1}$] & $\frac{\rd\sigma}{\rd\ptlep\text{(trailing)} }$ [pb/\GeVns{}] \\
\hline
$[20, 35]$ & (3.184 $\pm$ 0.008 $\pm$ 0.032) $ \times 10^{-2}$  & 0.361 $\pm$ 0.001 $\pm$ 0.024 \\
$[35, 50]$ & (1.916 $\pm$ 0.007 $\pm$ 0.02) $ \times 10^{-2}$  & 0.217 $\pm$ 0.001 $\pm$ 0.012 \\
$[50, 90]$ & (5.063 $\pm$ 0.021 $\pm$ 0.058) $ \times 10^{-3}$  & (5.74 $\pm$ 0.027 $\pm$ 0.345)  $ \times 10^{-2}$  \\
$[90, 140]$ & (5.493 $\pm$ 0.064 $\pm$ 0.124) $ \times 10^{-4}$  & (6.228 $\pm$ 0.073 $\pm$ 0.41)  $ \times 10^{-3}$  \\
$[140, 400]$ & (1.903 $\pm$ 0.058 $\pm$ 0.062) $ \times 10^{-5}$  & (2.158 $\pm$ 0.065 $\pm$ 0.157)  $ \times 10^{-4}$  \\
\end{tabular}
\label{tab:norm_particle17}
\end{table}

\clearpage

\begin{table}[!htpb]
\centering
\topcaption{The measured differential cross section and bin boundaries for each bin of the normalized and absolute measurements of the \ttbar differential cross section at particle level in the fiducial phase space as a function of \etalep are tabulated.}
\begin{tabular}{ c  c  c }
\etalep  & $\frac{1}{\sigma}$ $\frac{\rd\sigma}{\rd\etalep}$ & $\frac{\rd\sigma}{\rd\etalep}$ [pb] \\
\hline
$[-2.4, -2.1]$ & (6.907 $\pm$ 0.105 $\pm$ 0.27) $ \times 10^{-2}$  & 0.781 $\pm$ 0.012 $\pm$ 0.061 \\
$[-2.1, -1.8]$ & 0.101 $\pm$ 0.001 $\pm$ 0.003 & 1.143 $\pm$ 0.014 $\pm$ 0.086 \\
$[-1.8, -1.5]$ & 0.147 $\pm$ 0.002 $\pm$ 0.003 & 1.665 $\pm$ 0.017 $\pm$ 0.103 \\
$[-1.5, -1.2]$ & 0.193 $\pm$ 0.002 $\pm$ 0.003 & 2.186 $\pm$ 0.02 $\pm$ 0.141 \\
$[-1.2, -0.9]$ & 0.244 $\pm$ 0.002 $\pm$ 0.003 & 2.755 $\pm$ 0.022 $\pm$ 0.171 \\
$[-0.9, -0.6]$ & 0.283 $\pm$ 0.002 $\pm$ 0.004 & 3.203 $\pm$ 0.023 $\pm$ 0.193 \\
$[-0.6, -0.3]$ & 0.312 $\pm$ 0.002 $\pm$ 0.004 & 3.524 $\pm$ 0.024 $\pm$ 0.215 \\
$[-0.3, 0]$ & 0.318 $\pm$ 0.002 $\pm$ 0.003 & 3.598 $\pm$ 0.025 $\pm$ 0.224 \\
$[0, 0.3]$ & 0.317 $\pm$ 0.002 $\pm$ 0.003 & 3.58 $\pm$ 0.026 $\pm$ 0.224 \\
$[0.3, 0.6]$ & 0.311 $\pm$ 0.002 $\pm$ 0.003 & 3.518 $\pm$ 0.025 $\pm$ 0.218 \\
$[0.6, 0.9]$ & 0.28 $\pm$ 0.002 $\pm$ 0.002 & 3.167 $\pm$ 0.023 $\pm$ 0.192 \\
$[0.9, 1.2]$ & 0.246 $\pm$ 0.002 $\pm$ 0.004 & 2.777 $\pm$ 0.022 $\pm$ 0.171 \\
$[1.2, 1.5]$ & 0.191 $\pm$ 0.002 $\pm$ 0.003 & 2.163 $\pm$ 0.02 $\pm$ 0.142 \\
$[1.5, 1.8]$ & 0.151 $\pm$ 0.002 $\pm$ 0.003 & 1.709 $\pm$ 0.019 $\pm$ 0.107 \\
$[1.8, 2.1]$ & 0.104 $\pm$ 0.001 $\pm$ 0.003 & 1.18 $\pm$ 0.015 $\pm$ 0.084 \\
$[2.1, 2.4]$ & (6.531 $\pm$ 0.103 $\pm$ 0.226) $ \times 10^{-2}$  & 0.738 $\pm$ 0.012 $\pm$ 0.055 \\
\end{tabular}
\label{tab:norm_particle18}
\end{table}

\begin{table}[!htpb]
\centering
\topcaption{The measured differential cross section and bin boundaries for each bin of the normalized and absolute measurements of the \ttbar differential cross section at particle level in the fiducial phase space as a function of \etaalep are tabulated.}
\begin{tabular}{ c  c  c }
\etaalep  & $\frac{1}{\sigma}$ $\frac{\rd\sigma}{\rd\etaalep}$ & $\frac{\rd\sigma}{\rd\etaalep}$ [pb] \\
\hline
$[-2.4, -2.1]$ & (6.86 $\pm$ 0.106 $\pm$ 0.191) $ \times 10^{-2}$  & 0.775 $\pm$ 0.012 $\pm$ 0.055 \\
$[-2.1, -1.8]$ & 0.104 $\pm$ 0.001 $\pm$ 0.003 & 1.176 $\pm$ 0.014 $\pm$ 0.081 \\
$[-1.8, -1.5]$ & 0.151 $\pm$ 0.002 $\pm$ 0.003 & 1.704 $\pm$ 0.018 $\pm$ 0.109 \\
$[-1.5, -1.2]$ & 0.198 $\pm$ 0.002 $\pm$ 0.002 & 2.235 $\pm$ 0.02 $\pm$ 0.136 \\
$[-1.2, -0.9]$ & 0.244 $\pm$ 0.002 $\pm$ 0.003 & 2.758 $\pm$ 0.022 $\pm$ 0.172 \\
$[-0.9, -0.6]$ & 0.28 $\pm$ 0.002 $\pm$ 0.003 & 3.162 $\pm$ 0.023 $\pm$ 0.198 \\
$[-0.6, -0.3]$ & 0.306 $\pm$ 0.002 $\pm$ 0.002 & 3.454 $\pm$ 0.024 $\pm$ 0.212 \\
$[-0.3, 0]$ & 0.313 $\pm$ 0.002 $\pm$ 0.004 & 3.533 $\pm$ 0.025 $\pm$ 0.225 \\
$[0, 0.3]$ & 0.32 $\pm$ 0.002 $\pm$ 0.003 & 3.612 $\pm$ 0.026 $\pm$ 0.225 \\
$[0.3, 0.6]$ & 0.306 $\pm$ 0.002 $\pm$ 0.003 & 3.459 $\pm$ 0.025 $\pm$ 0.208 \\
$[0.6, 0.9]$ & 0.287 $\pm$ 0.002 $\pm$ 0.004 & 3.239 $\pm$ 0.023 $\pm$ 0.199 \\
$[0.9, 1.2]$ & 0.239 $\pm$ 0.002 $\pm$ 0.003 & 2.702 $\pm$ 0.022 $\pm$ 0.176 \\
$[1.2, 1.5]$ & 0.195 $\pm$ 0.002 $\pm$ 0.003 & 2.205 $\pm$ 0.02 $\pm$ 0.139 \\
$[1.5, 1.8]$ & 0.151 $\pm$ 0.002 $\pm$ 0.002 & 1.703 $\pm$ 0.018 $\pm$ 0.109 \\
$[1.8, 2.1]$ & 0.107 $\pm$ 0.001 $\pm$ 0.002 & 1.21 $\pm$ 0.015 $\pm$ 0.08 \\
$[2.1, 2.4]$ & (6.635 $\pm$ 0.104 $\pm$ 0.203) $ \times 10^{-2}$  & 0.75 $\pm$ 0.012 $\pm$ 0.055 \\
\end{tabular}
\label{tab:norm_particle19}
\end{table}

\begin{table}[!htpb]
\centering
\topcaption{The measured differential cross section and bin boundaries for each bin of the normalized and absolute measurements of the \ttbar differential cross section at particle level in the fiducial phase space as a function of \etalep (leading) are tabulated.}
\begin{tabular}{ c  c  c }
\etalep (leading)  & $\frac{1}{\sigma}$ $\frac{\rd\sigma}{\rd\etalep \text{(leading)}}$ & $\frac{\rd\sigma}{\rd\etalep \text{(leading)}}$ [pb] \\
\hline
$[-2.4, -2.1]$ & (6.102 $\pm$ 0.102 $\pm$ 0.211) $ \times 10^{-2}$  & 0.69 $\pm$ 0.012 $\pm$ 0.052 \\
$[-2.1, -1.8]$ & (9.44 $\pm$ 0.124 $\pm$ 0.304) $ \times 10^{-2}$  & 1.067 $\pm$ 0.014 $\pm$ 0.084 \\
$[-1.8, -1.5]$ & 0.146 $\pm$ 0.002 $\pm$ 0.002 & 1.655 $\pm$ 0.018 $\pm$ 0.102 \\
$[-1.5, -1.2]$ & 0.194 $\pm$ 0.002 $\pm$ 0.003 & 2.198 $\pm$ 0.02 $\pm$ 0.138 \\
$[-1.2, -0.9]$ & 0.243 $\pm$ 0.002 $\pm$ 0.003 & 2.749 $\pm$ 0.022 $\pm$ 0.17 \\
$[-0.9, -0.6]$ & 0.29 $\pm$ 0.002 $\pm$ 0.003 & 3.28 $\pm$ 0.024 $\pm$ 0.2 \\
$[-0.6, -0.3]$ & 0.318 $\pm$ 0.002 $\pm$ 0.004 & 3.591 $\pm$ 0.025 $\pm$ 0.219 \\
$[-0.3, 0]$ & 0.321 $\pm$ 0.002 $\pm$ 0.004 & 3.625 $\pm$ 0.025 $\pm$ 0.23 \\
$[0, 0.3]$ & 0.326 $\pm$ 0.002 $\pm$ 0.004 & 3.689 $\pm$ 0.026 $\pm$ 0.23 \\
$[0.3, 0.6]$ & 0.315 $\pm$ 0.002 $\pm$ 0.004 & 3.563 $\pm$ 0.026 $\pm$ 0.219 \\
$[0.6, 0.9]$ & 0.287 $\pm$ 0.002 $\pm$ 0.004 & 3.247 $\pm$ 0.024 $\pm$ 0.197 \\
$[0.9, 1.2]$ & 0.245 $\pm$ 0.002 $\pm$ 0.003 & 2.767 $\pm$ 0.023 $\pm$ 0.175 \\
$[1.2, 1.5]$ & 0.192 $\pm$ 0.002 $\pm$ 0.003 & 2.168 $\pm$ 0.021 $\pm$ 0.139 \\
$[1.5, 1.8]$ & 0.146 $\pm$ 0.002 $\pm$ 0.004 & 1.652 $\pm$ 0.019 $\pm$ 0.106 \\
$[1.8, 2.1]$ & (9.55 $\pm$ 0.129 $\pm$ 0.268) $ \times 10^{-2}$  & 1.08 $\pm$ 0.015 $\pm$ 0.082 \\
$[2.1, 2.4]$ & (5.913 $\pm$ 0.101 $\pm$ 0.159) $ \times 10^{-2}$  & 0.669 $\pm$ 0.012 $\pm$ 0.049 \\
\end{tabular}
\label{tab:norm_particle20}
\end{table}

\begin{table}[!htpb]
\centering
\topcaption{The measured differential cross section and bin boundaries for each bin of the normalized and absolute measurements of the \ttbar differential cross section at particle level in the fiducial phase space as a function of \etalep (trailing) are tabulated.}
\begin{tabular}{ c  c  c }
\etalep (trailing)  & $\frac{1}{\sigma}$ $\frac{\rd\sigma}{\rd \etalep \text{(trailing)}}$ & $\frac{\rd\sigma}{\rd \etalep \text{(trailing)}}$ [pb] \\
\hline
$[-2.4, -2.1]$ & (7.656 $\pm$ 0.113 $\pm$ 0.167) $ \times 10^{-2}$  & 0.866 $\pm$ 0.013 $\pm$ 0.059 \\
$[-2.1, -1.8]$ & 0.111 $\pm$ 0.001 $\pm$ 0.003 & 1.25 $\pm$ 0.015 $\pm$ 0.087 \\
$[-1.8, -1.5]$ & 0.152 $\pm$ 0.002 $\pm$ 0.003 & 1.715 $\pm$ 0.018 $\pm$ 0.112 \\
$[-1.5, -1.2]$ & 0.197 $\pm$ 0.002 $\pm$ 0.003 & 2.225 $\pm$ 0.021 $\pm$ 0.139 \\
$[-1.2, -0.9]$ & 0.245 $\pm$ 0.002 $\pm$ 0.004 & 2.765 $\pm$ 0.022 $\pm$ 0.173 \\
$[-0.9, -0.6]$ & 0.273 $\pm$ 0.002 $\pm$ 0.004 & 3.086 $\pm$ 0.023 $\pm$ 0.189 \\
$[-0.6, -0.3]$ & 0.3 $\pm$ 0.002 $\pm$ 0.002 & 3.388 $\pm$ 0.024 $\pm$ 0.209 \\
$[-0.3, 0]$ & 0.31 $\pm$ 0.002 $\pm$ 0.004 & 3.504 $\pm$ 0.025 $\pm$ 0.218 \\
$[0, 0.3]$ & 0.309 $\pm$ 0.002 $\pm$ 0.003 & 3.498 $\pm$ 0.026 $\pm$ 0.22 \\
$[0.3, 0.6]$ & 0.303 $\pm$ 0.002 $\pm$ 0.003 & 3.422 $\pm$ 0.025 $\pm$ 0.208 \\
$[0.6, 0.9]$ & 0.28 $\pm$ 0.002 $\pm$ 0.003 & 3.166 $\pm$ 0.024 $\pm$ 0.195 \\
$[0.9, 1.2]$ & 0.24 $\pm$ 0.002 $\pm$ 0.003 & 2.713 $\pm$ 0.023 $\pm$ 0.172 \\
$[1.2, 1.5]$ & 0.195 $\pm$ 0.002 $\pm$ 0.002 & 2.2 $\pm$ 0.021 $\pm$ 0.141 \\
$[1.5, 1.8]$ & 0.156 $\pm$ 0.002 $\pm$ 0.003 & 1.762 $\pm$ 0.019 $\pm$ 0.116 \\
$[1.8, 2.1]$ & 0.116 $\pm$ 0.001 $\pm$ 0.003 & 1.31 $\pm$ 0.016 $\pm$ 0.086 \\
$[2.1, 2.4]$ & (7.226 $\pm$ 0.112 $\pm$ 0.232) $ \times 10^{-2}$  & 0.817 $\pm$ 0.013 $\pm$ 0.06 \\
\end{tabular}
\label{tab:norm_particle21}
\end{table}

\clearpage

\begin{table}[!htpb]
\centering
\topcaption{The measured differential cross section and bin boundaries for each bin of the normalized and absolute measurements of the \ttbar differential cross section at particle level in the fiducial phase space as a function of \ptll are tabulated.}
\begin{tabular}{ c  c  c }
\ptll [\GeVns{}]  & $\frac{1}{\sigma}$ $\frac{\rd\sigma}{\rd\ptll}$ [\GeVns$^{-1}$] & $\frac{\rd\sigma}{\rd\ptll}$ [pb/\GeVns{}] \\
\hline
$[0, 10]$ & (1.691 $\pm$ 0.032 $\pm$ 0.068) $ \times 10^{-3}$  & (1.913 $\pm$ 0.036 $\pm$ 0.155)  $ \times 10^{-2}$  \\
$[10, 20]$ & (4.682 $\pm$ 0.049 $\pm$ 0.109) $ \times 10^{-3}$  & (5.297 $\pm$ 0.057 $\pm$ 0.371)  $ \times 10^{-2}$  \\
$[20, 40]$ & (7.593 $\pm$ 0.042 $\pm$ 0.076) $ \times 10^{-3}$  & (8.591 $\pm$ 0.052 $\pm$ 0.562)  $ \times 10^{-2}$  \\
$[40, 60]$ & (1.094 $\pm$ 0.005 $\pm$ 0.009) $ \times 10^{-2}$  & 0.124 $\pm$ 0.001 $\pm$ 0.008 \\
$[60, 100]$ & (9.554 $\pm$ 0.028 $\pm$ 0.05) $ \times 10^{-3}$  & 0.108 $\pm$ ${<}10^{-3}$ $\pm$ 0.006 \\
$[100, 150]$ & (3.04 $\pm$ 0.016 $\pm$ 0.031) $ \times 10^{-3}$  & (3.44 $\pm$ 0.019 $\pm$ 0.21)  $ \times 10^{-2}$  \\
$[150, 400]$ & (1.259 $\pm$ 0.015 $\pm$ 0.043) $ \times 10^{-4}$  & (1.425 $\pm$ 0.017 $\pm$ 0.108)  $ \times 10^{-3}$  \\
\end{tabular}
\label{tab:norm_particle22}
\end{table}

\begin{table}[!htpb]
\centering
\topcaption{The measured differential cross section and bin boundaries for each bin of the normalized and absolute measurements of the \ttbar differential cross section at particle level in the fiducial phase space as a function of \mll are tabulated.}
\begin{tabular}{ c  c  c }
\mll [\GeVns{}]  & $\frac{1}{\sigma}$ $\frac{\rd\sigma}{\rd \mll}$ [\GeVns$^{-1}$] & $\frac{\rd\sigma}{\rd \mll}$ [pb/\GeVns{}] \\
\hline
$[20, 30]$ & (3.44 $\pm$ 0.042 $\pm$ 0.14) $ \times 10^{-3}$  & (3.903 $\pm$ 0.048 $\pm$ 0.337)  $ \times 10^{-2}$  \\
$[30, 50]$ & (5.338 $\pm$ 0.033 $\pm$ 0.101) $ \times 10^{-3}$  & (6.058 $\pm$ 0.041 $\pm$ 0.419)  $ \times 10^{-2}$  \\
$[50, 76]$ & (7.656 $\pm$ 0.034 $\pm$ 0.144) $ \times 10^{-3}$  & (8.688 $\pm$ 0.042 $\pm$ 0.608)  $ \times 10^{-2}$  \\
$[76, 106]$ & (7.475 $\pm$ 0.037 $\pm$ 0.123) $ \times 10^{-3}$  & (8.482 $\pm$ 0.049 $\pm$ 0.482)  $ \times 10^{-2}$  \\
$[106, 130]$ & (5.446 $\pm$ 0.031 $\pm$ 0.042) $ \times 10^{-3}$  & (6.18 $\pm$ 0.038 $\pm$ 0.372)  $ \times 10^{-2}$  \\
$[130, 170]$ & (3.481 $\pm$ 0.019 $\pm$ 0.041) $ \times 10^{-3}$  & (3.95 $\pm$ 0.022 $\pm$ 0.233)  $ \times 10^{-2}$  \\
$[170, 260]$ & (1.342 $\pm$ 0.007 $\pm$ 0.02) $ \times 10^{-3}$  & (1.523 $\pm$ 0.009 $\pm$ 0.091)  $ \times 10^{-2}$  \\
$[260, 650]$ & (1.149 $\pm$ 0.01 $\pm$ 0.021) $ \times 10^{-4}$  & (1.304 $\pm$ 0.012 $\pm$ 0.081)  $ \times 10^{-3}$  \\
\end{tabular}
\label{tab:norm_particle23}
\end{table}

\begin{table}[!htpb]
\centering
\topcaption{The measured differential cross section and bin boundaries for each bin of the normalized and absolute measurements of the \ttbar differential cross section at particle level in the fiducial phase space as a function of \delphill are tabulated.}
\begin{tabular}{ c  c  c }
\delphill & $\frac{1}{\sigma}$ $\frac{\rd\sigma}{\rd\delphill}$ & $\frac{\rd\sigma}{\rd\delphill}$ [pb] \\
\hline
$[0, 0.4]$ & 0.194 $\pm$ 0.001 $\pm$ 0.003 & 2.191 $\pm$ 0.017 $\pm$ 0.14 \\
$[0.4, 0.78]$ & 0.247 $\pm$ 0.002 $\pm$ 0.003 & 2.793 $\pm$ 0.02 $\pm$ 0.179 \\
$[0.78, 1.14]$ & 0.278 $\pm$ 0.002 $\pm$ 0.003 & 3.141 $\pm$ 0.023 $\pm$ 0.196 \\
$[1.14, 1.48]$ & 0.299 $\pm$ 0.002 $\pm$ 0.003 & 3.381 $\pm$ 0.024 $\pm$ 0.205 \\
$[1.48, 1.8]$ & 0.326 $\pm$ 0.002 $\pm$ 0.002 & 3.688 $\pm$ 0.026 $\pm$ 0.225 \\
$[1.8, 2.1]$ & 0.352 $\pm$ 0.002 $\pm$ 0.003 & 3.98 $\pm$ 0.027 $\pm$ 0.247 \\
$[2.1, 2.38]$ & 0.374 $\pm$ 0.002 $\pm$ 0.003 & 4.235 $\pm$ 0.028 $\pm$ 0.269 \\
$[2.38, 2.64]$ & 0.393 $\pm$ 0.002 $\pm$ 0.003 & 4.449 $\pm$ 0.028 $\pm$ 0.273 \\
$[2.64, 2.89]$ & 0.42 $\pm$ 0.002 $\pm$ 0.004 & 4.749 $\pm$ 0.029 $\pm$ 0.296 \\
$[2.89, 3.142]$ & 0.418 $\pm$ 0.003 $\pm$ 0.004 & 4.724 $\pm$ 0.03 $\pm$ 0.301 \\
\end{tabular}
\label{tab:norm_particle24}
\end{table}

\begin{table}[!htpb]
\centering
\topcaption{The measured differential cross section and bin boundaries for each bin of the normalized and absolute measurements of the \ttbar differential cross section at particle level in the fiducial phase space as a function of \deletall are tabulated.}
\begin{tabular}{ c  c  c }
\deletall  & $\frac{1}{\sigma}$ $\frac{\rd\sigma}{\rd \deletall}$ & $\frac{\rd\sigma}{\rd\deletall}$ [pb] \\
\hline
$[-2.4, -1.7]$ & (3.516 $\pm$ 0.047 $\pm$ 0.083) $ \times 10^{-2}$  & 0.398 $\pm$ 0.005 $\pm$ 0.025 \\
$[-1.7, -1.2]$ & 0.135 $\pm$ 0.001 $\pm$ 0.002 & 1.526 $\pm$ 0.012 $\pm$ 0.091 \\
$[-1.2, -0.8]$ & 0.245 $\pm$ 0.002 $\pm$ 0.003 & 2.766 $\pm$ 0.02 $\pm$ 0.171 \\
$[-0.8, -0.4]$ & 0.352 $\pm$ 0.002 $\pm$ 0.003 & 3.98 $\pm$ 0.024 $\pm$ 0.248 \\
$[-0.4, 0]$ & 0.43 $\pm$ 0.002 $\pm$ 0.004 & 4.864 $\pm$ 0.026 $\pm$ 0.31 \\
$[0, 0.4]$ & 0.428 $\pm$ 0.002 $\pm$ 0.003 & 4.844 $\pm$ 0.026 $\pm$ 0.305 \\
$[0.4, 0.8]$ & 0.349 $\pm$ 0.002 $\pm$ 0.003 & 3.951 $\pm$ 0.023 $\pm$ 0.246 \\
$[0.8, 1.2]$ & 0.24 $\pm$ 0.002 $\pm$ 0.003 & 2.715 $\pm$ 0.018 $\pm$ 0.17 \\
$[1.2, 1.7]$ & 0.132 $\pm$ 0.001 $\pm$ 0.002 & 1.495 $\pm$ 0.012 $\pm$ 0.093 \\
$[1.7, 2.4]$ & (3.41 $\pm$ 0.046 $\pm$ 0.085) $ \times 10^{-2}$  & 0.386 $\pm$ 0.005 $\pm$ 0.025 \\
\end{tabular}
\label{tab:norm_particle25}
\end{table}

\begin{table}[!htpb]
\centering
\topcaption{The measured differential cross section and bin boundaries for each bin of the normalized and absolute measurements of the \ttbar differential cross section at particle level in the fiducial phase space as a function of $\pt^{\mathrm{b}} \text{(leading)}$ are tabulated.}
\begin{tabular}{ c  c  c }
\ptb (leading) [\GeVns{}]  & $\frac{1}{\sigma}$ $\frac{\rd\sigma}{\rd\ptb \text{(leading)}}$ [\GeVns$^{-1}$] & $\frac{\rd\sigma}{\rd\ptb \text{(leading)}}$ [pb/\GeVns{}] \\
\hline
$[30, 60]$ & (6.234 $\pm$ 0.041 $\pm$ 0.373) $ \times 10^{-3}$  & (7.039 $\pm$ 0.048 $\pm$ 0.553)  $ \times 10^{-2}$  \\
$[60, 95]$ & (1.163 $\pm$ 0.005 $\pm$ 0.014) $ \times 10^{-2}$  & 0.131 $\pm$ 0.001 $\pm$ 0.008 \\
$[95, 150]$ & (5.411 $\pm$ 0.025 $\pm$ 0.123) $ \times 10^{-3}$  & (6.109 $\pm$ 0.031 $\pm$ 0.433)  $ \times 10^{-2}$  \\
$[150, 230]$ & (1.134 $\pm$ 0.009 $\pm$ 0.042) $ \times 10^{-3}$  & (1.28 $\pm$ 0.01 $\pm$ 0.105)  $ \times 10^{-2}$  \\
$[230, 500]$ & (6.525 $\pm$ 0.128 $\pm$ 0.337) $ \times 10^{-5}$  & (7.367 $\pm$ 0.146 $\pm$ 0.671)  $ \times 10^{-4}$  \\
\end{tabular}
\label{tab:norm_particle26}
\end{table}

\begin{table}[!htpb]
\centering
\topcaption{The measured differential cross section and bin boundaries for each bin of the normalized and absolute measurements of the \ttbar differential cross section at particle level in the fiducial phase space as a function of \ptb (trailing) are tabulated.}
\begin{tabular}{ c  c  c }
$ \ptb \text{(trailing)}$ [\GeVns{}]  & $\frac{1}{\sigma}$ $\frac{\rd\sigma}{\rd\ptb \text{(trailing)}}$ [\GeVns$^{-1}$] & $\frac{\rd\sigma}{\rd\ptb \text{(trailing)}}$ [pb/\GeVns{}] \\
\hline
$[30, 45]$ & (2.48 $\pm$ 0.012 $\pm$ 0.096) $ \times 10^{-2}$  & 0.28 $\pm$ 0.001 $\pm$ 0.019 \\
$[45, 70]$ & (1.535 $\pm$ 0.008 $\pm$ 0.033) $ \times 10^{-2}$  & 0.173 $\pm$ 0.001 $\pm$ 0.011 \\
$[70, 110]$ & (4.691 $\pm$ 0.032 $\pm$ 0.148) $ \times 10^{-3}$  & (5.293 $\pm$ 0.039 $\pm$ 0.377)  $ \times 10^{-2}$  \\
$[110, 170]$ & (8.159 $\pm$ 0.101 $\pm$ 0.302) $ \times 10^{-4}$  & (9.206 $\pm$ 0.117 $\pm$ 0.7)  $ \times 10^{-3}$  \\
$[170, 500]$ & (2.304 $\pm$ 0.078 $\pm$ 0.205) $ \times 10^{-5}$  & (2.6 $\pm$ 0.088 $\pm$ 0.292)  $ \times 10^{-4}$  \\
\end{tabular}
\label{tab:norm_particle27}
\end{table}

\begin{table}[!htpb]
\centering
\topcaption{The measured differential cross section and bin boundaries for each bin of the normalized and absolute measurements of the \ttbar differential cross section at particle level in the fiducial phase space as a function of \etab (leading) are tabulated.}
\begin{tabular}{ c  c  c }
\etab (leading)  & $\frac{1}{\sigma}$ $\frac{\rd\sigma}{\rd \etab \text{(leading)}}$ & $\frac{\rd\sigma}{\rd \etab \text{(leading)}}$ [pb] \\
\hline
$[-2.4, -1.8]$ & (8.029 $\pm$ 0.099 $\pm$ 0.414) $ \times 10^{-2}$  & 0.908 $\pm$ 0.011 $\pm$ 0.087 \\
$[-1.8, -1.2]$ & 0.162 $\pm$ 0.001 $\pm$ 0.004 & 1.829 $\pm$ 0.015 $\pm$ 0.131 \\
$[-1.2, -0.6]$ & 0.269 $\pm$ 0.002 $\pm$ 0.004 & 3.046 $\pm$ 0.018 $\pm$ 0.178 \\
$[-0.6, 0]$ & 0.319 $\pm$ 0.002 $\pm$ 0.005 & 3.605 $\pm$ 0.02 $\pm$ 0.213 \\
$[0, 0.6]$ & 0.326 $\pm$ 0.002 $\pm$ 0.005 & 3.685 $\pm$ 0.021 $\pm$ 0.209 \\
$[0.6, 1.2]$ & 0.261 $\pm$ 0.002 $\pm$ 0.003 & 2.954 $\pm$ 0.019 $\pm$ 0.176 \\
$[1.2, 1.8]$ & 0.168 $\pm$ 0.001 $\pm$ 0.003 & 1.9 $\pm$ 0.015 $\pm$ 0.131 \\
$[1.8, 2.4]$ & (8.11 $\pm$ 0.093 $\pm$ 0.348) $ \times 10^{-2}$  & 0.917 $\pm$ 0.011 $\pm$ 0.082 \\
\end{tabular}
\label{tab:norm_particle28}
\end{table}

\begin{table}[!htpb]
\centering
\topcaption{The measured differential cross section and bin boundaries for each bin of the normalized and absolute measurements of the \ttbar differential cross section at particle level in the fiducial phase space as a function of \etab (trailing) are tabulated.}
\begin{tabular}{ c  c  c }
\etab (trailing)  & $\frac{1}{\sigma}$ $\frac{\rd\sigma}{\rd \etab\text{(trailing)}}$ & $\frac{\rd\sigma}{\rd\etab \text{(trailing)}}$ [pb] \\
\hline
$[-2.4, -1.8]$ & (9.96 $\pm$ 0.111 $\pm$ 0.339) $ \times 10^{-2}$  & 1.126 $\pm$ 0.013 $\pm$ 0.087 \\
$[-1.8, -1.2]$ & 0.171 $\pm$ 0.001 $\pm$ 0.004 & 1.937 $\pm$ 0.016 $\pm$ 0.136 \\
$[-1.2, -0.6]$ & 0.252 $\pm$ 0.002 $\pm$ 0.003 & 2.847 $\pm$ 0.019 $\pm$ 0.173 \\
$[-0.6, 0]$ & 0.301 $\pm$ 0.002 $\pm$ 0.005 & 3.409 $\pm$ 0.02 $\pm$ 0.2 \\
$[0, 0.6]$ & 0.31 $\pm$ 0.002 $\pm$ 0.005 & 3.504 $\pm$ 0.021 $\pm$ 0.204 \\
$[0.6, 1.2]$ & 0.258 $\pm$ 0.002 $\pm$ 0.003 & 2.916 $\pm$ 0.019 $\pm$ 0.171 \\
$[1.2, 1.8]$ & 0.178 $\pm$ 0.001 $\pm$ 0.003 & 2.014 $\pm$ 0.016 $\pm$ 0.138 \\
$[1.8, 2.4]$ & (9.679 $\pm$ 0.104 $\pm$ 0.434) $ \times 10^{-2}$  & 1.095 $\pm$ 0.012 $\pm$ 0.098 \\
\end{tabular}
\label{tab:norm_particle29}
\end{table}

\begin{table}[!htpb]
\centering
\topcaption{The measured differential cross section and bin boundaries for each bin of the normalized and absolute measurements of the \ttbar differential cross section at particle level in the fiducial phase space as a function of \ptbb are tabulated.}
\begin{tabular}{ c  c  c }
\ptbb [\GeVns{}]  & $\frac{1}{\sigma}$ $\frac{\rd\sigma}{\rd \ptbb}$ [\GeVns$^{-1}$] & $\frac{\rd\sigma}{\rd \ptbb }$ [pb/\GeVns{}] \\
\hline
$[0, 30]$ & (3.518 $\pm$ 0.029 $\pm$ 0.144) $ \times 10^{-3}$  & (3.976 $\pm$ 0.034 $\pm$ 0.26)  $ \times 10^{-2}$  \\
$[30, 60]$ & (7.185 $\pm$ 0.038 $\pm$ 0.216) $ \times 10^{-3}$  & (8.12 $\pm$ 0.047 $\pm$ 0.469)  $ \times 10^{-2}$  \\
$[60, 100]$ & (8.916 $\pm$ 0.034 $\pm$ 0.16) $ \times 10^{-3}$  & 0.101 $\pm$ ${<}10^{-3}$ $\pm$ 0.006 \\
$[100, 180]$ & (3.672 $\pm$ 0.015 $\pm$ 0.161) $ \times 10^{-3}$  & (4.15 $\pm$ 0.019 $\pm$ 0.367)  $ \times 10^{-2}$  \\
$[180, 400]$ & (1.298 $\pm$ 0.02 $\pm$ 0.117) $ \times 10^{-4}$  & (1.467 $\pm$ 0.023 $\pm$ 0.192)  $ \times 10^{-3}$  \\
\end{tabular}
\label{tab:norm_particle30}
\end{table}

\begin{table}[!htpb]
\centering
\topcaption{The measured differential cross section and bin boundaries for each bin of the normalized and absolute measurements of the \ttbar differential cross section at particle level in the fiducial phase space as a function of \mbb are tabulated.}
\begin{tabular}{ c  c  c }
\mbb [\GeVns{}]  & $\frac{1}{\sigma}$ $\frac{\rd\sigma}{\rd \mbb}$ [\GeVns$^{-1}$] & $\frac{\rd\sigma}{\rd \mbb}$ [pb/\GeVns{}] \\
\hline
$[0, 60]$ & (1.222 $\pm$ 0.011 $\pm$ 0.044) $ \times 10^{-3}$  & (1.377 $\pm$ 0.012 $\pm$ 0.097)  $ \times 10^{-2}$  \\
$[60, 120]$ & (4.964 $\pm$ 0.018 $\pm$ 0.09) $ \times 10^{-3}$  & (5.592 $\pm$ 0.023 $\pm$ 0.338)  $ \times 10^{-2}$  \\
$[120, 240]$ & (3.822 $\pm$ 0.011 $\pm$ 0.026) $ \times 10^{-3}$  & (4.306 $\pm$ 0.015 $\pm$ 0.261)  $ \times 10^{-2}$  \\
$[240, 650]$ & (4.152 $\pm$ 0.024 $\pm$ 0.138) $ \times 10^{-4}$  & (4.677 $\pm$ 0.029 $\pm$ 0.365)  $ \times 10^{-3}$  \\
\end{tabular}
\label{tab:norm_particle31}
\end{table}

\begin{table}[!htpb]
\centering
\topcaption{The measured differential cross section and bin boundaries for each bin of the normalized and absolute measurements of the \ttbar differential cross section at particle level in the fiducial phase space as a function of \Nj\ are tabulated.}
\begin{tabular}{ c  c  c }
\Nj\ & $\frac{1}{\sigma}$ $\frac{\rd\sigma}{\rd \Nj}$ & $\frac{\rd\sigma}{\rd \Nj}$ [pb] \\
\hline
$[1.5, 2.5]$ & 0.532 $\pm$ 0.001 $\pm$ 0.016 & 6.01 $\pm$ 0.02 $\pm$ 0.341 \\
$[2.5, 3.5]$ & 0.303 $\pm$ 0.001 $\pm$ 0.006 & 3.42 $\pm$ 0.017 $\pm$ 0.24 \\
$[3.5, 4.5]$ & 0.116 $\pm$ 0.001 $\pm$ 0.007 & 1.307 $\pm$ 0.012 $\pm$ 0.128 \\
$[4.5, 5.5]$ & (3.68 $\pm$ 0.06 $\pm$ 0.313) $ \times 10^{-2}$  & 0.416 $\pm$ 0.007 $\pm$ 0.052 \\
$[5.5, 6.5]$ & (1.035 $\pm$ 0.027 $\pm$ 0.104) $ \times 10^{-2}$  & 0.117 $\pm$ 0.003 $\pm$ 0.016 \\
$[6.5, 7.5]$ & (2.514 $\pm$ 0.137 $\pm$ 0.426) $ \times 10^{-3}$  & (2.84 $\pm$ 0.155 $\pm$ 0.571)  $ \times 10^{-2}$  \\
\end{tabular}
\label{tab:norm_particle32}
\end{table}

\clearpage

\section{Tables of \texorpdfstring{$\chi^{2}/\mathrm{dof}$ and $p$-values} {chi2/dof}}
\label{chi2:tables}

The $\chi^{2}/\mathrm{dof}$ and $p$-values between data and all theoretical predictions for all measured differential cross sections are  are tabulated in Tables \ref{tab:norm_parton}--\ref{tab:abs_particle}. The $\chi^{2}/\mathrm{dof}$ and $p$-value calculations take into account the inter-bin correlations of the data.

\renewcommand{\arraystretch}{1.3}
\setlength{\tabcolsep}{3pt}

\begin{table}[!htpb]
\topcaption{The \chidof and $p$-values quantifying the agreement between the three MC predictions and the measured, parton-level normalised cross sections are shown.}
\centering
\begin{tabular}{l >{\centering\arraybackslash}p{1.7cm} >{\centering\arraybackslash}p{1.7cm} >{\centering\arraybackslash}p{1.7cm} >{\centering\arraybackslash}p{1.7cm} >{\centering\arraybackslash}p{2.3cm} >{\centering\arraybackslash}p{2.3cm} }
& \multicolumn{2}{c}{\pwhgpy} &
\multicolumn{2}{c}{\pwhghpp} &
\multicolumn{2}{c}{\mgamcpy} \\
& \chidof & $p$-value  & \chidof & $p$-value  & \chidof & $p$-value \\
\hline
\pttop & 43/5 & $< 10^{-3}$ & 6/5 & 0.269 & 21/5 & $< 10^{-3}$ \\
\ptantitop & 35/5 & $< 10^{-3}$ & 7/5 & 0.257 & 18/5 & 0.003 \\
\pttop (leading) & 42/5 & $< 10^{-3}$ & 3/5 & 0.650 & 25/5 & $< 10^{-3}$ \\
\pttop (trailing) & 44/5 & $< 10^{-3}$ & 14/5 & 0.016 & 18/5 & 0.003 \\
\pttop (\ttbar RF) & 32/5 & $< 10^{-3}$ & 7/5 & 0.209 & 16/5 & 0.008 \\
\ytop & 6/9 & 0.723 & 6/9 & 0.758 & 5/9 & 0.825 \\
\yantitop & 3/9 & 0.976 & 3/9 & 0.974 & 3/9 & 0.966 \\
\ytop (leading) & 3/7 & 0.862 & 3/7 & 0.899 & 4/7 & 0.820 \\
\ytop (trailing) & 3/7 & 0.897 & 4/7 & 0.795 & 2/7 & 0.965 \\
\pttt & 24/5 & $< 10^{-3}$ & 83/5 & $< 10^{-3}$ & 16/5 & 0.007 \\
\ytt & 3/9 & 0.951 & 4/9 & 0.916 & 4/9 & 0.940 \\
\mtt & 17/6 & 0.009 & 2/6 & 0.882 & 6/6 & 0.382 \\
\delytt & 1/7 & 0.987 & 3/7 & 0.899 & 4/7 & 0.738 \\
\delphitt & 0/3 & 0.978 & 2/3 & 0.503 & 2/3 & 0.633 \\
\end{tabular}
\label{tab:norm_parton}
\end{table}

\begin{table}[!htpb]
\topcaption{The \chidof and $p$-values quantifying the agreement between the five theoretical predictions with \nnloew and \nnlonnllprime precision and the measured, parton-level normalised cross sections are shown.}
\centering
\begin{tabular}{ l  c  c  c  c  c  c  c  c  c  c }
\multirow{5}{*}{} &
\multicolumn{2}{c}{\parbox{2.3cm}{\centering \nnloew \\ (LUXQED17) \\ \mt~=~173.3\GeV}} &
\multicolumn{2}{c}{\parbox{2.3cm}{\centering \nnloew \\ (LUXQED17) \\ \mt~=~172.5\GeV}} &
\multicolumn{2}{c}{\parbox{2.3cm}{\centering \nnloew \\ (NNPDF3.1) \\ \mt~=~173.3\GeV}} &
\multicolumn{2}{c}{\parbox{2.3cm}{\centering \nnlonnllprime \\ (NNPDF3.1) \\ \mt~=~173.3\GeV}} &
\multicolumn{2}{c}{\parbox{2.3cm}{\centering \nnlonnllprime \\ (NNPDF3.1) \\ \mt~=~172.5\GeV}} \\
& \chidof & $p$-value  & \chidof & $p$-value  & \chidof & $p$-value  & \chidof & $p$-value  & \chidof & $p$-value \\
\hline
\pttop & 16/5 & 0.006 & 12/5 & 0.036 & 12/5 & 0.029 & 20/5 & 0.001 & 15/5 & 0.011 \\
\ptantitop & 16/5 & 0.007 & 12/5 & 0.041 & 13/5 & 0.027 & 17/5 & 0.005 & 13/5 & 0.026 \\
\ytop & 9/9 & 0.434 & 8/9 & 0.554 & 7/9 & 0.665 & \NA & \NA & \NA & \NA \\
\yantitop & 4/9 & 0.915 & 5/9 & 0.875 & 2/9 & 0.990 & \NA & \NA & \NA & \NA \\
\pttt & 26/5 & $< 10^{-3}$ & 25/5 & $< 10^{-3}$ & 25/5 & $< 10^{-3}$ & \NA & \NA & \NA & \NA \\
\ytt & 7/9 & 0.597 & 7/9 & 0.644 & 3/9 & 0.960 & \NA & \NA & \NA & \NA \\
\mtt & 34/6 & $< 10^{-3}$ & 24/6 & $< 10^{-3}$ & 28/6 & $< 10^{-3}$ & 30/6 & $< 10^{-3}$ & 19/6 & 0.004 \\
\delytt  & 5/7 & 0.608 & 6/7 & 0.537 & 2/7 & 0.966 & \NA & \NA & \NA & \NA \\
\end{tabular}
\label{tab:norm_parton_bnlo1}
\end{table}

\begin{table}[!htpb]
\topcaption{The \chidof and $p$-values quantifying the agreement between the two theoretical predictions with \annnlo and \annlo precision and the measured,
parton-level normalised cross sections are shown.}
\centering
\begin{tabular}{ l  c  c  c  c }
\multirow{2}{*}{} &
\multicolumn{2}{c}{\parbox{2.3cm}{\centering \annnlo \\ (NNPDF3.0) \\ \mt~=~172.5\GeV}} &
\multicolumn{2}{c}{\parbox{2.3cm}{\centering \annlo \\ (CT14NNLO) \\ \mt~=~172.5\GeV}} \\
& \chidof & $p$-value  & \chidof & $p$-value \\
\hline
\pttop & 7/5 & 0.242 & 51/5 & $< 10^{-3}$ \\
\ytop & 6/9 & 0.698 & 60/9 & $< 10^{-3}$ \\
\end{tabular}
\label{tab:norm_parton_bnlo2}
\end{table}

\begin{table}[!htpb]
\centering
\topcaption{The \chidof and $p$-values quantifying the agreement between the three MC predictions and the measured, parton-level absolute cross sections are shown.}
\begin{tabular}{l >{\centering\arraybackslash}p{1.7cm} >{\centering\arraybackslash}p{1.7cm} >{\centering\arraybackslash}p{1.7cm} >{\centering\arraybackslash}p{1.7cm} >{\centering\arraybackslash}p{2.3cm} >{\centering\arraybackslash}p{2.3cm} }
& \multicolumn{2}{c}{\pwhgpy} &
\multicolumn{2}{c}{\pwhghpp} &
\multicolumn{2}{c}{\mgamcpy} \\
& \chidof & $p$-value  & \chidof & $p$-value  & \chidof & $p$-value \\
\hline
\pttop & 51/6 & $< 10^{-3}$ & 8/6 & 0.239 & 18/6 & 0.007 \\
\ptantitop & 41/6 & $< 10^{-3}$ & 9/6 & 0.157 & 14/6 & 0.026 \\
\pttop (leading) & 47/6 & $< 10^{-3}$ & 4/6 & 0.627 & 20/6 & 0.003 \\
\pttop (trailing) & 38/6 & $< 10^{-3}$ & 16/6 & 0.012 & 9/6 & 0.150 \\
\pttop (\ttbar RF) & 40/6 & $< 10^{-3}$ & 11/6 & 0.077 & 13/6 & 0.046 \\
\ytop & 5/10 & 0.864 & 5/10 & 0.885 & 4/10 & 0.936 \\
\yantitop & 2/10 & 0.991 & 2/10 & 0.992 & 3/10 & 0.983 \\
\ytop (leading) & 3/8 & 0.948 & 2/8 & 0.966 & 3/8 & 0.924 \\
\ytop (trailing) & 3/8 & 0.956 & 3/8 & 0.912 & 2/8 & 0.976 \\
\delytt & 1/8 & 0.995 & 3/8 & 0.902 & 4/8 & 0.849 \\
\delphitt & 0/4 & 0.980 & 3/4 & 0.607 & 2/4 & 0.741 \\
\pttt & 22/6 & 0.001 & 36/6 & $< 10^{-3}$ & 12/6 & 0.054 \\
\ytt & 4/10 & 0.967 & 4/10 & 0.945 & 5/10 & 0.891 \\
\mtt & 12/7 & 0.109 & 3/7 & 0.906 & 6/7 & 0.533 \\
\end{tabular}
\label{tab:abs_parton}
\end{table}

\begin{table}[!htpb]
\centering
\topcaption{The \chidof and $p$-values quantifying the agreement between the five theoretical predictions with \nnloew and \nnlonnllprime precision and the measured, parton-level absolute cross sections are shown.}
\begin{tabular}{ l  c  c  c  c  c  c  c  c  c  c }
\multirow{5}{*}{} &
\multicolumn{2}{c}{\parbox{2.3cm}{\centering \nnloew \\ (LUXQED17) \\ \mt~=~173.3\GeV}} &
\multicolumn{2}{c}{\parbox{2.3cm}{\centering \nnloew \\ (LUXQED17) \\ \mt~=~172.5\GeV}} &
\multicolumn{2}{c}{\parbox{2.3cm}{\centering \nnloew \\ (NNPDF3.1) \\ \mt~=~173.3\GeV}} &
\multicolumn{2}{c}{\parbox{2.3cm}{\centering \nnlonnllprime \\ (NNPDF3.1) \\ \mt~=~173.3\GeV}} &
\multicolumn{2}{c}{\parbox{2.3cm}{\centering \nnlonnllprime \\ (NNPDF3.1) \\ \mt~=~172.5\GeV}} \\
& \chidof & $p$-value  & \chidof & $p$-value  & \chidof & $p$-value  & \chidof & $p$-value  & \chidof & $p$-value \\
\hline
\pttop & 14/6 & 0.026 & 12/6 & 0.071 & 10/6 & 0.115 & 17/6 & 0.010 & 14/6 & 0.032 \\
\ptantitop & 14/6 & 0.027 & 12/6 & 0.070 & 10/6 & 0.122 & 13/6 & 0.047 & 11/6 & 0.098 \\
\ytop & 9/10 & 0.510 & 7/10 & 0.694 & 6/10 & 0.787 & \NA & \NA & \NA & \NA \\
\yantitop & 5/10 & 0.912 & 5/10 & 0.877 & 3/10 & 0.990 & \NA & \NA & \NA & \NA \\
\pttt & 18/6 & 0.006 & 17/6 & 0.008 & 14/6 & 0.030 & \NA & \NA & \NA & \NA \\
\ytt & 9/10 & 0.548 & 8/10 & 0.666 & 4/10 & 0.937 & \NA & \NA & \NA & \NA \\
\mtt & 22/7 & 0.003 & 17/7 & 0.015 & 16/7 & 0.024 & 16/7 & 0.029 & 11/7 & 0.126 \\
\delytt  & 6/8 & 0.687 & 6/8 & 0.667 & 2/8 & 0.971 & \NA & \NA & \NA & \NA \\
\end{tabular}
\label{tab:abs_parton_bnlo1}
\end{table}

\begin{table}[!htpb]
\centering
\topcaption{The \chidof and $p$-values quantifying the agreement between the two theoretical predictions with \annnlo and \annlo precision and the measured, parton-level absolute cross sections are shown.}
\begin{tabular}{ l  c  c  c  c }
\multirow{2}{*}{} &
\multicolumn{2}{c}{\parbox{2.3cm}{\centering \annnlo \\ (NNPDF3.0) \\ \mt~=~172.5\GeV}} &
\multicolumn{2}{c}{\parbox{2.3cm}{\centering \annlo \\ (CT14NNLO) \\ \mt~=~172.5\GeV}} \\
& \chidof & $p$-value  & \chidof & $p$-value \\
\hline
\pttop & 8/6 & 0.220 & 306/6 & ${<}10^{-3}$ \\
\ytop & 6/10 & 0.838 & 1038/10 & ${<}10^{-3}$ \\
\end{tabular}
\label{tab:abs_parton_bnlo2}
\end{table}

\begin{table}[!htpb]
\centering
\topcaption{The \chidof and $p$-values quantifying the agreement between the three MC predictions and the measured, particle-level normalised cross sections are shown.}
\begin{tabular}{l >{\centering\arraybackslash}p{1.7cm} >{\centering\arraybackslash}p{1.7cm} >{\centering\arraybackslash}p{1.7cm} >{\centering\arraybackslash}p{1.7cm} >{\centering\arraybackslash}p{2.3cm} >{\centering\arraybackslash}p{2.3cm} }
& \multicolumn{2}{c}{\pwhgpy} &
\multicolumn{2}{c}{\pwhghpp} &
\multicolumn{2}{c}{\mgamcpy} \\
& \chidof & $p$-value  & \chidof & $p$-value  & \chidof & $p$-value \\
\hline
\pttop & 47/5 & $< 10^{-3}$ & 3/5 & 0.710 & 19/5 & 0.002 \\
\ptantitop & 41/5 & $< 10^{-3}$ & 3/5 & 0.630 & 18/5 & 0.003 \\
\pttop (leading) & 49/5 & $< 10^{-3}$ & 3/5 & 0.635 & 24/5 & $< 10^{-3}$ \\
\pttop (trailing) & 39/5 & $< 10^{-3}$ & 6/5 & 0.274 & 14/5 & 0.015 \\
\pttop (\ttbar RF) & 36/5 & $< 10^{-3}$ & 7/5 & 0.187 & 15/5 & 0.009 \\
\ytop & 6/9 & 0.701 & 9/9 & 0.443 & 7/9 & 0.639 \\
\yantitop & 3/9 & 0.961 & 3/9 & 0.952 & 3/9 & 0.945 \\
\ytop (leading) & 3/7 & 0.858 & 4/7 & 0.799 & 5/7 & 0.659 \\
\ytop (trailing) & 4/7 & 0.826 & 5/7 & 0.655 & 3/7 & 0.913 \\
\pttt & 28/5 & $< 10^{-3}$ & 104/5 & $< 10^{-3}$ & 15/5 & 0.010 \\
\ytt & 3/9 & 0.965 & 5/9 & 0.821 & 4/9 & 0.910 \\
\mtt & 12/6 & 0.058 & 29/6 & $< 10^{-3}$ & 5/6 & 0.606 \\
\delytt & 1/7 & 0.987 & 7/7 & 0.411 & 4/7 & 0.825 \\
\delphitt & 0/3 & 0.977 & 0/3 & 0.941 & 1/3 & 0.722 \\
\ptlep & 87/4 & $< 10^{-3}$ & 2/4 & 0.699 & 30/4 & $< 10^{-3}$ \\
\ptalep & 36/4 & $< 10^{-3}$ & 1/4 & 0.915 & 10/4 & 0.047 \\
\ptlep (leading) & 112/4 & $< 10^{-3}$ & 2/4 & 0.794 & 36/4 & $< 10^{-3}$ \\
\ptlep (trailing) & 32/4 & $< 10^{-3}$ & 4/4 & 0.396 & 10/4 & 0.034 \\
\etalep & 18/15 & 0.238 & 23/15 & 0.094 & 22/15 & 0.119 \\
\etaalep & 29/15 & 0.015 & 31/15 & 0.008 & 37/15 & 0.001 \\
\etalep (leading) & 13/15 & 0.582 & 13/15 & 0.565 & 21/15 & 0.142 \\
\etalep (trailing) & 22/15 & 0.098 & 32/15 & 0.007 & 27/15 & 0.028 \\
\ptll & 14/6 & 0.027 & 14/6 & 0.034 & 7/6 & 0.302 \\
\mll & 34/7 & $< 10^{-3}$ & 3/7 & 0.887 & 5/7 & 0.648 \\
\delphill & 31/9 & $< 10^{-3}$ & 16/9 & 0.063 & 12/9 & 0.233 \\
\deletall & 5/9 & 0.815 & 4/9 & 0.887 & 6/9 & 0.690 \\
\ptb (leading) & 31/4 & $< 10^{-3}$ & 14/4 & 0.006 & 14/4 & 0.006 \\
\ptb (trailing) & 27/4 & $< 10^{-3}$ & 18/4 & 0.001 & 11/4 & 0.029 \\
\etab (leading) & 10/7 & 0.186 & 13/7 & 0.082 & 8/7 & 0.295 \\
\etab (trailing) & 12/7 & 0.114 & 14/7 & 0.047 & 9/7 & 0.227 \\
\ptbb & 14/4 & 0.007 & 6/4 & 0.229 & 9/4 & 0.071 \\
\mbb & 3/3 & 0.393 & 17/3 & $< 10^{-3}$ & 1/3 & 0.753 \\
\Nj & 13/5 & 0.025 & 38/5 & $< 10^{-3}$ & 36/5 & $< 10^{-3}$ \\
\end{tabular}
\label{tab:norm_particle}
\end{table}

\begin{table}[!htpb]
\centering
\topcaption{The \chidof and $p$-values quantifying the agreement between the three MC predictions and the measured, particle-level absolute cross sections are shown.}
\begin{tabular}{l >{\centering\arraybackslash}p{1.7cm} >{\centering\arraybackslash}p{1.7cm} >{\centering\arraybackslash}p{1.7cm} >{\centering\arraybackslash}p{1.7cm} >{\centering\arraybackslash}p{2.3cm} >{\centering\arraybackslash}p{2.3cm} }
& \multicolumn{2}{c}{\pwhgpy} &
\multicolumn{2}{c}{\pwhghpp} &
\multicolumn{2}{c}{\mgamcpy} \\
& \chidof & $p$-value & \chidof & $p$-value & \chidof & $p$-value \\
\hline
\pttop & 52/6 & ${<}10^{-3}$ & 3/6 & 0.830 & 17/6 & 0.008 \\
\ptantitop & 44/6 & ${<}10^{-3}$ & 3/6 & 0.786 & 16/6 & 0.012 \\
\pttop (leading) & 50/6 & ${<}10^{-3}$ & 3/6 & 0.756 & 21/6 & 0.002 \\
\pttop (trailing) & 39/6 & ${<}10^{-3}$ & 5/6 & 0.576 & 11/6 & 0.099 \\
\pttop (\ttbar RF) & 38/6 & ${<}10^{-3}$ & 4/6 & 0.710 & 12/6 & 0.053 \\
\ytop & 6/10 & 0.785 & 8/10 & 0.627 & 6/10 & 0.795 \\
\yantitop & 3/10 & 0.989 & 3/10 & 0.981 & 3/10 & 0.983 \\
\ytop (leading) & 3/8 & 0.927 & 4/8 & 0.894 & 4/8 & 0.840 \\
\ytop (trailing) & 3/8 & 0.939 & 4/8 & 0.850 & 2/8 & 0.973 \\
\pttt & 34/6 & ${<}10^{-3}$ & 29/6 & ${<}10^{-3}$ & 17/6 & 0.011 \\
\ytt & 3/10 & 0.968 & 6/10 & 0.776 & 5/10 & 0.858 \\
\mtt & 12/7 & 0.095 & 11/7 & 0.135 & 5/7 & 0.676 \\
\delytt & 1/8 & 0.994 & 6/8 & 0.595 & 3/8 & 0.907 \\
\delphitt & 0/4 & 0.975 & 0/4 & 0.982 & 2/4 & 0.823 \\
\ptlep & 36/5 & ${<}10^{-3}$ & 3/5 & 0.771 & 11/5 & 0.059 \\
\ptalep & 35/5 & ${<}10^{-3}$ & 1/5 & 0.951 & 9/5 & 0.095 \\
\ptlep (leading) & 41/5 & ${<}10^{-3}$ & 3/5 & 0.731 & 12/5 & 0.039 \\
\ptlep (trailing) & 33/5 & ${<}10^{-3}$ & 5/5 & 0.404 & 6/5 & 0.318 \\
\etalep & 18/16 & 0.339 & 20/16 & 0.207 & 20/16 & 0.234 \\
\etaalep & 29/16 & 0.024 & 28/16 & 0.033 & 33/16 & 0.007 \\
\etalep (leading) & 11/16 & 0.781 & 12/16 & 0.766 & 17/16 & 0.388 \\
\etalep (trailing) & 21/16 & 0.194 & 29/16 & 0.025 & 24/16 & 0.092 \\
\ptll & 14/7 & 0.053 & 11/7 & 0.120 & 6/7 & 0.524 \\
\mll & 37/8 & ${<}10^{-3}$ & 3/8 & 0.948 & 4/8 & 0.810 \\
\delphill & 27/10 & 0.003 & 13/10 & 0.235 & 10/10 & 0.478 \\
\deletall & 6/10 & 0.848 & 4/10 & 0.930 & 6/10 & 0.842 \\
\ptb (leading) & 35/5 & ${<}10^{-3}$ & 10/5 & 0.085 & 17/5 & 0.004 \\
\ptb (trailing) & 27/5 & ${<}10^{-3}$ & 10/5 & 0.063 & 11/5 & 0.050 \\
\etab (leading) & 10/8 & 0.251 & 11/8 & 0.188 & 8/8 & 0.452 \\
\etab (trailing) & 11/8 & 0.204 & 12/8 & 0.151 & 8/8 & 0.456 \\
\ptbb & 12/5 & 0.030 & 9/5 & 0.096 & 7/5 & 0.232 \\
\mbb & 3/4 & 0.584 & 16/4 & 0.003 & 1/4 & 0.897 \\
\Nj & 14/6 & 0.035 & 37/6 & ${<}10^{-3}$ & 18/6 & 0.006 \\
\end{tabular}
\label{tab:abs_particle}
\end{table}

\clearpage
\cleardoublepage \section{The CMS Collaboration \label{app:collab}}\begin{sloppypar}\hyphenpenalty=5000\widowpenalty=500\clubpenalty=5000\vskip\cmsinstskip
\textbf{Yerevan Physics Institute, Yerevan, Armenia}\\*[0pt]
A.M.~Sirunyan, A.~Tumasyan
\vskip\cmsinstskip
\textbf{Institut f\"{u}r Hochenergiephysik, Wien, Austria}\\*[0pt]
W.~Adam, F.~Ambrogi, E.~Asilar, T.~Bergauer, J.~Brandstetter, M.~Dragicevic, J.~Er\"{o}, A.~Escalante~Del~Valle, M.~Flechl, R.~Fr\"{u}hwirth\cmsAuthorMark{1}, V.M.~Ghete, J.~Hrubec, M.~Jeitler\cmsAuthorMark{1}, N.~Krammer, I.~Kr\"{a}tschmer, D.~Liko, T.~Madlener, I.~Mikulec, N.~Rad, H.~Rohringer, J.~Schieck\cmsAuthorMark{1}, R.~Sch\"{o}fbeck, M.~Spanring, D.~Spitzbart, A.~Taurok, W.~Waltenberger, J.~Wittmann, C.-E.~Wulz\cmsAuthorMark{1}, M.~Zarucki
% ...
\vskip\cmsinstskip
% ...
1:  Also at Vienna University of Technology, Vienna, Austria\\
% ...
\end{sloppypar}
\end{document}